\documentclass[12pt,twoside,openright]{report}

\usepackage{epsf}

\usepackage{fancyheadings}
\usepackage{these}
\usepackage{emlines}

\def\Ref#1{(\ref{#1})}
\def\plabel#1{\label{#1}}

\newcommand{\be}{\begin{equation}}
\newcommand{\ee}{\end{equation}}
\newcommand{\beq}{\begin{eqnarray}}
\newcommand{\eeq}{\end{eqnarray}}
\newcommand{\bea}[2]{\be\label{#2}\begin{array}{#1}}
\newcommand{\eea}{\end{array}\ee}

\newcommand{\under}[2]{\mathop{#1}\limits_{#2}}

\def\lfig#1#2#3#4{
 \begin{figure}
 \refstepcounter{figure}
 \label{#4}
 \addtocounter{figure}{-1}
 \epsfxsize=#3
 \centerline{\epsfbox{#2}}
 {\bf \caption{{\rm #1}}}
 \end{figure}
}

\def\figlabel#1{\xdef#1{\the\figno}}
\def\encadremath#1{\vbox{\hrule\hbox{\vrule\kern8pt\vbox{\kern8pt
\hbox{$\displaystyle #1$}\kern8pt}
\kern8pt\vrule}\hrule}}

\def\ie{{\it i.e.\ }}
\newcommand{\etc}{{\it etc.}}

\def\Rb{{\rm \bf R}}
\def\Cb{{\rm \bf C}}
\def\Zb{{\rm \bf Z}}
\def\Tr{\,{\rm Tr}\, }
\def\det{\,{\rm det}\, }
\def\diag{{\rm diag}}

\def\adj{{\rm adj}}

\def\tr{\,{\rm tr}\, }

\def\Im{\,{\rm Im}\, }
\def\Re{\,{\rm Re}\, }

\def\rangl{\right\rangle   }
\def\langl{\left\langle  }
\def\({\left(}
\def\){\right)}
\def\[{\left[}
\def\]{\right]}
\def\p{\partial}

\def\11{1\!\! 1}
\def\tint{{\int\!\!\!\int}}

\def\hf{{1\over 2}}

\def\g{\gamma}

\def\e{\epsilon}
\def\eps{\varepsilon}

\def\l{\lambda}
\def\m{\mu}

\def\s{\sigma}

\def\u{\upsilon }
\def\z{\zeta }
\def\vp{\varphi}

\def\G{\Gamma}

\def\Sig{\Sigma}

\def\o{\omega }
\def\dd{\nabla}

   \def\CC {{\cal C}}
   \def\CD {{\cal D}}
   \def\CE {{\cal E}}
   \def\CF {{\cal F}}

   \def\CK {{\cal K}}
   \def\CL {{\cal L}}
   
   \def\CN {{\cal N}}
   \def\CO {{\cal O}}
   \def\CP {{\cal P}}
   
   \def\CR {{\cal R}}
   
   \def\CT {{\cal T}}
   
   \def\CV {{\cal V}}
   \def\CW {{\cal W}}

   \def\CZ {{\cal Z}}

\def\bV{ \bar V }

\def\bL{\bar L}
\def\bH{\bar H}
\def\bM{\bar M}
\def\bCW{\bar \CW}

\def\bm{\bar m}
\def\bn{\bar n}
\def\bw{\bar w}

\def\by{\bar y}
\def\bz{\bar z}
\def\bu{\bar u}

\def\hb{\hbar}
\def\bh{{1\over \hbar}}
\def\ibh{ {i\over \hb} }
\def\bZ{ Z^\dagger  }

\def\ho{\hat\o}

\def\tX{{\tilde X}}

\def\tV{{\tilde V}}
\def\tlf{\tilde f}

\def\tl{{\tilde \lambda}}
\def\teta{\tilde \eta}

\def\tom{\tilde \omega  }

\def\rvacs{|s\rangle }

\def\rvacl{|l\rangle }
\def\lvacl{\langle l| }
\def\Rvacl{|l\rangl }
\def\Lvacl{\langl l| }

\def\Pe{\Psi^{_{E}}}

\def\pse{ \psi^{_{E}} }
\def\pee{\psi^{_{E'}}}

\def\xp{x_{_{+}}}
\def\xm{x_{_{-}}}
\def\xpm{x_{_{\pm}}}
\def\xmp{x_{_{\mp}}}

\def\xkpm{x_{_{\pm, k}}}

\def\xipm#1{x_{_{\pm,#1}}}

\def\op{\omega_{_{+}}}

\def\bop{\omega^{}_{_{+}}}
\def\bom{\omega^{}_{_{-}}}
\def\bopm{\omega^{}_{_{\pm}}}

\def\tp{t_1}
\def\tm{t_{-1}}
\def\tpm{t_{\pm 1}}
\def\tmp{t_{\mp 1}}

\def\Xp{X_{+}}
\def\Xm{X_{-}}
\def\Xpm{X_{\pm }}

\def\psip{\psi_{_{+}}}
\def\psim{\psi_{_{-}}}

\def\Pep{\Pe_{+}}
\def\Pem{\Pe_{-}}
\def\Pepm{\Pe_{\pm }}

\def\psem{\pse_{_{-}}}
\def\psepm{\pse_{_{\pm }}}
\def\psemp{\pse_{_{\mp }}}

\def\peep{\pee_{_{+}}}

\def\Pip#1{\Psi^{_{#1}}_{+}}
\def\Pim#1{\Psi^{_{#1}}_{-}}
\def\Pipm#1{\Psi^{_{#1}}_{\pm}}

\def\Pspm{\Psi_{{\pm}}}

\def\Phipm{\Phi_{\pm}}

\def\Phpm#1{\Phi^{#1}_{\pm}}

\def\PPspm#1{\Psi'^{#1}_{\pm}}

\def\bff{\varphi}
\def\ff{\varphi_0}
\def\dl{\Phi}

\def\mt{m_{\eta}}
\def\oo{\o_{(0)}}

\def\op{\tilde\o}
\def\pp{\tilde p}

\def\xtp{\tilde x_t}

\def\tf{\tau}
\def\xf{q}

\def\dpi{{1\over 2\pi}}
\def\dd{\nabla}

\def\oR{{1\over R}}
\def\alp{\alpha'}
\def\pal{\alpha'^{-1}}

\def\w{w}

\def\ti{t}

\def\eao{e^{-{2\over R}X_0}}
\def\ebo{e^{-X_0}}
\def\eco{e^{-2{R-1\over R}X_0}}
\def\efo{e^{-{1\over R}X_0}}
\def\ego{e^{-{R-1\over R}X_0}}

\def\ea{e^{-{2\over R}X}}
\def\eb{e^{-{1\over R^2}X}}
\def\ec{e^{-2{1-R\over R^2}X}}

\def\Fsea{\rm ^{_{Fermi\ sea}}}

\def\ctf{\sqrt{\Lambda} }

\def\KK{K}

\def\vvp{\Phi}
\def\ctf{\sqrt{\Lambda} }

\def\mul{\mu_{_L}}

\def\Dc{D_{\rm cr}}
\def\gc{g_{\rm cl}}
\def\go{g_{\rm op}}
\def\gst{g_{\rm str}}
\def\str{\gamma_{\rm str}}
\def\cap{\kappa}
\def\eig{x}
\def\bfg{{\rm \bf g}}

\def\Phx{\Phi}
\def\Phy{\tilde \Phi}
\def\Lx{X}
\def\Ly{Y}
\def\Px{P}
\def\Py{Q}

\def\sing{{\rm (sing)}}
\def\adj{{\rm (adj)}}

\def\SUN{{\rm SU(N)}}
\def\HR{H^{(r)}}
\def\DR{D^{(r)}}
\def\dR{d_r}
\def\tauR{\tau^{(r)}}
\def\PhiR{\Phi^{(r)}}
\def\PsiR{\Psi^{(r)}}
\def\Phis{\Phi^{(\rm sing)}}
\def\Psis{\Psi^{(\rm sing)}}
\def\Psia{\Psi^{\adj}}
\def\taua{\tau^{\adj}}
\def\psa{\psi^{\adj}}
\def\Hadj{H^{\adj}}
\def\Padj{P^{\adj}}
\def\Qadj{Q^{\adj}}
\def\ads#1{|\tau_{#1}\rangle}

\def\HMQM{H_{_{\rm MQM}}}
\def\SMQM{S_{_{\rm MQM}}}
\def\SSL{S_{{\rm SL}}}

\def\ZM{\CZ^{^{_{\rm MQM}}}}
\def\ZN{\CZ^{^{_{\rm NMM}}}}
\def\taum{\tau}

\def\Zs{Z^{\sing}}

\def\Fm{\tilde F}
\def\Fs{F^{\sing}}
\def\fs{f^{\sing}}
\def\CFs{\CF^{\sing}}

\def\CFadj{\CF^{\adj}}
\def\FSL{\CF}

\chardef\tempcat=\the\catcode`\@ \catcode`\@=11
\def\cyracc{\def\u##1{\if \i##1\accent"24 i%
    \else \accent"24 ##1\fi }}
\newfam\cyrfam

\begin{document}
\renewcommand{\baselinestretch}{1.0} \normalsize
\pagenumbering{roman}

\topmargin=0in
\enlargethispage{3cm}
\thispagestyle{empty}
\vspace*{0cm}
\centerline{\scshape
\large Service de Physique Th{\'e}orique -- C.E.A.-Saclay}
\vskip 1.5cm
\centerline{\large \bfseries UNIVERSIT{\'E} PARIS XI}
\vskip 1.5cm
\centerline{\large \bf PhD Thesis}
\vskip 5.0cm
\centerline{\Huge \bfseries \itshape
Matrix Quantum Mechanics and }
\vskip .3cm
\centerline{\Huge \bfseries \itshape
Two-dimensional String Theory }
\vskip .3cm
\centerline{\Huge \bfseries \itshape in Non-trivial Backgrounds}
\vskip 2cm
\centerline{\Large \bf Sergei Alexandrov}

\newpage
\topmargin=2cm

\thispagestyle{empty}

\

\vskip 2cm 

\centerline{\bf \large \itshape Abstract }

\vskip 1cm

String theory is the most promising candidate for the theory unifying
all interactions including gravity. It has an extremely difficult
dynamics. Therefore, it is useful to study some its simplifications.
One of them is non-critical string theory which can be defined in low
dimensions. A particular interesting case is 2D string theory.
On the one hand, it has a very rich structure and, on the other hand,
it is solvable.
A complete solution of 2D string theory in the simplest linear dilaton
background was obtained using its representation as Matrix Quantum
Mechanics. This matrix model provides a very powerful technique
and reveals the integrability hidden in the usual CFT formulation.

This thesis extends the matrix model description of 2D string theory
to non-trivial backgrounds. We show how perturbations changing
the background are incorporated into Matrix Quantum Mechanics.
The perturbations are integrable and governed by Toda Lattice hierarchy.
This integrability is used to extract various information about
the perturbed system: correlation functions, thermodynamical behaviour,
structure of the target space. The results concerning these and
some other issues, like non-perturbative effects in non-critical string
theory, are presented in the thesis.

\newpage

\renewcommand{\baselinestretch}{1.1} \normalsize
\thispagestyle{empty}
\enlargethispage{5mm}

\

\vskip 1cm 

\centerline{\bf \large \itshape Acknowledgements } 

\vskip 1cm

This work was done at the Service de Physique Th\'eorique du
centre d'\'etudes de Saclay.
I would like to thank the laboratory for the excellent conditions
which allowed to accomplish my work. Also I am grateful to CEA
for the financial support during these three years.
Equally, my gratitude is directed to the Laboratoire de Physique
Th\'eorique de l'Ecole Normale Sup\'erieure where I also
had the possibility to work all this time.
I am thankful to all members of these two labs for
the nice stimulating atmosphere.

Especially, I would like to thank my scientific advisers,
Volodya Kazakov and Ivan Kostov who opened a new domain
of theoretical physics for me. Their creativity and deep knowledge
were decisive for the success of our work. Besides, their care
in all problems helped me much during these years of life in France.

I am grateful to all scientists with whom I had discussions and who
shared their ideas with me. In particular, let me express my gratitude to
Constantin Bachas, Alexey Boyarsky, Edouard Br\'ezin, Philippe Di Francesco, 
David Kutasov, Marcus Mari\~no,
Andrey Marshakov, Yuri Novozhilov,
Volker Schomerus, Didina Serban, Alexander Sorin, Cumrum Vafa,
Pavel Wiegmann, Anton Zabrodin, Alexey Zamolodchikov, Jean-Bernard Zuber
and, especially, to Dmitri Vassilevich.
He was my first advisor in Saint-Petersburg
and I am indebted to him for my first steps in physics as well as for
a fruitful collaboration after that.

Also I am grateful to the Physical Laboratory of Harvard University
and to the Max--Planck Institute of Potsdam University for
the kind hospitality during the time I visited there.

It was nice to work in the friendly atmosphere created 
by Paolo Ribeca and Thomas Quella at Saclay and
Nicolas Couchoud, Yacine Dolivet, Pierre Henry-Laborder, 
Dan Israel and Louis Paulot at ENS with whom I shared the office.

Finally, I am thankful to
Edouard Br\'ezin and Jean-Bernard Zuber who 
accepted to be the members of my jury
and to Nikita Nekrasov and Matthias Staudacher, who agreed to be
my reviewers, to read the thesis and helped me to improve it
by their corrections.


\renewcommand{\baselinestretch}{1.0} \normalsize
\tableofcontents
\newpage\thispagestyle{empty}

\cleardoublepage
\pagenumbering{arabic}
\begin{chapternon}{Introduction}

This thesis is devoted to application of the matrix model approach to
non-critical string theory.

More than fifteen years have passed since matrix models were first applied
to string theory. Although they have not helped to solve critical string
and superstring theory, they have taught us many things about
low-dimensional bosonic string theories. Matrix models have provided
so powerful technique that a lot of results which were obtained
in this framework are still inaccessible using the usual continuum approach.
On the other hand, those results that were reproduced turned out to be in
the excellent agreement with the results obtained by field theoretical
methods.

One of the main subjects of interest in the early years of the matrix model
approach was the $c=1$ non-critical string theory which is equivalent to
the two-dimensional critical string theory in the linear dilaton background.
This background is the simplest one for the low-dimensional theories.
It is flat and the dilaton field appearing in the low-energy target space
description is just proportional to one of the spacetime coordinates.

In the framework of the matrix approach this string theory
is described in terms of {\it Matrix Quantum Mechanics} (MQM).
Already ten years ago MQM gave a complete solution of the 2D string
theory. For example, the exact $S$-matrix of scattering processes
was found and many correlation functions were explicitly calculated.

However, the linear dilaton background is only one of the possible
backgrounds of 2D string theory. There are many other backgrounds
including ones with a non-vanishing curvature which contain a dilatonic
black hole. It was a puzzle during long time how to describe
such backgrounds in terms of matrices. And only recently some progress
was made in this direction.

In this thesis we try to develop the matrix model description of
2D string theory in non-trivial backgrounds.
Our research covers several possibilities to deform the initial simple
target space. In particular, we analyze winding and tachyon perturbations.
We show how they are incorporated into Matrix Quantum Mechanics and study
the result of their inclusion.

A remarkable feature of these perturbations is that they are exactly
solvable. The reason is that the perturbed theory is described by Toda
Lattice integrable hierarchy. This is the result obtained entirely
within the matrix model framework. So far this integrability has not been
observed in the continuum approach.
On the other hand, in MQM it appears quite naturally being a generalization
of the KP integrable structure of the $c<1$  models.
In this thesis we extensively use the Toda description because
it allows to obtain many exact results.

We tried to make the thesis selfconsistent. Therefore, we give a long
introduction into the subject. We begin by briefly reviewing the main
concepts of string theory. We introduce the Polyakov action for a bosonic
string, the notion of the Weyl invariance and the anomaly
associated with it. We show how the critical string theory emerges and
explain how it is generalized to superstring theory
avoiding to write explicit formulae. We mention also the modern view
on superstrings which includes D-branes and dualities.
After that we discuss the low-energy limit of bosonic string theories
and possible string backgrounds. A special attention is paid to the linear
dilaton background which appears in the discussion of non-critical
strings. Finally, we present in detail 2D string theory both in the linear 
dilaton and perturbed backgrounds. We elucidate its degrees of freedom and how
they can be used to perturb the theory.
In particular, we present a conjecture that relates 2D string theory
perturbed by windings modes to the same theory in a curved black hole background.

The next chapter is an introduction to matrix models. We explain
what the matrix models are and how they are related to various physical
problems and to string theory, in particular. The relation is established
through the sum over discretized surfaces and such important notions
as the $1/N$ expansion and the double scaling limit are introduced.
Then we consider the two simplest examples, the one-
and the two-matrix model. They are used to present two of the several 
known methods to solve matrix models. 
First, the one-matrix model is solved in the large
$N$-limit by the saddle point approach. Second, it is shown
how to obtain the solution of the two-matrix model by the technique
of orthogonal polynomials which works, in contrast to the
first method, to all orders in perturbation theory.
We finish this chapter giving an introduction to Toda hierarchy.
The emphasis is done on its Lax formalism.
Since the Toda integrable structure is the main tool of this thesis,
the presentation is detailed and may look too technical.
But this will be compensated by the power of this approach.

The third chapter deals with a particular matrix model ---
Matrix Quantum Mechanics. We show how it incorporates
all features of 2D string theory. In particular, we identify the tachyon
modes with collective excitations of the singlet sector of MQM and
the winding modes of the compactified string theory with
degrees of freedom propagating in the non-trivial representations
of the SU(N) global symmetry of MQM. We explain the free
fermionic representation of the singlet sector and present
its explicit solution both in the non-compactified and compactified cases.
Its target space interpretation is elucidated with the help
of the Das--Jevicki collective field theory.

Starting from the forth chapter, we turn to 2D string theory in non-trivial
backgrounds and try to describe it in terms of perturbations of Matrix
Quantum Mechanics. First, the winding perturbations of the compactified
string theory are incorporated into the matrix framework. We review the work
of Kazakov, Kostov and Kutasov where this was first done.
In particular, we identify the perturbed partition function with a
$\tau$-function of Toda hierarchy showing that the introduced perturbations
are integrable. The simplest case of the windings of the minimal charge
is interpreted as a matrix model for the 2D string theory in the black hole
background. For this case we present explicit results for the free energy.
Relying on these description, we explain our first work in this
domain devoted to
calculation of winding correlators in the theory with the simplest winding
perturbation. This work is little bit technical. Therefore, we concentrate
mainly on the conceptual issues.

The next chapter is about tachyon perturbations of 2D string theory
in the MQM framework. It consists from three parts representing our three
works. In the first one, we show how the tachyon perturbations should be
introduced. Similarly to the case of windings, we find that the
perturbations are integrable. In the quasiclassical limit we interpret them
in terms of the time-dependent Fermi sea of fermions of the singlet sector.
The second work provides a thermodynamical interpretation to
these perturbations. For the simplest case corresponding to
the Sine--Liouville perturbation, we are able to find all thermodynamical
characteristics of the system. However, many of the results
do not have a good explanation and remain to be mysterious for us.
In the third work we discuss how to obtain the structure of the
string backgrounds corresponding to the perturbations introduced
in the matrix model.

The sixth chapter is devoted to our fifth work where we establish
an equivalence between the MQM description of tachyon perturbations and
the so called Normal Matrix Model. We explain the basic features of
the latter and its relation to various problems in physics and mathematics.
The equivalence is interpreted as a kind of duality
for which a mathematical as well as a physical sense can be given.

In the last chapter we present our sixth work
on non-perturbative effects in matrix models and their relation
to D-branes. We calculate the leading non-perturbative
corrections to the partition function for both 
$c=1$ and $c<1$ string theories. In the beginning we present 
the calculation based on the matrix model formulation and then
we reproduce some of the obtained results from 
D-branes of Liouville theory.

We would like to say several words about the presentation. We tried to
do it in such a way that all the reported material would be connected
by a continuous line of reasonings. Each result is supposed to be a
more or less natural development of the previous ideas and results.
Therefore, we tried to give a motivation for each step leading to
something new. Also we explained various subtleties which occur
sometimes and not always can be found in the published articles.

Finally, we tried to trace all the coefficients and signs
and write all formulae in the once chosen normalization.
Their discussion sometimes may seem to be too technical for the reader.
But we hope he will forgive us because it is done to give the possibility
to use this thesis as a source for correct equations in the presented
domains.

\end{chapternon}





\chapter{String theory}
\label{chSTR}

String theory is now considered as the most promising candidate
to describe the unification of all interactions and quantum gravity.
It is a very wide subject of research possessing a very rich mathematical
structure. In this chapter we will give just a brief review
of the main ideas underlying string theory to understand
its connection with our work. For a detailed introduction to string theory,
we refer to the books \cite{GSW,Polbookf,Polbook}.

\section{Strings, fields and quantization}

\subsection{A little bit of history}

String theory has a very interesting history in which one can find
both the dark periods and remarkable breakthroughs of new ideas.
In the beginning it appeared as an attempt to describe
the strong interaction. In that time QCD was not yet known and
there was no principle to explain a big tower of particles
discovered in processes involving the strong interaction.
Such a principle was suggested by Veneziano \cite{Venec} in
the so called {\it dual} models.
He required that the sum of scattering amplitudes in $s$ and $t$
channels should coincide (see fig. \ref{dualamp}).

\lfig{Scattering amplitudes in dual models.}{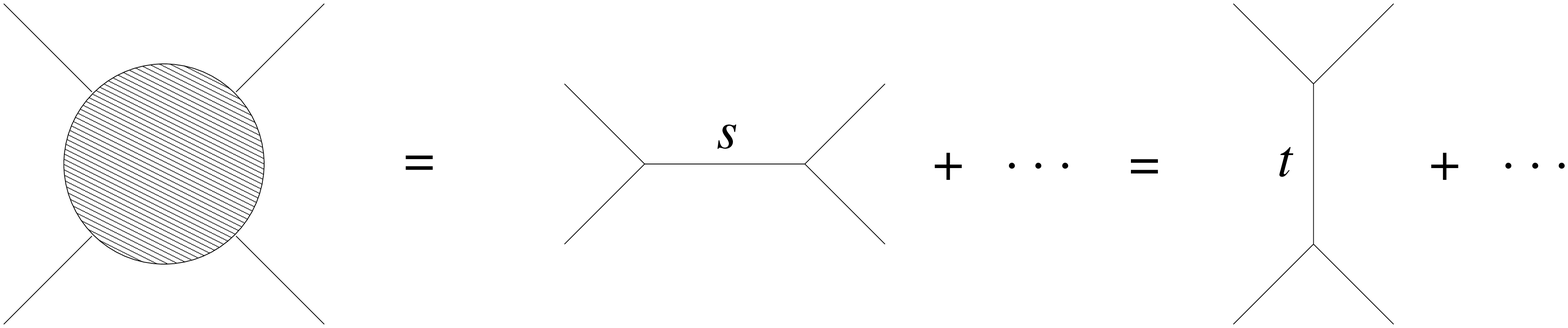}{9cm}{dualamp}

This requirement together with unitarity,
locality and {\it etc.} was strong enough
to fix completely the amplitudes.
Thus, it was possible to find them explicitly for the simplest cases
as well as to establish their general asymptotic properties.
In particular, it was shown that the scattering amplitudes
in dual models are much softer then the usual
field theory amplitudes, so that
the problems of field-theoretic divergences should be absent
in these models.

Moreover, the found amplitudes coincided with scattering amplitudes
of strings --- objects extended in one dimension \cite{NAMB,Nielsen,Suss}.
Actually, this is natural because for strings the property of
duality is evident: two channels can be seen as two degenerate
limits of the same string configuration (fig. \ref{dualstr}).
Also the absence of ultraviolet divergences got a natural explanation
in this picture. In field theory the divergences appear due to
a local nature of interactions related to the fact that
the interacting objects are thought to be pointlike.
When particles (pointlike objects) are replaced
by strings the singularity is smoothed out over the string world sheet.

\lfig{Scattering string amplitude can be seen in two ways.}
{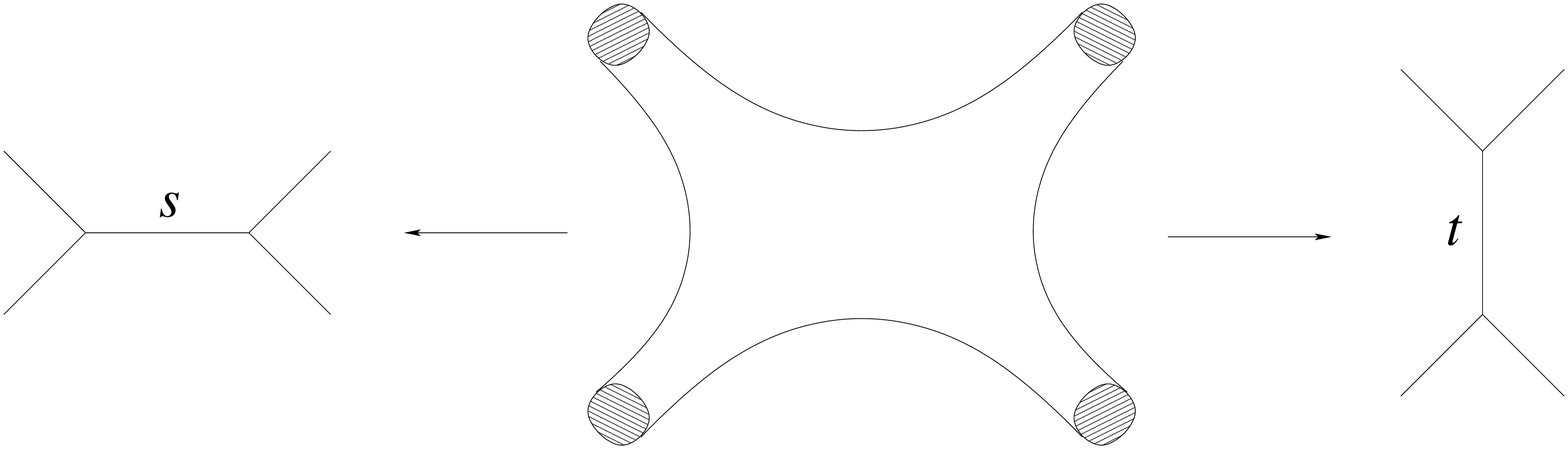}{9cm}{dualstr}

However, this nice idea was rejected by the discovery of QCD and
description of all strongly interacting particles as composite states
of fundamental quarks. Moreover, the exponential fall-off of
string amplitudes turned out to be inconsistent with
the observed power-like asymptotics. Thus, strings lost the initial
reason to be related to fundamental physics.

But suddenly another reason was found. Each string possesses
a spectrum of excitations. All of them can be interpreted as
particles with different spins and masses. For a closed string,
which can be thought just as a circle, the spectrum contains
a massless mode of spin 2.
But the graviton, quantum of gravitational interaction, has
the same quantum numbers. Therefore, strings might be used to
describe quantum gravity! If this is so, a theory based on strings
should describe the world at the very microscopic level,
such as the Planck scale, and should reproduce the standard model
only in some low-energy limit.

This idea gave a completely new status to string theory.
It became a candidate for the unified theory of all
interactions including gravity.
Since that time string theory has been developed into a rich theory
and gave rise to a great number of new physical concepts.
Let us have a look how it works.

\subsection{String action}

As is well known, the action for the relativistic particle is given by
the length of its world line. Similarly, the string action
is given by the area of its world sheet so that classical trajectories
correspond to world sheets of minimal area. The standard expression
for the area of a two-dimensional surface leads to the action
\cite{Nambu,Goto}
\be
S_{\rm NG}=-{1\over 2\pi \alp}
\int_\Sig  d\tau d\sigma\, \sqrt{- h}, \qquad h=\det h_{ab},
\plabel{SNG}
\ee
which is called the Nambu--Goto action. Here $\alp$ is a constant
of dimension of squared length. The matrix $h_{ab}$ is
the metric induced on the world sheet and can be represented as
\be
h_{ab}=G_{\mu\nu}\p_a X^{\mu}\p_b X^{\nu},
\plabel{indmet}
\ee
where $X^{\mu}(\tau,\sigma)$ are coordinates of a point $(\tau,\sigma)$
on the world sheet in the spacetime where the string moves.
Such a spacetime is called {\it target space} and $G_{\mu\nu}(X)$
is the metric there.

Due to the square root even in the flat target space
the action \Ref{SNG} is highly non-linear. Fortunately, there is
a much more simple formulation which is classically
equivalent to the Nambu--Goto action. This is the Polyakov action
\cite{Polyak}:
\be
S_{\rm P}=-{1\over 4\pi \alp}
\int_\Sig d\tau d\sigma \,
\sqrt{-h}\, G_{\mu\nu}h^{ab}\p_a X^{\mu}\p_b X^{\nu}.
\plabel{SPOL}
\ee
Here the world sheet metric is considered as a dynamical variable
and the relation \Ref{indmet} appears only
as a classical equation of motion. (More exactly, it is valid only up to
some constant multiplier.)
This means that we deal with a gravitational theory on the world sheet.
We can even add the usual Einstein term
\be
\chi={1\over 4\pi}\int_\Sig d\tau d\sigma\, \sqrt{-h}\, \CR.
\plabel{Eulerterm}
\ee
In two dimensions $\sqrt{-h}\CR$ is a total derivative.
Therefore, $\chi$ depends only on the topology of the surface $\Sig$,
which one integrates over, and produces its Euler characteristic.
In fact, any compact connected oriented two-dimensional surface
can be represented as a sphere with $g$ handles and $b$ boundaries.
In this case the Euler characteristic is
\be
\chi=2-2g-b.
\plabel{Euler}
\ee
Thus, the full string action reads
\be
S_{\rm P}=-{1\over 4\pi \alp}
\int_\Sig d\tau d\sigma \,
\sqrt{- h} \left( G_{\mu\nu}h^{ab}\p_a X^{\mu}\p_b X^{\nu}
+\alp \nu \CR \right),
\plabel{SPOLf}
\ee
where we introduced the coupling constant $\nu$.
In principle, one could add also a two-dimensional
cosmological constant. However, in this case the action
would not be equivalent to the Nambu--Goto action.
Therefore, we leave this possibility aside.

\lfig{Open and closed strings.}
{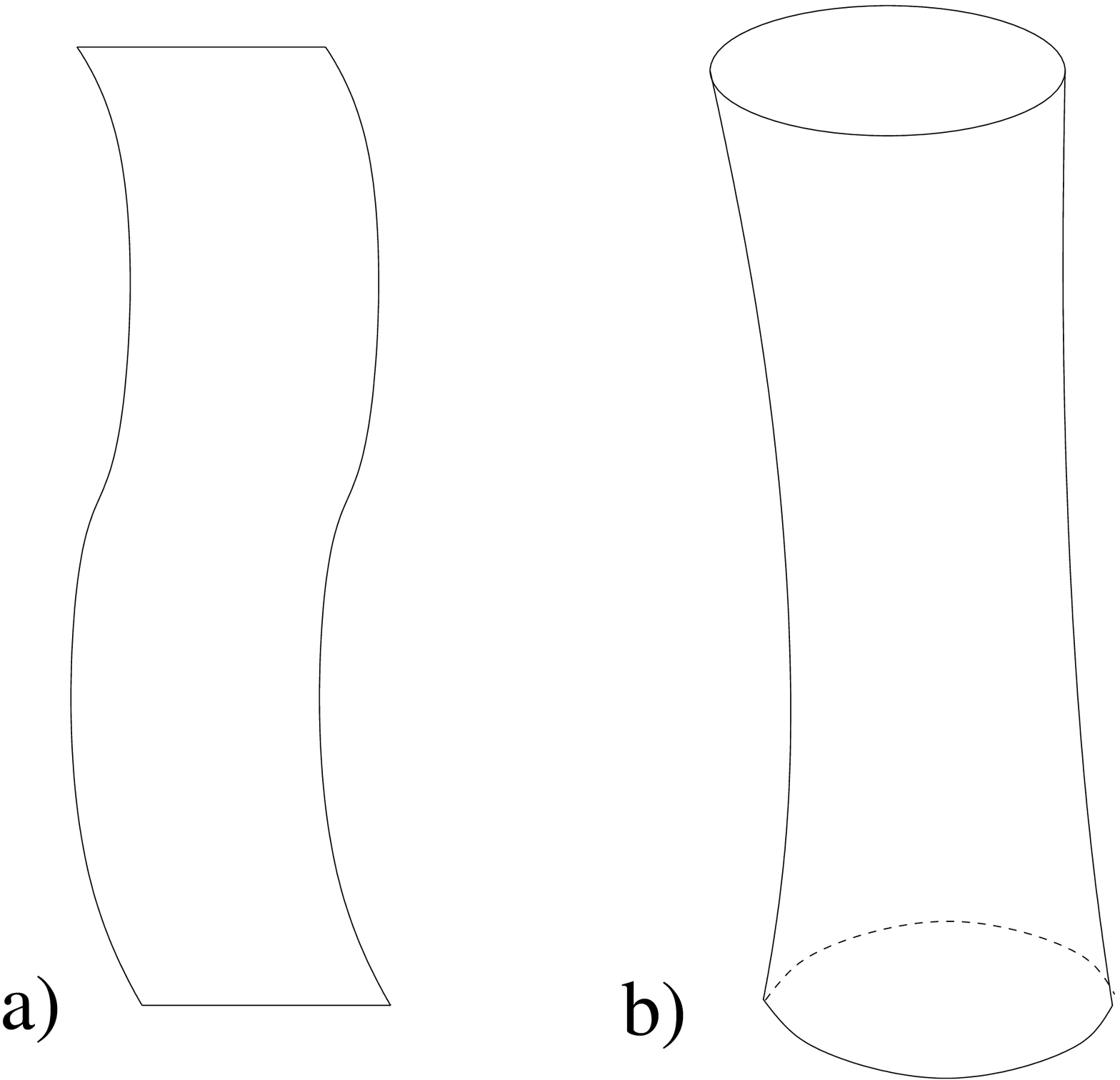}{5cm}{clop}

To completely define the theory, one should also impose some boundary
conditions on the fields $X^{\mu}(\tau,\sigma)$.
There are two possible choices corresponding to
two types of strings which one can consider.
The first choice is to take Neumann boundary conditions
$n^a\p_a X^{\mu}=0$ on $\p \Sig$, where $n^a$ is the normal
to the boundary. The presence of the boundary means that one considers
an {\it open} string with two ends (fig. \ref{clop}a).
Another possibility is given by periodic boundary conditions.
The corresponding string is called {\it closed} and it is
topologically equivalent to a circle (fig. \ref{clop}b).

\subsection{String theory as two-dimensional gravity}
\label{congr}

The starting point to write the Polyakov action was
to describe the movement of a string in a target space.
However, it possesses also an additional interpretation.
As we already mentioned, the two-dimensional metric $h_{ab}$
in the Polyakov formulation is a dynamical variable.
Besides, the action \Ref{SPOLf} is invariant under general coordinate
transformations on the world sheet. Therefore, the Polyakov action
can be equally considered as describing two-dimensional gravity
coupled with matter fields $X^{\mu}$.
The matter fields in this case are usual scalars.
The number of these scalars coincides with the dimension of
the target space.

Thus, there are two {\it dual} points of view:
target space and world sheet pictures. In the second one we can actually
completely forget about strings and consider it as
the problem of quantization of two-dimensional gravity in the presence
of matter fields.

It is convenient to do the analytical continuation to
the Euclidean signature on the world sheet $\tau\to -i\tau$.
Then the path integral over two-dimensional metrics
can be better defined, because the topologically non-trivial surfaces
can have non-singular Euclidean metrics,
whereas in the Minkowskian signature their metrics are always singular.
In this way we arrive at a statistical problem for which
the partition function is given by a sum over fluctuating
two-dimensional surfaces and quantum fields on
them\footnote{Note, that the Euclidean action $S_{\rm P}^{\rm (E)}$
differs by sign from the Minkowskian one.}

\be
Z=\sum\limits_{\rm surfaces \ \Sig}\int \CD X_{\mu}\,
e^{-S_{\rm P}^{\rm (E)}[X,\Sig]}.
\plabel{partfun}
\ee
The sum over surfaces should be understood as a sum over all possible
topologies plus a functional integral over metrics.
In two dimensions all topologies are classified. For example, for
closed oriented surfaces the sum over topologies corresponds to the sum
over genera $g$ which is the number of handles attached to a sphere.
In this case one gets
\be
\sum\limits_{\rm surfaces \ \Sig}=
\sum\limits_{g}\int \CD \varrho(h_{ab}).
\plabel{sumsurf}
\ee
On the contrary, the integral over metrics is yet to be defined.
One way to do this is to discretize surfaces and to replace the integral by
the sum over discretizations. This way leads to matrix models
discussed in the following chapters.

In string theory one usually follows another approach. It treats
the two-dimensional diffeomorphism invariance as an ordinary gauge
symmetry. Then the standard Faddeev--Popov gauge fixing procedure
is applied to make the path integral to be well defined.
However, the Polyakov action possesses an additional feature
which makes its quantization non-trivial.

\subsection{Weyl invariance}

The Polyakov action \Ref{SPOLf} is invariant under the local Weyl
transformations
\be
h_{ab} \longrightarrow e^{\phi}h_{ab},
\plabel{conftr}
\ee
where $\phi(\tau,\sigma)$ is any function on the world sheet.
This symmetry is very crucial because it allows to exclude one more
degree of freedom. Together with the diffeomorphism symmetry,
it leads to the possibility to express at the classical level
the world sheet metric in terms of
derivatives of the spacetime coordinates as in \Ref{indmet}.
Thus, it is responsible for the equivalence of the Polyakov and
Nambu--Goto actions.

However, the classical Weyl symmetry can be broken at
the quantum level. The reason can be found in the non-invariance
of the measure of integration over world sheet metrics.
Due to the appearance of divergences the measure should be regularized.
But there is no regularization preserving all symmetries including
the conformal one.

The anomaly can be most easily seen analyzing the energy-momentum tensor
$T_{ab}$. In any classical theory invariant under the Weyl
transformations the trace of $T_{ab}$ should be zero. Indeed,
the energy-momentum tensor is defined by
\be
T_{ab}=-{2\pi\over \sqrt{-h}}{\delta S\over \delta h^{ab}}.
\ee
If the metric is varied along eq. \Ref{conftr} ($\phi$ should be
taken infinitesimal), one gets
\be
T_a^{\ a}={2\pi\over \sqrt{-h}}{\delta S\over \delta \phi}=0.
\ee
However, in quantum theory $T_{ab}$ should be replaced by a renormalized
average of the quantum operator of the energy-momentum tensor.
Since the renormalization in general breaks the Weyl invariance,
the trace will not vanish anymore.

Let us restrict ourselves to the flat target space
$G_{\mu\nu}=\eta_{\mu\nu}$. Then explicit calculations
lead to the following anomaly
\be
\langl T_a^{\ a}\rangl_{\rm ren}=-{c\over 12}\CR.
\plabel{conam}
\ee

To understand the origin of the coefficient $c$, we
choose the flat gauge $h_{ab}=\delta_{ab}$.
Then the Euclidean Polyakov action takes the following form
\be
S_{\rm P}^{\rm (E)}=\nu\chi+{1\over 4\pi \alp}
\int_\Sig d\tau d\sigma \,
\delta^{ab}\p_a X^{\mu}\p_b X_{\mu}.
\plabel{EPOL}
\ee
This action is still invariant under conformal transformations
which preserve the flat metric. They are a special combination
of the Weyl and diffeomorphism transformations of the initial action.
Thus, the gauged fixed action \Ref{EPOL} represents a particular
case of conformal field theory (CFT).
Each CFT is characterized by a number $c$,
the so called {\it central charge}, which defines a quantum
deformation of the algebra of generators of conformal transformations.
It is this number that appears in the anomaly \Ref{conam}.

The central charge is determined by the field content of CFT.
Each bosonic degree of freedom contributes $1$ to the central charge,
each fermionic degree of freedom gives $1/2$, and ghost fields
which have incorrect statistics give rise to negative values of $c$.
In particular, the ghosts arising after a gauge fixation
of the diffeomorphism symmetry contribute $-26$.
Thus, if strings propagate in the flat spacetime of dimension $D$,
the central charge of CFT \Ref{EPOL} is
\be
c=D-26.
\plabel{centrch}
\ee

This gives the exact result for the Weyl anomaly.
Thus, one of the gauge symmetries of the classical theory
turns out to be broken. This effect can be seen also in
another approaches to string quantization.
For example, in the framework of canonical quantization
in the flat gauge one finds the breakdown of unitarity.
Similarly, in the light-cone quantization one encounters
the breakdown of global Lorentz symmetry in the target space.
All this indicates that the Weyl symmetry is extremely
important for the existence of a viable theory of strings.

\newpage

\section{Critical string theory}

\subsection{Critical bosonic strings}

We concluded the previous section with the statement that
to consistently quantize string theory we need to
preserve the Weyl symmetry. How can this be done?
The expression for the central charge \Ref{centrch} shows that it
is sufficient to place strings into spacetime of dimension $\Dc=26$
which is called {\it critical} dimension.
Then there is no anomaly and quantum theory is well defined.

Of course, our real world is four-dimensional. But now the idea
of Kaluza \cite{Kaluza} and Klein \cite{Klein} comes to save us.
Namely, one supposes that extra $22$ dimensions are compact
and small enough to be invisible at the usual scales.
One says that the initial spacetime is {\it compactified}.
However, now one has to choose some compact space to be used in
this compactification. It is clear that the effective
four-dimensional physics crucially depends on this choice.
But {\it a priori} there is no any preference and it seems to be
impossible to find the right compactification.

Actually, the situation is worse.
Among modes of the bosonic string, which are interpreted as
fields in the target space,
there is a mode with a negative squared mass that is a tachyon.
Such modes lead to instabilities of the vacuum and can break
the unitarity. Thus, the bosonic string theory in $26$ dimensions
is still a ``bad'' theory.

\subsection{Superstrings}

An attempt to cure the problem of the tachyon of bosonic strings
has led to a new theory where the role of fundamental objects is
played by superstrings. A superstring is a generalization of the
ordinary bosonic string including also fermionic degrees of freedom.
Its important feature is a supersymmetry.
In fact, there are two formulations of superstring theory with
the supersymmetry either in the target space or on the world sheet.

\subsubsection{Green--Schwarz formulation}

In the first formulation, developed by Green and Schwarz \cite{GS},
to the fields $X^{\mu}$ one adds
one or two sets of world sheet scalars $\theta^A$. They transform
as Maiorana--Weyl spinors with respect to the global Lorentz symmetry
in the target space. The number of spinors determines the number of
supersymmetric charges so that there are two possibilities to have
$\CN=1$ or $\CN=2$ supersymmetry. It is interesting that already at
the classical level one gets some restrictions on possible dimensions $D$.
It can be $3$, $4$, $6$ or $10$. However, the quantization
selects only the last possibility which is the critical dimension for
superstring theory.

In this formulation one has the explicit supersymmetry in the target
space.\footnote{Superstring can be interpreted as a string moving
in a superspace.}
Due to this, the tachyon mode cannot be present in the spectrum
of superstring and the spectrum starts with massless modes.

\subsubsection{RNS formulation}

Unfortunately, the Green--Schwarz formalism
is too complicated for real calculations. It is much more convenient
to use another formulation with a supersymmetry on the world sheet
\cite{Ramond,NS}.
It represents a natural extension of CFT \Ref{EPOL} being
a two-dimensional super-conformal field theory
(SCFT).\footnote{In fact, it is two-dimensional
supergravity coupled with superconformal matter.
Thus, in this formulation one has a supersymmetric generalization
of the interpretation discussed in section \ref{congr}.}
In this case the additional degrees of freedom are world sheet
fermions $\psi^{\mu}$ which form a vector under
the global Lorentz transformations in the target space.

Since this theory is a particular case of conformal theories,
the formula \Ref{conam} for the conformal anomaly remains valid.
Therefore, to find the critical dimension in this formalism,
it is sufficient to calculate the central charge.
Besides the fields discussed in the bosonic case,
there are contributions to the central charge from the world
sheet fermions and ghosts which arise after a gauge fixing
of the local fermionic symmetry. This symmetry is a superpartner of
the usual diffeomorphism symmetry and is a necessary part of
supergravity. As was mentioned, each fermion gives
the contribution $1/2$, whereas for the new superconformal ghosts
it is $11$. As a result, one obtains
\be
c=D-26+\hf D +11={3\over 2}(D-10).
\plabel{centrchf}
\ee
This confirms that the critical dimension for superstring theory is
$\Dc=10$.

To analyze the spectrum of this formulation, one should impose
boundary conditions on $\psi^{\mu}$.
But now the number of possibilities is doubled with respect
to the bosonic case. For example, since $\psi^{\mu}$ are fermions,
for the closed string not only periodic,
but also antiperiodic conditions can be chosen.
This leads to the existence of two independent sectors
called Ramond (R) and Neveu--Schwarz (NS) sectors.
In each sector superstrings have different spectra of modes.
In particular, from the target space point of view, R-sector
describes fermions and NS-sector contains bosonic fields.
But the latter suffers from the same problem as bosonic string theory
--- its lowest mode is a tachyon.

Is the fate of RNS formulation the same as that of the bosonic
string theory in $26$ dimensions? The answer is {\it not}.
In fact, when one calculates string amplitudes of perturbation theory,
one should sum over all possible spinor structures on the world sheet.
This leads to a special projection of the spectrum, which is called
Gliozzi--Scherk--Olive (GSO) projection \cite{GSO}.
It projects out the tachyon and several other modes.
As a result, one ends up with a well defined theory.

Moreover, it can be checked that after the projection
the theory possesses the global supersymmetry in the target space.
This indicates that actually GS and RNS formulations are equivalent.
This can be proven indeed and is related to some intriguing symmetries
of superstring theory in 10 dimensions.

\subsubsection{Consistent superstring theories}

Once we have constructed general formalism, one can ask
how many consistent theories of superstrings do exist?
Is it unique or not?

At the classical level it is certainly not unique.
One has open and closed, oriented and non-oriented,
$\CN=1$ and $\CN=2$ supersymmetric string theories. Besides,
in the open string case one can also introduce Yang--Mills gauge symmetry
adding charges to the ends of strings. It is clear that the gauge group
is not fixed anyhow. Finally, considering closed strings with
$\CN=1$ supersymmetry, one can construct the so called heterotic strings
where it is also possible to introduce a gauge group.

However, quantum theory in general suffers from anomalies arising
at one and higher loops in string perturbation theory. The requirement
of anomaly cancellation forces to restrict ourselves only to
the gauge group $SO(32)$ in the open string case and $SO(32)$ or
$E_8\times E_8$ in the heterotic case \cite{GSanom}.
Taking into account also restrictions on possible boundary conditions
for fermionic degrees of freedom,
one ends up with five consistent superstring theories.
We give their list below:
\begin{itemize}
\item type IIA: $\CN=2$ oriented non-chiral closed strings;
\item type IIB: $\CN=2$ oriented chiral closed strings;
\item type I: $\CN=1$ non-oriented open strings with the gauge group
$SO(32)$ $+$ non-oriented closed strings;
\item heterotic $SO(32)$: heterotic strings with the gauge group
$SO(32)$;
\item heterotic $E_8 \times E_8$: heterotic strings with the gauge group
$E_8 \times E_8$.
\end{itemize}

\subsection{Branes, dualities and M-theory}

\label{Mth}

Since there are five consistent superstring theories, the resulting
picture is not completely satisfactory. One should either choose
a correct one among them or find a further unification.
Besides, there is another problem. All string theories are defined
only as asymptotic expansions in the string coupling constant.
This expansion is nothing else but the sum over genera
of string world sheets in the closed case (see \Ref{sumsurf})
and over the number of boundaries in the open case.
It is associated with the {\it string loop expansion} since adding a handle
(strip) can be interpreted as two subsequent interactions:
a closed (open) string is emitted and then reabsorbed (fig. \ref{strint}).

\lfig{Interactions of open and closed strings.}
{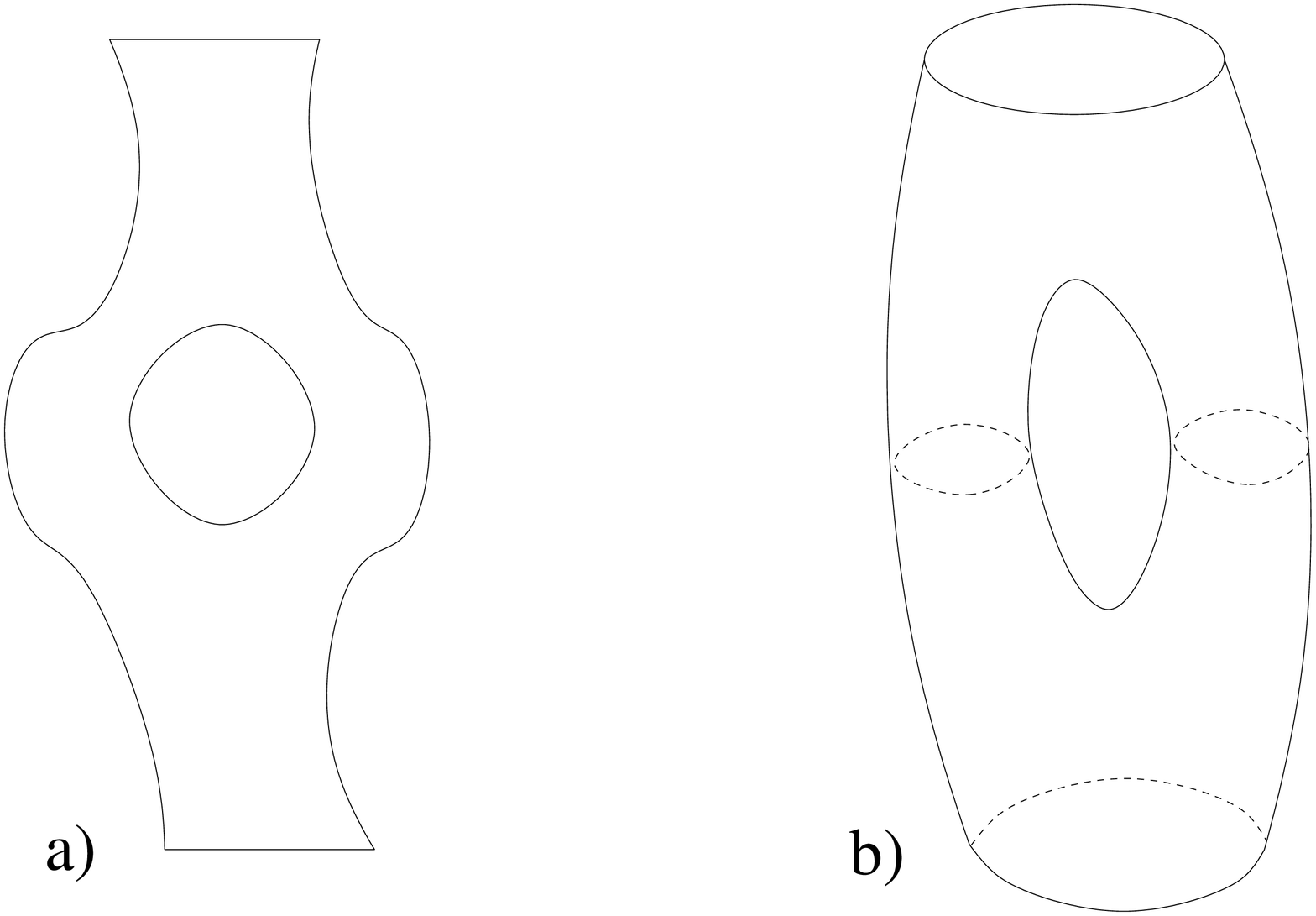}{6cm}{strint}

Note, that from the action \Ref{EPOL} it follows that
each term in the partition function \Ref{partfun} is weighted
by the factor $e^{-\nu\chi}$ which depends only on the topology of
the world sheet.
Due to this one can associate $e^{2\nu}$ with each handle
and $e^{\nu}$ with each strip. On the other hand, each interaction
process should involve a coupling constant. Therefore, $\nu$
determines the closed and open string coupling constants
\be
\gc\sim e^{\nu}, \qquad \go\sim e^{\nu/2}.
\plabel{cc}
\ee
Since string theories are defined as asymptotic expansions,
any finite value of $\nu$ leads to troubles. Besides, it looks like
a free parameter and there is no way to fix its value.

A way to resolve both problems came from the discovery
of a net of dualities relating different superstring theories.
As a result, a picture was found where different theories appear
as different vacua of a single (yet unknown) theory
which got the name ``M-theory''. A generic point in its moduli space
corresponds to an 11-dimensional vacuum. Therefore, one says
that the unifying M-theory is 11 dimensional. In particular,
it has a vacuum which is Lorentz invariant and described by 11-dimensional
flat spacetime. It is shown in fig. \ref{duality} as a circle labeled D=11.

\lfig{Chain of dualities relating all superstring theories.}
{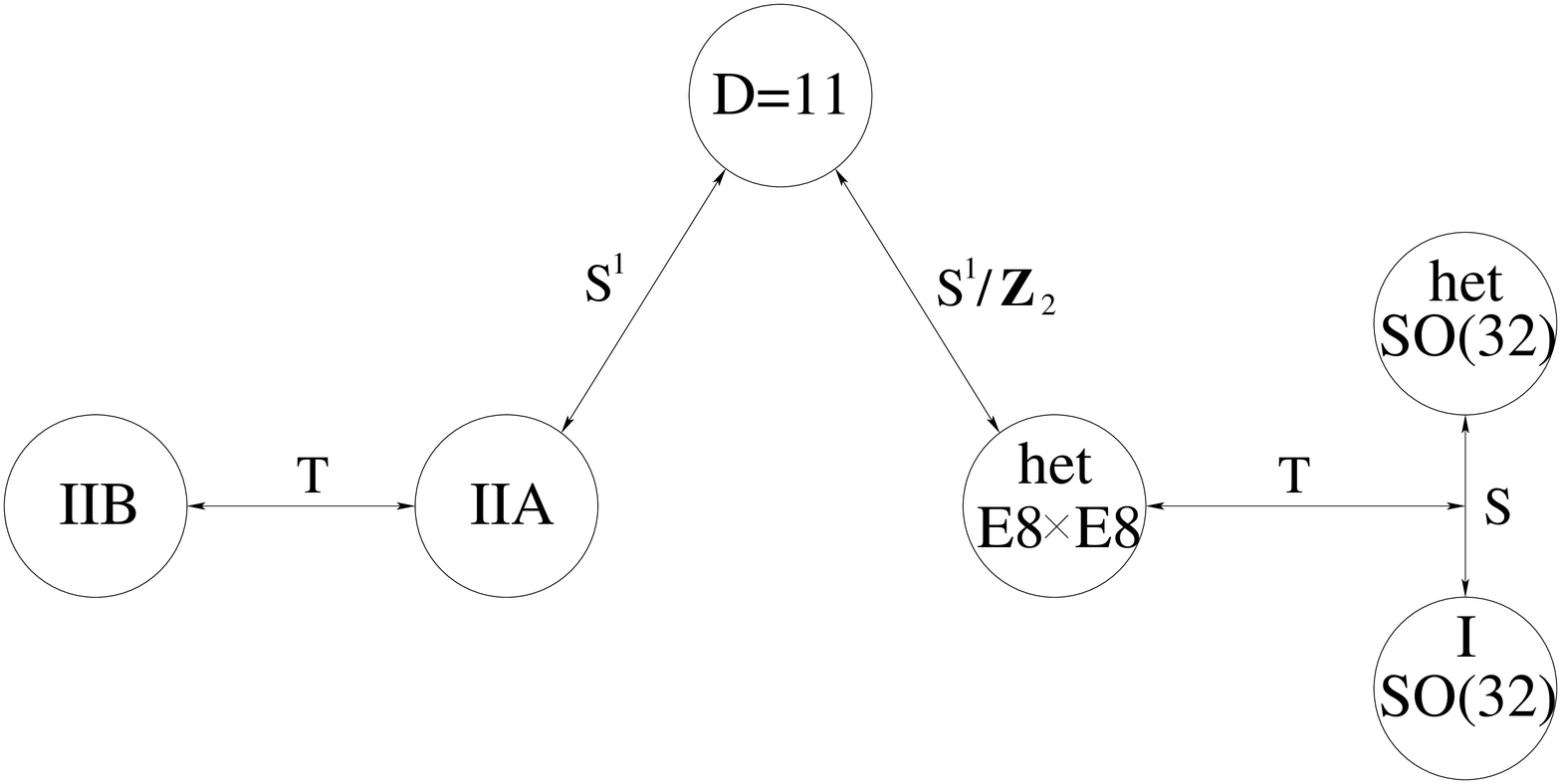}{11cm}{duality}

Other superstring theories can be obtained by different compactifications
of this special vacuum. Vacua with $\CN=2$ supersymmetry
arise after compactification on a torus, whereas
$\CN=1$ supersymmetry appears as a result of compactification
on a cylinder. The known superstring theories are
reproduced in some degenerate limits of the torus and cylinder.
For example, when one of the radii of the torus is much larger
than the other, so that one considers compactification on a circle,
one gets the IIA theory.
The small radius of the torus determines the string coupling constant.
The IIB theory is obtained when the two radii both vanish and
the corresponding string coupling is given by their ratio.
Similarly, the heterotic and type I theories appear in the same limits
for the radius and length of the cylinder.

This picture explains all existing relations between superstring
theories, a part of which is shown in fig. \ref{duality}.
The most known of them are given by T and S-dualities.
The former relates compactified theories with inverse
compactification radii and exchanges the windings around
compactified dimension with the usual momentum modes in this direction.
The latter duality says that the strong coupling limit of
one theory is the weak coupling limit of another.
It is important that T-duality has also a world sheet
realization: it changes sign of the right modes on
the string world sheet:
\be
X_L\to X_L, \qquad X_R\to -X_R.
\plabel{tdual}
\ee

The above picture indicates that the string coupling constant is always
determined by the background on which string theory is considered.
Thus, it is not a free parameter but one of the moduli of the underlying
M-theory.

It is worth to note that the realization of the dualities was possible
only due to the discovery of new dynamical objects in string theory
--- D branes \cite{PolchD}. They appear in several ways. On the one hand,
they are solitonic solutions of supergravity equations
determining possible string backgrounds. On the other hand,
they are objects where open strings can end. In this case
Dirichlet boundary conditions are imposed on the fields
propagating on the open string world sheet.
Already at this point it is clear that such objects must present
in the theory because the T-duality transformation \Ref{tdual}
exchanges the Neumann and Dirichlet boundary conditions.

We stop our discussion of critical superstring theories here.
We see that they allow for a nice unified picture of all interactions.
However, the final theory remains to be hidden from us
and we even do not know what principles should define it.
Also a correct way to compactify extra dimensions to get
the 4-dimensional physics is not yet found.

\newpage

\section{Low-energy limit and string backgrounds}
\label{lowen}

\subsection{General $\sigma$-model}

In the previous section we discussed string theory in
the flat spacetime. What changes if the target space is curved?
We will concentrate here only on the bosonic theory.
Adding fermions does not change much in the conclusions of this section.

In fact, we already defined an action for the string moving
in a general spacetime. It is given by the $\sigma$-model
\Ref{SPOLf} with an arbitrary $G_{\mu\nu}(X)$.
On the other hand, one can think about a non-trivial spacetime metric
as a coherent state of gravitons which appear in the closed string
spectrum. Thus, the insertion of the metric $G_{\mu\nu}$ into the world
sheet action is, roughly speaking, equivalent to summing of excitations
of this mode.

But the graviton is only one of the massless modes of the string spectrum.
For the closed string the spectrum contains also two other massless fields:
the antisymmetric tensor $B_{\mu\nu}$ and the scalar dilaton $\dl$.
There is no reason to turn on the first mode and to leave other modes
non-excited. Therefore, it is more natural to write a generalization of
\Ref{SPOLf} which includes also $B_{\mu\nu}$ and $\dl$.
It is given by the most general world sheet action which is invariant
under general coordinate transformations and renormalizable
\cite{CFMP}:\footnote{In the following, the world sheet metric is
always implied to be Euclidean.}
\be
S_\sigma={1\over4\pi\alp}\int d^2 \sigma\,\sqrt{h}
\left[\left(h^{ab}G_{\mu\nu}(X)+i\e^{ab}B_{\mu\nu}(X)\right)
\p_a X^{\mu} \p_b X^{\nu}
+\alp\CR \dl(X)\right],
\plabel{Smod}
\ee

In contrast to the Polyakov action in flat spacetime,
the action \Ref{Smod} is non-linear and represents an interacting
theory. The couplings of this theory are coefficients of
$G_{\mu\nu}$, $B_{\mu\nu}$ and $\dl$ of their expansion in $X^{\mu}$.
These coefficients are dimensionfull and the actual dimensionless
couplings are their combinations with the parameter $\alp$.
This parameter has dimension of squared length and
determines the {\it string scale}. It is clear that
the perturbation expansion of the world sheet quantum
field theory is an expansion in $\alp$ and, at the same time,
it corresponds to the long-range or
low-energy expansion in the target space.
At large distances compared to the string scale, the internal
structure of the string is not important and we should obtain
an effective theory. This theory is nothing else but
an effective field theory of massless string modes.

\subsection{Weyl invariance and effective action}
\label{weylback}

The effective theory, which appears in the low-energy limit,
should be a theory of fields in the target space.
On the other hand, from the world sheet point of view,
these fields represent an infinite set of couplings of a
two-dimensional quantum field theory. Therefore, equations
of the effective theory should be some constraints on the couplings.

What are these constraints? The only condition, which is not imposed by
hand, is that the $\sigma$-model \Ref{Smod} should define a consistent
string theory. In particular, this means
that the resulting quantum theory preserves the Weyl invariance.
It is this requirement that gives the necessary equations on the
target space fields.

With each field one can associate a $\beta$-function.
The Weyl invariance requires the vanishing of all $\beta$-functions
\cite{FriedSIG}.
These are the conditions we were looking for.
In the first order in $\alp$ one can find the following equations
\beq
\beta_{\mu\nu}^G &=&R_{\mu\nu}+2\dd_{\mu}\dd_{\nu} \dl-
{1\over 4}H_{\mu\lambda\sigma}{H_{\nu}}^{\lambda\sigma}+O(\alp)=0,
\nonumber \\
\beta_{\mu\nu}^{B}&=&-\hf \dd_{\lambda}{H_{\mu\nu}}^{\lambda}+
{H_{\mu\nu}}^{\lambda}\dd_{\lambda}\dl+O(\alp)=0,
\plabel{eqss} \\
\beta^{\dl}&=& {D-26 \over 6\alp }-{1\over 4}R- \dd^2 \dl+
(\dd\dl)^2+{1\over 48}H_{\mu\nu\lambda}H^{\mu\nu\lambda}+O(\alp)=0,
\nonumber
\eeq
where
\be
H_{\mu\nu\lambda}=\p_{\mu} B_{\nu\lambda}+\p_{\lambda} B_{\mu\nu}
+\p_{\nu} B_{\lambda\mu}
\ee
is the field strength for the antisymmetric tensor $B_{\mu\nu}$.

A very non-trivial fact which, on the other hand, can be considered
as a sign of consistency of the approach, is that the equations
\Ref{eqss} can be derived from the spacetime action \cite{CFMP}
\be
S_{\rm eff}=\hf \int d^D X \, \sqrt{-G}\, e^{-2\dl }
\left[ -{2(D-26)\over 3\alp}+
R+4(\dd\dl)^2-{1\over 12}H_{\mu\nu\lambda}H^{\mu\nu\lambda}\right].
\plabel{Seff}
\ee
All terms in this action are very natural representing the simplest
Lagrangians for symmetric spin-2, scalar, and antisymmetric spin-2 fields.
The first term plays the role of the cosmological constant.
It is huge in the used approximation since it is proportional to
$\alp^{-1}$. But just in the critical dimension it vanishes identically.

The only non-standard thing is the presence of the factor $e^{-2\dl}$
in front of the action. However, it can be removed by rescaling the metric.
As a result, one gets the usual Einstein term what means that
in the low-energy approximation string theory reproduces Einstein gravity.

\subsection{Linear dilaton background}

Any solution of the equations \Ref{eqss} defines a consistent string
theory. In particular, among them one finds the simplest flat,
constant dilaton background
\be
G_{\mu\nu}=\eta_{\mu\nu}, \qquad B_{\mu\nu}=0, \qquad \dl=\nu,
\ee
which is a solution of the equations of motion only in $\Dc=26$ dimensions
reproducing the condition we saw above.

There are also solutions which do not require any restriction
on the dimension of spacetime. To find them it is enough
to choose a non-constant dilaton to cancel the first term in $\beta^\dl$.
Strictly speaking, it is not completely satisfactory
because the first term has another order in $\alp$ and,
if we want to cancel it, one has to take into account contributions
from the next orders. Nevertheless, there exist
{\it exact} solutions which do not involve the higher orders.
The most important solution is the so called {\it linear
dilaton background}
\be
G_{\mu\nu}=\eta_{\mu\nu}, \qquad B_{\mu\nu}=0, \qquad \dl=l_{\mu}X^{\mu},
\plabel{ldl}
\ee
where
\be
l_{\mu}l^{\mu}={26-D \over 6\alp}.
\plabel{vector}
\ee

Note that the dilaton is a generalization of the coupling constant $\nu$
in \Ref{SPOLf}. Therefore, from \Ref{cc} it is clear that
this is the dilaton that defines the string coupling constant
which can now vary in spacetime
\be
\gc\sim e^\dl.
\plabel{ccc}
\ee
But then for the solution \Ref{ldl} there is a region where
the coupling diverges and the string perturbation theory fails.
This means that such background does not define a satisfactory
string theory.
However, there is a way to cure this problem.

\subsection{Inclusion of tachyon}
\label{intach}

When we wrote the renormalizable $\sigma$-model \Ref{Smod}, we actually
missed one possible term which is a generalization of
the two-dimensional cosmological constant
\be
S_{\sigma}^{T}= {1\over4\pi\alp}\int d^2 \sigma\,\sqrt{h}\, T(X).
\plabel{sigtach}
\ee
From the target space point of view, it introduces a tachyon field
which is the lowest mode of bosonic strings.
One can repeat the analysis of section \ref{weylback} and calculate
the contributions of this term to the $\beta$-functions.
Similarly to the massless modes, all of them can be deduced from
the spacetime action which should be added to \Ref{Seff}
\be
S_{\rm tach}=-\hf \int d^D X \, \sqrt{-G}\, e^{-2\dl }
\left[ (\dd T)^2-{4\over \alp}T^2\right].
\plabel{Stach}
\ee

\lfig{String propagation in the linear dilaton background in
the presence of the tachyon mode. The non-vanishing tachyon produces
a wall prohibiting the penetration into the region of
a large coupling constant.}
{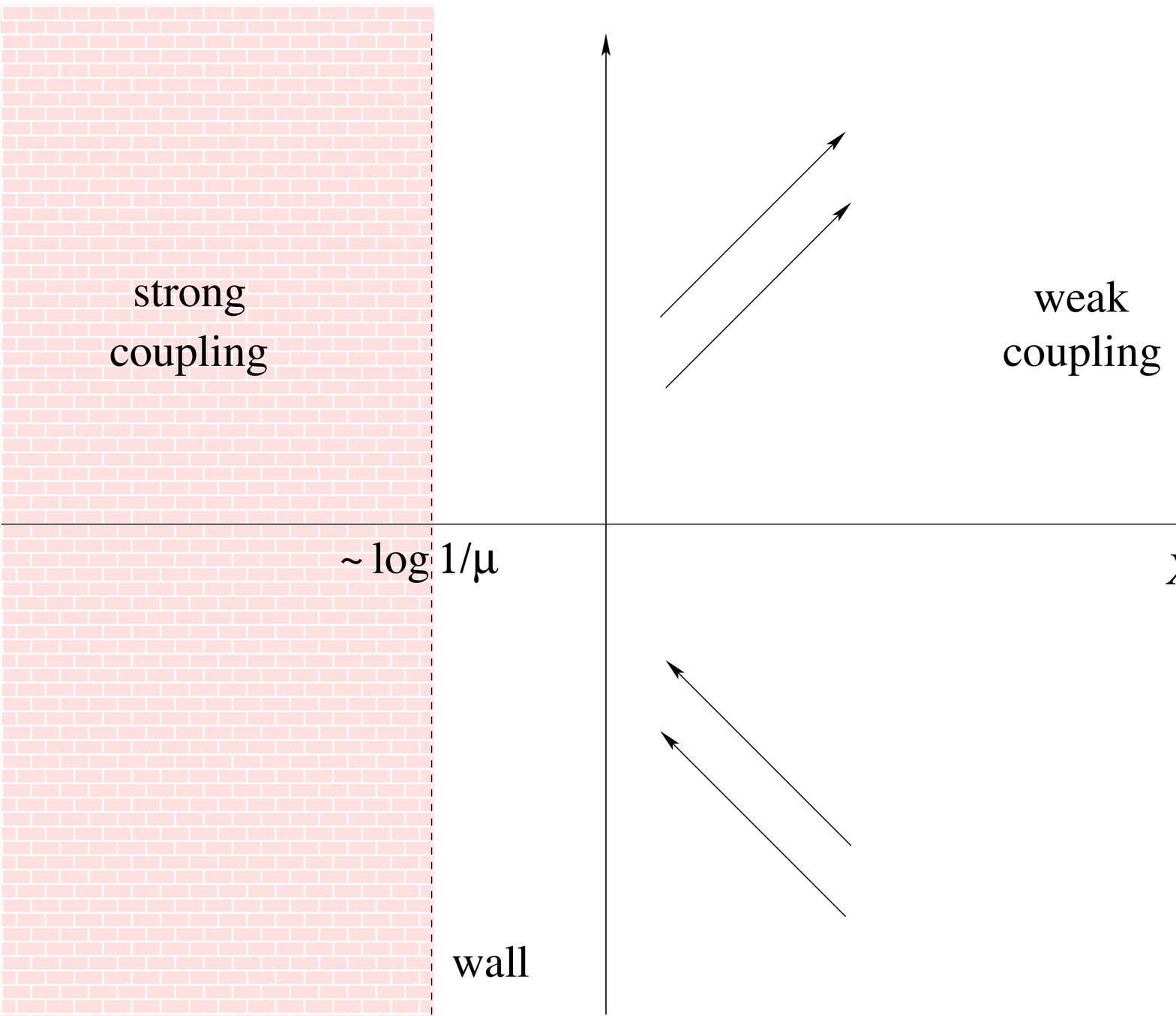}{8cm}{ldlwall}

Let us consider the tachyon as a field moving in the fixed linear
dilaton background. Substituting \Ref{ldl}
into the action \Ref{Stach}, one obtains the following equation of
motion
\be
\p^2 T-2l^{\mu}\p_{\mu}T+{4\over \alp}T=0.
\plabel{eqtach}
\ee
It is easy to find its general solution
\be
T=\mu\exp(p_{\mu}X^{\mu}), \qquad
(p-l)^2={2-D \over 6\alp}.
\plabel{soltach}
\ee
Together with \Ref{ldl}, \Ref{soltach} defines a generalization
of the linear dilaton background.
Strictly speaking, it is not a solution of the equations of motion
derived from the common action $S_{\rm eff}+S_{\rm tach}$.
However, this action includes only the first order in $\alp$, whereas,
in general, as we discussed above, one should take into account
higher order contributions. The necessity to do this is seen from
the fact that the background fields \Ref{ldl} and \Ref{soltach}
involve $\alp$ in a non-trivial way. The claim is that they give
an {\it exact} string background. Indeed, in this background
the complete $\sigma$-model action takes the form
\be
S^{\rm l.d.}_{\sigma}= {1\over4\pi\alp}\int d^2 \sigma\,\sqrt{h}
\left[h^{ab}\p_a X^{\mu} \p_b X_{\mu}
+\alp\CR l_{\mu}X^{\mu}+\mu e^{p_{\mu}X^{\mu}} \right].
\plabel{ldlsigma}
\ee
It can be checked that it represents an exact CFT
and, consequently, defines a consistent string theory.

Why does the introduction of the non-vanishing tachyonic mode make
the situation better? The reason is that this mode gives rise to
an exponential potential, which suppresses the string propagation
into the region where the coupling constant $\gc$ is large. It acts
as an effective wall placed at
$X^{\mu}\sim {p^{\mu}\over p^{2}}\log(1/\mu)$.
The resulting qualitative picture is shown in fig. \ref{ldlwall}.
Thus, we avoid the problem to consider strings at strong coupling.

\newpage

\section{Non-critical string theory}

In the previous section we saw that,
if to introduce non-vanishing expectation
values for the dilaton and the tachyon, it is possible to
define consistent string theory not only in the spacetime of
critical dimension $\Dc=26$.
Still one can ask the question: is there any sense for
a theory where the conformal anomaly is not canceled?
For example, if we look at the $\sigma$-model just as a statistical
system of two-dimensional surfaces embedded into $d$-dimensional
space and having some internal degrees of freedom, there is no reason
for the system to be Weyl-invariant. Therefore, even in the presence
of the Weyl anomaly, the system should possess some interpretation.
It is called {\it non-critical string theory}.

When one uses the interpretation we just described,
even at the classical level one can introduce terms breaking the Weyl
invariance such as the world sheet cosmological constant. Then
the conformal mode of the metric becomes a dynamical field
and one should gauge fix only the world sheet diffeomorphisms.
It can be done, for example, using the conformal gauge
\be
h_{ab}=e^{\phi(\sigma)}{\hat h}_{ab}.
\plabel{confg}
\ee
As a result, one obtains an effective action where, besides
the matter fields, there is a contribution depending on $\phi$
\cite{Polyak}. Let us work in the flat target space.
Then, after a suitable rescaling of $\phi$ to get the right kinetic term,
the action is written as
\be
S_{\rm CFT}={1\over4\pi\alp}\int d^2 \sigma\, \sqrt{\hat h}
\left[ {\hat h}^{ab}\partial_a X^{\mu} \partial_b X_{\mu}
+{\hat h}^{ab}\partial_a\phi\partial_b\phi
-\alp Q{\hat\CR}\phi + \mu e^{\gamma\phi}+{\rm ghosts}\right].
\label{confstr}
\ee
The second and third terms, which give dynamics to the conformal mode,
come from the measure of integration over all fields due to
its non-invariance under the Weyl transformations.
The coefficient $Q$ can be calculated from the conformal anomaly
and is given by
\be
Q=\sqrt{25-d\over 6\alp}.
\plabel{QQ}
\ee
The coefficient $\gamma$ is fixed by the condition that the theory
should depend only on the full metric $h_{ab}$. This means that
the effective action \Ref{confstr} should be invariant under
the following Weyl transformations
\be
{\hat h}_{ab}(\sigma)\longrightarrow e^{\rho(\sigma)}{\hat h}_{ab}(\sigma),
\qquad
\phi(\sigma)\longrightarrow \phi(\sigma)-\rho(\sigma).
\plabel{Ws}
\ee
This implies that the action \Ref{confstr} defines CFT.
This is indeed the case only if
\be
\gamma=-{1\over \sqrt{6\alp}}\left(\sqrt{25-d}-\sqrt{1-d}\right).
\plabel{gg}
\ee
The CFT \Ref{confstr} is called Liouville theory coupled with $c=d$
matter. The conformal mode $\phi$ is the Liouville field.

The comparison of the two CFT actions \Ref{confstr} and \Ref{ldlsigma}
shows that they are equivalent if one takes $D=d+1$,
$p_{\mu}\sim l_{\mu}$ and identifies $X^D=\phi$.
Then all coefficients also coincide as follows from
\Ref{vector}, \Ref{soltach}, \Ref{QQ} and \Ref{gg}.
Thus, the conformal mode of the world sheet metric can be interpreted as
an additional spacetime coordinate. With this interpretation
non-critical string theory in the flat $d$-dimensional spacetime
is seen as critical string theory in the $d+1$-dimensional
linear dilaton background. The world sheet cosmological constant $\mu$
is identified with the amplitude of the tachyonic mode.

\newpage

\section{Two-dimensional string theory}

In the following we will concentrate on the particular case
of 2D bosonic string theory. It represents
the main subject of this thesis. I hope to convince the reader
that it has a very rich and interesting structure and, at the same
time, it is integrable and allows for many detailed
calculations.\footnote{There is the so called $c=1$ barrier which
coincides with 2D string theory.
Whereas string theories with $c\le 1$ are solvable, we cannot say much
about $c>1$ cases.}
Thus, the two-dimensional case looks to be special and
it is a particular realization of a very universal structure.
It appears in the description of different physical and mathematical
problems. We will return to this question in the last chapters of
the thesis. Here we just mention two interpretations which, as
we have already seen, are equivalent to the critical string theory.

From the point of view of non-critical
strings, 2D string theory is a model of fluctuating two-dimensional surfaces
embedded into 1-dimensional time. The second space coordinate arises
from the metric on the surfaces.

Another possible interpretation
of this system described in section \ref{congr}
considers it as two-dimensional gravity coupled with the $c=1$ matter.
The total central charge vanishes since
the Liouville field $\phi$, arising due to the conformal anomaly,
contributes $1+6\alp Q^2$, where $Q$ is given in \Ref{QQ},
and cancels the contribution of matter and ghosts.

\subsection{Tachyon in two-dimensions}
\label{tachtwo}

To see that the two-dimensional case is indeed very special,
let us consider the effective action \Ref{Stach}
for the tachyon field in the linear dilaton background
\be
S_{\rm tach}=-\hf \int d^D X \,  e^{2Q\phi }
\left[ (\p T)^2-{4\over \alp}T^2\right],
\plabel{Stachldl}
\ee
where $\phi$ is the target space coordinate coinciding with
the gradient of the dilaton. According to the previous section,
it can be considered as the conformal mode of the world sheet metric
of non-critical strings. After the redefinition $T=e^{-Q\phi}\eta$,
the tachyon action becomes an action of a scalar field
in the flat spacetime
\be
S_{\rm tach}=-\hf \int d^D X \,
\left[ (\p \eta)^2+ m_{\eta}^2 \eta^2\right],
\plabel{Stachr}
\ee
where
\be
m_{\eta}^2=Q^2-{4\over \alp}={2-D\over 6\alp}
\plabel{mtach}
\ee
is the mass of this field. For $D>2$ the field $\eta$ has an imaginary mass
being a real tachyon. However, for $D=2$ it becomes massless.
Although we will still call this mode ``tachyon'', strictly speaking,
it represents a good massless field describing the stable vacuum
of the two-dimensional bosonic string theory.
As always, the appearance in the spectrum
of the additional massless field
indicates that the theory acquires some special properties.

In fact, the tachyon is the only field theoretic degree of freedom of
strings in two dimensions. This is evident in the light cone gauge
where there are physical excitations associated with
$D-2$ transverse oscillations and the motion of the string center of
mass. The former are absent in our case and the latter is identified
with the tachyon field.

To find the full spectrum of states and the corresponding
vertex operators, one should investigate the CFT \Ref{confstr}
with one matter field $X$.
The theory is well defined when the kinetic term for the $X$ field
enters with the $+$ sign so that $X$ plays the role of a space
coordinate. Thus, we will consider the following CFT
\be
S_{\rm CFT}={1\over4\pi}\int d^2 \sigma\, \sqrt{\hat h}
\left[ {\hat h}^{ab}\partial_a X \partial_b X
+{\hat h}^{ab}\partial_a\phi\partial_b\phi
-2{\hat\CR}\phi + \mu e^{-2\phi}+{\rm ghosts}\right],
\plabel{constr}
\ee
where we chose $\alp=1$ and took into account that in two dimensions
$Q=2$, $\gamma=-2$.
This CFT describes the Euclidean target space.
The Minkowskian version is defined by the analytical continuation
$X\to it$.

The CFT \Ref{constr} is a difficult interacting theory
due to the presence of the Liouville term $\mu e^{-2\phi}$.
Nevertheless, one can note that
in the region $\phi \to \infty$ this interaction is
negligible and the theory becomes free. Since the interaction is
arbitrarily weak in the asymptotics, it cannot
create or destroy states concentrated in this region.
However, it removes from the spectrum all states concentrated
at the opposite side of the Liouville direction. Therefore,
it is sufficient to investigate the spectrum of the free theory
with $\mu=0$ and impose the so called {\it Seiberg bound}
which truncates the spectrum by half \cite{Seibbound}.

The (asymptotic form of) vertex operators of the tachyon
have already been found in \Ref{soltach}.
If $l_{\mu}=(0,-Q)$ and $p_{\mu}=(p_X,p_{\phi})$,
one obtains the equation
\be
p_X^2+(p_{\phi}+Q)^2=0
\plabel{speqq}
\ee
with the general solution ($Q=2$)
\be
p_X=ip, \qquad p_{\phi}=-2\pm |p|, \qquad p\in \Rb.
\plabel{speqsol}
\ee
 Imposing the Seiberg bound, which forbids the operators growing
at $\phi\to -\infty$, we have to choose the $+$ sign in \Ref{speqsol}.
Thus, the tachyon vertex operators are
\be
V_p=\int d^2\sigma\, e^{ipX}e^{(|p|-2)\phi}.
\plabel{tachver}
\ee
Here $p$ is the Euclidean momentum of the tachyon. When we go to
the Minkowskian signature, the momentum should also be continued
as follows
\be
X\to it, \qquad p\to -ik.
\plabel{Mcont}
\ee
As a result, the vertex operators take the form
\bea{rcl}{Mtachver}
V^{-}_k&=&\int d^2\sigma\, e^{ik(t-\phi)}e^{-2\phi}, \\
V^{+}_k&=&\int d^2\sigma\, e^{-ik(t+\phi)}e^{-2\phi},
\eea
where $k>0$. The two types of operators describe
outgoing right movers and incoming left movers, respectively.
They are used to calculate the scattering of tachyons
off the Liouville wall.

\subsection{Discrete states}

Although the tachyon is the only target space
field in 2D string theory, there are also physical states
which are remnants of the transverse excitations of the string
in higher dimensions.
They appear at special values of momenta and
they are called {\it discrete states} \cite{disGKN,disPol,disLZ,WITTENGR}.

To define their vertex operators, we introduce
the {\it chiral} fields
\beq
W_{j,m}&=&\CP_{j,m}(\p X, \p^2 X, \dots) e^{2imX_L}e^{2(j-1)\phi_L},
\plabel{discrch}\\
\bar W_{j,m}&=&\CP_{j,m}(\bar \p X, \bar \p^2 X, \dots)
e^{2imX_R}e^{2(j-1)\phi_R},
\eeq
where $j=0,\hf,1,\dots$, $m=-j,\dots,j$ and
we used the decomposition of the world sheet fields into
the chiral (left and right) components
\be
X(\tau,\sigma)= X_L(\tau+i\sigma)+X_R(\tau-i\sigma)
\plabel{Xdec}
\ee
and similarly for $\phi$.
$\CP_{j,m}$ are polynomials in the chiral derivatives of $X$.
Their dimension is $j^2-m^2$. Due to this,
$\CP_{j,\pm j}=1$.
For each fixed $j$, the set of operators $W_{j,m}$ forms
an SU(2) multiplet of spin $j$.
Altogether, the operators \Ref{discrch} form $W_{1+\infty}$ algebra.

With the above definitions,
the operators creating the discrete states are given by
\be
V_{j,m}=\int d^2\sigma\, W_{j,m}\bar W_{j,m},
\plabel{discr}
\ee
Thus, the discrete states appear at the following momenta
\be
p_X=2im,
\qquad
p_\phi=2(j-1).
\plabel{mdiscr}
\ee
It is clear that the lowest and highest components $V_{j,\pm j}$
of each multiplet are just special cases of the vertex
operators \Ref{tachver}. The simplest non-trivial discrete state
is the zero-momentum dilaton
\be
V_{1,0}=\int d^2\sigma\, \p X\bar \p X.
\plabel{disdil}
\ee

\subsection{Compactification, winding modes and T-duality}
\label{comwin}

So far we considered 2D string theory in the usual flat
Euclidean or Minkowskian spacetime. The simplest thing which
we can do with this spacetime is to compactify it.
Since there is no translational invariance in the Liouville
direction, it cannot be compactified. Therefore, we do
compactification only for the Euclidean ``time'' coordinate $X$. We
require
\be
X\sim X+\beta, \qquad \beta=2\pi R,
\plabel{compX}
\ee
where $R$ is the radius of the compactification.
Because it is the time direction that is compactified,
we expect the resulting Minkowskian theory be equivalent to
a thermodynamical system at temperature $T=1/\beta$.

The compactification restricts the allowed tachyon momenta to discrete
values $p_n=n/R$ so that we have only a discrete set of vertex operators.
Besides, depending on the radius, the compactification can create or destroy
the discrete states. Whereas for rational values of the radius some discrete
states are present in the spectrum, for general irrational radius
there are no discrete states.

But the compactification also leads to the existence of new
physical string states. They correspond to configurations where
the string is wrapped around the compactified dimension.
Such excitations are called {\it winding modes}.
To describe these configurations in the CFT terms, one should use
the decomposition \Ref{Xdec} of
the world sheet field $X$ into the left and right moving components.
Then the operators creating the winding modes,
the {\it vortex} operators, are defined in terms of the dual field
\be
\tX(\tau,\sigma)= X_L(\tau+i\sigma)-X_R(\tau-i\sigma).
\plabel{tXdec}
\ee
They also have a discrete spectrum, but with the inverse frequency:
$q_m=mR$. In other respects they are similar to the vertex operators
\Ref{tachver}
\be
\tilde V_q=\int d^2\sigma\, e^{iq\tX}e^{(|q|-2)\phi}.
\plabel{vort}
\ee

The vertex and vortex operators are related by T-duality, which
exchanges the radius of compactification $R\leftrightarrow 1/R$ and
the world sheet fields corresponding to the compactified direction
$X\leftrightarrow \tX$ (cf. \Ref{tdual}). Thus, from the CFT point of view
it does not matter whether vertex or vortex operators are used to perturb the free
theory. For example, the correlators of tachyons at the radius $R$
should coincide with the correlators of windings at the radius
$1/R$.\footnote{In fact, one should also change the cosmological
constant $\mu\to R\mu$ \cite{KLEBANOV}. This change is equivalent to
a constant shift of the dilaton which is necessary to preserve
the invariance to all orders in the genus expansion.}

Note that the self-dual radius $R=1$ is distinguished by a higher symmetry
of the system in this case. As we will see, its mathematical description
is especially simple.

\newpage

\section{2D string theory in non-trivial backgrounds}
\label{nontrback}

\subsection{Curved backgrounds: Black hole}
\label{curback}

In the previous section we described the basic properties of string theory
in two-dimensions in the linear dilaton background.
In this thesis we will be interested in more general backgrounds.
In the low-energy limit all of them  can be described
by an effective theory. Its action can be extracted from \Ref{Seff}
and \Ref{Stach}. Since there is no antisymmetric 3-tensor in two
dimensions, the $B$-field does not contribute and we remain with
the following action
\be
S_{\rm eff}=\hf \int d^2 X \, \sqrt{-G}\, e^{-2\dl }
\left[ {16\over\alp}+
R+4(\dd\dl)^2 -(\dd T)^2+{4\over \alp}T^2\right].
\plabel{Sefftwo}
\ee

It is a model of dilaton gravity non-minimally coupled with
a scalar field, the tachyon $T$. It is known to have solutions with
non-vanishing curvature. Moreover, without the tachyon its general
solution is well known and is written as \cite{MSWbh} ($X^{\mu}=(t,r)$,
$Q=2/\sqrt{\alp}$)
\be
ds^2= -\left(1-e^{-2Qr}\right)dt^2+{1\over 1-e^{-2Qr}}dr^2,
\qquad  \dl=\vp_0-Qr.
\plabel{bhsol}
\ee
In this form the solution resembles the radial part of
the Schwarzschild metric for a spherically symmetric black hole.
This is not a coincidence since the spacetime \Ref{bhsol} does
correspond to a two-dimensional black hole. At $r=-\infty$
the curvature has a singularity and at $r=0$ the metric has a coordinate
singularity corresponding to the black hole horizon.
There is only one integration constant $\vp_0$ which can be related
to the mass of black hole
\be
M_{\rm bh}=2Q e^{-2\vp_0}.
\plabel{mass}
\ee

As the usual Schwarzschild black hole, this black hole emits the Hawking
radiation at the temperature $T_H={Q\over 2\pi}$ \cite{CGHS}
and has a non-vanishing
entropy \cite{GP,NP}. Thus, 2D string theory incorporates all problems of
the black hole thermodynamics and represents a model to approach their
solution. Compared to the quantum field theory analysis
on curved spacetime, in string theory the situation is better since
it is a well defined theory. Therefore, one can hope to solve the issues
related to physics at Planck scale, such as microscopic description
of the black hole entropy, which are inaccessible by the usual methods.

To accomplish this task, one needs to know the background not only in
the low-energy limit but also at all scales. Remarkably, an exact CFT,
which reduces in the leading order in $\alp$
to the world sheet string action in the black hole
background \Ref{bhsol}, was constructed \cite{WITbh}.
It is given by the so called $[{\rm SL}(2,\Rb)]_k/{\rm U}(1)$
coset $\sigma$-model where $k$ is the level of the representation
of the current algebra. Relying on this CFT, the exact form of the
background \Ref{bhsol}, which ensures the Weyl invariance in all
orders in $\alp$, was found \cite{DVV}. We write it in the following
form
\beq
&ds^2= -l^2(x) dt^2+dx^2, \qquad
l(x)=\frac{(1-p)^{1/2}\tanh Qx}{(1-p\tanh^2 Qx)^{1/2}}, &
\plabel{bhmet} \\
& \dl=\vp_0 - \log \cosh Qx - \frac14 \log(1-p\tanh^2 Qx), &
\plabel{bhdil}
\eeq
where $p$, $Q$ and the level $k$ are related by
\be
p=\frac{2\alp Q^2}{1+2\alp Q^2}, \qquad k={2\over p}=2+{1\over\alp Q^2}
\plabel{bhparam}
\ee
so that in our case $p=8/9$, $k=9/4$.
To establish the relation with the background \Ref{bhsol},
one should change the radial coordinate
\be
Qr = \ln\left[\frac{ \sqrt{1-p}}{1+\sqrt{1-p}}
\left(\cosh Qx +
\sqrt{ \cosh^2 Qx +\frac{p}{1-p}}\right)\right]
\ee
and take $p\to 0$ limit. This exact solution possesses the same properties
as the approximate one. However, it is difficult to extract
its quantitative thermodynamical characteristics such as mass, entropy,
and free energy. The reason is that we do not know any action
for which the metric \Ref{bhmet} and the dilaton \Ref{bhdil}
give a solution.\footnote{It is worth to mention
the recent result that \Ref{bhmet}, \Ref{bhdil} cannot be solution
of any dilaton gravity model with only second derivatives \cite{VGno}.}
The existing attempts to derive
these characteristics rely on some assumptions and lead to ambiguous
results \cite{KZ}.

The form \Ref{bhmet} of the solution is convenient for the continuation
to the Euclidean metric. It is achieved by $t=-iX$ what changes
sign of the first term.
The resulting space can be represented by a smooth manifold if to take
the time coordinate $X$ be periodic with the period
\be
\beta={2\pi\over Q\sqrt{1-p}}.
\plabel{peri}
\ee
The manifold looks as a cigar (fig. \ref{cigar})
and the choice \Ref{peri} ensures the absence
of conical singularity at the tip.
It is clear that this condition reproduces the Hawking temperature
in the limit $p\to 0$ and generalizes it to all orders in $\alp$.
The function $l(x)$ multiplied by $R=\sqrt{\alp k}$ 
plays the role of the radius of the compactified
dimension. It approaches the constant value $R$ at infinity
and vanishes at the tip so that this point represents
the horizon of the Minkowskian black hole. Thus, the cigar describes only
the exterior of the black hole.

\lfig{The Euclidean black hole.}
{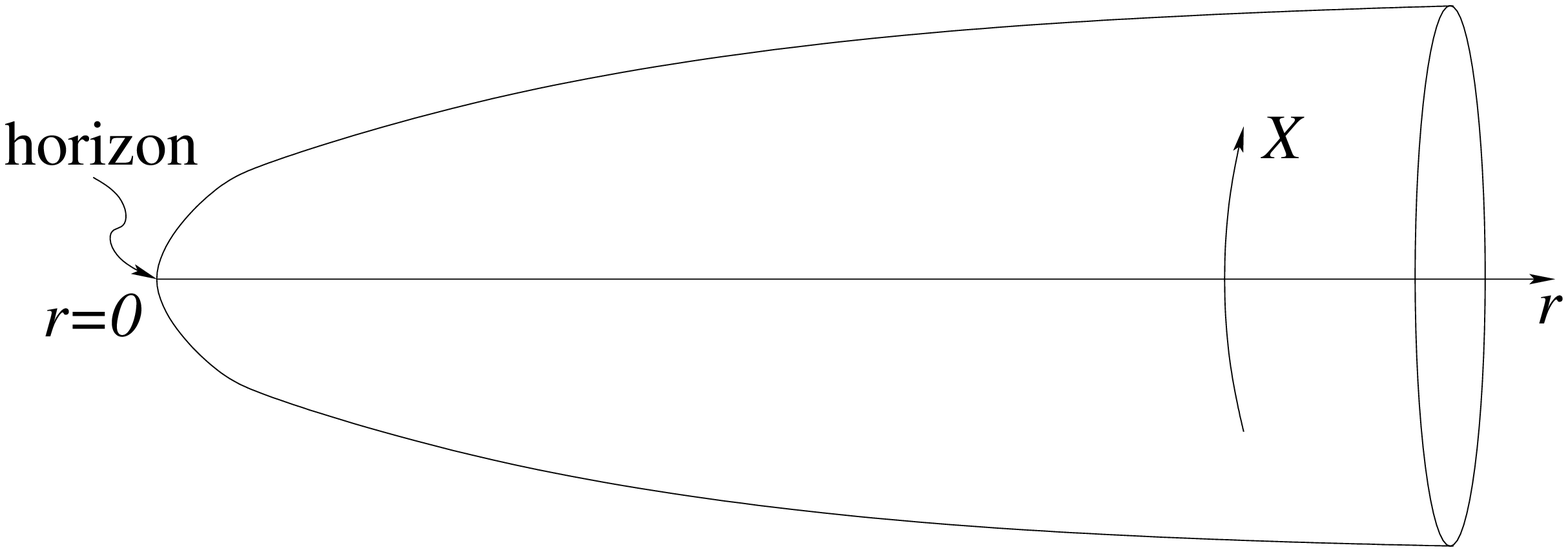}{7cm}{cigar}

Note that the CFT describing this Euclidean continuation is represented by 
the coset
\be
[H^+_3]_k/{\rm U}(1), \quad H^+_3\equiv {{\rm SL}(2,\Cb)\over {\rm SU}(2)},
\label{coset}
\ee
where $H^+_3$ can be thought as Euclidean $AdS_3$.

Using the coset CFT, two and three-point correlators
of tachyons and windings on the black hole background
were calculated \cite{DVV,Tesch1,Tesch2,Tesch3}.
By T-duality they coincide with winding and tachyon correlators,
respectively, on a dual spacetime, which is called {\it trumpet}
and can be obtained replacing $\cosh$ and $\tanh$ in
\Ref{bhmet}, \Ref{bhdil} by $\sinh$ and $\coth$. This dual spacetime
describes a naked (without horizon) black hole of a negative mass
\cite{DVV}.
In fact, it appears as a part of the global analytical continuation of
the initial black hole spacetime.

\subsection{Tachyon and winding condensation}
\label{twback}

In the CFT terms, string theory on the curved background considered above
is obtained as a $\sigma$-model. If one chooses the dilaton as
the radial coordinate, then the $\sigma$-model looks as CFT \Ref{constr}
where the kinetic term is coupled with the black hole metric $G_{\mu\nu}$
and there is no Liouville exponential interaction.
The change of the metric can be represented as a perturbation of
the linear dilaton background by
the gravitational vertex operator. Note that this operator creates one of
the discrete states.

It is natural to consider also perturbations by another relevant operators
existing in the initial CFT \Ref{constr} defined in
the linear dilaton background. First of all, these are the tachyon vertex
operators $V_p$ \Ref{tachver}. Besides, if we consider the Euclidean
theory compactified on a circle, there exist the vortex operators
$\tilde V_q$
\Ref{vort}. Thus, both types of operators can be used to perturb
the simplest CFT \Ref{constr}
\be
 S=S_{\rm CFT}+\sum_{n\ne 0}( t_n V_n +\tilde t_n\tilde V_n),
\plabel{PERTS}
\ee
where we took into account that tachyons and windings have
discrete spectra in the compactified theory.

What backgrounds of 2D string theory do these perturbations correspond to?
The couplings $t_n$ introduce a non-vanishing vacuum expectation value of
the tachyon. Thus, they simply change the background value of $T$.
Note that, in contrast to the cosmological constant term
$\mu e^{-2\phi}$, these tachyon condensates are time-dependent.
For the couplings $\tilde t_n$ we cannot give such a simple picture.
The reason is that the windings do not have a local target space interpretation.
Therefore, it is not clear which local characteristics of the background
change by the introduction of a condensate of winding modes.

A concrete proposal has been made for the simplest case
$\tilde \tpm \ne 0$, which is called {\it Sine-Liouville CFT}.
It was suggested that this CFT is equivalent to the
$H^+_3$/U(1) $\sigma$-model describing string theory on the black
hole background \cite{FZZ}. This conjecture was justified by the coincidence of
spectra of the two CFTs as well as of two- and three-point correlators
as we discuss in the next paragraph.
Following this idea, it is natural to suppose that any general 
winding perturbation changes the target space metric.

Note, that the world sheet T-duality relates the CFT \Ref{PERTS} with
one set of couplings $(t_n,\tilde t_n)$ and radius of compactification $R$
to the similar CFT, where the couplings are exchanged $(\tilde t_n, t_n)$
and the radius is inverse $1/R$.
However, these two theories should not describe the same background
because the target space interpretations of tachyons and windings
are quite different. T-duality allows to relate their correlators,
but it says nothing how their condensation changes the target space.

\subsection{FZZ conjecture}
\label{conjFZZ}

In this paragraph we give the precise formulation of the conjecture proposed by 
V. Fateev, A. Zamolodchikov and Al. Zamolodchikov \cite{FZZ}. 
It states that the coset CFT $H^+_3$/U(1), 
describing string theory on the Euclidean black hole background,
at arbitrary level $k$ is equivalent to the CFT given by the following action
\be
\SSL={1\over4\pi\alp}\int d^2 \sigma\,
\left[ (\partial X)^2+(\partial \phi)^2
-\alp Q{\hat\CR}\phi +\lambda e^{\rho\phi}\cos(R\tX) \right],
\plabel{SLCFTbh}
\ee
where the field $X$ is compactified at radius $R$ and
the parameters are expressed through the level $k$ 
\be
Q={1\over \sqrt{\alp (k-2)}}, \qquad 
R=\sqrt{\alp k}, \qquad  
\rho=-\sqrt{k-2 \over \alp}.
\plabel{bhpark}
\ee

These identifications can be understood as follows.
First, the duality requires the coincidence of the central 
charges of the two theories
\be
c_{\rm bh}={3k\over k-2}-1  \quad {\rm and} \quad
c_{\rm SL}=2+6\alp Q^2.  
\label{centrtwo}
\ee
This gives the first condition. 
The second equation in \Ref{bhpark} allows to 
identify the two CFTs in the free asymptotic region 
$\phi\to \infty$ ($r\to \infty$). Indeed, in both cases the target space looks as
a cylinder of the radius $R$ so that the world sheet field $X$ 
coincides with the radial coordinate on the cigar (see fig. \ref{cigar}).
Comparing the expressions for the dilaton, one also concludes
that $r\sim \phi$. 
Finally, the last formula in \Ref{bhpark} follows from the requirement that 
the scaling dimension of the interaction term is equal to one and 
from the first two identifications.

The first evidence for the equivalence is the coincidence of the spectra
of the two theories. In both cases the observables $V_{j,n,m}$ are labeled by
three indices: $j$ related to representations of SL(2,$\Rb$),
$n\in \Zb$ measuring the momentum along the compactified direction, and
$m\in\Zb$ associated with the winding number. 
In the free asymptotic region they have the form
\be
V_{j,n,m}\sim e^{ip_LX_L+ip_RX_R +2Qj\phi}
\label{obsas}
\ee
and their scaling dimensions agree
\bea{c}{scdimbh}
\Delta_{j,n,m}={\alp p_L^2 \over 4}-\alp Q^2 j(j+1)
={n_L^2\over k}-{j(j+1) \over k-2}, \\
\bar\Delta_{j,n,m}={\alp p_R^2 \over 4}-\alp Q^2 j(j+1)
={n_R^2\over k}-{j(j+1) \over k-2},
\eea  
where
\bea{cc}{mnnn}
p_L={n\over R}+{mR\over \alp} \quad & 
\quad p_R={n\over R}-{mR\over \alp}, \\
n_L=\hf(n+km), \quad & \quad n_R=-\hf(n-km).
\eea

Note also that in both theories there is a conservation of the momentum 
$n$, and the winding number $m$ is not conserved. But the reason for that is different.
Whereas in the cigar CFT the winding modes can slip off the tip of the cigar,
in the Sine--Liouville CFT \Ref{SLCFTbh} the winding conservation is broken explicitly
by the interaction term.  

The next essential piece of evidence in favour of the FZZ conjecture is
provided by the analysis of correlators in the two models.
The two-point correlators on the cigar in the spherical approximation 
are written as follows \cite{DVV}
\be
\langle V_{j,n,m}V_{j,-n,-m}\rangle=(k-2)
[\nu(k)]^{2j+1}{\Gamma(1-{2j+1\over k-2})\Gamma(-2j-1)\Gamma(j-n_L+1)
\Gamma(1+j+n_R)\over\Gamma({2j+1\over k-2})\Gamma(2j+2)
\Gamma(-j-n_L)\Gamma(n_R-j)},
\label{twobh}
\ee
where
\be
\nu(k)\equiv{1\over\pi} {\Gamma(1+{1\over k-2})\over
\Gamma(1-{1\over k-2})}.
\ee
It was shown that they agree with the same correlators calculated in the Sine--Liouville
theory \cite{FZZ}.
Besides, the same statement was established also for
the three point correlators.

Of course, this does not give a proof of the conjecture yet. But
this represents a very non-trivial fact which is hardly believed to be accidental.
Moreover, there is a supersymmetric generalization of this conjecture
proposed in \cite{GivKut}. It relates the $\CN=1$ superconformal coset model 
${\rm SL}(2,\Rb)/{\rm U}(1)$ to the $\CN=2$ Liouville theory. The former theory
has an accidental $\CN=2$ supersymmetry which is a special case of the 
Kazama--Suzuki construction \cite{KazSuz}. Therefore, the proposed relation is not quite
surprising. As it often happens, supersymmetry simplifies the problem and,
in contrast to the original bosonic case, this 
conjecture was explicitly proven \cite{kapust}.   

Finally, one remark is in order. 
The FZZ conjecture was formulated for arbitrary level $k$ and radius $R$.
However, it is relevant for two-dimensional string theory only when
the central charge is equal to 26. Therefore, in our case we have to fix
all parameters
\be
Q=2/\sqrt{\alp}, \qquad R=3\sqrt{\alp}/2, \qquad k=9/4. 
\label{fixval}
\ee
This means that there is only one point in the moduli space 
where we can apply the described duality.




\chapter{Matrix models}
\label{chMM}

In this chapter we introduce a powerful mathematical technique,
which allows to solve many physical problems. Its main feature
is the use of matrices of a large size. Therefore, the models
formulated using this technology are called {\it matrix models}.
Sometimes a matrix formulation is not only a useful mathematical
description of a physical system, but it also sheds light on its
fundamental degrees of freedom.

We will be interested mostly in application of matrix models to string
theory. However, in the beginning we should explain their relation
to physics, their general properties, and basic methods
to solve them (for an extensive review, see \cite{Mehta}).
This is the goal of this chapter.

\section{Matrix models in physics}

Working with matrix models, one usually considers the situation
when the size of matrices is very large. Moreover,
these models imply integration over matrices or averaging over
them taking all matrix elements as independent variables.
This means that one deals with systems where some random processes
are expected. Indeed, this is a typical behaviour for the systems
described by matrix models.

\subsubsection{Statistical physics}

Historically, for the first time matrix models appeared
in nuclear physics. It was discovered by Wigner \cite{Wigner} that
the energy levels of large atomic nuclei are distributed according
to the same law, which describes the spectrum of eigenvalues
of one Hermitian matrix in the limit where the size of the matrix
goes to infinity. Already this result showed the important feature
of universality: it could be applied to any nucleus and
did not depend on particular characteristics of this nucleus.

Following this idea, one can generalize the matrix description
of statistics of energy levels to any system, which either
has many degrees of freedom and is too complicated for an exact
description, or possesses a random behaviour. A typical example of
systems of the first type is given by mesoscopic physics,
where one is interested basically only in macroscopic characteristics.
The second possibility is realized, in particular, in chaotic systems.

\subsubsection{Quantum chromodynamics}

Another subject, where matrix models gave a new method of calculation,
is particle physics. The idea goes back to the work of 't Hooft
\cite{thoft} where he suggested to use the $1/N$ expansion for calculations
in gauge theory with the gauge group SU(N). Initially, he suggested
this expansion for QCD as an alternative to the usual perturbative
expansion, which is valid only in the weak coupling region and fails
at low energies due to the confinement.
However, in the case of QCD it is not well justified
since the expansion parameter equals $1/3^2$ and
is not very small.

Nevertheless,\ 't Hooft realized several important facts about the $1/N$
expansion. First, SU(N) gauge theory can be considered as a model
of $N\times N$ unitary matrices since the gauge fields are operators
in the adjoint representation. Then the $1/N$ expansion corresponds to
the limit of large matrices.
Second, it coincides with the topological expansion where all Feynman
diagrams are classified according to their topology, which one can associate
if all lines in Feynman diagrams are considered as double lines.
This gives the so called {\it fat graphs}. In the
limit $N\to \infty$ only the {\it planar} diagrams survive.
These are the diagrams which can be drawn on the 2-sphere
without intersections.
Thus, with each matrix model one can associate a diagrammatic expansion
so that the size of matrices enters only as a prefactor for
each diagram.

Although this idea has not led to a large progress in QCD,
it gave rise to new developments, related with matrix models,
in two-dimensional quantum gravity and string theory
\cite{fdavid,vkaz,KKM,BKKM,KAZMIG}.
In turn, there is still a hope to find a connection between string theory
and QCD relying on matrix models \cite{MakMig}. Besides, recently they
were applied to describe supersymmetric gauge theories \cite{DVc}.

\subsubsection{Quantum gravity and string theory}

The common feature of two-dimensional quantum gravity and string theory
is a sum over two-dimensional surfaces. It turns out that it also
has a profound connection with matrix models ensuring their relevance
for these two theories. We describe this connection in detail
in the next section because it deserves a special attention.
Here we just mention that the reformulation of string theory in terms
of matrix models has lead to significant results in the low-dimensional
cases such as 2D string theory. Unfortunately, this reformulation
has not helped much in higher dimensions.

It is worth to note that there are matrix models of M-theory
\cite{BFSS,IKKT}, which is thought to be a unification of all
string theories (see section \ref{chSTR}.\ref{Mth}). 
(See also \cite{Halp,dWHN,Town} for earlier attempts to describe 
the quantum mechanics of the supermembranes.)
They claim to be fundamental non-perturbative and background independent
formulations of Planck scale physics.
However, they are based on the ideas different from the ``old'' matrix models
of low-dimensional string theories.

Also matrix models appear in the so called {\it spin foam} approach to
3 and 4-dimensional quantum gravity \cite{spinf,spinfp}.
Similarly to the models of M--theory,
they give a non-perturbative and background
independent formulation of quantum gravity but do not help
with calculations.

\newpage

\section{Matrix models and random surfaces}

\label{sdissurf}

\subsection{Definition of one-matrix model}

Now it is time to define what a matrix model is.
In the most simple case of {\it one-matrix model} (1MM), one considers
the following integral over $N\times N$ matrices
\be
Z=\int dM\, \exp\[ -N \tr V(M)\] ,
\plabel{mint}
\ee
where
\be
V(M)=\sum\limits_{k>0}{g_k \over k}M^k
\plabel{mpot}
\ee
is a potential and the measure $dM$ is understood as a product of
the usual differentials of all independent matrix elements
\be
dM=\prod\limits_{i,j} dM_{ij}.
\plabel{mmes}
\ee
The integral \Ref{mint} can be interpreted as the partition function
in the canonical ensemble of a statistical model.
Also it appears as a generating function for the correlators of
the operators $\tr M^k$, which are obtained differentiating $Z$ with respect
to the couplings $g_k$.
For general couplings, the integral \Ref{mint} is divergent
and should be defined by analytical continuation.

Actually, one can impose some restrictions on the matrix $M_{ij}$
which reflect the symmetry of the problem. Correspondingly,
there exist 3 ensembles of random matrices:
\begin{itemize}
\item ensemble of hermitian matrices with the symmetry group U(N);
\item ensemble of real symmetric matrices with the symmetry group O(N);
\item ensemble of quaternionic matrices with the symmetry group Sp(N).
\end{itemize}
We will consider only the first ensemble. Therefore, our systems
will always possess the global U(N) invariance under the transformations
\be
M\longrightarrow \Omega^{\dagger}M\Omega
\quad (\Omega^{\dagger}\Omega=I).
\plabel{mtrun}
\ee

In general, the integral \Ref{mint} cannot be evaluated exactly
and one has to use its perturbative expansion in the coupling constants.
Following the usual methodology of quantum field theories,
each term in this expansion can be represented
as a Feynman diagram. Its ingredients, propagator and vertices,
can be extracted from the potential \Ref{mpot}. The main difference
with the case of a scalar field is that the propagator is represented
by an oriented double line what reflects the index structure carried
by matrices.

\input{propag.pic}

\noindent
The expression which is associated with the propagator can be obtained
as an average of two matrices $\langle M_{ij}M_{kl}\rangle_0$ with
respect to the Gaussian part of the potential. The coupling of indices
corresponds to the usual matrix multiplication law.
The vertices come from the terms of the potential of third and higher
powers. They are also composed from the double lines and are given by
a product of Kronecker symbols.

\input{vertex.pic}

\vskip - 0.8cm
\noindent
Note that each loop in the diagrams gives the factor $N$ coming from
the sum over contracted indices. As a result, the partition function
\Ref{mint} is represented as
\be
Z=\sum\limits_{\rm diagrams}{1\over s}
\left({1\over Ng_2}\right)^{E}N^{L} \prod\limits_k (-Ng_k)^{n_k},
\plabel{mintexp}
\ee
where the sum goes over all diagrams constructed from the drawn
propagators and vertices (fat graphs)
and we introduced the following notations:
\begin{itemize}
\item
$n_k$ is the number of vertices with $k$ legs in the diagram,
\item
$V=\sum_k n_k$ is the total number of vertices,
\item
$L$ is the number of loops,
\item
$E=\hf \sum_k kn_k$ is the number of propagators,
\item
$s$ is the symmetry factor given by the order of the discrete
group of symmetries of the diagram.
\end{itemize}
Thus, each diagram contributes to the partition function
\be
{1\over s} N^{V-E+L}g_2^{-E}
\prod\limits_k (-g_k)^{n_k}.
\plabel{mdig}
\ee

\subsection{Generalizations}
\label{Sgener}

Generalizations of the one-matrix model can be obtained by
increasing the number of matrices. The simplest generalization
is {\it two-matrix model} (2MM). In a general case it is defined by
the following integral
\be
Z=\int dA\, dB\, \exp\[ -N W(A,B)\] ,
\plabel{twomint}
\ee
where $W(A,B)$ is a potential invariant under the global unitary
transformations
\be
A\longrightarrow \Omega^{\dagger}A\Omega,
\qquad
B\longrightarrow \Omega^{\dagger}B\Omega.
\plabel{twomtrun}
\ee
The structure of this model is already much richer than the structure
of the one-matrix model. It will be important for us in the study of
2D string theory.

Similarly, one can consider 3, 4, {\it etc.} matrix models.
Their definition is the same as \Ref{twomint}, where one requires
the invariance of the potential under the simultaneous unitary
transformation of all matrices. A popular choice for the potential is
\be
W(A_1,\dots,A_n)=\sum\limits_{k=1}^{n-1}c_k\tr (A_kA_{k+1})-
\sum\limits_{k=1}^{n}\tr V_k(A_k).
\plabel{twompot}
\ee
It represents a linear {\it matrix chain}. Other choices are also possible.
For example, one can close the chain into a circle adding the term
$\tr(A_kA_1)$. This crucially changes the properties of the model, since
it loses integrability which is present in the case of the chain.

When the number of matrices increases to infinity, one can change
the discrete index by a continuous argument. Then one considers
a one-matrix integral, but the matrix is already a function.
Interpreting the argument as a time variable, one obtains a quantum
mechanical problem. It is called {\it matrix quantum mechanics} (MQM).
The most part of the thesis is devoted to its investigation.
Therefore, we will discuss it in detail in the next chapters.

Further generalizations include cases when one adds new arguments and
discrete indices to matrices and combines them in different ways.
One can even consider grassmanian, rectangular and other types of
matrices. Also multitrace terms can be included into the potential.

\subsection{Discretized surfaces}

A remarkable fact, which allows to make contact between
matrix models and two-dimensional
quantum gravity and string theory, is that the matrix integral \Ref{mint}
can be interpreted as a sum over discretized surfaces 
\cite{fdavid,vkaz,KKM}.
Each Feynman diagram represented by a fat graph is dual to
some triangulation of a two-dimensional surface as shown in fig.
\ref{dsdual}. To construct the dual surface, one associates
a $k$-polygon with each $k$-valent vertex and joins them along edges
intersecting propagators of the Feynman diagram.

\lfig{Duality between Feynman graphs and discretized surfaces.}
{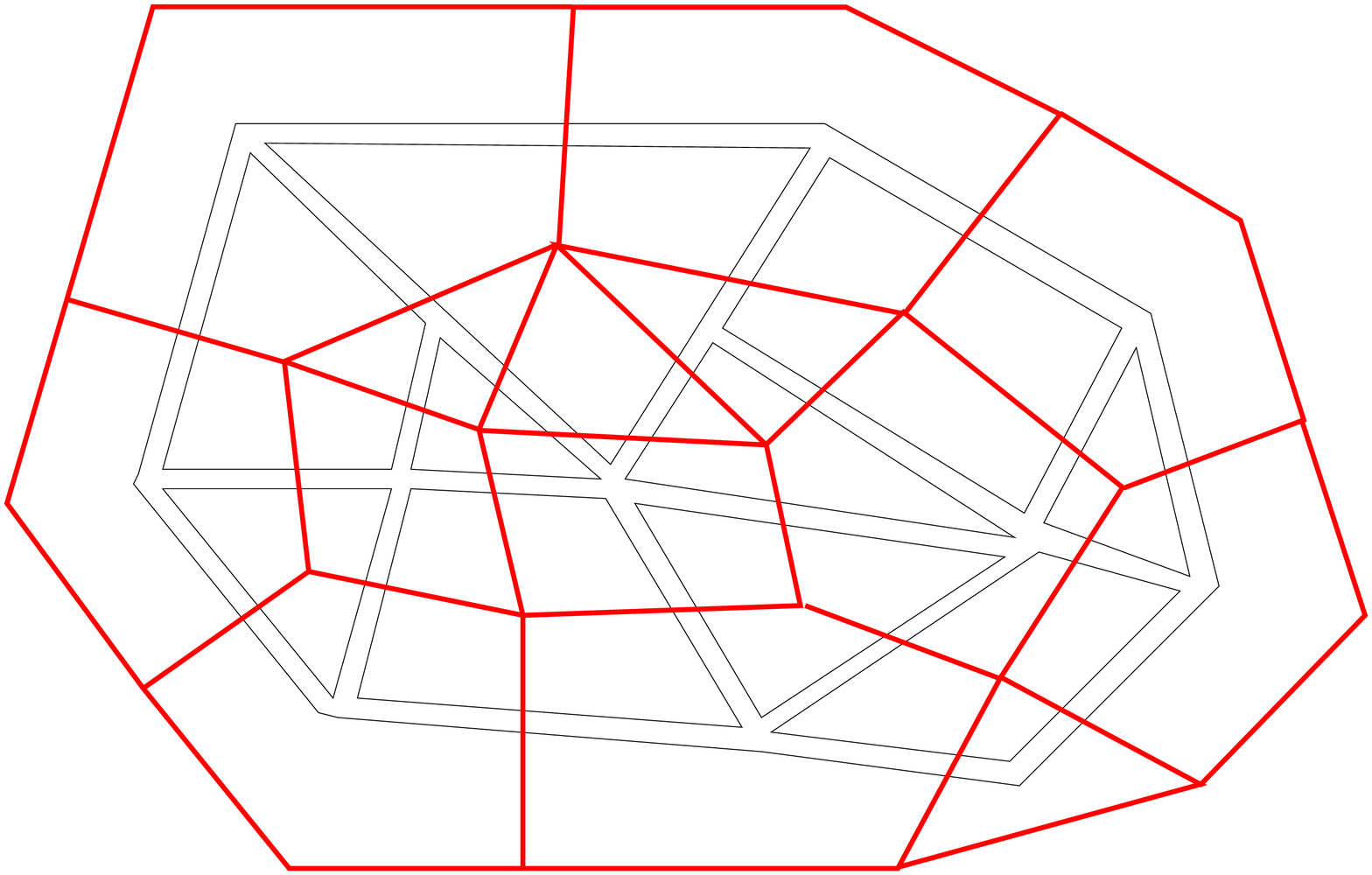}{8cm}{dsdual}

Note that the partition function is a sum over both connected and
disconnected diagrams. Therefore, it gives rise to both connected and
disconnected surfaces.
If we are interested, as in quantum
gravity, only in the connected surfaces, one should consider
the free energy, which is the logarithm of the partition function,
$F=\log Z$.
Thus, taking into account \Ref{mdig},
the duality of matrix diagrams and discretized surfaces leads
to the following representation
\be
F=\sum\limits_{\rm surfaces}
{1\over s} N^{V-E+L}g_2^{-E}
\prod\limits_k (-g_k)^{n_k}.
\plabel{mdisc}
\ee
where we interpret:
\begin{itemize}
\item
$n_k$ is the number of $k$-polygons used in the discretization,
\item
$V=\sum_k n_k$ is the total number of polygons (faces),
\item
$L$ is the number of vertices,
\item
$E=\hf \sum_k kn_k$ is the number of edges,
\item
$s$ is the order of the group of automorphisms of the
discretized surface.
\end{itemize}
It is clear that the relative numbers of $k$-polygons are controlled by
the couplings $g_k$. For example, if one wants to use only triangles,
one should choose the cubic potential in the corresponding matrix model.

The sum over discretizations \Ref{mdisc}
(more exactly, its continuum limit)
can be considered as a definition
of the sum over surfaces appearing in \Ref{sumsurf}.
Each discretization induces a curvature on the surface, which is
concentrated at vertices where several polygons are joint to each other.
For example, if at $i$th vertex there are $n^{(i)}_k$ $k$-polygons,
the discrete counterpart of the curvature is
\be
\CR_i=2\pi\frac{\left(2-\sum\limits_k {k-2\over k}n^{(i)}_k\right) }
{\sum\limits_k{1\over k} n^{(i)}_k}.
\plabel{curv}
\ee
It counts the deficit angle at the given vertex. In the limit
of large number of vertices, the discretization approximates some
continuous geometry. Varying discretization,
one can approximate any continuous distribution
of the curvature with any given accuracy.

Note, that the discretization encodes only the information about
the curvature which is diffeomorphism invariant. Therefore,
the sum over discretizations realizes already a gauge fixed version of
the path integral over geometries. Due to this, one does not need to deal
with ghosts and other problems related to gauge fixing.

If one considers generalizations of the one-matrix model, the dual
surfaces will carry additional structures. For example, the Feynman
diagrams of the two-matrix model are drawn using two types of lines
corresponding to two matrices. All vertices are constructed from
the lines of a definite type since they come from the potential
for either the first or the second matrix.
Therefore, with each face of the dual discretization one can associate
a discrete variable taking two values, say $\pm 1$. Summing over all
structures, one obtains Ising model on a random lattice
\cite{KazIzing,KBIzing}. Similarly,
it is possible to get various fields living on two-dimensional
dynamical surfaces or, in other words, coupled with two-dimensional
gravity.

\subsection{Topological expansion}

From \Ref{mdisc} one concludes that each surface enters with the weight
$N^{V-E+L}$. What is the meaning of the parameter $N$ for surfaces?
To answer this question, we note that the combination $V-E+L$ gives
the Euler number $\chi=2-2g$ of the surface.
Indeed, the Euler number is defined
as in \Ref{Eulerterm}. Under discretization the curvature turns into
\Ref{curv}, the volume element at $i$th vertex becomes
\be
\sqrt{h_i}={\sum\limits_k n^{(i)}_k/k},
\plabel{volem}
\ee
and the integral over the surface is replaced by the sum over
the vertices. Thus, the Euler number for the discretized surface
is defined as
\be
\chi={1\over 4\pi}\sum\limits_{i} \sqrt{h_i}\CR_i=
\hf\sum\limits_{i}
\left(2-\sum\limits_k{k-2\over k}n^{(i)}_k\right)=
L-\hf\sum\limits_k (k-2)n_k=L-E+V.
\plabel{Eulerdis}
\ee
Due to this,
we can split the sum over surfaces in \Ref{mdisc} into the sum over
topologies and the sum over surfaces of a given topology, which
imitates the integral over metrics,
\be
F=\sum\limits_{g=0}^{\infty} N^{2-2g} F_g(g_k).
\plabel{mtop}
\ee
Thus, $N$ allows to distinguish surfaces of different topology.
In the large $N$ limit only surfaces of the spherical topology
survive. Therefore, this limit is called also the {\it spherical limit}.

\lfig{Diagrams of different genera.}
{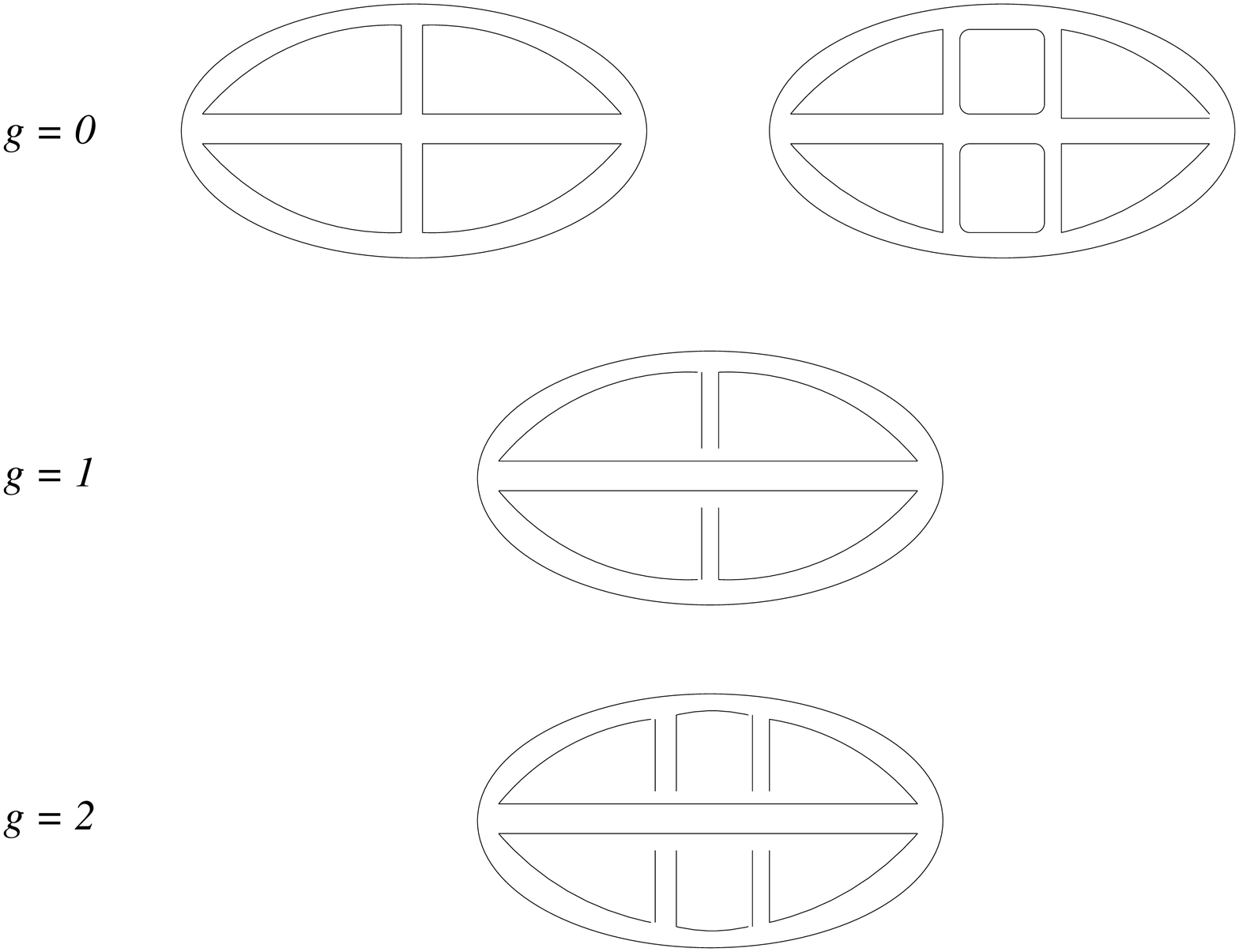}{8cm}{diagr}

In terms of fat graphs of the matrix model, this classification
by topology means the following. The diagrams appearing with
the coefficient $N^{2-2g}$, which correspond to the
surfaces of genus $g$, can be drawn on such surfaces without
intersections.
In particular, the leading diagrams coupled with $N^2$
are called {\it planar} and can be drawn on a 2-sphere or on a plane.
For other diagrams, $g$ can be interpreted as the minimal number of
intersections which are needed to draw the diagram on a plane
(see fig. \ref{diagr}).

\subsection{Continuum and double scaling limits}
\label{conlim}

Our goal is to relate the matrix integral to the sum over Riemann
surfaces. We have already completed the first step reducing the matrix
integral to the sum over discretized surfaces. And the sum over
topologies was automatically included. It remains only to extract a
continuum limit.

To complete this second step, let us work for simplicity with
the cubic potential
\be
V(M)=\hf M^2-{\lambda \over 3}M^3.
\plabel{mpotc}
\ee
Then \Ref{mdisc} takes the form
\be
F=\sum\limits_{g=0}^{\infty} N^{\chi}
\sum\limits_{\rm genus\ g \atop triangulations} {1\over s} \lambda^V.
\plabel{mcub}
\ee
We compare this result
with the partition function of two-dimensional quantum gravity
\be
Z_{\rm QG}=\sum\limits_{g=0}^{\infty} \int \CD \varrho(h_{ab})
e^{-\nu\chi-\mu A},
\plabel{twogr}
\ee
where $g$ is the genus of the surface,
$A=\int d^2\sigma\,\sqrt{h}$ is its area, and $\nu$ and $\mu$ are
coupling constants.
First of all, we see that one can identify
$N=e^{-\nu}$. Then, if one assumes that all triangles have unit area,
the total area is given by the number of triangles $V$.
Due to this, \Ref{mcub} implies $\lambda=e^{-\mu}$.

However, this was a formal identification because one needs the
coincidence of the partition function $Z_{\rm QG}$
and the free energy of the matrix model $F$. It is possible only
in a continuum limit where we sum over the same set of
continuous surfaces.
In this limit the area of triangles used in
the discretization should vanish. Since we fixed their area,
in our case the continuum limit implies that the number of triangles
should diverge $V\to \infty$.

In quantum theory one can speak only
about expectation values. Hence we are interested in the behaviour of
$\langle V\rangle$. 
In fact, for the spherical topology
this quantity is dominated by non-universal contributions.
To see the universal behaviour related to the continuum limit,
one should consider more general correlation functions 
$\langle V^n\rangle$.
From \Ref{mcub},
for these quantities one obtains a simple expression
\be
\langle V^n\rangle=\(\lambda{\p\over \p \lambda}\)^n\log F.
\plabel{avarea}
\ee
The typical form of the contribution of genus $g$ to the
free energy is
\be
F_g\sim F_g^{\rm (reg)}+(\lambda_c-\lambda)^{(2-\str)\chi/2},
\plabel{aspf}
\ee
where $\str$ is the so called {\it string susceptibility},
which defines the critical behaviour, $\lambda_c$ is some
critical value of the coupling constant, and we took into account
the non-universal contribution $F_g^{\rm (reg)}$. 
The latter leads to the fact 
that the expectation value $\langle V\rangle$
remains finite for $\chi>0$. But for all $n>1$,
one finds
\be
{\langle V^n\rangle \over \langle V^{n-1}\rangle}
\sim{1\over \lambda_c- \lambda}.
\plabel{avare}
\ee
This shows that in the limit $\lambda\to \lambda_c$,
the sum \Ref{mcub} is dominated by
triangulations with large number of triangles.
Thus, the continuum limit is obtained taking
$\lambda\to \lambda_c$. One can renormalize the area and
the couplings so that they remain finite in this limit.

Now we encounter the following problem.
According to \Ref{aspf}, the (universal part of the)
free energy in the continuum limit either
diverges or vanishes depending on the genus.
On the other hand, in the natural limit $N\to \infty$ only the spherical
contribution survives. How can one obtain contributions for all genera?
It turns out that taking both limits not independently, but together
in a correlated manner, one arrives at the desired result
\cite{BrKz,DgSh,GrMg}.
Indeed, we introduce the ``renormalized'' string coupling
\be
\cap^{-1}=N(\lambda_c-\lambda)^{(2-\str)/2},
\plabel{coupl}
\ee
and consider the limit $N\to \infty$, $\lambda\to\lambda_c$,
where $\cap$ is kept fixed. In this limit
the free energy is written as an asymptotic expansion for $\cap\to 0$
\be
F=\sum\limits_{g=0}^{\infty} \cap^{-2+2g}f_g.
\plabel{dblsc}
\ee
Thus, the described limit allows to keep all genera in the expansion
of the free energy. It is called the {\it double scaling limit}.
In this limit the free energy of the one-matrix model reproduces
the partition function of two-dimensional quantum gravity.
This correspondence holds also for various correlators
and is extended to other models. In particular, the double scaling
limit of MQM, which we describe in the next chapter,
gives 2D string theory.

The fact that nothing depends on the particular form of the potential,
which is equivalent to the independence of the type of polygons used
to discretize surfaces (triangles, quadrangles, \etc), is known as
{\it universality} of matrix models.
All of them can be splitted into classes of universality.
Each class is associated with some continuum theory
and characterized by the limiting behaviour of a matrix model near
its critical point \cite{Kcrit}.

\newpage

\section{One-matrix model: saddle point approach}

In the following two sections we review two basic methods to solve
matrix models. This will be done relying on explicit examples
of the simplest matrix models, 1MM and 2MM.
This section deals with the first model, which is defined by
the integral over one hermitian matrix
\be
Z=\int dM\, \exp\[ -N \tr V(M)\] .
\plabel{mone}
\ee
The potential was given in \Ref{mpot}.
Since the matrix is hermitian, its independent matrix elements
are $M_{ij}$ with $i\le j$ and the diagonal elements are real.
Therefore, the measure $dM$ is given by
\be
dM=\prod\limits_{i} dM_{ii}\prod\limits_{i<j} d\Re M_{ij}\, d\Im M_{ij}.
\plabel{mesone}
\ee

\subsection{Reduction to eigenvalues}

Each hermitian matrix can be diagonalized by a unitary transformation
\be
M=\Omega^{\dagger}\eig\Omega, \qquad
\eig=\diag(\eig_1,\dots,\eig_N),\ \Omega^{\dagger}\Omega=I.
\plabel{mdiag}
\ee
Therefore, one can change variables from the matrix elements
$M_{ij}$ to the eigenvalues $\eig_k$ and elements of the unitary matrix
$\Omega$ diagonalizing $M$. This change produces a Jacobian.
To find it, one considers a hermitian matrix which is obtained
by an infinitesimal unitary transformation $\Omega=I+d\omega$
of the diagonal matrix $\eig$, where $d\omega$ is antisymmetric. 
In the first order in $\omega$, one
obtains
\be
dM_{ij}\approx \delta_{ij}d\eig_j+[\eig,d\omega]_{ij}=
\delta_{ij}d\eig_j+(\eig_i-\eig_j)d\omega_{ij}.
\plabel{untr}
\ee
This leads to the following result
\be
dM=[d\Omega]_{SU(N)}\prod\limits_{k=1}^N d\eig_k \, \Delta^2(x),
\plabel{mesur}
\ee
where $[d\Omega]_{SU(N)}$ is the Haar measure on SU(N) and
\be
\Delta(x)=\prod\limits_{i<j}(\eig_i-\eig_j)=
\mathop{\det}\limits_{i,j}\eig_i^{j-1}
\plabel{Vand}
\ee
is the Vandermonde determinant.

Due to the U(N)-invariance of the potential, after
the substitution \Ref{mdiag} into the integral \Ref{mone}
the unitary matrix $\Omega$ decouples and can be integrated out.
As a result, one arrives at the following representation
\be
Z={\rm Vol}(SU(N))
\int \prod\limits_{k=1}^N d\eig_k \,  \Delta^2(x)
\exp\[ -N \sum\limits_{k=1}^N V(\eig_k)\] .
\plabel{moned}
\ee
The volume of the SU(N) group is a constant depending only
on the matrix size $N$. It is not relevant for the statistics of
eigenvalues, although it is important for the dependence of the
free energy on $N$.

The representation \Ref{moned} is very important because it reduces
the problem involving $N^2$ degrees of freedom to the problem of
only $N$ eigenvalues. This can be considered as a ``generalized
integrability'': models allowing such reduction are in a sense
``integrable''.

\subsection{Saddle point equation}

The integral \Ref{moned} can also be presented in the form
\be
Z=\int \prod\limits_{k=1}^N d\eig_k \,
e^{ -N \CE }, \qquad
\CE=\sum\limits_{k=1}^N V(\eig_k)
-{2\over N}\sum\limits_{i<j}\log|\eig_i-\eig_j|,
\plabel{intcul}
\ee
where we omitted the irrelevant constant factor.
It describes a system of $N$ particles interacting by the
two-dimensional repulsive Coulomb law in the common potential $V(\eig)$.
In the limit $N\to \infty$, one can apply the usual saddle point
method to evaluate the integral \Ref{intcul} \cite{BIPZ}. It says that
the main contribution comes from configurations of the eigenvalues
satisfying the classical equations of motion $\p \CE/\p\eig_k=0$.
Thus, one obtains the following system of $N$ algebraic equations
\be
V'(\eig_k)={2\over N}\sum\limits_{j\ne k}{1\over \eig_k-\eig_j}.
\plabel{sadeq}
\ee

If we neglected the Coulomb force, all eigenvalues would sit at the minima
of the potential $V(x)$. Due to the Coulomb repulsion they are
spread around these minima and fill some finite intervals
as shown in fig. \ref{filcut}.
In the large $N$ limit their distribution is characterized 
by the density function defined as follows
\be
\rho(\eig)={1\over N}\langl\tr \delta(\eig-M) \rangl.
\plabel{dens}
\ee
The density contains an important information about the system. 
For example, the spherical limit of the free energy
$F_0=\lim\limits_{N\to \infty}N^{-2}\log Z$
is given by
\be
F_0=-\int d\eig \, \rho(\eig) V(\eig)
+\tint d\eig dy\, \rho(\eig)\rho(y)\log|\eig-y|.
\plabel{fre}
\ee

\lfig{In the large $N$ limit the eigenvalues fill finite intervals
around the minima of the potential.}
{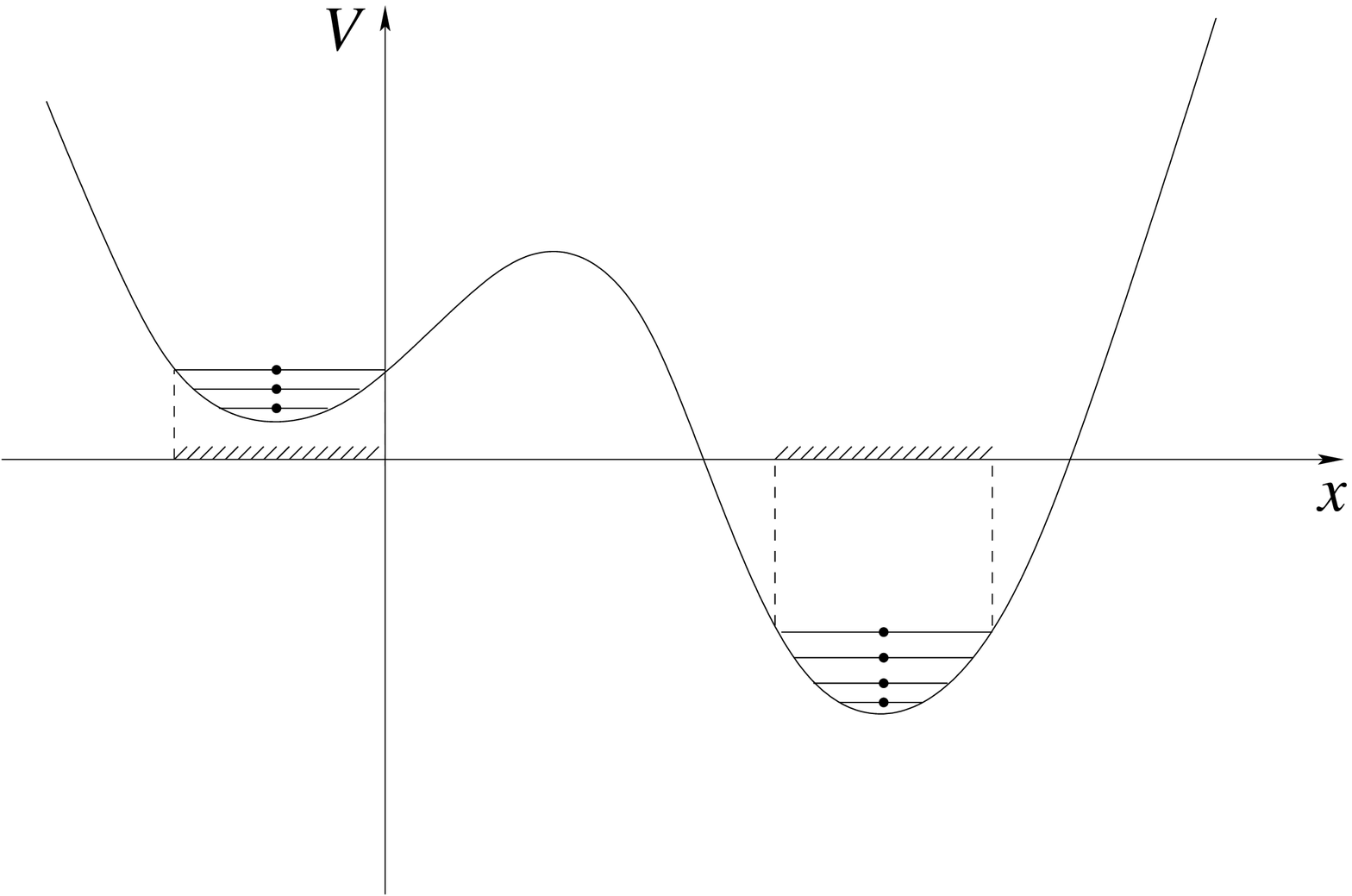}{7cm}{filcut}

The system of equations \Ref{sadeq} can be also rewritten as one integral
equation for $\rho(\eig)$
\be
V'(\eig)=2{\CP.v.}\int{\rho(y)\over \eig-y}dy,
\plabel{sadeqq}
\ee
where ${\CP.v.}$ indicates the principal value of the integral.
To solve this equation, we introduce the following function
\be
\omega (z) =\int {\rho(\eig)\over z-\eig}d\eig.
\plabel{resolv}
\ee
Substitution of the definition \Ref{dens} shows that it is
the resolvent of the matrix $M$
\be
\omega(z)={1\over N}\langl\tr {1\over z-M} \rangl.
\plabel{resl}
\ee
It is an analytical function on the whole complex plane
except the intervals filled by the eigenvalues. At these intervals it has
a discontinuity given by the density $\rho(\eig)$
\be
\omega(\eig+i0)-\omega(\eig-i0)=-2\pi i\rho(\eig),
\qquad \eig\in {\rm sup}[\rho].
\plabel{discon}
\ee
On the other hand, the real part of the resolvent on the support of
$\rho(\eig)$ coincides with the principal value integral as in
\Ref{sadeqq}. Thus, one obtains
\be
\omega(\eig+i0)+\omega(\eig-i0)=V'(\eig),
\qquad \eig\in {\rm sup}[\rho].
\plabel{sdeq}
\ee
This equation is already simple enough to be solved explicitly.

\subsection{One cut solution}

In a general case the potential $V(\eig)$ has several minima and
all of them can be filled by the eigenvalues. It means that
the support of the density $\rho(\eig)$ will have several disconnected
components.
To understand the structure of the solution, let us consider
the case when the support consists of only one interval $(a,b)$.
Then the resolvent $\omega(z)$ should be an analytical function on
the complex plane with one cut along this interval.
On this cut the equations \Ref{discon} and \Ref{sdeq} must hold.
They are sufficient to fix the general form of $\omega(z)$
\be
\omega(z)=\hf\left(V'(z)-P(z)\sqrt{(z-a)(z-b)}\right).
\plabel{grone}
\ee
$P(z)$ is an analytical function which is fixed by the
asymptotic condition
\be
\omega(z)\under{\sim}{z\to\infty} 1/z
\plabel{asres}
\ee
following from \Ref{resolv} and normalization of the density
$\int \rho(x)dx=1$.
If $V(z)$ is a polynomial of degree $n$, $P(z)$ should be a polynomial of
degree $n-2$. In particular, for the case of the Gaussian potential,
it is a constant.
The same asymptotic condition \Ref{asres} fixes the boundaries of the cut,
$a$ and $b$.

The density of eigenvalues is found as the discontinuity of the
resolvent \Ref{grone} on the cut and is given by
\be
\rho(\eig)={P(\eig)\over 2\pi} \sqrt{(\eig-a)(b-\eig)}.
\plabel{denone}
\ee
If the potential is quadratic, then $P(\eig)=const$ and one obtains the
famous semi-circle law of Wigner for the distribution of
eigenvalues of random matrices in the Gaussian ensemble.

The free energy is obtained by substitution of \Ref{denone} into
\Ref{fre}. Note that
the second term is one half of the first one due to \Ref{sadeqq}.
Therefore, one has
\be
F_0=-\hf \int_a^b d\eig \, \rho(\eig) V(\eig).
\plabel{freone}
\ee

\subsection{Critical behaviour}
\label{critomm}

The result \Ref{denone} shows that near the end of the support
the density of eigenvalues does not depend on the potential
and always behaves as a square root. This is a manifestation
of {\it universality} of matrix models mentioned in the end
of section \ref{sdissurf}.

However, there are degenerate cases when this behaviour is violated.
This can happen if the polynomial $P(\eig)$ vanishes at $\eig=a$ or
$\eig=b$. For example, if $a$ is a root of $P(\eig)$ of degree $m$,
the density behaves as $\rho(\eig)\sim(\eig-a)^{m+1/2}$ near this point.
It is clear that this situation is realized only for special
values of the coupling constants $g_k$. They correspond to
the critical points of the free energy discussed in connection
with the continuum limit in paragraph \ref{conlim}.
Indeed, one can show that near these configurations the
spherical free energy \Ref{freone} looks as \Ref{aspf} with $\chi=2$
and $\str=-1/(m+1)$ \cite{Kcrit,Stcrit}.

Thus, each critical point with a given $m$ defines a class of
universality. All these classes correspond to continuum theories which
are associated with a special discrete series of CFTs living on dynamical
surfaces. This discrete series is a part of the so called {\it minimal
conformal theories} which possess only finite number of primary fields
\cite{BPZ}.
The minimal CFTs are characterized by two relatively prime integers
$p$ and $q$ with the following central charge and string susceptibility
\be
c=1-6{(p-q)^2\over pq}, \qquad
\str=-{2\over p+q-1}.
\plabel{mincc}
\ee
Our case is obtained when $p=2m+1$, $q=2$. Thus, for $m=1$ the central
charge vanishes and we describe the pure two-dimensional quantum gravity.
The critical points with other $m$ describe
the two-dimensional quantum gravity coupled to the matter characterized by
the rational central charge $c=1-3{(2m-1)^2\over 2m+1}$.

\lfig{Critical point in one-matrix model.}
{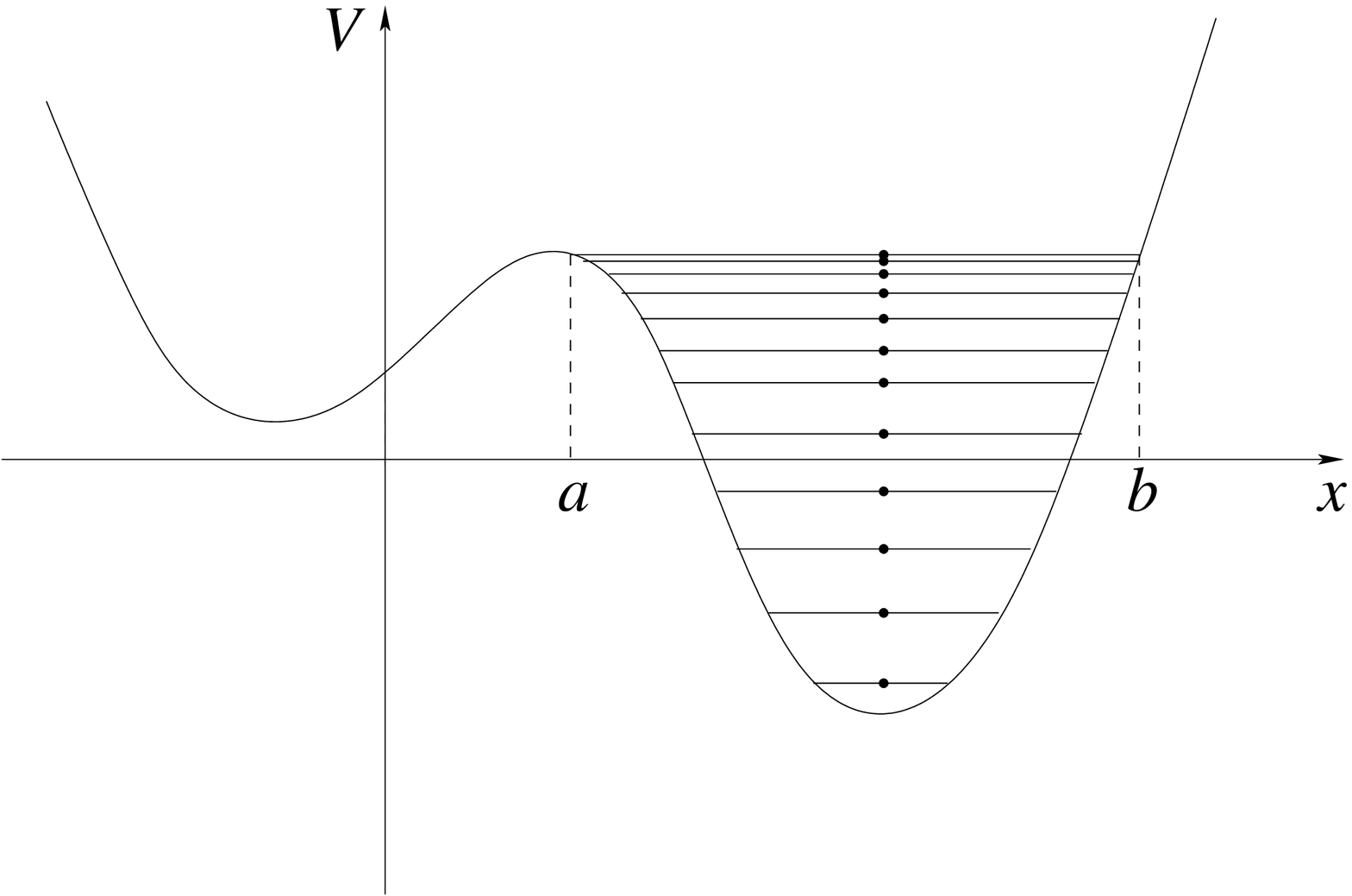}{7cm}{critbeh}

From the matrix model point of view, there is a clear interpretation
for the appearance of the critical points.
Usually, in the large $N$ limit the matrix eigenvalues fill consecutively
the lowest energy levels in the well around a minimum of the 
effective potential. The critical behaviour arises 
when the highest filled level reaches an extremum
of the effective potential.
Generically, this happens as shown in fig. \ref{critbeh}.
By fine tuning of the coupling constants, one can get more special
configurations which will describe the critical behaviour with $m>1$.

\subsection{General solution and complex curve}

So far we considered the case
where the density is concentrated on one interval.
The general form of the solution can be obtained if we note that
equation \Ref{sdeq} can be rewritten as the following equation
valid in the whole complex plane
\be
y^2(z)=\tilde Q(\z), \qquad y=\omega(z)-\hf V'(z),
\plabel{sdecp}
\ee
where $\tilde Q(z)$ is some analytical function. 
Its form is fixed by
the condition \Ref{asres} which leads to $\tilde Q(z)=V'^2(z)+Q(z)$,
where $Q(z)$ is a polynomial of degree $n-2$. Thus, the solution reads
\be
y(z)=\sqrt{V'^2(z)+Q(
z)}.
\plabel{sdsol}
\ee
In a general case, the polynomial $\tilde Q(z)$ has $2(n-1)$ roots
so that the solution is
\be
y(z)=\sqrt{\prod\limits_{k=1}^{n-1}(z-a_k)(z-b_k)},
\plabel{sdsoll}
\ee
where we imply the ordering $a_1<b_1<a_2<\dots<b_{n-1}$.
The intervals $(a_k,b_k)$ represent the support of the density.
When $a_k$ coincides with $b_{k}$, the corresponding interval collapses
and we get a factor $(z-a_k)$ in front of the square root.
When $n-2$ intervals collapse, we return to the one cut solution
\Ref{grone}.
Note that at the formal level the eigenvalues can appear
around each extremum, not only around minima, of the potential.
However, the condensation of the eigenvalues around maxima is not physical
and cannot be realized as a stable configuration.

Thus, we see that the solution of 1MM in the large $N$ limit is completely
determined by the resolvent $\omega(z)$ whose general structure is given
in \Ref{sdsol}.
It is an analytical function with at most $n-1$ cuts having the square
root structure.
Therefore one can consider a Riemann surface associated with this function.
It consists from two sheets joint by all cuts (fig. \ref{rsur}).
Similarly, it can be viewed as a genus $n_c$ complex curve
where $n_c$ is the number of cuts.

\lfig{Riemann surface associated with the solution of 1MM.}
{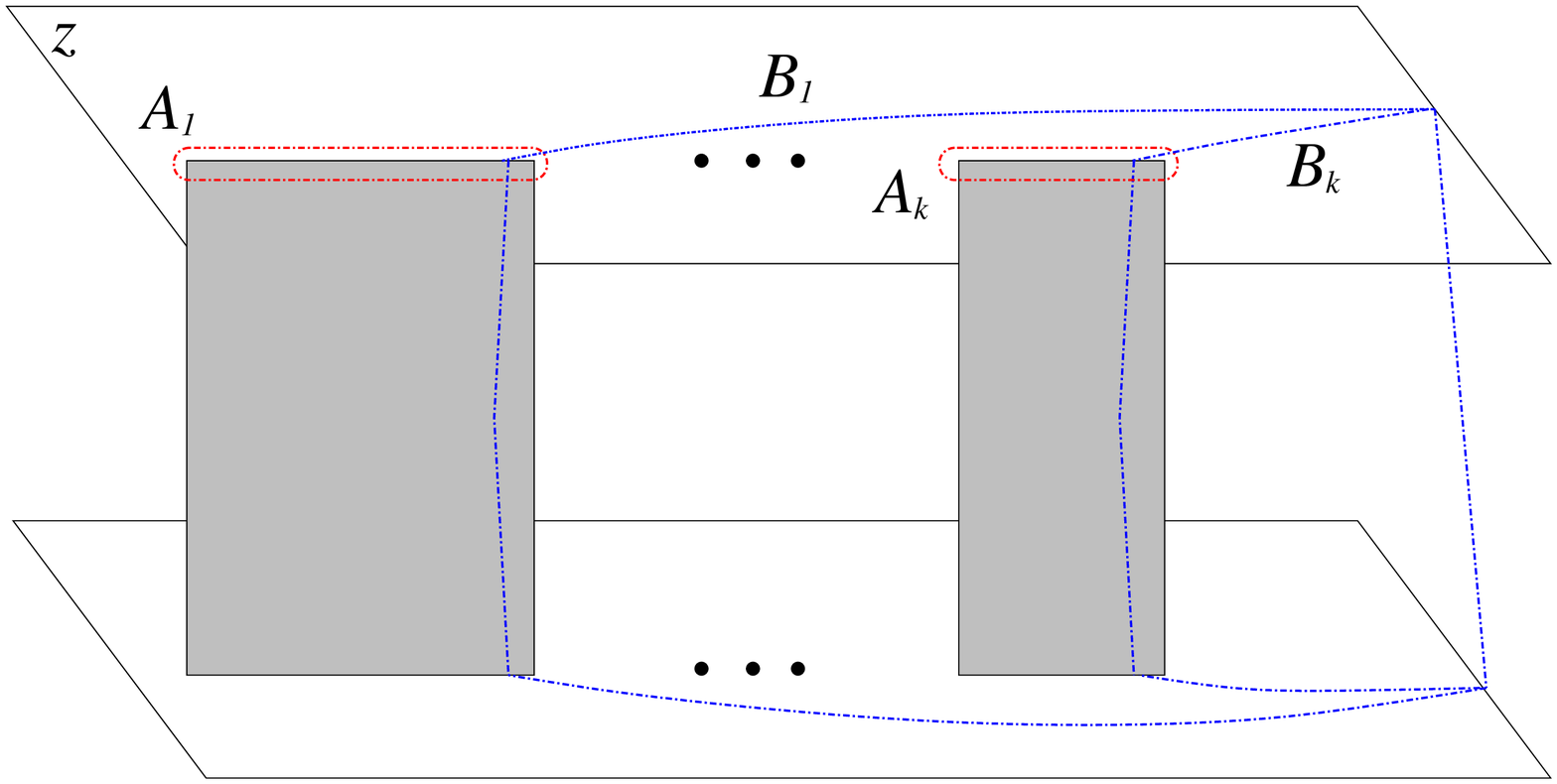}{11cm}{rsur}

On such curve there are $2n_c$ independent cycles.
$n_c$ compact cycles $A_k$ go around cuts and $n_c$ non-compact
cycles $B_k$ join cuts with infinity. The integrals of a holomorphic
differential along these cycles can be considered as the moduli of the curve.
In our case the role of such differential is played by $y(z)dz$.
From the definition \Ref{sdecp} it is clear that the integrals along
the cycles $A_k$ give the relative numbers of eigenvalues in each cut
\be
{1\over 2\pi i} \oint_{A_k} y(z)dz=
\int_{a_k}^{b_k}\rho(\eig)d\eig \stackrel{\rm def}{=} n_k.
\plabel{cycla}
\ee
By definition $\sum_{k=1}^{n_c}n_k=1$. The integrals along the cycles
$B_k$ can also be calculated and are given by the derivatives of
the free energy \cite{DAVID,DavidEy}
\be
 \int_{B_k} y(z)dz={\p F_0\over \p n_k}.
\plabel{cyclb}
\ee
Thus, the solution of the one-matrix model in the large $N$ limit
is encoded in the complex
structure of the Riemann surface associated with the resolvent of
the matrix.

\newpage

\section{Two-matrix model: method of orthogonal polynomials}
\label{twomm}

In this section we consider the two-matrix model \Ref{twomint} which
describes the Ising model on a random lattice.
We restrict ourselves to the simplest case of the potential of
the type \Ref{twompot}.  Thus, we are interested in the
following integral over two hermitian matrices
\be
Z=\int dA\, dB\, \exp\[-N \tr\left(AB-V(A)-\tV(B)\right)\] ,
\plabel{twoint}
\ee
where the potentials $V(A)$ and $\tV(B)$ are some polynomials and
the measures are the same as in \Ref{mesone}.
We will solve this model by the method of orthogonal polynomials.
This approach can be also applied to 1MM where it looks even simpler.
However, we would like to illustrate the basic features of 2MM
and the technique of orthogonal polynomials is quite convenient for
this.

\subsection{Reduction to eigenvalues}

Similarly to the one-matrix case, one can diagonalize the matrices
and rewrite the integral in terms of their eigenvalues.
However, now one has to use two unitary matrices,
which are in general different, for the diagonalization
\be
A=\Omega_A^{\dagger}x\Omega_A, \qquad
B=\Omega_B^{\dagger}y\Omega_B.
\plabel{twodig}
\ee
At the same time, the action in
\Ref{twoint} is invariant only under the common unitary transformation
\Ref{twomtrun}. Therefore, only one of the two unitary matrices
is canceled. As a result, one arrives at the following representation
\be
Z=C_N\int \prod\limits_{k=1}^N dx_k\, dy_k\,
e^{N\left(V(x_k)+\tV(y_k)\right)}
\Delta^2(x) \Delta^2(y)  I(x,y),
\plabel{twointoo}
\ee
where
\be
I(x,y)=\int [d\Omega]_{SU(N)}
\exp\[ -N \tr\left(\Omega^{\dagger}x\Omega y\right)\]
\plabel{IZint}
\ee
and $\Omega=\Omega_A\Omega_B^{\dagger}$. The integral \Ref{IZint}
is known as Itzykson--Zuber--Charish-Chandra 
integral and can be calculated explicitly
\cite{IZ}
\be
I(x,y)=\tilde C_N {\det e^{-N x_i y_j}\over \Delta(x)\Delta(y)},
\plabel{IZin}
\ee
where $\tilde C_N$ is some constant. Substitution of this result into
\Ref{twointoo} leads to the cancellation of a half of the Vandermonde
determinants. The remaining determinants make the integration measure
antisymmetric under permutations $x_k$ and $y_k$. Due to this
antisymmetry, the determinant $\det e^{Nx_i y_j}$ can be replaced
by the the product of diagonal terms. The final result reads
\be
Z=C'_N\int \prod\limits_{k=1}^N d\mu(x_k, y_k)\,
\Delta(x) \Delta(y),
\plabel{twointo}
\ee
where we introduced the measure
\be
d\mu(x,y)=dx\, dy\, e^{-N\left(xy-V(x)-\tV(y)\right)}.
\plabel{twome}
\ee

\subsection{Orthogonal polynomials}

Let us introduce the system of polynomials orthogonal with respect
to the measure \Ref{twome}
\be
\int d\mu(x,y) \Phx_n(x)\Phy_m(y)=\delta_{nm}.
\plabel{orpolcon}
\ee
It is easy to check that they are given by the following expressions
\beq
\Phx_n(x)&=&{1\over n!\sqrt{h_n}}\int \prod\limits_{k=1}^n
{d\mu(x_k,y_k)\over h_{k-1}}\Delta(x)\Delta(y)
\prod\limits_{k=1}^n (x-x_k),
\plabel{orpol}
\\
\Phy_n(y)&=&{1\over n!\sqrt{h_n}}\int \prod\limits_{k=1}^n
{d\mu(x_k,y_k)\over h_{k-1}}\Delta(x)\Delta(y)
\prod\limits_{k=1}^n (y-y_k),
\plabel{orpolt}
\eeq
where the coefficients $h_n$ are fixed by
the normalization condition in \Ref{orpolcon}.
They can be calculated recursively by the relation
\be
h_n={1\over (n+1)!}\(\prod\limits_{k=1}^{n} h_{k-1}\)^{-1}
\int {\prod\limits_{k=1}^{n+1}
d\mu(x_k,y_k)}\Delta(x)\Delta(y).
\plabel{hn}
\ee
Due to this, the general form of the polynomials is the following
\beq
\Phx_n(x)&=&{1\over \sqrt{h_n}}x^n+\sum\limits_{k=0}^{n-1}c_{n,k}x^k,
\plabel{orpl}
\\
\Phy_n(y)&=&{1\over \sqrt{h_n}}y^n+\sum\limits_{k=0}^{n-1}d_{n,k}y^k.
\plabel{orplt}
\eeq

Using these polynomials, one can rewrite the partition function
\Ref{twointo}. Indeed, due to the antisymmetry the Vandermonde
determinants \Ref{Vand} can be replaced by determinants of the orthogonal
polynomials multiplied by the product of the normalization coefficients.
Then one can apply the orthonormality relation \Ref{orpolcon} so that
\be
Z=C'_N \(\prod\limits_{k=0}^{N-1} h_k\)
\int \prod\limits_{k=1}^N d\mu(x_k, y_k)\,
\mathop{\det}\limits_{ij}\left(\Phx_{j-1}(x_i)\right)
\mathop{\det}\limits_{ij}\left(\Phy_{j-1}(y_i)\right)=
C'_N N!\prod\limits_{k=0}^{N-1} h_k.
\plabel{intpl}
\ee
Thus, we reduced the problem of calculation of the partition function
to the problem of finding the orthogonal polynomials and their
normalization coefficients.

\subsection{Recursion relations}

To find the coefficients $h_n$, $c_{n,k}$ and $d_{n,k}$ of
the orthogonal polynomials, one uses recursions relations
which can be obtained for $\Phx_n$ and $\Phy_n$.
They are derived using two pairs of conjugated operators which are
the operators of multiplication and derivative \cite{MMDouglas}.
We introduce them as the following matrices
representing these operators in the basis of
the orthogonal polynomials
\beq
& x\Phx_n(x)=\sum\limits_{m} \Lx_{nm}\Phx_m(x),
\qquad
{1\over N}{\p\over \p  x}\Phx_n(x)=\sum\limits_{m} \Px_{nm}\Phx_m(x), &
\plabel{deflax}\\
& y\Phy_n(y)=\sum\limits_{m}  \Phy_m(y)\Ly_{mn},
\qquad
{1\over N}{\p\over \p  y}\Phy_n(y)=\sum\limits_{m} \Phy_m(y)\Py_{mn}. &
\plabel{deflaxy}
\eeq
Integrating by parts, one finds the following relations
\be
\Px_{nm}=\Ly_{nm}-[V'(\Lx)]_{nm}, \qquad
\Py_{nm}=\Lx_{nm}-[\tV'(\Ly)]_{nm}.
\plabel{relLP}
\ee
If one takes the potentials $V(x)$ and $\tV(y)$ of degree $p$ and $q$,
correspondingly, the form of the orthogonal polynomials \Ref{orpl},
\Ref{orplt} and relations \Ref{relLP} imply the following properties
\bea{c}{proplax}
\Lx_{n,n+1}=\Ly_{n+1,n}=\sqrt{h_{n+1}/h_{n}},
\\
\Lx_{nm}=0, \quad m>n+1 {\rm \ and\ } m<n-q+1,
\\
\Ly_{mn}=0, \quad m>n+1 {\rm \ and\ } m<n-p+1,
\\
\Px_{n,n-1}=\Py_{n-1,n}={n\over N}\sqrt{h_{n-1}/h_{n}},
\\
\Px_{nm}=\Py_{nm}=0, \quad m>n-1 {\rm \ and\ } m<n-(p-1)(q-1).
\eea
They indicate that it is more convenient to work with redefined indices
\be
\Lx_k(n/N)=\Lx_{n,n-k}, \qquad \Ly_k(n/N)=\Ly_{n-k,n}
\plabel{redind}
\ee
and similarly for $\Px$ and $\Py$.
Then each set of functions can be organized into one operator
as follows
\beq
&\hat\Lx(s)=\sum\limits_{k=-1}^{q-1}\Lx_k(s)\ho^{-k}, \qquad
\hat\Px(s)=\sum\limits_{k=1}^{(p-1)(q-1)}\Px_k(s)\ho^{-k}, &
\plabel{xlax} \\
&\hat\Ly(s)=\sum\limits_{k=-1}^{p-1}\ho^k\Ly_k(s), \qquad
\hat\Py(s)=\sum\limits_{k=1}^{(p-1)(q-1)}\ho^k\Py_k(s), &
\plabel{ylax}
\eeq
where we denoted $s=n/N$ and introduced the shift operator
$\ho=e^{{1\over N}{\p\over\p s}}$.
These operators satisfy
\beq
& \hat\Lx(n/N)\Phx_n(x)=x\Phx_n(x), \qquad
\hat\Px(n/N)\Phx_n(x)={1\over N}{\p\over \p x}\Phx_n(x), &
\plabel{Lortx} \\
& \hat\Ly^{\dagger}(n/N)\Phy_{n}(x)=y\Phy_{n}(y), \qquad
\hat\Py^{\dagger}(n/N)\Phy_n(y)={1\over N}{\p\over \p y}\Phy_n(y), &
\plabel{Lorty}
\eeq
where the conjugation is defined as $(\ho a(s))^{\dagger}=a(s)\ho^{-1}$.
With these definitions
the relations \Ref{relLP} become
\be
\hat\Px(s)=\hat\Ly(s)-V'(\hat\Lx(s)), \qquad
\hat\Py(s)=\hat\Lx(s)-\tV'(\hat\Ly(s)).
\plabel{relLPop}
\ee

Substitution of the expansions \Ref{xlax} and \Ref{ylax}
into \Ref{relLPop} gives a system of finite-difference
algebraic equations, which are obtained
comparing the coefficients in front of powers of $\ho$.
Actually, one can restrict the attention only to negative powers.
Then the reduced system contains only equations on
the functions $\Lx_k$ and $\Ly_k$, because the left hand side of
\Ref{relLPop} is expanded only in positive powers of $\ho$.
This is a triangular system and for each given potential it can be
solved by a recursive procedure.
The free energy can be reproduced from the function
$R_{n+1}\stackrel{\rm def}{=}\Lx_{-1}^2(n/N)=\Ly_{-1}^2(n/N)$.
The representation \Ref{intpl} implies
\be
Z=C'_N N!\prod\limits_{k=0}^{N-1} R_k^{N-k}.
\plabel{pfR}
\ee
This gives the solution for all genera.

In fact, to find the solution of \Ref{relLPop},
one has to use the perturbative expansion in $1/N$.
Then the problem is reduced to a hierarchy of differential equations.
The hierarchy appearing here is Toda lattice hierarchy which
will be described in detail in the next section.
The use of methods of Toda theory simplifies the problem
and allows to find explicit differential, and even algebraic equations
directly for the free energy.

\subsection{Critical behaviour}

We saw in section \ref{critomm} that the multicritical points of
the one-matrix model correspond to a one-parameter family of the
minimal conformal theories.
The two-matrix model is more general than the one-matrix model.
Therefore, it can encompass a larger class of continuous models.
It turns out that all minimal models \Ref{mincc} can be obtained by
appropriately adjusting the matrix model potentials \cite{MMDouglas,DKK}.

There is an infinite set of critical potentials for each $(p,q)$ point.
The simplest one is when one of the potentials has degree $p$ and the other
has degree $q$. Their explicit form has been constructed in \cite{DKK}.
The key fact of the construction is that
the momentum function  $\Px(\o,s)$
(up to an analytical piece) is identified with the resolvent \Ref{resl}
of the corresponding matrix $\Lx$.
Comparing with the one-matrix case, one concludes that the $(p,q)$ point
is obtained when the resolvent behaves near its singularity
as $(x-x_c)^{p/q}$. Thus, it is sufficient to take the operators
with the following behaviour at $\o\to 1$
\be
X-X_c\sim (\log\o)^q, \qquad P-P_c\sim (\log\o)^p.
\plabel{crbh}
\ee
Note that the singular asymptotics of $\Ly$ and $\Py$ follow from
\Ref{relLPop} and lead to the dual $(q,p)$ point.
Then one can take some fixed $\Lx(\w)$ and $\Ly(\w)$ with the necessary
asymptotics at $\o\to 1$ and $\o\to \infty$ and solve the equations
\Ref{relLPop} with respect to the potentials $V(X)$ and $\tV(Y)$.
The resulting explicit formulae can be found in \cite{DKK}.

\subsection{Complex curve}
\label{cctwo}

As in the one-matrix model, the solution of 2MM in the large $N$ limit
can be represented in terms of a complex curve \cite{kkvwz,KM}.
However, there is a difference between these two cases.
Whereas in the former case the curve coincides with the Riemann surface
of the resolvent, in the latter case the origin of the curve is different.
To illuminate it, let us consider how the solution of the model in the large
$N$ limit arises.

In this approximation one can apply the saddle point approach described
in the previous section. It leads to the following
two equations on the resolvents of matrices $X$ and $Y$
\be
y=V'(x)+\o(x), \qquad x=\tV'(y)+\tom(y).
\plabel{eqtmm}
\ee
In fact, these equations are nothing else but the classical limit
of the relations \Ref{relLPop} obtained using the orthogonal polynomials.
They coincide due to the identification mentioned above of
the momentum operators $\Px$ and $\Py$ with the resolvents $\o$ and $\tom$,
respectively.

The equations \Ref{eqtmm} can be considered as definitions of
the multivalued analytical functions
$y(x)$ and $x(y)$. It is clear that they must be mutually inverse.
This is actually a non-trivial restriction which, together with 
the asymptotic condition \Ref{asres}, fixes the resolvents
and gives the solution. 
The complex curve, one should deal with, is the Riemann surface
of one of these functions.

The general structure of this complex curve was studied in \cite{KM}.
It was established that if the potentials are of degree $n+1$
and $\tilde n+1$, the maximum genus of the curve is $n\tilde n -1$.
In this most general case
the Riemann surface is represented by $n+1$ sheets. One of them is the
``physical sheet'' glued with each ``unphysical'' one along $\tilde n$ cuts
and all ``unphysical sheets'' join at infinity by the
$n$th order branch point (see fig. \ref{curven}).

\lfig{Generic curve of 2MM as a cover of $x$-plane. Each fat line
consists from $\tilde n$ cuts.}
{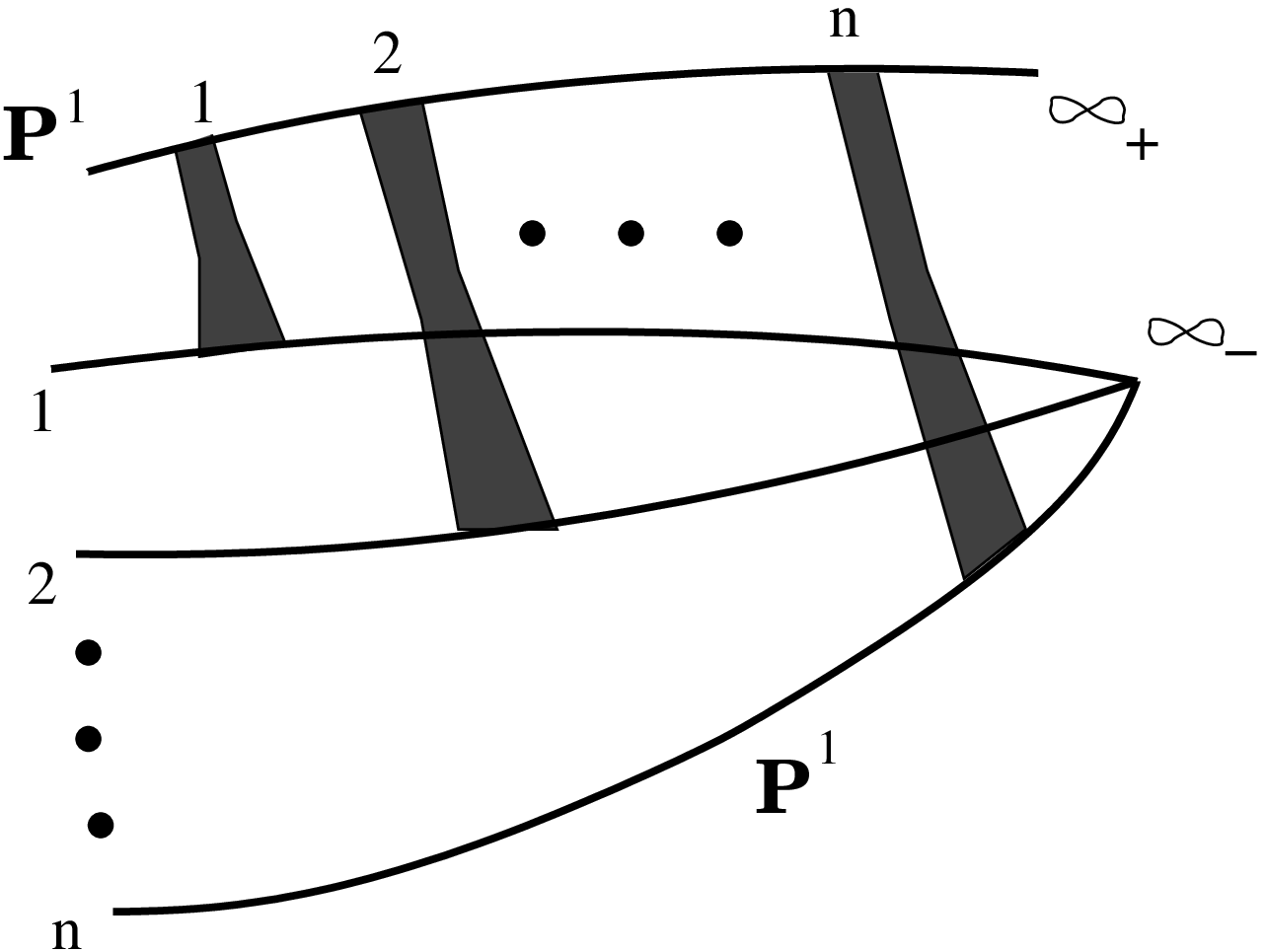}{6cm}{curven}

One can prove the analog of the formulae \Ref{cycla} and \Ref{cyclb}
\cite{KM}. We still integrate around two conjugated sets of
independent cycles $A_k$ and $B_k$ on the curve. The role of
the holomorphic differential is played again by $y(z)dz$ where
$y(x)$ is a solution of \Ref{eqtmm}.

\lfig{Eigenvalue plane of 2MM. The eigenvalues fill several spots
which contain information about the density.}
{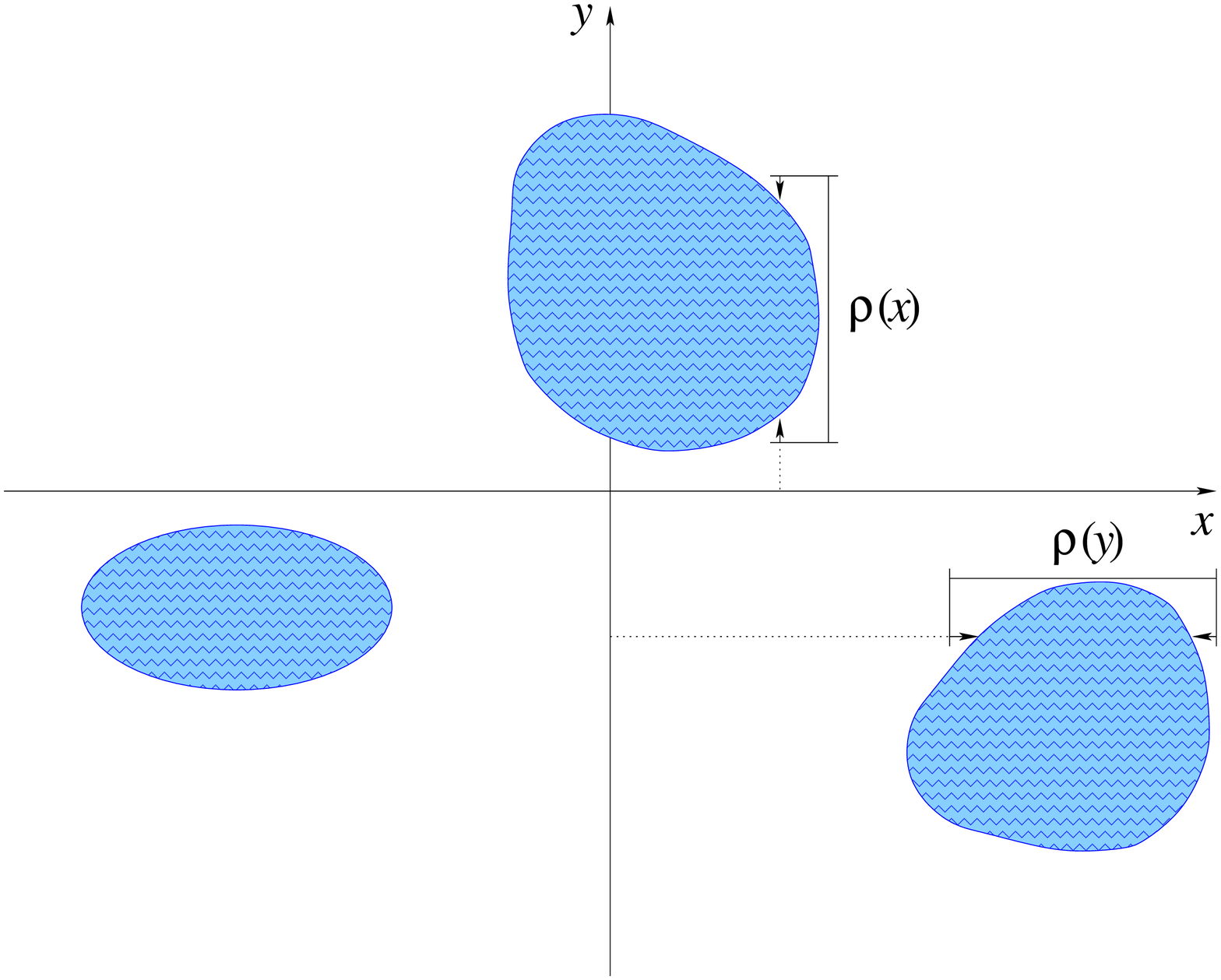}{9cm}{eigtwo}

As earlier, the cycles $A_k$ surround the cuts of the resolvent
on the physical sheet. This is the place where the eigenvalues of
the matrix $X$ live. However, in this picture it is not clear
how to describe the distribution of eigenvalues of $Y$. Of course,
it is enough to invert the function $y(x)$ to find it.
But the arising picture is nevertheless non-symmetric with respect 
to the exchange of $x$ and $y$.

A more symmetric picture can be obtained considering the so called
``double'' \cite{KM}. To introduce this notion, note that since
the two matrices $X$ and $Y$ are of the same size, with each
eigenvalue $x_i$ one can associate an eigenvalue $y_i$ of the second
matrix. Thus, one obtains $N$ pairs $(x_i,y_i)$ which can be put on a
plane. In the large $N$ limit, the eigenvalues are distributed
continuously so that on the plane $(x,y)$ their distribution
appears as several disconnected regions. Each region corresponds
to a cut of the resolvent. Its width in the $y$ direction
at a given point $x$ is nothing else but the density $\rho(x)$
and {\it vice versa}. Thus, one arrives at the two-dimensional picture shown
on fig. \ref{eigtwo}.
From this point of view, the functions $y(x)$ and $x(y)$,
which are solutions of \Ref{eqtmm},
determine the boundaries of the spots of eigenvalues.
The fact that they are inverse means that they define the same
boundary.

Now to define the double, we take two copies of the $(x,y)$ plane,
cut off the spots, and glue the two planes along the boundaries. The resulting
surface shown in fig. \ref{double} is smooth and can be considered as
a genus $n_c-1$ surface with two punctures (corresponding to two infinities)
where $n_c$ is the number of spots on the initial surface equal to the number
of cuts of the resolvent $\o(x)$.

\lfig{Complex curve associated with the solution of 2MM viewed as
a ``double'' of $(x,y)$-plane.}
{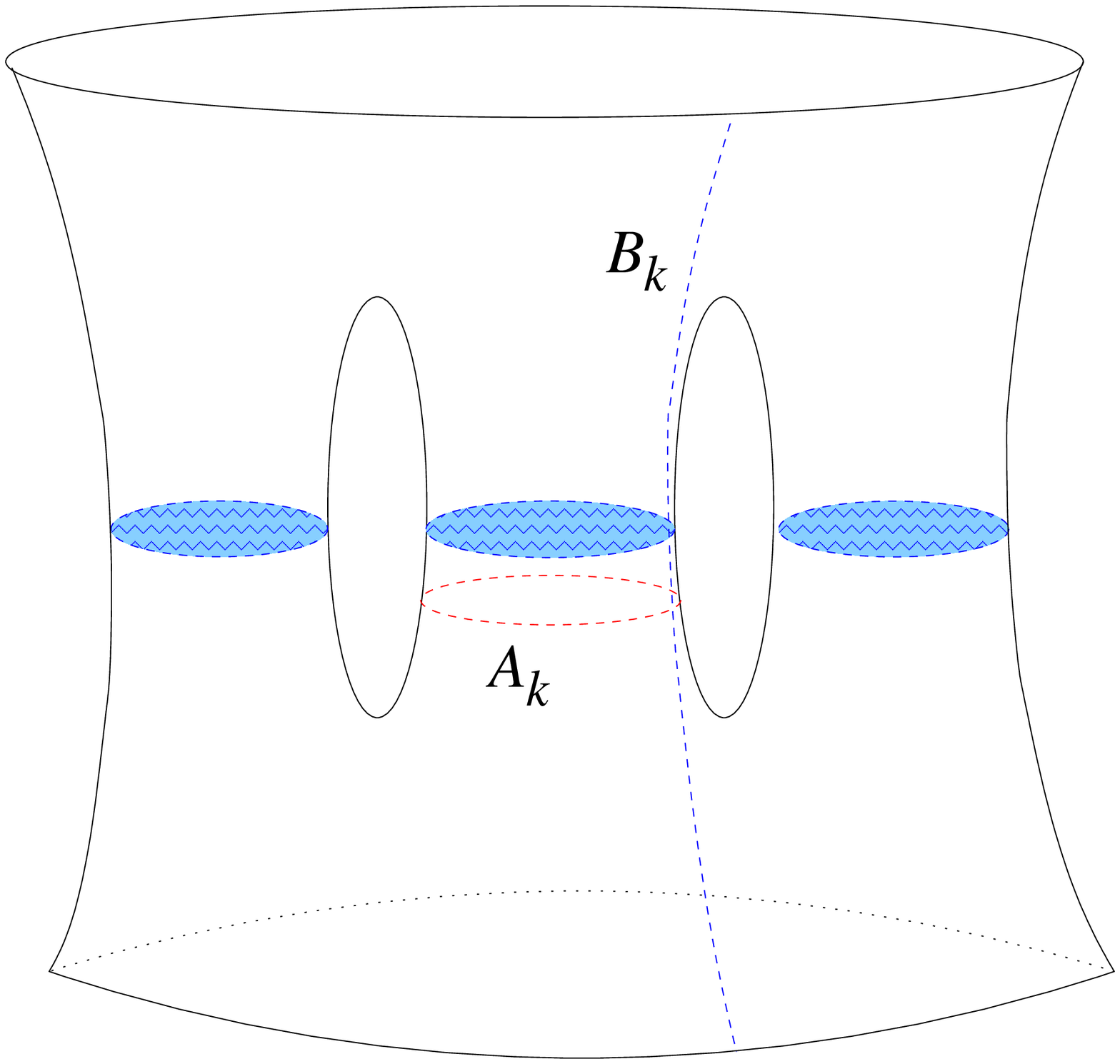}{7cm}{double}

All cycles $A_k$ and $B_k$ exist also in this picture
and the integration formulae \Ref{cycla} and \Ref{cyclb} 
along them hold as well.
Note that the integrals along $A_k$ cycles can be rewritten as
two-dimensional integrals over the $(x,y)$ plane with the density
equal to 1 inside the spots and vanishing outside them.
Such behaviour of the density is characteristic for fermionic systems.
And indeed, 2MM can be interpreted as a system of free fermions.

\subsection{Free fermion representation}
\label{fermm}

Let us identify the functions
\be
\psi_n(x)=\Phx_n(x)e^{NV(x)},
 \qquad
\tilde\psi_n(y)=\Phy_n(y)e^{N\tV(y)}.
\plabel{BAfun}
\ee
with fermionic wave functions.
The functions $\psi_n$ and $\tilde\psi_n$
can be considered as two representations of the same state
similarly to the coordinate and momentum representations.
The two representations are related by a kind of Fourier transform
and the scalar product between functions
of different representations is given by the following integral
\be
\langl \tilde\psi|\psi\rangl=\int dx\,dy\,
e^{-Nxy}\tilde\psi(y) \psi(x).
\plabel{scprod}
\ee

The second-quantized fermionic fields
are defined as
\be
\psi(x)=\sum\limits_{n=0}^{\infty}a_n\psi_n(x),
\qquad
\tilde\psi(y)=\sum\limits_{n=0}^{\infty}a_n^{\dagger}\tilde\psi_n(y).
\plabel{fermf}
\ee
Due to the orthonormality of the wave functions \Ref{BAfun} with respect to
the scalar product \Ref{scprod},
the creation and annihilation operators satisfy the following
anticommutation relations
\be
\{ a_n,a_m^{\dagger} \}=\delta_{n,m}.
\plabel{comab}
\ee
The fundamental state of $N$ fermions is defined by
\be
a_n|N\rangle=0, \quad n\ge N,
\qquad
a_n^{\dagger}|N\rangle=0, \quad n< N.
\plabel{fundst}
\ee
Its wave function can be represented by the Slater determinant.
For example, in the $x$-representation it looks as
\be
\Psi_N(x_1,\dots,x_N)=\mathop{\det}\limits_{i,j}\psi_i(x_j).
\plabel{fundfun}
\ee

The key observation which establishes the equivalence of the two
systems is that the correlators of matrix operators coincide with
expectation values of the corresponding fermionic operators in
the fundamental state \Ref{fundfun}:
\be
\langl \prod\limits_j \tr B^{m_j}\prod\limits_i \tr A^{n_i}\rangl=
\langl N| \prod\limits_j \hat y^{m_j}\prod\limits_i \hat x^{n_i}
|N\rangl.
\plabel{matfer}
\ee
Here a second-quantized operator $\hat \CO (x,y)$ is defined as follows
\be
\hat \CO (x,y)=\int dx\, dy\, e^{-Nxy} \tilde\psi(y)\CO(x,y)\psi(x).
\plabel{secop}
\ee
The proof of the statement \Ref{matfer} relies on the properties of
the orthogonal polynomials. For example, for the one-point correlator
we have
\beq
\langl \tr A^{n}\rangl&\equiv&
Z^{-1}\int dA\, dB\, \tr A^{n}\, e^{- N \tr\left(AB-V(A)-\tV(B)\right)}
\nonumber \\
&=& \left( N!\prod\limits_{k=0}^{N-1} h_k\right)^{-1}
\int \prod\limits_{k=1}^N d\mu(x_k, y_k)\,
\Delta(x) \Delta(y) \sum\limits_{i=1}^N x_i^n
\nonumber \\
&=& \frac{1}{ (N-1)!}
\int \prod\limits_{k=1}^N d\mu(x_k, y_k)\,
\mathop{\det}\limits_{ij}\left(\Phx_{j-1}(x_i)\right)
\mathop{\det}\limits_{ij}\left(\Phy_{j-1}(y_i)\right) x_1^n
\nonumber \\
&=& \sum\limits_{j=0}^{N-1} \int d\mu(x, y)\,
\Phx_{j}(x)\Phy_{j}(y) x^n
=\langl N| \hat x^{n}|N\rangl.
\plabel{matferproof}
\eeq

This equivalence supplies us with a powerful technique
for calculations. For example, all correlators of the type \Ref{matfer}
can be expressed through the two-point function
\be
K_N(x,y)=\langl N|\tilde\psi(y)\psi(x)|N\rangl
\plabel{twopo}
\ee
and it is sufficient to study this quantity. In general, the
existence of the representation in terms of free fermions
indicates that the system is integrable.
This, in turn, is often related to the possibility to introduce
orthogonal polynomials.

We will see that the similar structures appear in matrix quantum
mechanics. Although MQM has much richer physics,
it turns out to be quite similar to 2MM.

\newpage

\section{Toda lattice hierarchy}
\label{TODA}

\subsection{Integrable systems}
\label{intsys}

One and two-matrix models considered above are examples of {\it integrable
systems}. There exists a general theory of such systems.
A system is considered as integrable if it has
an infinite number of commuting Hamiltonians. Each Hamiltonian generates
an evolution along some direction in the parameter space of the model.
Their commutativity means that there is an infinite number of
conserved quantities associated with them and, at least in principle, it
is possible to describe any point in the parameter space.

Usually, the integrability implies the existence of a {\it hierarchy} of
equations on some specific quantities characterizing the system.
The hierarchical structure means that the equations can be solved
one by one, so that the solution of the first equation should be
substituted into the second one and \etc\
This recursive procedure allows to
reproduce all information about the system.

The equations to be solved are most often of the finite-difference type.
In other words, they describe a system on a lattice.
Introducing a parameter measuring the spacing between nodes
of the lattice, one can organize a perturbative expansion in this parameter.
Then the finite-difference equations are replaced by an infinite set
of partial differential equations. They also form a hierarchy
and can be solved recursively. The equations appearing at the first
level, corresponding to the vanishing spacing,
describe a closed system which is considered as a continuum or
classical limit of the initial one.
In turn, starting with the classical system
describing by a hierarchy of differential equations, one can construct
its quantum deformation arriving at the full hierarchy.

The integrable systems can be classified according to the type of
hierarchy which appears in their description. In this section we consider
the so called {\it Toda hierarchy} \cite{UT}.
It is general enough to include
all integrable matrix models relevant for our work. In particular,
it describes 2MM and some restriction of MQM, whereas 1MM
corresponds to its certain reduction.

\subsection{Lax formalism}
\label{laxform}

There are several ways to introduce the Toda hierarchy. The most convenient
for us is to use the so called {\it Lax formalism}.

Take two semi-infinite
series\footnote{In fact, the first coefficients in the expansions
\Ref{Lax} can be chosen arbitrarily. Only their product has a sense
and how it is distributed between the two operators can be considered
as a choice of some gauge.
In particular, often one uses the gauge where one of
the coefficients equals 1 \cite{TakTak}.
We use the symmetric gauge which agrees with the choice of the
orthogonal polynomials \Ref{orpolcon} normalized to the Kronecker symbol
in 2MM.}
\beq
L= r(s)\ho+\sum\limits_{k=0}^{\infty} u_k(s)\ho^{-k},
\qquad
\bL= \ho^{-1}r(s)+\sum\limits_{k=0}^{\infty} \ho^{k}\bu_{k}(s),
\plabel{Lax}
\eeq
where $s$ is a discrete variable labeling the nodes of an infinite
lattice and $\ho=e^{\hb\p/\p s}$ is the shift operator in $s$.
The Planck constant $\hbar$ plays the role of
the spacing parameter. The operators \Ref{Lax} are called
Lax operators.
The coefficients $r$, $u_k$ and $\bu_{k}$ are also functions
of two infinite sets of ``times'' $\{t_{\pm k}\}_{k=1}^{\infty}$.
Each time variable gives rise to an evolution along its direction.
This system represents Toda hierarchy if the evolution associated with each
$t_{\pm k}$ is generated by Hamiltonians $H_{\pm k}$
\bea{rclcrcl}{evolH}
 \hb{\p L\over\p t_k} &=&[H_k,L], &\qquad&
 \hb{\p \bL\over \p t_k} &=&[H_k,\bL], \\
 \hb{\p L\over \p t_{-k}} &=&[H_{-k},L], &\qquad&
 \hb{\p \bL\over \p t_{-k}} &=&[H_{-k},\bL],
\eea
which are expressed through the Lax operators \Ref{Lax} as
follows\footnote{\label{foottn}
Sometimes the second set of Hamiltonians is defined
with the opposite sign. This corresponds to the change of sign of
$t_{-k}$. Doing both these replacements, one can establish
the full correspondence with the works using this sign convention.}
\be
H_k=(L^k)_> +\hf(L^k)_0,\qquad
H_{-k}=(\bL^k)_< +\hf(\bL^k)_0,
\plabel{laxh}
\ee
where the symbol $(\ )_{\stackrel{>}{<}}$ means the positive (negative)
part of the series in the shift operator $\ho$ and $(\ )_0$ denotes
the constant part.
Thus, Toda hierarchy is a collection of non-linear equations of
the finite-difference type in $s$ and differential with respect to $t_k$
for the coefficients $r(s,t)$, $u_k(s,t)$ and $\bu_k(s,t)$.

From the commutativity of the second derivatives, it is easy to obtain
that the Lax--Sato equations \Ref{evolH} are equivalent to
the zero-curvature condition for the Hamiltonians
\be
\hb {\p H_k \over \p t_l}-\hb {\p H_l \over \p t_k}+[H_k,H_l]=0.
\plabel{zerocur}
\ee
It shows that the system possesses an infinite set of commutative flows
$\hb{\p \over \p t_k}-H_k$ and, therefore, Toda hierarchy is integrable.

One can get another equivalent formulation if one considers
the following eigenvalue problem
\be
x\Psi=L\Psi(x;s),\qquad
\hb {\p \Psi \over \p t_k}=H_k\Psi(x;s),\qquad
\hb {\p \Psi \over \p t_{-k}}=H_{-k}\Psi(x;s).
\plabel{eigprob}
\ee
The previous equations \Ref{evolH} and \Ref{zerocur} appear as the
integrability condition for \Ref{eigprob}.
Indeed, differentiating the first equation, one reproduces the evolution
law of the Lax operators \Ref{evolH} and the second and third
equations lead to the zero-curvature condition \Ref{zerocur}.
The eigenfunction $\Psi$ is known as Baker--Akhiezer function.
It is clear that it contains all information about the system.

Note that equations of Toda hierarchy allow a representation in terms of
semi-infinite matrices. Then the Baker--Akhiezer function is
a vector whose elements correspond to different values of the discrete
variable $s$. The positive/negative/constant parts of the series in
$\ho$ are mapped to upper/lower/diagonal triangular parts of matrices.

The equations to be solved are either the equations
\Ref{eigprob} on $\Psi$ or the Lax--Sato equations \Ref{evolH}
on the coefficients of the Lax operators.
Their hierarchic structure is reflected in the fact
that one obtains a closed equation on the first coefficient $r(s,t)$
and its solution provides the necessary information for the following
equations. This first equation is derived considering the Lax--Sato
equations \Ref{evolH} for $k=1$. They give
\beq
& \hb {\p \log r^{2}(s)\over \p t_1}=u_0(s+\hb)-u_0(s), \qquad
 \hb {\p \log r^{2}(s)\over \p t_{-1}}=\bu_0(s)-\bu_0(s+\hb), &
\plabel{prtt}\\
& \hb {\p \bu_0(s)\over \p t_{1}}=r^2(s)-r^2(s-\hb), \qquad
\hb {\p u_0(s)\over \p t_{-1}}=r^2(s-\hb)-r^2(s). &
\eeq
Combining these relations, one finds the so called Toda equation
\be
\hb^2 {\p^2 \log r^{2}(s)\over \p t_1 \p_{-1}}=
2r^{2}(s)-r^{2}(s+\hb)-r^{2}(s-\hb).
\plabel{todaeq}
\ee

Often it is convenient to introduce the following Orlov--Shulman operators
\cite{orlshu}
\bea{rcl}{ORSH}
M&=&\sum\limits_{k\ge 1}kt_k L^k+s+\sum\limits_{k\ge 1}v_k L^{-k}, \\
\bM&=&-\sum\limits_{k\ge 1}kt_{-k} \bL^k+s-\sum\limits_{k\ge 1}v_{-k} \bL^{-k}.
\eea
The coefficients $v_{\pm k}$ are fixed by the condition
on  their commutators with the Lax operators
\be
[L,M]=\hb L,\qquad [\bL,\bM]=-\hb \bL.
\plabel{laxorsh}
\ee
The main application of these operators is that they can be considered
as perturbations of the simple operators of multiplication by
the discrete variable $s$. Indeed, if one requires that $v_{\pm k}$
vanish when all $t_{\pm k}=0$, then in this limit $M=\bM=s$.
Similarly the Lax operators reduce to the shift operator.
The perturbation leading to the general expansions \Ref{Lax} and
\Ref{ORSH} can be described by the dressing operators $\CW$ and $\bCW$
\bea{ccc}{dreslax}
L=\CW \ho \CW^{-1}, &\qquad &
 M=\CW s \CW^{-1},
 \\
\bL=\bCW \ho^{-1} \bCW^{-1},&\qquad &
\bM=\bCW s \bCW^{-1}.
\eea
The commutation relations \Ref{laxorsh} are nothing else but
the dressed version of the evident relation
\be
[\ho,s]=\hb\ho.
\plabel{osos}
\ee

To produce the expansions \Ref{Lax} and \Ref{ORSH}, the dressing operators
should have the following general form
\bea{rcl}{dress}
\CW&=& e^{{1\over 2\hb}\phi}\left( 1+\sum\limits_{k\ge 1}w_k\ho^{-k}\right)
\exp\( {1\over \hb}\sum\limits_{k\ge 1}t_k\ho^k\) ,\\
\bCW&=& e^{-{1\over 2\hb}\phi}\left( 1+\sum\limits_{k\ge 1}\bw_k\ho^{k}\right)
\exp\({1\over \hb}\sum\limits_{k\ge 1}t_{-k}\ho^{-k}\),
\eea
where the zero mode $\phi(s)$ is related to the coefficient $r(s)$
as
\be
r^2(s)=e^{\bh(\phi(s)-\phi(s+\hb))}.
\plabel{phir}
\ee
However, the coefficients in this expansion are not arbitrary and
the dressing operators should be subject of some additional
condition. It can be understood considering evolution along the times
$t_{\pm k}$. Differentiating \Ref{dreslax} with respect to $t_{\pm k}$,
one finds the following expression of the Hamiltonians
in terms of the dressing operators
\bea{c}{HW}
H_k=\hb (\p_{t_k} \CW)\CW^{-1}, \qquad
H_{-k}=\hb(\p_{t_{-k}} \CW)\CW^{-1},
\\
\bH_k=\hb(\p_{t_k} \bCW)\bCW^{-1}, \qquad
\bH_{-k}=\hb(\p_{t_{-k}} \bCW)\bCW^{-1},
\eea
Here $H_{\pm k}$ generate evolution of $L$ and $\bH_{\pm k}$ are
Hamiltonians for $\bL$. However, \Ref{evolH} implies that for both
operators one should use the same Hamiltonian. This imposes the condition
that $H_{\pm k}=\bH_{\pm k}$ which relates two dressing operators.
This condition can be rewritten in a more explicit way.
Namely, it is equivalent to the requirement that $\CW^{-1} \bCW$
does not depend on times $t_{\pm k}$ \cite{UT,Takebe}.

Studying the evolution laws of the Orlov--Shulman operators, one can find
that \cite{TakTak}
\be
{\p v_k\over \p t_l}={\p v_l\over \p t_k}.
\plabel{vttv}
\ee
It means that there exists a generating function $\tau_s[t]$
of all coefficients $v_{\pm k}$
\be
v_k(s)=\hb^2\, {\p \log \tau_s[t]\over \p t_k}.
\plabel{tauvk}
\ee
It is called {\it $\tau$-function} of Toda hierarchy. It also allows
to reproduce the zero mode $\phi$ and, consequently,
the first coefficient in the expansion of the Lax operators
\be
e^{\bh \phi(s)}={\tau_{s}\over \tau_{s+\hb}},
\qquad
r^2(s-\hb)={\tau_{s+\hb}\tau_{s-\hb}\over\tau_s^2}.
\plabel{taur}
\ee
The $\tau$-function is usually the main subject of interest in the systems
described by Toda hierarchy. The reason is that it coincides with
the partition function of the model. Then the coefficients $v_{k}$
are the one-point correlators of the operators generating the
commuting flows $H_{ k}$. We will show a concrete realization
of these ideas in the end of this section and in the next chapters.

\subsection{Free fermion and boson representations}
\label{ferbosrep}

To establish an explicit connection with physical systems, it is sometimes
convenient to use a representation of Toda hierarchy in terms of
second-quantized free chiral fermions or bosons.
The two representations are related
by the usual bosonization procedure.

\subsubsection{Fermionic picture}

To define the fermionic representation, let us consider
the chiral fermionic fields with the following expansion
\be
\psi(z)=\sum\limits_{r\in \Zb+\hf}\psi_r z^{-r-\hf} , \qquad
\psi^*(z)=\sum\limits_{r\in \Zb+\hf}\psi^*_{-r} z^{-r-\hf}
\plabel{expfer}
\ee
Their two-point function
\be
\langle l| \psi(z)\psi^*(z')|l\rangle={(z'/z)^l\over z-z'}
\plabel{twofer}
\ee
leads to the following commutation relations for the modes
\be
\{\psi_r,\psi^*_s\}=\delta_{r,s}.
\plabel{comfer}
\ee
The fermionic vacuum of charge $l$ is defined by
\be
\psi_r\rvacl=0, \quad r>l, \qquad \psi^*_{r}\rvacl=0, \quad r<l.
\plabel{vacfer}
\ee
Also we need to introduce the current
\be
J(z)=\psi^*(z)\psi(z)=
{\hat p} z^{-1} +\sum\limits_{n\ne 0} H_n z^{-n-1}
\plabel{curfer}
\ee
whose components $H_n$ are associated with the Hamiltonians generating
the Toda flows. In terms of the fermionic modes
they are represented as follows
\be
H_n=\sum\limits_{r\in\Zb+\hf}\psi^*_{r-n}\psi_r
\plabel{Hamfer}
\ee
and for any $l$ they satisfy
\be
H_n\rvacl=\lvacl H_{-n}=0, \quad n>0.
\plabel{Hnvac}
\ee
Finally, we introduce an operator of $GL(\infty)$ rotation
\be
\bfg=\exp\left(\bh\sum\limits_{r,s\in \Zb+\hf} A_{rs}\psi_r\psi^*_s\right).
\plabel{glfer}
\ee

With these definitions the $\tau$-function of Toda hierarchy
is represented as the following vacuum expectation value
\be
\tau_{l\hb}[t]=\Lvacl e^{\bh H_+[t]}\bfg e^{-\bh H_-[t]} \Rvacl,
\plabel{taufer}
\ee
where
\be
H_+[t]=\sum\limits_{k> 0} t_k H_k , \qquad
H_-[t]=\sum\limits_{k< 0} t_k H_k.
\plabel{hphm}
\ee
It is clear that each solution of Toda hierarchy is
characterized in the unique way by the choice of the matrix $A_{rs}$.

\subsubsection{Bosonic picture}

The bosonic representation now follows from the bosonization
formulae
\be
\psi(z)=:e^{\vp(z)}:, \qquad
\psi^*(z)=:e^{-\vp(z)}:, \qquad
\p\vp(z)=:\psi^*(z)\psi(z):.
\plabel{bosoniz}
\ee
Since the last expression in \Ref{bosoniz} is the current \Ref{curfer},
the Hamiltonians $H_n$ appear now
as the coefficients in the mode expansion of the free bosonic field
\be
\vp(z)=\hat q+\hat p \log z +\sum\limits_{n\ne 0} {1\over n}H_n z^{-n}.
\plabel{freebos}
\ee
From \Ref{Hamfer} and \Ref{comfer} one finds
the following commutation relations
\be
 [\hat p,\hat q]=1, \qquad [H_n,H_m]=n\delta_{m+n,0},
\plabel{commbos}
\ee
which lead to the two-point function of the free boson
\be
\langle \vp(z)\vp(z')\rangle=\log(z-z').
\plabel{twobos}
\ee
The bosonic vacuum is defined by the Hamiltonians and
characterized by the quantum number
$s$, which is the eigenvalue of the momentum operator,
\be
\hat p\rvacs=s\rvacs, \qquad  H_n\rvacs=0, \quad (n>0).
\plabel{vacbos}
\ee

To rewrite the operator $\bfg$ \Ref{glfer} in the bosonic terms, we
introduce the vertex operators
\be
V_{\alpha}(z)=:e^{\alpha\vp(z)}:.
\plabel{verbos}
\ee
Here the normal ordering is defined by putting all $H_n$, $n>0$ to the right
and $H_n$, $n<0$ to the left.
Besides $:\hat q\hat p:=:\hat p\hat q:=\hat q\hat p$.
Then the operator of $GL(\infty)$ rotation is given by
\be
\bfg=\exp\left(-{1\over 4\pi^2\hb}
\oint dz  \oint dw\, A(z,w) V_{1}(z)V_{-1}(w)\right),
\plabel{glbos}
\ee
where
\be
A(z,w)=\sum\limits_{r,s}A_{rs}z^{r-\hf} w^{-s-\hf}.
\plabel{Anm}
\ee
The substitution of \Ref{glbos} into \Ref{taufer} and replacing the
fermionic vacuum $\rvacl$ by the bosonic one $\rvacs$
gives the bosonic representation of the $\tau$-function $\tau_s[t]$.

\subsubsection{Connection with the Lax formalism}

The relation of these two representations with the objects
considered in paragraph \ref{laxform} is based on the realization
of the Baker--Akhiezer function $\Psi(z;s)$ as
the expectation value of the one-fermion field
\be
\Psi(z;s)=
\tau^{-1}_{s}[t]\langl\hb^{-1}s| e^{\bh H_+[t]}\psi(z)\bfg e^{-\bh H_-[t]}
|\hb^{-1}s\rangl.
\plabel{BAferm}
\ee
One can show that it does satisfy the relations \Ref{eigprob} with
$H_k$ defined as in \Ref{laxh} and the Lax operators having the form
\Ref{Lax}.

\subsection{Hirota equations}

The most explicit manifestation of the hierarchic
structure of the Toda system is a set of equations on the $\tau$-function,
which can be obtained from the fermionic representation introduced above.
One can show \cite{JM} that the ensemble of
the $\tau$-functions of the Toda hierarchy with different charges satisfies
a set of bilinear equations.
They are known as Hirota equations and can be
written in a combined way as follows
\begin{eqnarray}
&\oint_{C_{\infty}} dz\, z^{l-l'}
\exp\left(\bh\sum\limits_{k> 0}(t_k-t'_k)z^k\right)
\tau_{l}[t-\tilde\zeta_+]\tau_{l'}[t'+\tilde\zeta_+] =
& \nonumber \\
&\oint_{C_{0}} dz\, z^{l-l'}
\exp\left(\bh\sum\limits_{k< 0}(t_k-t'_k)z^{k}\right)
\tau_{l+1}[t-\tilde\zeta_-]\tau_{l'-1}[t'+\tilde\zeta_-],
& \plabel{hiro}
\end{eqnarray}
where
\be
\tilde\zeta_{+}/\hb =(\dots,0,0,z^{-1}, z^{-2}/2,z^{-3}/3,\dots),
\qquad
\tilde\zeta_{-}/\hb =(\dots,z^{3}/3, z^{2}/2,z,0,0,\dots)
\ee
and we omitted $\hb$ in the index of the $\tau$-function.
The proof of \Ref{hiro} relies on the representation \Ref{taufer}
of the $\tau$-function with $\bfg$ taken from \Ref{glfer}.
The starting point is the fact that the following operator
\be
\CC=\sum\limits_{r\in\Zb+\hf}\psi^*_r\otimes\psi_r=\oint {dz\over 2\pi i}\,
\psi^*(z)\otimes\psi(z)
\plabel{casfer}
\ee
plays the role of the Casimir operator for the diagonal subgroup of
$GL(\infty)\otimes GL(\infty)$. This means that it commutes with
the tensor product of two $\bfg$ operators
\be
\CC\, (\bfg \otimes \bfg)=(\bfg \otimes \bfg)\, \CC.
\plabel{gcom}
\ee
Multiplying this relation by
$\langle l+1|e^{\bh H_+[t]}\otimes \langle l'-1|e^{\bh H_+[t']}$
from the left and by
$e^{-\bh H_-[t]}|l\rangle\otimes e^{-\bh H_-[t']}|l'\rangle$ from the right,
one can commute the fermion operators until they hit
the left (right) vacuum.
The final result is obtained using the following relations
\beq
& \langle l+1| e^{\bh H_+[t]}\psi^*(z)\bfg e^{-\bh H_-[t]}|l\rangle
=z^l \exp\left( \bh\sum\limits_{n>0}t_n z^n\right)
 \langl l| e^{\bh H_+[t-\tilde\zeta_+]}\bfg e^{-\bh H_-[t]}|l\rangl ,& \\
& \langle l+1| e^{\bh H_+[t]}\bfg\psi^*(z) e^{-\bh H_-[t]}|l\rangle
=z^l \exp\left( \bh\sum\limits_{n<0}t_n z^n\right)
 \langl l+1| e^{\bh H_+[t]}\bfg e^{-\bh H_-[t-\tilde\zeta_-]}|l+1\rangl &
\eeq
together with the similar relations for $\psi(z)$. They can be proven
in two steps. First, one commutes the fermionic fields with the
perturbing operators
\bea{c}{ferham}
e^{-\bh H_\pm[t]}\psi^*(z) e^{\bh H_\pm[t]} =
\exp\(-\bh \sum\limits_{n{>}0}t_{\pm n} z^{\pm n}\) \psi^*(z),
\\
e^{\bh H_{\pm}[t]}\psi(z) e^{-\bh H_{\pm}[t]} =
\exp\(-\bh \sum\limits_{n{>}0}t_{\pm n} z^{\pm n} \) \psi(z).
\eea
After this it remains to show that, for example, $\psi^*(z)|l\rangle=z^l
e^{\bh H_-[\tilde \zeta_-]}|l+1\rangle$. The easiest way to do it is to
use the bosonization formulae \Ref{bosoniz} and \Ref{freebos}.

The identities \Ref{hiro} can be rewritten in a more explicit form.
For this we introduce the Schur polynomials $p_j$ defined by
\be
\sum\limits_{k=0}^{\infty}p_k[t]x^k=
\exp\left(\sum\limits_{k=1}^{\infty}t_n x^n\right)
\ee
and the following notations
\beq
y_{\pm} &=& (y_{\pm 1}, y_{\pm 2},y_{\pm 3},\dots), \\
\tilde D_{\pm} &=& (D_{\pm 1}, D_{\pm 2}/2,D_{\pm 3}/3,\dots),
\plabel{hirop}
\eeq
where $D_{\pm n}$ represent the Hirota's bilinear operators
\be
D_n f[t]\cdot g[t]=\left. \frac{\partial}{\partial
x}f(t_n+x)g(t_n-x)\right|_{x=0}.
\ee
Then identifying $y_n={1\over 2\hb}(t'_n-t_n)$,
one obtains a hierarchy of partial differential equations
\begin{eqnarray}
&\sum\limits_{j=0}^{\infty}p_{j+i}(-2y_+)p_j(\hb\tilde D_+)
\exp \left( \hb\sum\limits_{k\ne 0}y_kD_k\right)
\tau_{l+i+1}[t]\cdot\tau_l[t] =
& \nonumber \\
&\sum\limits_{j=0}^{\infty}p_{j-i}(-2y_-)p_j(\hb\tilde D_-)
\exp \left( \hb\sum\limits_{k\ne 0}y_kD_k\right)
\tau_{l+i}[t]\cdot\tau_{l+1}[t].&
\label{toda}
\end{eqnarray}

The Hirota equations lead to a triangular system of nonlinear
difference-differential equations for the $\tau$-function.
Since the derivatives of the $\tau$-function are identified with
correlators, the Hirota equations are also
equations for the correlators of operators generating the Toda flows.
The first equation of the hierarchy is obtained by taking $i=-1$
and extracting the coefficient in front of $y_{-1}$
\be
\hb^2\, \tau_l{\p^2\tau_l\over \p t_1\p t_{-1}}-
\hb^2\, {\p\tau_l\over \p t_1}{\p\tau_l\over \p t_{-1}}+
\tau_{l+1}\tau_{l-1}=0.
\plabel{todaeqq}
\ee
Rewriting this equations as
\be
\hb^2\, {\p^2\log\tau_l\over \p t_1\p t_{-1}}+
{\tau_{l+1}\tau_{l-1}\over \tau_l^2}=0,
\plabel{todaqq}
\ee
one can recognize the Toda equation \Ref{todaeq} if one
takes into account the identification \Ref{taur}.

\subsection{String equation}

Above we considered the general structure of the Toda hierarchy.
However, the equations of the hierarchy, for example, the Hirota
equations \Ref{toda}, have many solutions. A particular solution is
characterized by initial condition. The role of such condition can be
played by the partition function of a non-perturbed system.  If one
requires that it should be equal to the $\tau$-function at vanishing
times and coincides with the full $\tau$-function after the
perturbation, the perturbed partition function can be found by means
of the hierarchy equations with the given initial condition.

However, the Toda equations involve partial differential
equations of high orders and require
to know not only the $\tau$-function at vanishing times but also
its derivatives. Therefore, it is not always clear whether the
non-perturbed function provides a sufficient initial information.
Fortunately, there is another way to select a unique solution of
Toda hierarchy. It uses some equations on the operators, usually,
the Lax and Orlov--Shulman operators. The corresponding equations
are called {\it string equations}.

The string equations cannot be arbitrary because they should preserve
the structure of Toda hierarchy. For example, if they are given by two
equations of the following type
\be
\bL=f(L,M),\qquad \bM=g(L,M),
\plabel{streq}
\ee
the operators defined by the functions $f$ and $g$ must satisfy
\be
[f(\ho,s),g(\ho,s)]=-\hb f.
\plabel{constreq}
\ee
This condition appears since $\bL$ and $\bM$ commute in the same way.

The advantage of use of string equations is that they represent, in a
sense, already a partially integrated version of the hierarchy
equations. For example, as we will see, instead of differential
equations of the second order, they produce algebraic and first order
differential equations and make the problem of finding the
$\tau$-function much simpler.

\subsection{Dispersionless limit}
\label{dispers}

The classical limit of Toda hierarchy is obtained in the limit
where the parameter measuring the lattice spacing vanishes.
In our notations this parameter is the Planck constant $\hb$.
Putting it to zero, as usual, one replaces all operators by
functions. In particular, as in the usual quantum mechanical systems,
the classical limit of the derivative operator is a variable
conjugated to the variable with respect to which one differentiates.
In other words, one should consider the phase space consisting
from $s$ and $\o$, which is the classical limit of the shift operator.
The Poisson structure on this phase space is defined by the
Poisson bracket induced from \Ref{osos}
\be
\{\o,s\}=\o.
\plabel{oss}
\ee
All operators now become functions of $s$ and $\o$ and commutators
are replaced by the corresponding Poisson brackets defined through
\Ref{oss}.

All equations including the Lax--Sato equations \Ref{evolH},
zero curvature condition \Ref{zerocur}, commutators
with Orlov--Shulman operators \Ref{laxorsh} can be rewritten in
the new terms. Thus, the general structure of Toda hierarchy is
preserved although it becomes much simpler. The resulting structure
is called {\it dispersionless Toda hierarchy} and the classical limit
is also known as {\it dispersionless limit}.

As in the full theory, a solution of the dispersionless Toda hierarchy
is completely characterized by a dispersionless $\tau$-function.
In fact, one should consider the free energy since
it is the logarithm of the full $\tau$-function that
can be represented as a series in $\hb$
\be
\log\tau=\sum\limits_{n\ge 0}\hb^{-2+2n}F_n.
\plabel{sertauh}
\ee
Thus, the dispersionless limit is extracted as follows
\be
F_0=\mathop{\lim}\limits_{\hb\to 0}\hb^2\log\tau.
\plabel{disfr}
\ee
The dispersionless free energy $F_0$ satisfies the classical limit
of Hirota equations \Ref{toda} and selected from all solutions
by the same string equations \Ref{streq} as in the quantum case.

Since the evolution along the times $t_k$ is now generated by
the Hamiltonians through the Poisson brackets, it can be seen as
a canonical transformation in the phase space defined above.
This fact is reflected also in the commutation relations \Ref{laxorsh}.
Since the Lax and Orlov--Shulman operators are dressed versions of
$\o$ and $s$, correspondingly, and have the same Poisson brackets,
one can say that the dispersionless Toda hierarchy describe
a canonical transformation from the canonical pair $(\o,s)$ to $(L,M)$.
The free energy $F_0$ plays the role of the generating function
of this transformation.

\subsection{2MM as $\tau$-function of Toda hierarchy}
\label{tmmtau}

In this paragraph we show how all abstract ideas described above get
a realization in the two-matrix model. Namely, we identify
the partition function \Ref{twointo} with a particular
$\tau$-function of Toda hierarchy.
This can be done in two ways using either the fermionic representation
or the Lax formalism and its connection with the orthogonal polynomials.
However, the fermionic representation which arises in this case
is not exactly the same as in paragraph \ref{ferbosrep}, although
it still gives a $\tau$-function of Toda hierarchy.
The difference is that one should use two types of fermions \cite{IKunp}.
In fact, they can be reduced to the fermions
appearing in the fermionic representation of 2MM presented in section
\ref{fermm}. They differ only by the basis
which is used in the mode expansions \Ref{expfer} and \Ref{fermf}.

We will use the approach based on the Lax formalism.
Following this way, one should identify the Lax operators in
the matrix model and prove that they satisfy the Lax-Sato equations
\Ref{evolH}. Equivalently, one can obtain the Baker--Akhiezer function
satisfying \Ref{eigprob} where the Hamiltonians $H_k$ are related to
the Lax operators through \Ref{laxh}.

First of all, the Lax operators coincide with the operators
$\hat\Lx$ and $\hat\Ly$ defined in \Ref{xlax} and \Ref{ylax}.
Due to the first equation in \Ref{proplax}, the operators have
the same expansion as required in \Ref{Lax} where
the first coefficient is
\be
r(n\hb)=\sqrt{h_{n+1}/h_{n}}.
\plabel{fcoef}
\ee
The Baker--Akhiezer function is obtained as a semi-infinite
vector constructed from the functions $\psi_n(x)$ \Ref{BAfun}
\be
\Psi(x;n\hb)=\Phx_n(x)e^{NV(x)}.
\plabel{bafun}
\ee
Due to \Ref{Lortx}, $L\Psi=x\Psi$. Thus, it remains to consider
the evolution of $\Psi$ in the coupling constants.
We fix their normalization choosing the potentials as follows
\be
V(x)=\sum\limits_{n>0}t_n x^n, \qquad \tV(x)=-\sum\limits_{n>0}t_{-n} y^n.
\plabel{MMpot}
\ee
Then the differentiation of the orthogonality condition
\Ref{orpolcon} with respect to the coupling constants leads to the
following evolution laws
\bea{rcl}{relatP}
{1\over N}{\p \Phx_{n}(x) \over \p t_k}&=&
-\sum\limits_{m=0}^{n-1}(\Lx^k)_{nm}\Phx_{m}(x)
-\hf  (\Lx^k)_{nn}\Phx_{n}(x),  \\
{1\over N} {\p \Phx_{n}(x) \over \p t_{-k}}&=&
\sum\limits_{m=0}^{n-1}(\Ly^k)_{nm}\Phx_{m}(x)
+\hf  (\Ly^k)_{nn}\Phx_{n}(x).
\eea
Using these relations one finds
\be
{1\over N} {\p \Psi \over \p t_k} = H_k\Psi,
\qquad
{1\over N} {\p \Psi \over \p t_{-k}} =H_{-k}\Psi,
\plabel{evolBA}
\ee
where $H_{\pm k}$ are defined as in \Ref{laxh}. As a result, if
one identifies $1/N$ with $\hbar$, one reproduces all equations for
the Baker--Akhiezer function. This means that the dynamics of 2MM
with respect to the coupling constants is governed
by Toda hierarchy.

Combining \Ref{taur} and \Ref{fcoef}, one finds
\be
h_n={\tau_{n+1}\over\tau_{n}}.
\plabel{tauhh}
\ee
Then the representation \Ref{intpl} implies
\be
Z(N)\sim{\tau_{N}\over\tau_{0}}.
\plabel{Ztauo}
\ee
The factor $\tau_{0}$ does not depend on $N$ and appears as a
non-universal contribution to the free energy.
Therefore, it can be neglected and, choosing the appropriate
normalization, one can identify the partition function of 2MM
with the $\tau$-function of Toda hierarchy
\be
Z(N)={\tau_{N}}.
\plabel{Ztau}
\ee

Moreover, one can find string equations uniquely characterizing
the $\tau$-function. First, we note that the Orlov--Shulman operators
\Ref{ORSH} are given by
\be
M=\hat\Lx\(V'(\hat\Lx)+\hat\Px\)-\hb, \qquad
\bM=\hat\Ly\(\tV'(\hat\Ly)+\hat\Py\).
\plabel{OrShMM}
\ee
Then the relations \Ref{relLPop} imply
\be
L\bL=M+\hb, \qquad \bL L=\bM.
\plabel{strMM}
\ee
Multiplying the first equation by $L^{-1}$ from the left and by $L$
from the right and taking into account \Ref{laxorsh}, one obtains
that
\be
M=\bM.
\plabel{strMMM}
\ee
This result together with the second equation in \Ref{strMM}
gives one possible form of the string equations.
It leads to the following functions $f$ and $g$ from \Ref{streq}
\be
f(\ho,s)=s\ho^{-1},\qquad g(\ho,s)=s.
\plabel{streqMM}
\ee
It is easy to check that they satisfy the condition \Ref{constreq}.
Combining \Ref{strMM} and \Ref{strMMM}, one arrives at another very
popular form of the string equation
\be
[L,\bL]=\hb.
\plabel{strcom}
\ee

The identification \Ref{Ztauo} allows to use the powerful machinery of
Toda hierarchy to find the partition function of 2MM. For example,
one can write the Toda equation \Ref{todaqq} which, together with some
initial condition, gives the dependence of $Z(N)$ on the first times
$t_{\pm 1}$. In the dispersionless limit this equation simplifies
to a partial differential equation and sometimes it becomes even
an ordinary differential equation (for example,
when it is known that the partition
function depends only on the product of the coupling constants
$t_{1}t_{-1}$). Finally, the string equation \Ref{strcom}
can replace the initial condition for the differential equations of
the hierarchy and produce equations of lower orders.


\chapter{Matrix Quantum Mechanics}
\label{chMQM}

Now we approach the main subject of the thesis which
is {\it Matrix Quantum Mechanics}.
This chapter is devoted to the introduction to this model
and combines the ideas discussed in the previous two chapters.
The reader will see how the technique of matrix models allows
to solve difficult problems related to string theory.

\section{Definition of the model and its interpretation}

Matrix Quantum Mechanics is a natural generalization of the matrix chain
model presented in section \ref{chMM}.\ref{Sgener}.
It is defined as an integral
over hermitian $N\times N$ matrices
whose components are functions of one real
variable which is interpreted as ``time''. Thus, it represents the path
integral formulation of a quantum mechanical system with $N^2$ degrees
of freedom. We will choose the time to be Euclidean so that the matrix
integral takes the following form
\be
Z_N(g)=\int \CD M(t) \exp \[ -N\tr\int dt\, \( \hf {\dot M}^2+V(M)\) \],
\plabel{MQMint}
\ee
where the potential $V(M)$ has the form as in \Ref{mpot}.
As it was required for all (hermitian) matrix models, this integral
is invariant under the global unitary transformations
\be
M(t)\longrightarrow \Omega^{\dagger}M(t)\Omega
\quad (\Omega^{\dagger}\Omega=I).
\plabel{MQMmtrun}
\ee

The range of integration over the time variable in \Ref{MQMint}
can be arbitrary depending on the problem we are interested in.
In particular, it can be finite or infinite, and the possibility
of a special interest is the case when the time is compact
so that one considers MQM on a circle. The latter choice will be
important later and now we will concentrate on the simplest
case of the infinite time interval.

In section \ref{chMM}.\ref{sdissurf} it was shown that the free energy
of matrix models gives a sum over discretized two-dimensional
surfaces. In particular, its special {\it double scaling limit}
corresponds to the continuum limit for the discretization and
reproduces the sum over continuous surfaces, which is the path integral
for two-dimensional quantum gravity. For the case of the simple one-matrix
integral (if we do not tune the potential to a multicritical
point), the surfaces did not carry any additional structure,
whereas we argued that the multi-matrix case should correspond to
quantum gravity coupled to matter. Since the MQM integral goes over
a continuous set of matrices, we expect to obtain quantum gravity
coupled with one scalar field. In turn, such a system can be interpreted
as the sum over surfaces (or strings) embedded into one dimension 
\cite{KAZMIG}.

Let us show how it works.
As in section \ref{chMM}.\ref{sdissurf},
one can construct a Feynman expansion of the integral \Ref{MQMint}.
It is the same as in the one-matrix case except that the propagator
becomes time-dependent.

\input{propagM.pic}

\noindent
Then the expansion \Ref{mdisc} is generalized to
\be
F=\sum\limits_{g=0}^{\infty}
N^{2-2g}
\sum\limits_{{\rm genus}\ g\ {\rm connected} \atop {\rm diagrams}}
g_2^{-E}
\prod\limits_k (-g_k)^{n_k}
\prod\limits_{i=1}^V \int\limits_{-\infty}^{\infty}dt_i
\prod\limits_{\langl ij\rangl}G(t_i-t_j),
\plabel{MQMexp}
\ee
where $\langl ij\rangl$ denotes the edge connecting
$i$th and $j$th vertices and $G(t)=e^{-|t|}$.
As usual, each Feynman diagram is dual to
some discretized surface and the sum \Ref{MQMexp} is interpreted as the
sum over all discretizations. The new feature is the appearance of
integrals over real variables $t_i$ living at the vertices of the
Feynman diagrams or at the centers of the faces of triangulated surfaces.
They represent a discretization of the functional integral over
a scalar field $t(\sigma)$. The action for the scalar field can be restored
from the propagator $G(t)$. Its discretized version is given by
\be
-\sum\limits_{\langl ij\rangl} \log G(t_i-t_j).
\plabel{actdis}
\ee
Taking into account the exact form of the propagator,
in the continuum limit, where the lattice spacing goes to zero,
one finds that the action becomes
\be
\int d^2\sigma\sqrt{h}\,| h^{ab} \p_a t\p_b t|^{1/2}.
\plabel{actcont}
\ee
It is not the standard action for a scalar field in two dimensions.
The usual one would be obtained if we took the Gaussian propagator
$G(t)=e^{-t^2}$.  Does it mean that MQM does not describe two-dimensional
gravity coupled with $c=1$ matter?

In \cite{KAZMIG} it was argued that it does describe by the following reason.
The usual scalar field propagator in the momentum space has the Gaussian
form $G^{-1}(p)\sim e^{p^2}$.
Its leading small momentum behaviour coincides with $G^{-1}(p)\sim 1+{p^2}$
which is the momentum representation of the propagator for MQM.
Thus, the replacement of one propagator by another affects only
the short distance physics which is non-universal. The critical properties
of the model surviving in the continuum limit should not depend on
this choice. This suggestion was confirmed by a great number of
calculations which showed full agreement of the results obtained by the
CFT methods and in the framework of MQM.

Finally, we note that 2D gravity coupled with $c=1$ matter
can be interpreted as a non-critical string embedded in one dimension.
The latter is equivalent to 2D critical string theory in the linear
dilaton background. Thus Matrix Quantum Mechanics is an alternative
description of 2D string theory. As it will be shown, it allows to
manifest the integrability of this model and to address many questions
inaccessible in the usual CFT formulation.

\newpage

\section{Singlet sector and free fermions}

In this section we review the general structure of Matrix Quantum Mechanics
and present its solution in the so called {\it singlet sector}
of the Hilbert space \cite{BIPZ}.
The solution is relied on the interpretation
of MQM as a quantum mechanical system of fermions.
Therefore, we will consider $t$ as a real Minkowskian time to have a good
quantum mechanical description.

\subsection{Hamiltonian analysis}

To analyze the dynamics of MQM,
as in the one-matrix case we change
the variables from the matrix elements $M_{ij}(t)$ to the
eigenvalues and the angular degrees of freedom
\be
M(t)=\Omega^{\dagger}(t)\eig(t)\Omega(t), \qquad
\eig=\diag(\eig_1,\dots,\eig_N),\ \Omega^{\dagger}\Omega=I.
\plabel{MQMdiag}
\ee
Since the unitary matrix depends on time it is not canceled
in the action of MQM. The kinetic term gives rise to an
additional term
\be
\tr {\dot M}^2=\tr {\dot x}^2+\tr[x,\dot\Omega\Omega^{\dagger}]^2.
\plabel{MQMkinter}
\ee
The matrix $\dot\Omega\Omega^{\dagger}$ is anti-hermitian
and can be considered as an element of the $su(n)$ algebra.
Therefore, it can be decomposed in terms of the SU(N) generators
\be
\dot\Omega\Omega^{\dagger}=
\sum\limits_{i=1}^{N-1} \dot\alpha_{i}H_i+
{i\over \sqrt{2}}\sum\limits_{i<j}
\( \dot\beta_{ij}T_{ij}+\dot\gamma_{ij}\tilde T_{ij}\),
\plabel{MQMang}
\ee
where $H_i$ are the diagonal generators of the Cartan subalgebra
and the other generators are represented by the following matrices:
$(T_{ij})_{kl}=\delta_{ik}\delta_{jl}+\delta_{il}\delta_{jk}$
and $(\tilde T_{ij})_{kl}=
-i(\delta_{ik}\delta_{jl}-\delta_{il}\delta_{jk})$.
The MQM degrees of freedom are described now by $x_i$, $\alpha_i$,
$\beta_{ij}$ and $\gamma_{ij}$.
The Minkowskian action in terms of these variables takes the form
\be
\SMQM=\int dt\, \[ \sum\limits_{i=1}^N
\( \hf{\dot x}_i^2 -V(x_i) \)+
\hf\sum\limits_{i<j}(x_i-x_j)^2({\dot\beta}_{ij}^2+{\dot\gamma}_{ij}^2)\].
\plabel{MQMacd}
\ee
We did not included the overall multiplier $N$ into the action.
It plays the role of the Planck constant so that in the following
we denote $\hb=1/N$.

To understand the structure of the corresponding quantum theory we pass
to the Hamiltonian formulation (see review \cite{KLEBANOV}).
It is clear that the Hamiltonian is given by
\be
\HMQM=\sum\limits_{i=1}^N
\( {1\over 2}{p}_i^2 +V(x_i) \)+
{1\over 2}\sum\limits_{i<j}{{\Pi}_{ij}^2+{\tilde\Pi}_{ij}^2
\over (x_i-x_j)^2},
\plabel{MQMham}
\ee
where $p_i$, ${\Pi}_{ij}$ and ${\tilde\Pi}_{ij}$ are momenta conjugated to
$x_i$, ${\beta}_{ij}$ and ${\gamma}_{ij}$, respectively.
Besides, since the action \Ref{MQMacd} does not depend on $\alpha_i$,
we have the constraint that its momentum should vanish $\Pi_i=0.$

In quantum mechanics all these quantities should be realized as operators.
If we work in the coordinate representation,
${\Pi}_{ij}$ and ${\tilde\Pi}_{ij}$ are the usual derivatives.
But to find $\hat p_i$, one should take into account the Jacobian appearing
in the path integral measure after the change of variables \Ref{MQMdiag}.
The Jacobian is the same as in \Ref{mesur}.
To see how it affects the momentum operator,
we consider the scalar product in the Hilbert space of MQM.
The measure of the scalar product in the coordinate representation
coincides with the path integral measure and contains the same Jacobian.
It can be easy understood because the change \Ref{MQMdiag} can be done
directly in the scalar product so that
\be
\langl \Phi|\Phi'\rangl=\int dM\, \overline{\Phi(M)}\Phi'(M)=
\int d\Omega\int \prod\limits_{i=1}^N dx_i\, \Delta^2(x)
\overline{\Phi(x,\Omega)}\Phi'(x,\Omega).
\plabel{MQMsc}
\ee
Due to this the map to the momentum representation, where the
measure is trivial, is given by
\be
\Phi(p,\Omega)=\int \prod\limits_{i=1}^N \(dx_i\, e^{-{i\over\hb}p_ix_i}\)
\Delta(x)
\Phi(x,\Omega).
\plabel{MQMmc}
\ee
Then in the coordinate representation the momentum is realized as the
following operator
\be
\hat p_i={-i \hb\over\Delta(x)} {\p\over \p x_i}\Delta(x).
\plabel{MQMmom}
\ee
As a result, we obtain that the Hamiltonian \Ref{MQMham} is represented by
\be
\hat \HMQM=\sum\limits_{i=1}^N
\( -{\hb^2\over 2\Delta(x)} {\p^2\over \p x^2_i}\Delta(x) +N V(x_i) \)+
{1\over 2}\sum\limits_{i<j}{{\hat\Pi}_{ij}^2+
\hat{\tilde\Pi^{\lefteqn{_{2}}}_{\lefteqn{_{ij}}}}
\over (x_i-x_j)^2}.
\plabel{MQMhamop}
\ee
The wave functions are characterized by the Schr\"odinger and constraint
equations
\be
i\hb{\p \Phi(x,\Omega)\over \p t}=\hat H_{\rm MQM} \Phi(x,\Omega),
\qquad
\hat \Pi_i\Phi(x,\Omega)=0.
\plabel{MWMeq}
\ee

\subsection{Reduction to the singlet sector}

Using the Hamiltonian derived in the previous paragraph,
the partition function \Ref{MQMint} can be rewritten as follows
\be
Z_N=\Tr e^{-T\hb^{-1} \hat\HMQM},
\plabel{ZHMQM}
\ee
where $T$ is the time interval we are interested in.
If one considers the sum over surfaces embedded in the infinite real line,
the interval should also be infinite. In this limit
only the ground state of the Hamiltonian contributes to
the partition function and we have
\be
F=\mathop{\lim}\limits_{T\to\infty}{\log Z_N\over T}=-E_0/\hb.
\plabel{Ztinf}
\ee

Thus, we should look for an eigenfunction of the Hamiltonian \Ref{MQMhamop}
which realizes its minimum.
It is clear that the last term representing the angular degrees of freedom
is positive definite and should annihilate this eigenfunction.
To understand the sense of this condition, let us note that
the angular argument $\Omega$ of the wave functions belongs to
SU(N). Hence, the wave functions are functions on the group and
can be decomposed in its irreducible representations
\be
\Phi(x,\Omega)=\sum\limits_{r}\sum\limits_{a,b=1}^{\dR}
\DR_{ba}(\Omega)\PhiR_{ab}(x),
\plabel{MQMdecom}
\ee
where $r$ denotes an irreducible representation, $\dR$ is its dimension
and $\DR_{ba}(\Omega)$ is the representation matrix of an element
$\Omega\in \SUN$ in the representation $r$. The coefficients are functions
of only the eigenvalues $x_i(t)$. On the other hand, the operators
${\hat\Pi}_{ij}$ and $\hat{\tilde\Pi}_{ij}$ are generators of
the left rotations $\Omega\to U\Omega$. It is clear that in 
the sum \Ref{MQMdecom} the only term remaining invariant under 
this transformation corresponds to the trivial, or {\it singlet}, 
representation.
Thus, the condition ${\hat\Pi}_{ij}\Phi=\hat{\tilde\Pi}_{ij}\Phi=0$
restricts us to the sector of the Hilbert space where the wave functions
do not depend on the angular degrees of freedom
$\Phi(x,\Omega)=\Phis(x)$.

In this singlet sector the Hamiltonian reduces to
\be
\hat \HMQM^{(\rm sing)}=\sum\limits_{i=1}^N
\( -{\hb^2\over 2\Delta(x)} {\p^2\over \p x^2_i}\Delta(x) +V(x_i) \).
\plabel{MQMhamred}
\ee
Its form indicates that it is convenient to redefine the wave functions
\be
\Psis(x)=\Delta(x)\Phis(x).
\plabel{redfun}
\ee
In terms of these functions the Hamiltonian becomes the sum of the
one particle Hamiltonians
\be
\hat \HMQM^{(\rm sing)}=\sum\limits_{i=1}^N \hat h_i,
\qquad
\hat h_i=-{\hb^2\over 2} {\p^2\over \p x^2_i} +V(x_i).
\plabel{MQMhamone}
\ee
Moreover, since a permutation of eigenvalues is also a unitary
transformation, the singlet wave function $\Phis(x)$
should not change under such permutations and,
therefore, it is symmetric. Then the redefined wave function $\Psis(x)$
is completely antisymmetric. Taking into account
the result \Ref{MQMhamone}, we conclude that the problem involving
$N^2$ bosonic degrees of freedom has been reduced to a system of
$N$ non-relativistic free fermions moving in the potential $V(x)$
\cite{BIPZ}. This fact is at the heart of the integrability
of MQM and represents an interesting and still not well understood
equivalence between 2D critical string theory and free fermions.

\subsection{Solution in the planar limit}

According to the formula \Ref{Ztinf} and the fermionic interpretation
found in the previous paragraph,
one needs to find the ground state energy of the system of $N$
non-interacting fermions. All states of such system are described
by Slater determinants and characterized by the filled energy levels
of the one-particle Hamiltonian $\hat h$ \Ref{MQMhamone}
\be
\Psi_{n_1,\dots,n_N}(x)={1\over \sqrt{N!}}\mathop{\det}\limits_{k,l}
\psi_{n_k}(x_l),
\plabel{SLdet}
\ee
where $\psi_n(x)$ is the eigenfunction at the $n$th level
\be
\hat h\psi_{n}(x)=\e_n\psi_n(x).
\plabel{heig}
\ee
The ground state is obtained by filling the lowest $N$ levels
so that the corresponding energy is given by
\be
E_0=\sum\limits_{n=1}^N\e_n.
\plabel{engr}
\ee

Note, that if the cubic potential is chosen, the system is non-stable.
The same conclusion can be made for any unbounded potential.
Therefore, strictly speaking, there is no ground state in such situation.
However, we are interested only in the perturbative expansion in $1/N$,
which corresponds to the expansion in the Planck constant.
On the other hand, the amplitudes of tunneling from an unstable vacuum
are exponentially suppressed as $\sim e^{-1/\hb}$. Thus, these
effects are not seen in the perturbation theory and we can simplify
the life considering even unstable potentials forgetting about
the instabilities. All what we need is to separate the perturbative effects
from the non-perturbative ones.

\lfig{The ground state of the free fermionic system. The fermions
fill the first $N$ levels up to the Fermi energy. At the critical
point where the Fermi level touches the top of the potential the energy
levels condensate and the density diverges.}
{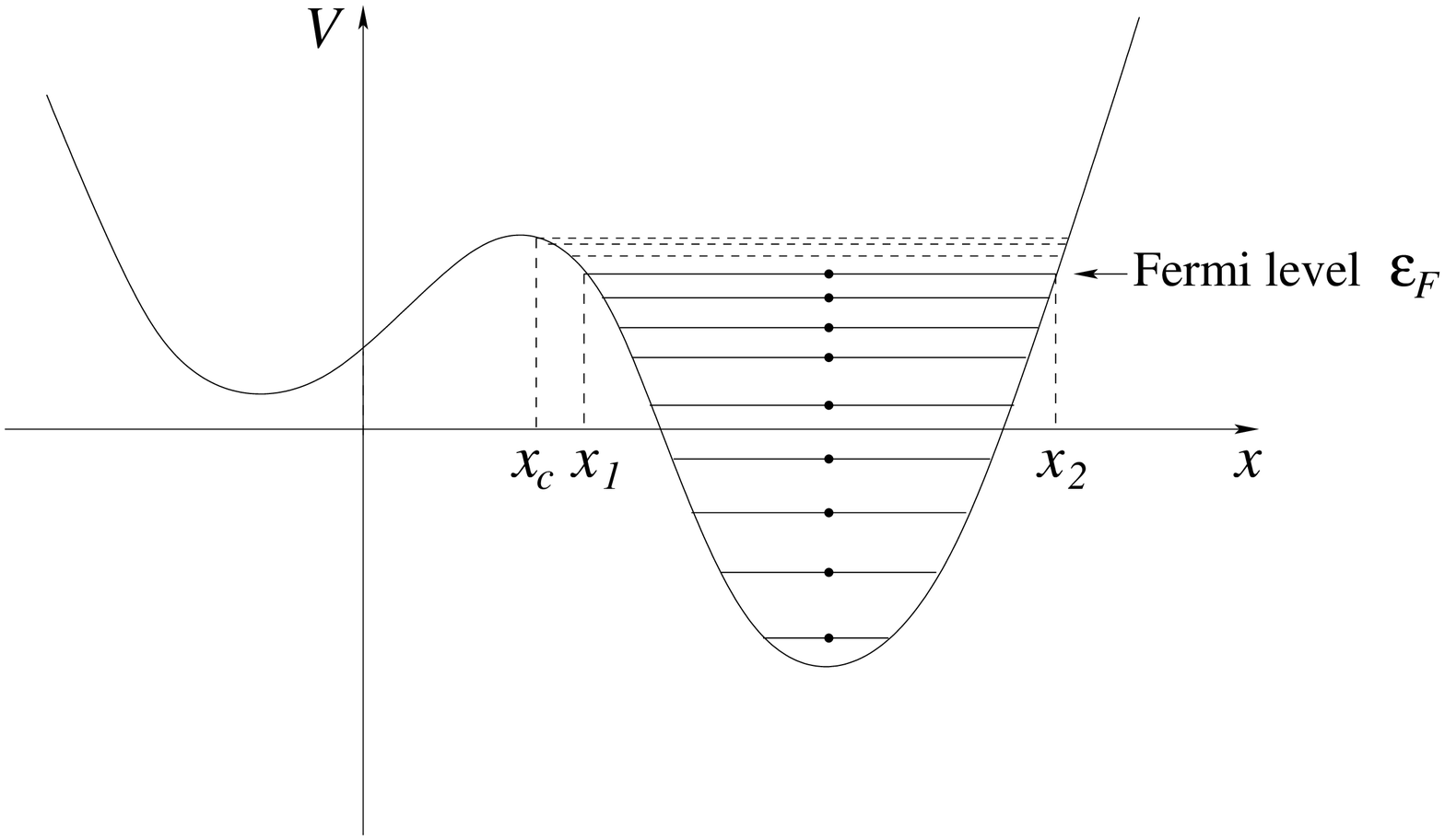}{10cm}{kolodets}

As a result, we get the picture presented in fig. \ref{kolodets}.
Let us consider this system in the large $N$ limit.
Since we identified $1/N$ with the Planck constant $\hb$,
$N\to \infty$ corresponds to the classical limit.
In this approximation the energy becomes continuous and particles are
characterized by their coordinates in the phase space.
In our case the phase space is two-dimensional and each particle
occupies the area $2\pi\hb$. Moreover, due to the fermionic nature,
two particles cannot take the same place.
Thus, the total area occupied by $N$ particles is $2\pi$. 
Due to the Liouville
theorem it is preserved in the time evolution. Therefore,
the classical description of $N$ free fermions is the same as that of
an incompressible liquid.

For us it is important now only that
the ground state corresponds to a configuration where
the liquid fills a connected region (Fermi sea)
with the boundary given by
the following equation
\be
h(x,p)=\hf p^2+V(x)=\e_F,
\plabel{Ferb}
\ee
where $\e_F=\e_N$ is the energy at the Fermi level.
Then one can write
\beq
N&=&\tint {dxdp\over 2\pi\hb}\theta(\e_F-h(x,p)),
\plabel{FSnorm} \\
E_0&=&\tint {dxdp\over 2\pi\hb}h(x,p)\theta(\e_F-h(x,p)).
\plabel{FSen}
\eeq
Differentiation with respect to $\e_F$ gives
\beq
\hb{\p N\over \p \e_F}&\stackrel{def}{=}&\rho(\e_F)=
\tint {dxdp\over 2\pi}\delta(\e_F-h(x,p))=
{1\over \pi}\int\limits_{x_1}^{x_2} {dx\over \sqrt{2(\e_F-V(x))}},
\plabel{FSnormd} \\
\hb{\p E_0\over \p \e_F}&=&
\tint {dxdp\over 2\pi}h(x,p)\delta(\e_F-h(x,p))
=\e_F\rho(\e_F),
\plabel{FSend}
\eeq
where $x_1$ and $x_2$ are turning points of the classical trajectory at the
Fermi level. These equations determine the energy in an inexplicit way.

To find the energy in terms of the coupling constants, one
should exclude the Fermi level $\e_F$ by means of the
normalization condition \Ref{FSnorm}.
For some simple potentials the integral in \Ref{FSnorm} can be calculated
explicitly, but in general this cannot be done.
However, the universal information related to the sum over continuous
surfaces and 2D string theory is contained only in the singular part
of the free energy. The singularity appears when the Fermi level
reaches the top of the potential similarly to the one-matrix case
(cf. figs. \ref{kolodets} and \ref{critbeh}).
Near this point the density diverges
and shows together with the energy a non-analytical behaviour.

From this it is clear that the singular contribution to
the integral \Ref{FSnormd} comes from the
region of integration around the maxima of the potential. Generically,
the maxima are of the quadratic type. Thus, up to analytical terms
we have
\be
\int\limits_{x_1}^{x_2} {dx\over \sqrt{2(\e_F-V(x))}}
\sim -\hf\log(\e_c-\e_F),
\plabel{intnonan}
\ee
where $\e_c$ is the critical value of the Fermi level.
Denoting $\e=\e_c-\e_F$, one finds \cite{KAZMIG}
\be
\rho(\e)= -{1\over2\pi}\log(\e/\Lambda),
\qquad F_0={1\over 4\pi\hb^2}\e^2\log(\e/\Lambda),
\plabel{CRden}
\ee
where we introduced a cut-off $\Lambda$ related to the
non-universal contributions.

\subsection{Double scaling limit}
\label{doubscl}

In the previous paragraph we reproduced the free energy and the density of states
in the planar limit. To find them in all orders in the genus expansion,
one needs to consider the double scaling limit as it was explained in section
\ref{chMM}.\ref{conlim}. For this one should correlate the large $N$ limit
with the limit where the coupling constants approach their critical values.
In our case this means that one should introduce coordinates describing
the region near the top of the potential.  Let $x_c$ is the coordinate
of the maximum and $y={1\over\sqrt{\hb}}(x-x_c)$.
Then the potential takes the form
\be
V(y)=\e_c-{\hb\over 2}y^2+{\hb^{3/2}\l\over 3}y^3+\cdots,
\plabel{potresc}
\ee
where the dots denote the terms of higher orders in $\hb$.
The Schr\"odinger equation for the eigenfunction at the Fermi level can be
rewritten as follows
\be
\left(-\hf{\p^2\over \p y^2}-{1\over 2}y^2+{\hb^{1/2}\l\over 3}y^3+\cdots\right)
\psi_N(y)=-\hb^{-1}\e\psi_{N}(y).
\plabel{SEQresc}
\ee
It shows that it is natural to define the rescaled energy variable
\be
\mu=\hb^{-1}(\e_c-\e_F).
\plabel{MQMdbs}
\ee
This relation defines the double scaling limit of MQM, which is obtained as
$N=\hb^{-1}\to \infty$, $\e_F\to\e_c$ and keeping $\mu$ to be fixed
\cite{BRKA,GIZI,PARISI,GRMI}.

Note that the double scaling limit \Ref{MQMdbs} differs from
the naive limit expected from 1MM where we kept fixed the product
of $N$ and some power of $\l_c-\l$ \Ref{coupl}.
In our case the latter is renormalized
in a non-trivial way. To get this renormalization, note that writing
the relation \Ref{FSnormd} we actually decoupled $N$ and $\hb$. This means
that in fact we rescaled the argument of the potential
so that we moved the coupling constant from
the potential to the coefficient in front of the action.
For example, we can do this with the cubic coupling constant
$\l$ by rescaling $x\to x/\l$.
In this normalization the overall coefficient should be
multiplied by $\l^{-2}$ what changes the relation between $N$
and the Planck constant to $\hb=\l^2/N$.
Then for $\Delta={2\pi \over\hb}(\l_c^2-\l^2)$,
\Ref{FSnormd} and \Ref{MQMdbs} imply
\be
{\p\Delta\over \p \mu}=2\pi\rho(\mu).
\plabel{MQMdmu}
\ee
Integrating this equation, one finds a complicated relation between two scaling variables.
In the planar limit this relation reads
\be
\Delta=-\mu\log(\mu/\Lambda).
\plabel{MQMdmup}
\ee

The remarkable property of the double scaling limit \Ref{MQMdbs} is that
it reduces the problem to the investigation of free fermions in the inverse
oscillator potential $V_{\rm ds}(x)=-\hf x^2$
\be
-\hf\left({\p^2\over \p x^2}+x^2\right)
\psi_{\e}(x)=\e\psi_{\e}(x),
\plabel{SEQdbs}
\ee
where we returned to the notations $x$ and $\e$ for already rescaled
matrix eigenvalues and energy.
All details of the initial potential disappear in this limit because after
the rescaling the cubic and higher terms suppressed by positive powers
of $\hb$.
This fact is the manifestation of the universality of MQM showing the
independence of its results of the form of the potential.

The equation \Ref{SEQdbs} for eigenfunctions has an explicit solution in terms of
the parabolic cylinder functions. They have a complicated form and we
do not give their explicit expressions. However, the density of states at the Fermi
level can be calculated knowing only their asymptotics at large $x$.
The density follows from the WKB quantization condition
\be
\left. \(\Phi_{\e_{n+1}}-\Phi_{\e_n}\)\right|_{\sqrt{\Lambda}}=2\pi,
\plabel{WKBq}
\ee
where $\Phi_{\e}$ is the phase of the wave function
$\psi_{\e}(x)={C\over \sqrt{x}}e^{{i}\Phi_{\e}(x)}$
and the difference is calculated at the cut-off
$x\sim\sqrt{\Lambda}\sim\sqrt{N}$. The asymptotic form of the phase is \cite{GinMoore}
\beq
\Phi_{\e}(x)&\approx& {1\over 2}x^2+\e\log x-\phi(\e),
\label{ASPH}
\\
\phi(\e)&=&{\pi\over 4}-
{i\over 2}\log{\Gamma(\hf+i\e)\over\Gamma(\hf-i\e)}.
\plabel{ASph}
\eeq
In the WKB approximation the index $n$ becomes continuous variable
and the density of states is defined as its derivative
\be
\rho(\e)\stackrel{\rm def}{=}
{\p n\over \p \e}={1\over 2\pi}\log\Lambda-{1\over 2\pi}{d\phi\over d\e}
={1\over 2\pi}\log\Lambda-{1\over 2\pi}\Re\psi(\hf+i\e),
\plabel{MQMdens}
\ee
where $\psi(\e)={d\over d\e}\log\Gamma(\e)$.
Neglecting the cut-off dependent term and expanding the digamma function
in $1/\mu$ ($\mu=-\e$), one finds the following result
\be
\rho(\mu)={1\over 2\pi}\left( -\log\mu +\sum\limits_{n=1}^{\infty}(2^{2n-1}-1)
{|B_{2n}|\over n}(2\mu)^{-2n}\right),
\plabel{MQMdensexp}
\ee
where $B_{2n}$ are Bernoulli numbers.
Integrating \Ref{FSend}, one obtains the expansion of the free energy
\be
F(\mu)={1\over 4\pi}\left( \mu^2\log\mu -{1\over 12}\log\mu+
\sum\limits_{n=1}^{\infty}{(2^{2n+1}-1)
|B_{2n+2}|\over 4n(n+1)}(2\mu)^{-2n}\right).
\plabel{MQMfrexp}
\ee

In fact, to compare this result with the genus expansion of the partition
function of 2D string theory, one should reexpand \Ref{MQMfrexp}
in terms of the renormalized string coupling.
Its role, as usual, is played by $\kappa=\Delta^{-1}$ and its relation to
$\mu$ is determined by \Ref{MQMdmu}. With $\rho(\mu)$ taken from
\Ref{MQMdensexp}, one can solve this equation with respect to
$\Delta(\mu)$. Then it is sufficient to make substitution
into \Ref{MQMfrexp} to get the following answer
\be
F(\Delta)={1\over 4\pi}\left( {\Delta^2\over\log\Delta} -
{1\over 12}\log\Delta+
\sum\limits_{n=1}^{\infty}{(2^{2n+1}-1)
|B_{2n+2}|\over 4n(n+1)(2n+1)}\({2\Delta\over \log\Delta}\)^{-2n}\right),
\plabel{MQMfrdl}
\ee
where terms $O(\log^{-1}\Delta)$ were neglected because they contain
the cut-off and vanish in the double scaling limit.

Some remarks related to the expansion \Ref{MQMfrdl} are in order.
First, the coefficients associated with genus $g$ grow as $(2g)!$.
This behaviour is characteristic for closed string theories where
the sum over genus-$g$ surfaces exhibits the same growth.
Besides, we observe a new feature in comparison with the one-matrix model.
Although the couplings were renormalized, the sums over spherical
and toroidal surfaces logarithmically diverge. This also can be
explained in the context of the CFT approach. Finally, comparing \Ref{MQMfrdl}
with \Ref{aspf}, we find that the string susceptibility for MQM
vanishes $\str=0$. This is again in the excellent agreement with
the continuum prediction of \cite{KPZ}.

Thus, all results concerning the free energy of MQM
coincide with the corresponding results for the partition function
of 2D string theory. Therefore, it is tempting to claim that
they are indeed equivalent theories.
But we know that 2D string theory possesses dynamical degrees of freedom:
the tachyon, winding modes and discrete states.
To justify the equivalence further, we should show how all of them
are realized in Matrix Quantum Mechanics.

\newpage

\section{Das--Jevicki collective field theory}
\label{dasjev}

The fermionic representation presented in the previous section
gives a microscopic description of 2D string theory. As usual,
the macroscopic description, which has a direct interpretation in terms
of the target space fields of 2D string theory, is obtained
as a theory of effective degrees of freedom. These degrees of freedom
are collective excitations of the fermions of the singlet sector of MQM
and identified with the tachyonic modes of 2D strings.
Their dynamics is governed by a {\it collective field theory}
\cite{JEVSAC}, which in the given case was developed by Das and Jevicki
\cite{DasJev}. This theory encodes all interactions of strings
in two dimensions and, therefore, it gives an example of
string field theory formulated directly in the target space.

\subsection{Effective action for the collective field}

A natural collective field in MQM is the density of eigenvalues
\be\label{cmf}
\bff(x,t)=\tr\delta\left(x-M(t)\right).
\ee
In the double scaling limit, which implies $N\to \infty$, it becomes
a continuous field. Its dynamics can be derived directly
from the MQM action with the inverse oscillator potential
\cite{DasJev}
\be
S=\hf \int dt\,\tr\left({\dot M}^2+M^2\right).
\plabel{mmL}
\ee
However, it is much easier to use the Hamiltonian formulation.
The Hamiltonian of MQM in the singlet representation is given by the energy
of the Fermi sea similarly to the ground state energy \Ref{FSen}.
The difference is that now the Fermi sea can have an arbitrary profile
which can differ from the trajectory of one fermion \Ref{Ferb}.
Besides, to the expression \Ref{FSen} one should add a term
which fixes the Fermi level and allows to vary the number of fermions.
Otherwise the density would be subject of some normalization condition.
Thus, the full double scaled Hamiltonian reads
\be
H_{\rm coll}=\mathop{\tint}\limits_{\rm Fermi \atop sea}
{dxdp\over 2\pi}(h(x,p)+\mu),
\plabel{FSendb}
\ee
where
\be
h(x,p)=\hf p^2+V(x), \quad V(x)=-\hf x^2.
\plabel{hham}
\ee

We restrict ourselves to the case where the boundary of the Fermi sea can
be represented by two functions, say $p_+(x,t)$ and $p_-(x,t)$ satisfying
the boundary condition $p_+(x_*,t)=p_-(x_*,t)$, where $x_*$ is the
leftmost point of the sea (fig. \ref{fers}a). It means that we forbid
the situations shown on fig, \ref{fers}b. In the fermionic picture they
do not cause any problems, but in the bosonic description they require
a special attention.

\lfig{The Fermi sea of the singlet sector of MQM. The first picture
shows the situation where the profile of the Fermi sea can be described by
a two-valued function. The second picture presents
a more general configuration.}
{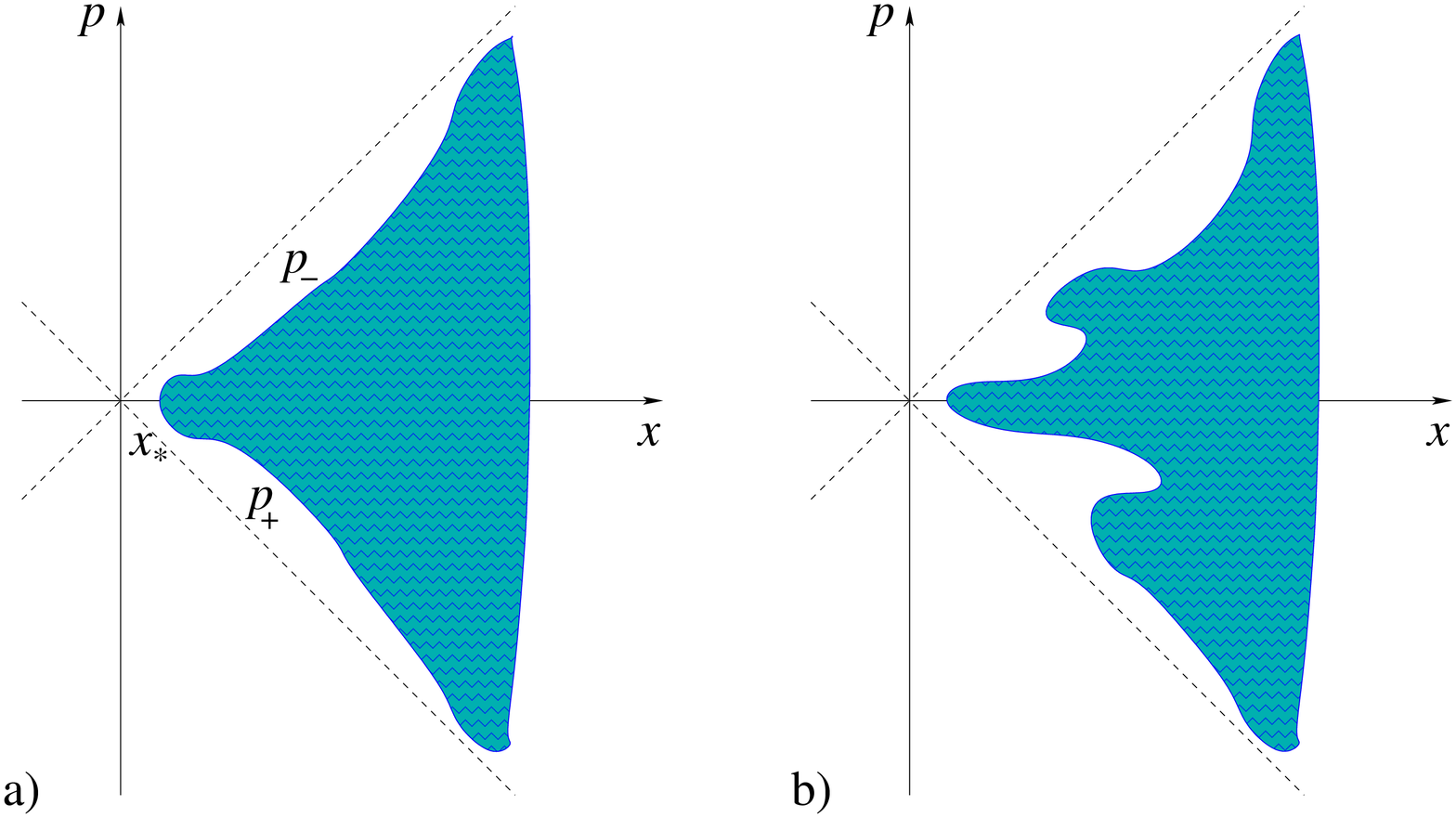}{10cm}{fers}

In this restricted situation one can take the integral over the momentum
in \Ref{FSendb}. The result is
\be
H_{\rm coll}=
\int {dx\over 2\pi}\, \left( {1\over 6} (p_+^3-p_-^3)+
(V(x)+\mu)(p_+-p_-) \right).
\plabel{FSendd}
\ee
It is clear that the difference of $p_+$ and $p_-$ coincides with the
density \Ref{cmf}, whereas their sum plays the role of a conjugate variable.
The right identification is the following \cite{JEVICKI}:
\be
p_\pm(x,t) = \p_x\Pi \pm \pi \bff(x,t),
\plabel{chf}
\ee
where the equal-time Poisson brackets are defined as
\be
\{ \bff(x), \Pi(y)\}=\delta(x-y).
\plabel{imp}
\ee
Substitution of \Ref{chf} into \Ref{FSendd} gives
\be
H_{\rm coll}=\int dx \left(
{1\over 2} \bff(\p_x\Pi)^2 +{\pi^2 \over 6}\bff^3
+(V(x)+\mu)\bff \right).
\plabel{Hcoll}
\ee
One can exclude the momentum $\Pi(x,t)$ by means of the equation of motion
\be
-\p_x \Pi={1\over \bff}\int dx \p_t \bff
\plabel{cleq}
\ee
what leads to the following collective field theory action
\be
S_{\rm coll}=\int dt \int dx\,
\left({1\over 2\bff}\left(\int dx \p_t\bff\right)^2-{\pi^2 \over 6}\bff^3
-(V(x)+\mu)\bff\right).
\plabel{Scol}
\ee

This action can be considered as a background independent
formulation of string field theory. It contains a cubic interaction
and a linear tadpole term.
The former describes the effect of splitting and joining strings
and the latter represents a process of string annihilation into the vacuum.
The important point is that the dynamical field $\bff(x,t)$
is two-dimensional. The dimension additional to the time $t$
appeared from the matrix eigenvalues. This shows that the target space
of the corresponding string theory is also two-dimensional in agreement
with our previous conclusion.

Another observation is that whereas the initial matrix model
in the inverse oscillator potential was simple with the linear equations
of motion
\be
\mathop{M}\limits^{..}(t)-M(t)=0,
\plabel{MATeq}
\ee
the resulting collective field theory is non-linear.
Thus, MQM provides a solution of a complicated
non-linear theory through the transformation of variables \Ref{cmf}.
Nevertheless, the integrability of MQM is present also in the
effective theory \Ref{Scol}. 
Indeed, consider the equations of motion for the fields $p_+$ and $p_-$.
The equations \Ref{chf} and \Ref{imp} imply the following Poisson brackets
\be
\{p_\pm(x),p_\pm(y)\}=\mp 2\pi \p_x\delta(x-y)
\plabel{Ppp}
\ee
so that the Hamiltonian \Ref{FSendd} gives
\be
\p_t p_\pm+p_\pm \p_x p_\pm+\p_x V(x)=0.
\plabel{Hopfeq}
\ee
This equation is a KdV type equation which is integrable.
This indicates that the whole theory is also exactly solvable.
In fact, one can write an infinite set of conserved commuting quantities
\cite{JEVICKI}
\be
H_n=\mathop{\tint}\limits_{\rm Fermi \atop sea}
{dxdp\over 2\pi}(p^2-x^2)^n.
\plabel{MQMhh}
\ee
It is easy to check that up to surface terms they satisfy
\be
\{H_n,H_m\}=0 \quad {\rm and} \quad {d\over dt}H_n=0.
\plabel{prophh}
\ee
The quantities $H_n$ can be considered as Hamiltonians generating 
some perturbations. Since all of them are commuting, 
according to the definition given in section
\ref{chMM}.\ref{intsys}, we conclude that the system is integrable.

\subsection{Identification with the linear dilaton background}
\label{identld}

Now we choose a particular background of string theory. This will allow
to identify the tachyon field and target space coordinates with
the corresponding quantities of the collective field theory.
In terms of the collective theory the choice of a background
means to consider the perturbation theory around some classical solution of
\Ref{Scol}. We choose the solution describing the ground state of MQM.
It is obtained  from \Ref{Ferb} and can be written as
\be
\pi \bff_0=\pm p_{\pm}=\sqrt{x^2-2\mu}.
\plabel{grstsol}
\ee
This solution is distinguished by the fact that it is static and
that the boundary of the Fermi
sea coincides with a trajectory of one fermion.

Taking \Ref{grstsol} as a background, we are interested in small fluctuations
of the collective field around this background solution
\be
\bff(x,t)=\bff_0(x)+{1\over \sqrt{\pi}}\p_x\eta(x,t).
\plabel{expf}
\ee
The dynamics of these fluctuations is described by the following action
obtained by substitution of \Ref{expf} into \Ref{Scol}
\be
S_{\rm coll}=\hf\int dt \int dx\,
\left({(\p_t\eta)^2\over (\pi\bff_0+{\sqrt{\pi}}\p_x\eta)}
-\pi\bff_0(\p_x\eta)^2-{\sqrt{\pi}\over 3}(\p_x\eta)^3
\right).
\plabel{Scolex}
\ee
The expansion of the denominator in the first term gives rise to
an infinite number of vertices of increasing order with the field $\eta$.
A compact version of this interacting theory would be obtained
if we worked with the Hamiltonian instead of
the action. Then only cubic interaction terms would appear.

Let us consider the quadratic part of the action \Ref{Scolex}. It is given by
\be
S_{(2)}=\hf\int dt \int dx\,
\left({(\p_t\eta)^2\over \pi\bff_0}-(\pi\bff_0)(\p_x\eta)^2\right).
\plabel{Scoltwo}
\ee
Thus, $\eta(x,t)$ can be interpreted as a massless field propagating
in the background metric
\be
g^{(0)}_{\mu\nu}=
\pmatrix{-\pi\bff_0 & 0 \cr
           0 & (\pi\bff_0)^{-1}}.
\plabel{met}
\ee
However, the non-trivial metric can be removed by a coordinate transformation.
It is enough to introduce the time-of-flight coordinate
\be
\xf(x)=\int\limits^x {dx\over\pi\bff_0(x) }.
\plabel{tfcor}
\ee
The change of coordinate \Ref{tfcor} brings the action to the form
\be
S_{\rm coll}=\hf\int dt \int d{\xf}\,
\left((\p_t\eta)^2
-(\p_{\xf}\eta)^2-{1\over 3\pi\sqrt{\pi}\bff_0^2}\( (\p_{\xf}\eta)^3+
3(\p_{\xf}\eta)(\p_{t}\eta)^2\)+\cdots
\right),
\plabel{Scolexq}
\ee
where we omitted the terms of higher orders in $\eta$.
The action \Ref{Scolexq} describes a massless field in the flat Minkowski
spacetime with a spatially dependent coupling constant
\be
\gst(\xf)={1\over(\pi\bff_0(\xf))^2 }.
\plabel{coupst}
\ee
Using the explicit formula \Ref{grstsol} for the background solution, one can
obtain
\be
x(\xf)=\sqrt{2\mu}\cosh\xf, \qquad
p(\xf)=\sqrt{2\mu}\sinh\xf.
\plabel{xpq}
\ee
Thus, the coupling constant behaves as
\be
\gst(\xf)={1\over 2\mu\sinh^2\xf}\sim {1\over \mu}e^{-2\xf},
\plabel{coupstq}
\ee
where the asymptotics is given for $\xf\to \infty$.

Comparing \Ref{coupstq} with \Ref{ccc}, we see that the collective
field theory action \Ref{Scolexq} describes 2D string theory in the linear
dilaton background. In the asymptotic region of large $\xf$
the flat coordinates $(t,\xf)$ can be identified with the coordinates
of the target space of string theory coming from the $c=1$ matter $X$
and the Liouville field $\phi$ on the world sheet.
The identification reads as follows
\be
it\leftrightarrow X,
\qquad
\xf\leftrightarrow \phi.
\plabel{idencoor}
\ee
Thus, the time of MQM and the time-of-flight coordinate, which is
a function of the matrix eigenvalue variable, form the flat
target space of the linear dilaton background.
It is also clear that the two-dimensional massless collective field
$\eta(\xf,t)$ coincides with the redefined tachyon $\eta=e^{2\phi}T$.

In fact, the above identification is valid only asymptotically.
When we go to the region of small $\xf$, one should include into account
the Liouville exponent $\mu e^{-2\phi}$ in the CFT action \Ref{constr}.
The tachyon field can be considered as a wave function describing
the lowest eigenstate of the Hamiltonian of this CFT. Therefore,
instead of Klein--Gordon equation, it should satisfy the Liouville
equation
\be
\left( \p^2_X+\p^2_{\phi}+4\p_\phi+4-\mu e^{-2\phi} \right) T(\phi,X)=0.
\plabel{Liouv}
\ee
Hence the Liouville mode does not coincide exactly with the
collective field coordinate $\xf$. The correct identification is obtained
as follows \cite{MooreSei}. We return to the eigenvalue variable $x$
and consider its conjugated momentum $p=-i\p/\p x$. The Fourier
transform of the Klein--Gordon equation written
in the metric \Ref{met} gives
\be
\left( \p^2_t-\sqrt{x^2-2\mu}\,\p_{x}\,\sqrt{x^2-2\mu}\,\p_x
\right) \eta(x,t)=0
\Rightarrow
\left( \p^2_t-(p\p_p)^2-2p\p_p-1-2\mu p^2 \right) \teta(p,t)=0.
\plabel{Liouvf}
\ee
where $\teta$ is the Fourier image of $\eta$.
Finally, the change of variables
\be
it=X,
\qquad
ip={1\over \sqrt{2}} e^{-\phi}
\plabel{idenL}
\ee
together with $T(\phi,X)\sim p^3\teta(p,t)$
brings \Ref{Liouvf} to the Liouville equation \Ref{Liouv}.
This shows that, more precisely, the Liouville coordinate is identified
with the logarithm of the momentum conjugated to the matrix eigenvalue.

The meaning of this rule becomes more clear after realizing that the Fourier
transform with an imaginary momentum of the collective field $\bff$
is the Wilson loop operator
\be
W(l,t)=\tr \( e^{-lM}\)=\int dx \, e^{-lx}\bff(x,t).
\plabel{WL}
\ee
This operator inserts a loop of the length $l$ into the world sheet.
Therefore, it has a direct geometrical interpretation and its parameter
$l$ is related to the scale of the metric, which is governed
by the Liouville mode $\phi$. Thus, it is quite natural that
$l$ and $\phi$ are identified through \Ref{idenL} where one should take
$ip=l$.  The substitution of the expansion \Ref{expf} gives
\be
W(l,t)=W_0+{1\over \sqrt{\pi}}\int dx \, e^{-lx}\p_x\eta(x,t)=
W_0+{l\over \sqrt{\pi}}\teta(-il,t),
\plabel{WLphi}
\ee
where
\be
W_0={\sqrt{2\mu}\over l}K_1(\sqrt{2\mu}l)
\plabel{WLzero}
\ee
is the genus zero one-point function of the density.
Using this representation, it is easy to check that the Wilson loop
operator satisfies the following Wheeler--DeWitt equation \cite{MSS}
\be
\left( \p^2_t-(l\p_l)^2+2\mu l^2 \right) W(l,t)=0.
\plabel{WLl}
\ee
Thus, in the $l$-representation it is the Wilson loop operator that
is the analog of the free field for which the derivative terms
have standard form. Therefore, we should identify the field
$\eta$ from section \ref{chSTR}.\ref{tachtwo} with $W$ rather than with
$\teta$ defined in \Ref{Liouvf}. The precise relation between the tachyon field
and the Wilson loop operator is the following
\be
T(\phi,X)=e^{-2\phi}W(l(\phi),-iX)=e^{-2\phi}W_0+
e^{-2\phi}\int\limits_0^{\infty}d\xf\,
\exp\left[-\sqrt{\mu} e^{-\phi}\cosh\xf \right]\p_{\xf}\eta.
\plabel{WLT}
\ee

The integral transformation \Ref{WLT} expresses solutions of the non-linear
Liouville equation through solutions of the Klein--Gordon equation.
This reduces the problem of calculating the tachyon scattering amplitudes
in the linear dilaton background
to calculation of the $S$-matrix for the collective field theory of
the Klein--Gordon field $\eta$. As we saw above, this theory is integrable.
Therefore, the scattering problem in 2D string theory
can be exactly solved.
Before to show that, we should introduce the operators creating
the asymptotic states, {\it i.e.}, the tachyon vertex operators.

\subsection{Vertex operators and correlation functions}
\label{VerCorFun}

The vertex operators of the tachyon field were constructed in section
\ref{chSTR}.\ref{tachtwo}. Their Minkowskian form is given by
\Ref{Mtachver} and describes the left and right movers. Note that
the representation for the operators was written only in the asymptotic
region $\phi\to \infty$ where the Liouville potential can be ignored.
Therefore, we can use the simple identification \Ref{idencoor}
to relate the matrix model quantities with the target space objects.
Then one should find operators that behave as left and right movers
in the space of $t$ and $\xf$.

First, the $t$-dependence of a matrix model operator
is completely determined by the inverse
oscillator potential, which leads to the following simple Heisenberg
equation
\be
{\p\over \p t}\hat A(t)=i\left[\hf(\hat p^2-\hat x^2),\hat A(t)\right].
\plabel{tdep}
\ee
Its solution is conveniently represented in the basis of the chiral operators
\be
\hat\xpm(t)\stackrel{\rm def}{=}
{\hat x(t)\pm \hat p(t)\over \sqrt{2}}=\hat \xpm(0)e^{\pm t}.
\plabel{xpm}
\ee
Therefore, the time-independent operators, which should be used
in the Schr\"odinger representation, are 
$e^{\mp t}\hat \xpm(t)$.
This suggests that the vertex operators can be constructed
from powers of $\xpm$. Indeed, it was argued \cite{JEVICKI} that
their matrix model realization is given by
\be
T^{\pm}_n=e^{\pm nt}\tr (M\mp P)^n.
\plabel{Tpm}
\ee

To justify further this choice, let us consider the collective field
representation of the operators \Ref{Tpm}
\be
T^{\pm}_n=e^{\pm nt}\int {dx\over 2\pi}\int\limits^{p_+(x,t)}_{p_-(x,t)} dp\,
(x\mp p)^n=
\left. {e^{\pm nt}\over n+1}
\int {dx\over 2\pi} (x\mp p)^{n+1}\right|^{p_+}_{p_-},
\plabel{Tpmcol}
\ee
where $p_{\pm}$ are second quantized fields satisfying equations
\Ref{Hopfeq}. We shift these fields by the classical solution
\be
p_{\pm}(x,t)=\pm \pi\bff_0(x) + {\alpha_{\mp}(x,t)\over \pi\bff_0(x)}.
\plabel{ppm}
\ee
Then the linearized equations of motion for
the quantum corrections $\alpha_{\pm}$ coincide with the conditions
for chiral fields
\be
(\p_t\mp\p_{\xf})\alpha_{\pm}=0 \Rightarrow
\alpha_{\pm}=  \alpha_{\pm}(t\pm \xf).
\ee
In the asymptotics $\xf\to\infty$, $\pi\bff_0\approx x\approx \hf e^{\xf}$.
Therefore, in the leading approximation the operators \Ref{Tpmcol}
read
\be
T^{\pm}_n\approx
 \int {d\xf\over 2\pi}e^{n(\xf\pm t)} \alpha_{\pm}.
\plabel{Tpmas}
\ee
The integral extracts the operator creating
the component of the chiral field which
behaves as $e^{n(\xf\pm t)}$.
This shows that the matrix operators \Ref{Tpm} do possess the necessary
properties.  The factor $e^{-2\phi}$, which is present in the definition of
the vertex operators \Ref{Mtachver} and absent in the matrix case,
can be seen as coming from the measure of integration over the
world sheet or, equivalently, as a result of
the redefinition of the tachyon field \Ref{WLT}.
In fact, it is automatically restored by the matrix model.

The matrix operators \Ref{Tpm} correspond to the Minkowskian vertex
operators with imaginary momenta $k=in$.
After the continuation of time to the Euclidean region $t\to -iX$,
they also can be considered as vertex operators with
Euclidean momenta $p=\mp n$.
To realize other momenta, one should analytically continue from this
discrete set to the whole complex plane. In particular, the vertex
operators of Minkowskian real momenta are obtained as
\be
V^{\pm}_k\sim T^{\pm}_{-ik} =e^{\mp ikt }\tr (M\mp P)^{-ik}.
\plabel{vermom}
\ee

Using these operators, one can construct and calculate
scattering amplitudes of tachyons. The result has been obtained
from both the collective field theory formalism
\cite{Polch,JevRod,MP,POLCHINSKI}
and the fermionic representation \cite{KLEBANOV,Mooredb,MPR,DMP,GinMoore}.
Moreover, the generating functional for all
$S$-matrix elements has been constructed \cite{DMP}.
It takes an especially transparent
form when the $S$-matrix is represented
as a composition of three processes: fermionization of incoming
tachyon modes, scattering in the free fermion theory and reverse
bosonization of the scattered fermions
\be
S_{TT}=\iota_{f\to b}\circ S_{FF} \circ \iota_{b\to f}.
\plabel{Smat}
\ee
The fermionic $S$-matrix $S_{FF}$ was explicitly
calculated from the properties
of the parabolic cylinder functions \cite{MPR}.
We do not give more details since these results will be reproduced
in much simpler way from the formalism which we develop in the next
chapters.

We restrict ourselves to two remarks. The first one is that
in those cases where the scattering amplitudes in 2D string theory
can be calculated by the CFT methods, the results
coincided with the corresponding calculations in MQM \cite{DiFK}.
The only thing to be done to ensure the complete agreement
is a local redefinition of the vertex operators.
It turns out that the exact relation between the tachyon operators
\Ref{Mtachver} and their matrix model realization \Ref{Tpm}
include the so called {\it leg-factors} \cite{GRKLleg}
\be
V^{+}_k={\Gamma(-ik)\over \Gamma(ik)} T^{+}_{-ik},
\qquad
V^{-}_k={\Gamma(ik)\over \Gamma(-ik)} T^{-}_{-ik}.
\plabel{legf}
\ee
This redefinition is not surprising because the matrix model gives only a
discrete approximation to the local vertex operators and in the continuum
limit the operators can be renormalized. Therefore, one should expect the
appearance of such leg-factors in any matrix/string correspondence.
Note that the Minkowskian leg-factors \Ref{legf} are pure phases. Thus,
they represent a unitary transformation and do not affect the amplitudes.
However, they are relevant for the correct spacetime physics,
in particular, for the gravitational scattering of tachyons
\cite{POLCHINSKI}. In fact, the leg-factors can be associated with a field
redefinition given by the integral transformation \Ref{WLT}.
Written in the momentum space for $\xf$, it gives rise to
additional factors for the left and right components whose ratio
produces the leg-factor.

The second remark is that in the case when the Euclidean momenta
of the incoming and outgoing
tachyons belong to an equally spaced lattice (as in a compactified theory),
the generating functional for $S$-matrix elements
has been shown to coincide with a $\tau$-function of
Toda hierarchy \cite{DMP}. However, this fact has not been used
to address other problems like scattering in presence of a tachyon
condensate. We will show that with some additional information added,
it allows to solve many interesting questions related to 2D string
theory in non-trivial backgrounds.

\subsection{Discrete states and chiral ring}

Finally, we show how the discrete states of 2D string theory appear in
MQM. They are created by a natural generalization of the
matrix operators \Ref{Tpm} \cite{AvJev}
\be
T_{n,\bn}=e^{(\bn-n)t}\tr \( (M+P)^n (M-P)^{\bn}\)
\plabel{Tdis}
\ee
which have the following Euclidean momenta
\be
p_X=i(n-\bn), \qquad p_{\phi}=n+\bn-2.
\plabel{momdisM}
\ee
Comparing with the momenta of the discrete states \Ref{mdiscr},
one concludes that
\be
m={n-\bn\over 2}, \qquad j={n+\bn\over 2}.
\plabel{numdis}
\ee
Taking into account that $n$ and $\bn$ are integers, one finds
that the so defined pair $(j,m)$ spans all discrete states.

It is remarkable that the collective field theory approach
allows to unveil the presence of a large symmetry group
\cite{AvJev,MooreSei,WITTENGR}.
Indeed, the operators \Ref{Tdis} are realized as
\be
T_{j,m}=e^{-2mt}\int {dx\over 2\pi}\int\limits^{p_+}_{p_-}
dp\,  (x+p)^{j+m} (x-p)^{j-m},
\plabel{Tdisst}
\ee
where we changed indices of the operator from $n,\bn$ to $j,m$.
One can check that they obey the commutation relations
of $w_{\infty}$ algebra
\be
\{ T_{j_1,m_1}, T_{j_2,m_2} \} =
4i\( j_1 m_2-j_2 m_1 \) T_{j_1+j_2-1,m_1+m_2}.
\plabel{winf}
\ee
In particular, the operators $T_{j,m}$ are eigenstates of the Hamiltonian
$H=-\hf T_{1,0}$
\be
\{ H, T_{j,m} \} =
-2im T_{j,m}
\plabel{disham}
\ee
what means that they generate its spectrum.
When we replace the Poisson brackets of the classical collective
field theory by the quantum commutators, the $w_{\infty}$ algebra
is promoted to a $W_{1+\infty}$ algebra.

Each generator \Ref{Tdis} gives rise to an element of
the ground ring of the $c=1$ CFT \cite{WITTENGR} which plays an important
role in many physical problems.
First, we introduce the so called {\it chiral ground ring}.
It consists of chiral ghost number zero, conformal spin zero
operators $\CO_{JM}$ which are closed under the operator product
$\CO\cdot\CO'\sim\CO''$ up to BRST commutators.
The entire chiral ring can be generated from the basic operators
\beq
\CO_{0,0}&=&1, \nonumber \\
y\stackrel{\rm def}{=}\CO_{\hf, \hf}&=&
 \(cb+ i\p X-\p\phi \) e^{ iX+\phi},
\plabel{genrin} \\
w\stackrel{\rm def}{=}\CO_{\hf,- \hf}&=&
 \(cb- i\p X-\p\phi \) e^{- iX+\phi}.
\nonumber
\eeq
The {\it ground ring} is constructed from products of the chiral
and antichiral operators. We consider the case of the theory compactified
at the self-dual radius $R=1$ in the absence of the cosmological constant
$\mu$. Then the ground ring contains
the following operators
\be
\CV_{j,m,\bm}=\CO_{j,m}\bar \CO_{j,\bm}.
\plabel{gring}
\ee
The ring has four generators
\be
a_1=y\by,\quad
a_2=w\bw,\quad
a_3=y\bw,\quad
a_4=w\by.
\plabel{genring}
\ee
These generators obey one obvious relation which determines the ground ring
of the $c=1$ theory
\be
a_1a_2-a_3a_4=0.
\plabel{relgr}
\ee

It has been shown \cite{WITTENGR}
that the symmetry algebra mentioned above is realized
on this ground ring as the algebra of diffeomorphisms of
the three dimensional cone \Ref{relgr} preserving the volume form
\be
\Theta={da_1da_2da_3\over a_3}.
\plabel{volcon}
\ee
Furthermore, it was argued that the inclusion of perturbations by marginal
operators deforms the ground ring to
\be
a_1a_2-a_3a_4=M(a_1,a_2),
\plabel{relgrdef}
\ee
where $M$ is an arbitrary function.
In particular, to introduce the cosmological constant, one should take
$M$ to be constant. As a result, one removes
the conic singularity and obtains a smooth manifold
\be
a_1a_2-a_3a_4=\mu.
\plabel{relgrmu}
\ee

In the limit of uncompactified theory only the generators $a_1$ and $a_2$
survive.
The symmetry of volume preserving diffeomorphisms is reduced to
some abelian transformations plus
area preserving diffeomorphisms of the plane $(a_1,a_2)$ that leave
fixed the curve
\be
a_1a_2=\mu
\plabel{aazero}
\ee
or its deformation according to \Ref{relgrdef}.
The suggestion was to identify the plane $(a_1,a_2)$ with the
eigenvalue phase space of MQM. Namely, one has the following relations
\be
a_1=x+p, \qquad a_2 =x-p.
\plabel{idrin}
\ee
Then it is clear that equation \Ref{aazero} corresponds to
the Fermi surface of MQM. The mentioned abelian transformations
are associated with time translations. And the operators \Ref{Tdis}
are identified with $a_1^na_2^{\bn}$.

The sense of the operators $a_3$ and $a_4$ existing in the compactified
theory is also known. They correspond to the winding modes of strings
which we are going to consider in the next section.
The relation \Ref{relgrmu} satisfied by these operators
is very important because it shows
how to describe the theory containing both the tachyon and winding modes.
However, it has not been yet understood 
how to obtain this relation directly from MQM.

\newpage

\section{Compact target space and winding modes in MQM}

In the previous section we showed that the modes of
2D string theory in the linear dilaton background are described
by the collective excitations of the singlet sector of Matrix
Quantum Mechanics. If we compactify the target space of
string theory, then there appear additional states --- windings
of strings around the compactified dimension. In this section
we demonstrate how to describe these modes in the matrix language.

\subsection{Circle embedding and duality}

In matrix models the role of the target space of non-critical string
theory is played by the parameter space of matrices.
Therefore, to describe 2D string theory
with one compactified dimension one should consider MQM on a circle.
The partition function of such matrix model is given by the integral
\Ref{MQMint} where the integration is over a finite Euclidean
time interval $t\in [0,\beta]$ with the two ends identified.
The length of the interval is $\beta=2\pi R$ where $R$ is
the radius of the circle. The identification $t\sim t+\beta$
requires to impose the periodic boundary condition $M(0)=M(\beta)$.
Thus, one gets
\be
Z_N(R,g)=\int\limits_{M(0)=M(\beta)} \CD M(t)
\exp \[ -N\tr\int\limits_{0}^{\beta} dt\, \( \hf {\dot M}^2+V(M)\) \].
\plabel{MQMintR}
\ee

As usual, one can write the Feynman expansion of this integral. It is the
same as in \Ref{MQMexp} with the only difference that the propagator
should be replaced by the periodic one
\be
G(t)=\sum\limits_{m=-\infty}^{\infty} e^{-|t+m\beta|}.
\plabel{propR}
\ee
We see that for large $\beta$ the term $m=0$ dominates and
we return to the uncompactified case. But for finite $\beta$
the sum should be retained and this leads to important phenomena
related with the appearance of vortices on the discretized world sheet
\cite{Berez,KT,Vill,GRKLa}.

We mentioned in section \ref{chSTR}.\ref{comwin} that the
$c=1$ string theory compactified at radius $R$ is T-dual to the same
theory at radius $1/R$. 
Does this duality appear in the sum over discretized surfaces regularizing
the sum over continuous geometries? To answer this question, we
perform the duality transformation of the Feynman expansion of
the matrix integral \Ref{MQMintR} 
\be
F=\sum\limits_{g=0}^{\infty}
N^{2-2g}
\sum\limits_{{\rm connected} \atop {\rm diagrams} \ \Gamma_g}
\lambda^{V}
\prod\limits_{i=1}^V \int\limits_{0}^{\beta}dt_i
\prod\limits_{\langl ij\rangl}
\sum\limits_{m_{ij}=-\infty}^{\infty} e^{-|t_i-t_j+\beta m_{ij}|},
\plabel{MQMexpR}
\ee
where we have chosen for simplicity the cubic potential \Ref{mpotc}.
The duality transformation is obtained applying the Poisson formula
to the propagator \Ref{propR}
\be
G(t_i-t_j)={1\over \beta} \sum\limits_{k_{ij}=-\infty}^{\infty}
e^{i{2\pi  \over \beta}k_{ij}(t_i-t_j)}\tilde G(k_{ij})=
{1\over \beta} \sum\limits_{k_{ij}=-\infty}^{\infty}
e^{i{2\pi  \over \beta}k_{ij}(t_i-t_j)}
{2\over 1+\({2\pi\over \beta} k_{ij}\)^2}.
\plabel{tfGG}
\ee
The substitution of \Ref{tfGG} into \Ref{MQMexpR} allows to integrate
over $t_i$ which gives the momentum conservation constraint at each
vertex
\be
k_{ij_1}+k_{ij_2}+k_{ij_3}=0.
\plabel{condkkk}
\ee
This reduces the number of independent variables from $E$ to $E-V+1$.
(One additional degree of freedom appears due to the zero mode which is
canceled in all $t_i-t_j$.) By virtue of the Euler theorem this equals to
$L-1+2g$. According to this, we attach a momentum $p_I$ to each
elementary loop (face) of the graph (one of $p_I$ is fixed) and
define remaining $2g$ variables as momenta $l_a$
running along independent non-contractable loops.
Thus, one arrives at the following representation
\be
F=\sum\limits_{g=0}^{\infty}
\({N\beta \over \lambda^2}\)^{2-2g} \!\!\!\!\!\!
\sum\limits_{{\rm connected} \atop {\rm diagrams} \ \tilde\Gamma}
 \!\!\({\lambda^2\over\beta }\)^L
\(\prod\limits_{I=1}^{L-1} \sum\limits_{p_I=-\infty}^{\infty}\)
\(\prod\limits_{a=1}^{2g} \sum\limits_{l_a=-\infty}^{\infty}\)
\prod\limits_{\langl IJ\rangl}
\tilde G\(p_I-p_J+\sum\limits_{a=1}^{2g}l_a\e^a_{IJ}\),
\plabel{MQMexpRd}
\ee
where the sum goes over the dual graphs (triangulations) with $L$ dual
vertices and we introduced the matrix $\e^a_{IJ}$ equal
$\pm 1$ when a dual edge
$\langl IJ\rangl$ crosses an edge belonging to $a$th non-contractable
cycle (the sign depends on the mutual orientation)
and zero otherwise.

The transformation \Ref{tfGG} changes $R\to 1/R$ which is seen from the form
of the propagators. But the result \Ref{MQMexpRd} does not seem to
be dual to the original representation \Ref{MQMexpR}. Actually, at the
spherical level, instead of
describing a compact target space of the inverse radius, it corresponds
to the embedding into the discretized real line with lattice
spacing $1/R$ \cite{KLEBANOV}. This is natural because the variables $p_I$
live in the momentum space of the initial theory which is discrete.

Thus, even in the continuum limit, the sum over discretized surfaces
embedded in a circle cannot be identical to its continuum analog.
The reason is that it possesses additional degrees of freedom which
are ignored in the naive continuum limit.
These are the vortex configurations. Indeed, in the continuum geometry
the simplest vortex of winding number $n$ is described by the field
$X(\theta)=nR\theta$ where $\theta$ is the azimuth angle.
However, this is a singular configuration and should be disregarded.
In contrast, on a lattice the singularity is absent and such configurations
are included into the statistical sum.
For example, in the notations of \Ref{MQMexpR} the number
of vortices associated with a face $I$ is given by
\be
w_I=\sum\limits_{\langl ij\rangl \in I}m_{ij}.
\plabel{numvor}
\ee
It is clear that it coincides with the number of times the string is
wrapped around the circle. In other words, the vortices are world sheet
realizations of windings in the target space. Thus, MQM with compactified
time intrinsically contains winding string configurations. But just due to
this fact, it fails to reproduce the partition function of
compactified 2D string theory.

It is clear that to obtain the sum over continuous
surfaces possessing selfduality one should somehow exclude
the vortices. This can be done restricting the sum over $m_{ij}$
in \Ref{MQMexpR}. The distribution $m_{ij}$ can be seen as an
abelian gauge field defined on links of a graph. Then the quantity
\Ref{numvor} is its field strength. We want that this strength vanishes.
With this condition only the ``pure gauge'' configurations of $m_{ij}$ are
admissible. They are represented as
\be
m_{ij}=m_i-m_j+\sum\limits_{a=1}^{2g}\tilde\e^a_{ij}\tilde l_a,
\plabel{gaugem}
\ee
where integers $\tilde l_a$ are associated with non-contractable
loops of the dual graph and $\tilde\e^a_{ij}$ is the matrix dual
to $\e^a_{IJ}$.
If we change the sum over all $m_{ij}$ by the sum over these
pure gauge configurations, the free energy \Ref{MQMexpR} is rewritten as
follows
\be
\Fm=\beta \sum\limits_{g=0}^{\infty}
N^{2-2g} \!\!\!\!
\sum\limits_{{\rm connected} \atop
{\rm diagrams} \ \Gamma_g}
\!\!\!\! \lambda^{V}
\prod\limits_{i=1}^{V-1} \int\limits_{-\infty}^{\infty}dt_i
\(\prod\limits_{a=1}^{2g} \sum\limits_{\tilde l_a=-\infty}^{\infty}\)
\prod\limits_{\langl ij\rangl}
\exp\left[-\large|t_i-t_j+
\beta\sum\limits_{a=1}^{2g}\tilde\e^a_{ij}\tilde l_a
\large| \right],
\plabel{MQMR}
\ee
where the overall factor $\beta$ arises from the integration over the zero
mode. The sum over $m_i$ resulted in the extension of
the integrals over $t_i$ to the whole line.
Due to this the dual transformation gives rise to
integrals over momenta rather than discrete sums. Repeating the steps
which led to \Ref{MQMexpRd} and renormalizing the momenta, one obtains
\beq
& \Fm={2\pi\over R}\sum\limits_{g=0}^{\infty}
\({N\beta \over \lambda^2}\)^{2-2g} \!\!\!\!\!\!
\sum\limits_{{\rm connected} \atop {\rm diagrams} \ \tilde\Gamma}
\!\! \({\lambda^2\over 4\pi^2 }\)^L
\(\prod\limits_{I=1}^{L-1} \int\limits_{-\infty}^{\infty} dp_I\) \times &
\nonumber \\
& \times \(\prod\limits_{a=1}^{2g} \sum\limits_{l_a=-\infty}^{\infty}\)
\prod\limits_{\langl IJ\rangl}
{2\over 1+{1\over 4\pi^2}\(p_I-p_J+
{2\pi\over R}\sum\limits_{a=1}^{2g}l_a\e^a_{IJ}\)^2}. &
\plabel{MQMRd}
\eeq

The only essential difference between two representation \Ref{MQMR}
and \Ref{MQMRd} is the propagator. However, the universality of
the continuum limit implies that the results in the macroscopic scale
do not depend on it. Moreover, if we choose the Gaussian propagator,
which follows from the usual Polyakov action,
its Fourier transform coincides with the original one.
Due to this one can neglect this discrepancy. Then the two representations
are dual to each other with the following matching of the arguments
\be
R \to 1/R, \qquad N\to RN.
\plabel{argd}
\ee
Thus, the exclusion of vortices allowed to make the sum over discretized
surfaces selfdual and they are those degrees of freedom that are responsible
for the breaking of this duality in the full matrix integral.

We succeeded to identify and eliminate the vortices in the sum over
discretized surfaces. How can this be done directly in the compactified
Matrix Quantum Mechanics defined by the integral \Ref{MQMintR}?
In other words, what matrix degrees of freedom describe the vortices?
Let us see how MQM in the Hamiltonian formulation
changes after compactification. The partition function is represented by
the trace in the Hilbert space of the theory of the evolution operator
as in \Ref{ZHMQM}
\be
Z_N(R)=\Tr e^{-{\beta\over\hb} \hat\HMQM}.
\plabel{ZHMQMR}
\ee
Now the time interval coincides with $\beta$ which
can also be considered as the inverse temperature. It is finite
so that one should consider the finite temperature partition function.
This fact drastically complicates the problem because one should take into
account the contributions of all states and not only the ground state.
In particular, all representations of the SU(N) global symmetry group
come to the game. The states associated with these representations are those
additional states which appear in the compactified theory.
Therefore, it is natural to expect that they correspond
to vortices on the discretized world sheet or windings of the string
\cite{GRKLa,GRKLb}.

The first check of this expectation which can be done is to verify
the duality \Ref{argd}. Since we expect vortices only in the non-singlet
representations of SU(N), to exclude them one should restrict
oneself to the singlet sector as in the uncompactified theory.
Then we still have a powerful description in terms of free fermions.
In the double scaling limit, we are interested in, the fermions
move in the inverse oscillator potential.
Thus we return to the problem stated by
equation \Ref{SEQdbs}. However, the presence of a finite temperature
gives rise to a big difference with the previous situation.
Due to the thermal fluctuations, the Fermi surface cannot be defined
in the ensemble with fixed number of particles $N$.
The solution is to pass to the grand canonical ensemble with
the following partition function
\be
\CZ(\mu,R)=\sum\limits_{N=0}^{\infty}e^{-\beta \mu N}Z_N(R).
\plabel{gcZ}
\ee
The chemical potential $\mu$ is exactly the Fermi level which
is considered now as the basic variable while $N$ becomes an operator.
The grand canonical free energy $\CF=\log\CZ$ can be expressed through the
density of states. In the singlet sector the relation reads as follows
\be
\CFs=\int\limits_{-\infty}^{\infty} d\e\, \rho(\e)
\log\left( 1+e^{-\beta(\e+\mu)}\right),
\plabel{gcF}
\ee
where the energy $\e$ is rescaled by the Planck constant to be of the same
order as $\mu$. The compactification does not affect the density
which is therefore given by equation \Ref{MQMdens}. There is a nice
integral representation of this formula
\be
\rho(\e)={1\over 2\pi}\Re\int\limits_{\Lambda^{-1}}^{\infty}
d\tau\, {e^{i\e\tau} \over 2\sinh{\tau\over 2}}
\Rightarrow {\p \rho(\e)\over \p \e}=
-{1\over 2\pi}\Im\int\limits_{0}^{\infty}
d\tau\, e^{i\e\tau}\, {\tau/2\over \sinh{\tau/ 2}}.
\plabel{intdens}
\ee
Taking the first derivative makes
the integral well defined and allows to remove the cut-off.
By the same reason, let us consider the third derivative of the free energy
\Ref{gcF} with respect to $\mu$. It can be written as
\be
{\p^3\CFs\over \p\mu^3}=-\beta\int\limits_{-\infty}^{\infty} d\e\,
{\p^2\rho\over\p\e^2}
{1\over  1+e^{\beta(\e+\mu)}}.
\plabel{gcFmu}
\ee
Then the substitution of \Ref{intdens} and taking the integral over $\e$
by residues closing the contour in the upper half plane gives \cite{GRKLa}
\be
{\p^3\CFs\over \p\mu^3}={\beta\over 2\pi}\Im\int\limits_{0}^{\infty} d\tau\,
e^{-i\mu\tau}{{\tau\over 2}\over \sinh{\tau\over 2}} \,
{{\pi\tau\over\beta}\over \sinh{\pi\tau\over \beta}}.
\plabel{gcFthree}
\ee
This representation possesses the explicit duality symmetry
\be
R\to 1/R, \qquad \mu\to R\mu,
\plabel{dsym}
\ee
where one should take into account that $\p^3_\mu\CFs\to R^{-3}\p^3_\mu\CFs$.
Thus, the grand canonical free energy of the singlet sector of
MQM compactified on a circle is indeed selfdual.
This fact can be verified also from the expansion of the free energy
in $1/\mu$. The result reads \cite{GRKLa}
\be
\CFs(\mu,R)=
 -{R\over 2}\mu^2\log\mu -{1\over 24}\(R+{1\over R}\)\log\mu+
\sum\limits_{n=1}^{\infty}\fs_{n+1}(R)(\mu\sqrt{R})^{-2n},
\plabel{FRexp}
\ee
where the coefficients are selfdual finite series in $R$
\be
\fs_n(R)={2^{-2n-1}(2n-2)! \over n-1}
\sum\limits_{k=0}^{n} |2^{2k}-2| |2^{2(n-k)}-2|
{|B_{2k}||B_{2(n-k)}|\over (2k)![2(n-k)]!} R^{n-2k}.
\plabel{fnR}
\ee

To return to the canonical ensemble, one should take the Laplace transform
of $\CF(\mu)$. But already analyzing equation \Ref{gcZ}, one can conclude
that the canonical free energy is also selfdual. This is because
$\beta N\mu$ is selfdual under the simultaneous change of $R$, $\mu$ and
$N$ according to \Ref{dsym} and \Ref{argd}.
The same result can be obtained by the direct calculation which shows
that the expansion of the canonical free energy in
${1\over \Delta}\log\Delta$
differs from the expansion of the grand canonical one in $1/\mu$
only by the sign of the first term
\cite{GRKLa,KLEBANOV}. These results confirm the expectation that
the singlet sector of MQM does not contain vortices and that the latters
are described by higher SU(N) representations.

\subsection{MQM in arbitrary representation: Hamiltonian analysis}
\label{MQMhrep}

We succeeded to describe the partition function of the
compactified MQM in the singlet sector which does not contain
winding excitations of the corresponding string theory.
Also it is possible to calculate
correlation functions of the tachyon modes in this case \cite{KLL}.
However, if we want to understand the dynamics of windings,
we should study MQM in the non-trivial SU(N) representations \cite{BULKA}.

The dynamics in the sector of the Hilbert space corresponding to
an irreducible representation $r$ is described by the projection
of the Hamiltonian \Ref{MQMhamop} on this subspace.
As in the case of the singlet representation, it is convenient
to redefine the wave function, which is now represented as a matrix,
by the Vandermonde determinant
\be
\PsiR_{ab}(x)=\Delta(x)\PhiR_{ab}(x).
\plabel{redfunr}
\ee
Then the action of the Hamiltonian $\hat\HMQM$ on the wave functions
$\PsiR_{ab}$ is given by the following matrix-differential operator
\be
\hat\HR_{ab}=\sum\limits_{c,d=1}^{\dR} P^{(r)}_{ac}\left[\,
\delta_{cd}\sum_{i=1}^N\left(
 -{\hb^2 \over 2}{\partial^2 \over \partial x_i{}^2}+V(x_i)\right)
 +{\hb^2 \over 4}\sum_{i\neq j}
 {\left(Q^{(r)}_{ij}\right)_{cd}\over (x_i-x_j)^2}\,\right]
P^{(r)}_{db},
\plabel{HR}
\ee
where
\be
  Q^{(r)}_{ij}\equiv\tauR_{ij}\tauR_{ji}+\tauR_{ji}\tauR_{ij}
\plabel{Qgen}
\ee
and we introduced the representation matrix of $u(N)$ generators
$\tau_{ij}\in u(N)$:
$\left(\tauR_{ij}\right)_{ab}=D^R_{ab}(\tau_{ij})$,
 which satisfies
\be
  [\tau_{ij},\tau_{kl}]
    =\delta_{jk}\tau_{il}-\delta_{il}\tau_{kj}.
\plabel{taucomm}
\ee
$P^{(r)}$ is the projector to the subspace
consisting of the wave functions that satisfy
\be
  \(\tauR_{mm}\PsiR\)_{ab}=
\sum\limits_{c=1}^{\dR}\left(\tauR_{mm}\right)_{ac}
\PsiR_{cb}=0, \quad m=1,\dots,N.
\plabel{mmcond}
\ee
The explicit form of $P^{(r)}$ is given by
\be
  P^{(r)}=\int\limits_0^{2\pi}\prod\limits_{m=1}^N
{d\theta_m\over 2\pi}\, e^{i\sum\limits_{m=1}^N\theta_m\tauR_{mm}}.
\plabel{proj}
\ee

The Hilbert structure is induced by the scalar product \Ref{MQMsc}
which has the following decomposition
\be
 \langle\Psi |\Psi'\rangle=\sum\limits_r{1\over \dR}
  \int\prod\limits_{i=1}^N dx_i \,\sum\limits_{a,b=1}^{\dR}
\overline{\Psi^{(r)}_{ab}(x)} \Psi'{}^{(r)}_{ab}(x).
\plabel{Rsc}
\ee
This shows that the full Hilbert space is indeed a direct sum
of the Hilbert spaces corresponding to irreducible representations of
SU(N). In the each subspace the scalar product is given by the
corresponding term in the sum \Ref{Rsc}.

Note that in the Schr\"odinger equation
\be
  i\hb{\p \PsiR_{ab}\over \p \ti}=
\sum\limits_{c=1}^{\dR}\hat\HR_{ac}\PsiR_{cb}.
\plabel{SHreq}
\ee
the last index $b$ is totally free and thus can be neglected.
In other words, one should consider the eigenvalue problem
given by the equation \Ref{SHreq} with the constraint \Ref{mmcond}
where the matrix $\PsiR_{ab}$ is replaced by the vector $\PsiR_a$ and
each solution is degenerate with multiplicity $\dR$.

\subsubsection{Example: adjoint representation}

Let us consider how the above construction works on the simplest
non-trivial example of the adjoint representation.
The representation space in this case is spanned by $\ads{ij}$.
The $u(N)$ generators act on these states as
\be
\tau_{ij}\ads{mn}\equiv |[\tau_{ij},\tau_{mn}]\rangle=
\delta_{jm}\ads{in}-\delta_{in}\ads{mj}
\plabel{tautau}
\ee
what means that their representation matrices are given by
\be
\(\taua_{ij}\)_{kl,mn}=
\delta_{ik}\delta_{jm}\delta_{ln}-\delta_{in}\delta_{jl}\delta_{km}.
\plabel{tauadj}
\ee
The operator $\Qadj_{ij}$ and the projector $\Padj$
are found to be
\beq
\(\Qadj_{ij}\)_{kl,mn}&=&
(\delta_{ik}+\delta_{jl})\delta_{km}\delta_{ln}-
(\delta_{ik}\delta_{jm}+\delta_{im}\delta_{jk})\delta_{kl}\delta_{mn}
+(i\leftrightarrow j),
\plabel{Qadj}   \\
\Padj_{kl,mn}&=&\delta_{kl}\delta_{nk}\delta_{ln}.
\plabel{Padj}
\eeq
The projector \Ref{Padj} leads to that only the diagonal components
of the adjoint wave function survive
\be
\Psia_{kl}=\sum\limits_{m,n=1}^N \Padj_{kl,mn}\Psia_{mn}=
\delta_{kl}\Psia_{kk}.
\plabel{psiproj}
\ee
Due to this it is natural to introduce the functions $\psa_k=\Psia_{kk}$
on which the operator $\Qadj_{ij}$ simplifies further
\be
\(\Qadj_{ij}\)_{kk,mm}=
2\left[ (\delta_{ik}+\delta_{jk})\delta_{km}-
(\delta_{ik}\delta_{jm}+\delta_{im}\delta_{jk})\right].
\plabel{Qadjj}
\ee
As a result, one obtains a set of $N$ coupled equations on $\psa_k(x,t)$
\be
\left\{i{\p\over \p t}+\sum_{i=1}^N\left(
 {1 \over 2N}{\partial^2 \over \partial x_i{}^2}-N V(x_i)\right)\right\}
\psa_k(x,t) -{1 \over N}\sum_{l(\neq k)}
 {\psa_k-\psa_l\over (x_k-x_l)^2}=0.
\plabel{Schradj}
\ee
This shows that instead by a system of free fermions
the adjoint representation is described by an interacting system
and we lose the integrability that allows to solve exactly
the singlet case.

\subsection{MQM in arbitrary representation: partition function}
\label{MQMpf}

Since the full Hilbert space is decomposed into the direct sum,
the partition function \Ref{ZHMQMR} of the compactified MQM
can be represented as the sum of contributions
from different representations
\be
Z_N(R)=\sum\limits_{r}\dR Z^{(r)}_N(R)=
\sum\limits_{r}\dR \Tr_{(r)}\, e^{-{\beta\over \hb}\hat \HR},
\plabel{ZNRr}
\ee
where the Hamiltonian is defined in \Ref{HR} and the trace is over the
subspace of the $r$th irreducible representation.
An approach to calculate the partition functions $Z^{(r)}_N$
was developed in \cite{BULKA}. It is based on the introduction of a new
object, the so called {\it twisted partition function}.
It is obtained by rotating the final state by a unitary transformation
with respect to the initial one
\be
Z_N(\Omega)=\Tr\( e^{-{\beta\over \hb}\hat \HMQM} \hat\Theta(\Omega)\),
\plabel{ZNRtw}
\ee
where $\hat\Theta(\Omega)$ is the rotation operator.
The partition functions in a given SU(N) representation can be obtained
by projecting the twisted partition function with help of
the corresponding character $\chi^{(r)}(\Omega)$
\be
Z^{(r)}_N=\int [d\Omega]_{SU(N)} \chi^{(r)}(\Omega)Z_N(\Omega).
\plabel{ZNRchar}
\ee
The characters for different representations are orthogonal to each other
\be
\int [d\Omega]_{SU(N)} \chi^{(r_1)}(\Omega^\dagger)
\chi^{(r_2)}(\Omega\cdot U) =\delta_{r_1,r_2}\chi^{(r_1)}(U)
\plabel{orchar}
\ee
and are given by the Weyl formula
\be
\chi^{(r)}(\Omega)\stackrel{\rm def}{=}\tr\left[ \DR(\Omega)\right]
={\mathop{\det}\limits_{i,j}\( e^{il_i\theta_j}\) \over
\Delta\( e^{i\theta}\) },
\plabel{defchar}
\ee
where $z_i=e^{i\theta_i}$ are eigenvalues of $\Omega$, $\Delta$ is
the Vandermonde determinant and the ordered set of integers
$l_1>l_2>\cdots >l_N$ is defined in terms of the components
of the highest weight $\{ m_k\}$: $l_i=m_i+N-i$
of the given representation.

Thus, the twisted partition function plays the role of the generating
functional for the set of partition functions $Z^{(r)}_N$.
The characters for the unitary group are well known objects.
Therefore, the main problem is to find $Z_N(\Omega)$.
It was done in the double scaling limit where the potential $V(x)$
becomes the potential of the inverse oscillator. Then it can be related
by analytical continuation to the usual harmonic oscillator potential
where the twisted partition function can be trivially found.
The derivation is especially simple when one uses the matrix
Green function defined by the following initial value problem
\be
\left\{ {\p\over \p\beta}-\hf\tr\( {\p^2\over \p M^2}-
\omega^2M^2\)\right\}G(\beta,M,M')=0,
\quad
G(0,M,M')=\delta^{(N^2)}(M-M').
\plabel{fGreen}
\ee
The solution is well known to be
\be
G(\beta,M,M')=\( {\omega\over 2\pi \sinh(\omega\beta)}\)^{\hf N^2}
\exp\left[-{\omega\over 2}\coth(\omega\beta)\tr(M^2+M^{'2})+
{\omega\over \sinh(\omega\beta)}\tr(MM') \right].
\plabel{Green}
\ee
It is clear that the twisted partition function is obtained from
this Green function as follows
\be
Z_N(\Omega)=\int dM \, G(\beta,M,\Omega^{\dagger}M\Omega).
\plabel{ZNOG}
\ee
One can easily perform the simple Gaussian integration over $M$ and
find the following result
\be
Z_N(\Omega;\omega)=
2^{-\hf N^2}\(2\sinh^2{\omega\beta\over 2}\)^{-N/2}\prod\limits_{i>j}
{1\over \cosh(\omega\beta)-\cos(\theta_i-\theta_j)}.
\plabel{ZNOo}
\ee
The answer for the inverse oscillator is obtained by the analytical
continuation to the imaginary frequency $\omega\to i$.
It can be also represented (up to $(-1)^{N/2}$) in the following form
\be
Z_N(\Omega)=q^{\hf N^2}\prod\limits_{i,j=1}^N
{1\over 1-qe^{i(\theta_i-\theta_j)}}.
\plabel{ZNO}
\ee
where $q=e^{i\beta}$. Remarkably, the partition function \Ref{ZNO}
depends only on the eigenvalues of the twisting matrix $\Omega$.
Due to this, the integral \Ref{ZNRchar} is rewritten as follows
\be
Z^{(r)}_N={1\over N!}\int \limits_0^{2\pi}
\prod\limits_{k=1}^N {d\theta_k\over 2\pi}
\left|\Delta(e^{i\theta})\right|^2 \chi^{(r)}(e^{i\theta})Z_N(\theta).
\plabel{ZNRth}
\ee

In fact, it is not evident that the analytical continuation of
the results obtained for the usual oscillator gives the correct
answers for the inverse oscillator. The latter is complicated
by the necessity to introduce a cut-off since otherwise
it would represent an unstable system. Because of that the analytical
continuation should be performed in a way that avoids the problems
related with arising divergences.
In this respect, the presented derivation is not rigorous.
However, the validity of the final result \Ref{ZNO} was confirmed
by the reasonable physical conclusions which were derived relying on it.
Besides, in \cite{BULKA} an alternative derivation
based on the density of states, which are eigenstates of the inverse
oscillator Hamiltonian with the twisted boundary conditions,
was presented. It led to the same formula as in \Ref{ZNO}.

\subsection{Non-trivial SU(N) representations and windings}
\label{MQMhrw}

The technique developed in the previous paragraph allows to study
the compactified MQM in the non-trivial representations in detail.
In particular, it was shown \cite{BULKA} that the partition function
associated with some representation of SU(N)
corresponds to the sum over surfaces in the presence of pairs
of vortices and anti-vortices of charge defined by this representation.
In terms of 2D string theory this means that $\log Z^{(r)}_N$
gives the partition function of strings among which there are
strings wrapped around the compactified
dimension $n$ times in one direction and the same number of strings
wrapped $n$ times in the opposite one.

This result can be established considering the diagrammatic expansion of
$Z^{(r)}_N$. The expansion is found using a very important fact
that the non-trivial representations are associated with correlators
of matrix operators which are traces of matrices taken in different
moments of time. For example, the two-point correlators describe
the propagation of states belonging to the adjoint representation
\be
\langl \tr \( e^{\alpha_1M(0)}e^{\alpha_2M(\beta)}\)\rangl=
\sum\limits_{i,j=1}^N
\langl 0| e^{\alpha_1 x_i} \(e^{-{\beta\over \hb}\hat \Hadj}\)_{ij}
e^{\alpha_2 x_j}|0\rangl.
\plabel{adjcor}
\ee
The diagrammatic expansion for the correlators is known and it gives
an expansion for the partition functions $Z^{(r)}_N$ after
a suitable identification of the legs of the Feynman graphs.

Another physical consequence of the previous analysis is that
there is a large energy gap between the singlet and the adjoint representations.
It was found to be \cite{GRKLb,BULKA}
\be
\delta=\CFadj-\CFs \mathop{\sim}\limits_{\mu\to \infty}
-{\beta\over 2\pi}\log(\mu/\Lambda).
\plabel{gap}
\ee
Due to this gap the contribution of few vortices to the partition
function is negligible and they seem to be suppressed.
However, the vortices have a large entropy related to the degeneracy
factor $\dR$ in the sum \Ref{ZNRr}. For the adjoint representation it
equals $d_{\adj}=N^2-1$. Therefore, there is a competition between
the two factors.
It leads to the existence of a phase transition when
the radius of compactification becomes sufficiently small \cite{GRKLb}.
Indeed, from \Ref{ZNRr} one finds
\be
Z_N(R)\approx \Zs_N(R)\(1+d_{\adj}e^{-\delta}+\cdots \)
\approx
\exp\left[ \Fs+{\rm const}\cdot N^2\, (\mu/\Lambda)^{R} \right].
\plabel{gappf}
\ee
Since $\Lambda\sim N$, we see that for large radii the second term
in the exponent is very small and is irrelevant with respect to the
first one. However, at $R_c=2$ the situation changes and now
the contribution of entropy dominates. Physically this means that
the vortex-antivortex pairs become dynamically more preferable and
populate densely the string world sheet. This effect is called the vortex
condensation and the change of behaviour at $R_c$
is known as the Berezinski--Kosterlitz--Thouless
phase transition \cite{Berez,KT}. For radii $R<R_c$ MQM does not describe
anymore the $c=1$ CFT. Instead it describes $c=0$ theory corresponding to
the pure two-dimensional gravity. This fact can be easily understood from
the MQM point of view because at very small radii we expect
the usual dimension reduction. The dimension reduction of MQM is
the simple one-matrix model which is known to describe pure gravity
as it was shown in section \ref{chMM}.\ref{sdissurf}.

Of course, this phase transition is seen also in the continuum formalism.
There it is related to the fact whether the operator creating
vortex-antivortex pairs is relevant or not. These operators were
introduced in \Ref{vort}. The simplest such operator has the form
\be
\int d^2\sigma\, e^{(R-2)\phi}\cos(R\tX).
\plabel{svor}
\ee
It is relevant if it decreases in the asymptotics $\phi\to\infty$.
This happens when $R<2$ exactly as it was predicted from the matrix model.
This gives one more evidence that we correctly identified the
winding modes of string theory with the
states of MQM arising in the non-trivial SU(N) representations.

\

\centerline{* \hspace{1cm} * \hspace{1cm} *}

\

We conclude that Matrix Quantum Mechanics successfully describes
2D string theory in the linear dilaton background. All excitations
of 2D string theory were identified with
appropriate degrees of freedom of MQM
and all continuum results were reproduced by the matrix model
technique. Moreover, MQM in the singlet sector represents an integrable
system which allowed to exactly solve the corresponding (tachyon) sector
of string theory.

Once the string physics in the linear dilaton background has been
understood and solved, it is natural to turn our attention to
other backgrounds. We have in our hands two main tools to obtain
new backgrounds: to consider either a tachyon or winding condensation
since their vertex operators are well known.
But the most interesting problem is string theory in
curved backgrounds. We know that 2D string theory does possess such
a background which describes the two-dimensional dilatonic black hole.
Therefore, we expect that the Matrix Quantum Mechanics is also able to
describe it and may be to provide its exact solution.



\chapter{Winding perturbations of MQM}
\label{chWIND}

\section{Introduction of winding modes}
\label{windmodel}

\subsection{The role of the twisted partition function}

In this chapter we consider one of the two possibilities to
change the string background in Matrix Quantum Mechanics.
We introduce non-perturbative sources of windings which perturb
the theory and give rise to a winding condensation.
From the previous chapter we know that the winding modes of string
theory are described in MQM by the non-trivial SU(N) representations.
However, the problem is that we do not have a control on them.
Namely, we do not know how to introduce a portion of windings of
charge 1, another portion of windings of charge 2, \etc\
In other words one should have an analog of the couplings $\tilde t_n$
which are associated with the perturbations by the vortex operators
in the CFT \Ref{PERTS}. Such couplings would allow to construct
a generating functional for all correlators of windings.

Actually, we already encountered one generating functional
of quantities related to windings. This was the twisted partition
function \Ref{ZNRtw}
which generated the partition functions of MQM in different representations.
It turns out that this is the object we are looking for because
it can be considered as the generating functional of vortex operators
with couplings being the moments of the twisting matrix.
In the following two sections we review the work \cite{KKK}
where this fact has been established and exploited to get a matrix model
for 2D string theory on the black hole background.

By definition the twisted partition function describes MQM with 
twisted boundary condition. Therefore, it can be represented by
the following matrix integral
\be
Z_N(\Omega)=\int\limits_{M(\beta)=\Omega^{\dagger}M(0)\Omega} \CD M(t)
\exp \[ -N\tr\int\limits_{0}^{\beta} dt\, \( \hf {\dot M}^2+V(M)\) \].
\plabel{MQMinto}
\ee
Hence it has the usual representation as the sum over Feynman diagrams
of the matrix model or as the sum over discretized two-dimensional
surfaces embedded in one-dimensional space. Since the target space is
compactified, we expect to obtain something like the expansion given
in \Ref{MQMexpR}. However, the presence of the twisting matrix introduces
new ingredients.

The only thing which can change is the propagator

\input{propagMO.pic}

\noindent
As it is seen from its definition through the two-point correlator,
it should satisfy the periodic boundary condition
\be
G_{ij,kl}(t+\beta,t')=
{\Omega^{\dagger}}_{ii'}\, G_{i'j',kl}(t,t')\, \Omega_{j'j}.
\plabel{Gperiod}
\ee
Due to this, the propagator is given by a generalization of the
simple periodic solution \Ref{propR}. In contrast to previous cases,
its index structure is not anymore described by the Kronecker symbols
but by the twisting matrix
\be
G_{ij,kl}(t,t')={1\over N}\sum\limits_{m=-\infty}^{\infty}
(\Omega^m)_{il}(\Omega^{-m})_{jk} e^{-|t-t'+m\beta|}.
\plabel{GOO}
\ee
As a result, the contraction of indices along each loop in a given graph
gives the factor $\tr\Omega^{w_I}$ where the integer field $w_I$
was defined in \Ref{numvor}. It is equal to the winding number of the loop
around the target space circle (see fig. \ref{planar}).
We will call this field {\it vorticity}.

\lfig{A piece of a discretized world sheet. The twisted boundary condition
associates with each loop of the dual graph a moment of
the twisting matrix corresponding to the winding number of the loop.}
{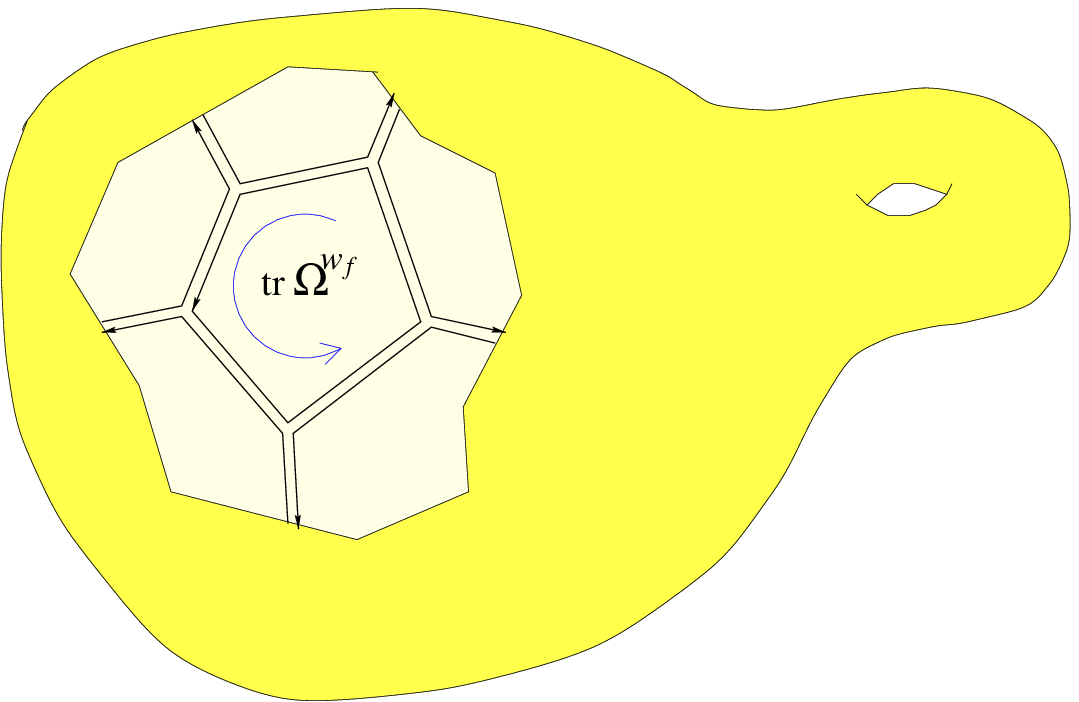}{8cm}{planar}

Thus, the only effect of the introduction of the twisted boundary condition
is that the factors $N$, which were earlier associated
with each loop, are replaced by the factors $\tr\Omega^{w_I}$.
The rest remains the same as in \Ref{MQMexpR}.
In particular, one has the sum over distributions $m_{ij}$. This
sum can be splitted into the sum over vorticities $w_I$ and
the sum over the ``pure gauge'' configurations \Ref{gaugem}.
The latter can be removed at the cost of extending
the integrals over time to the whole real axis.
We conclude that each term in the resulting sum is characterized by a
graph with particular distributions of times $t_i$ at vertices
and vorticity at faces (loops) $w_I$.
It enters the sum with the following coefficient
\be
N^{2-2g} \lambda^{V}
\prod\limits_{I=1}^{L}{tr\Omega^{w_I}\over N}
\prod\limits_{\langl ij\rangl} e^{-|t_i-t_j+\beta m_{ij}|},
\plabel{MQMexpO}
\ee

As usual, in the double scaling limit the sum over discretizations
becomes the sum over continuous geometries and
the twisted partition function \Ref{MQMinto} can be
interpreted as the partition function of 2D string theory including
the sum over vortex insertions.
The vortices of charge $m$ are coupled to the $m$th moment
of the twisting matrix so that the moments control the probability
to find vortices of a given vorticity.

To use this interpretation to extract some information concerning
a particular vortex configuration (for instance, to study
the dependence of the theory perturbed by vortices of charge 1
of the coupling constant), one should express the twisted partition
function as a function of moments $s_m=\tr \Omega^m$.
However, it turns out to be a quite difficult problem.
Therefore, one should find an alternative way to describe
the perturbed system.

\subsection{Vortex couplings in MQM}

The main problem with the twisted partition function is that its
natural argument is a matrix. Moreover, in the large $N$ limit
its size goes to infinity.
(Although $Z_N(\Omega)$ depends actually only on the eigenvalues,
as it was shown in section \ref{chMQM}.\ref{MQMpf},
this does not help much.)
On the other hand, usually one integrates over
matrices of a large size. For example, the partition functions of
MQM in different representations were represented as integrals
over the twisting matrix \Ref{ZNRchar}.
The measure of the integration was given by the characters of
irreducible representations. But the characters are not related
to the coupling constants of vortices in an explicit way.

We can generalize this construction and integrate the twisted partition
function with an arbitrary measure. Then the vortex coupling constants will
be associated with some parameters of the measure. Thus, the problem
can be reformulated as follows: what choice of the measure gives
the most convenient parameterization of the generating functional of
vortices?

The answer is as simple as it can be. Indeed, as usual, we require
from the measure the invariance under the unitary transformations
\be
\Omega \to U^{\dagger} \Omega U \quad (U^{\dagger} U=I).
\plabel{Oun}
\ee
Then, as in the one-matrix model, the most natural choice of
the measure is given by
the exponential of a potential. Since the matrix is unitary,
in contrast to 1MM, now both positive
and negative powers of the twisting matrix are allowed.
Thus, we define the following functional \cite{KKK}
\be
Z_N[\lambda]=\int [d\Omega]_{SU(N)}\,
\exp\(\sum\limits_{n\ne 0}\lambda_n\tr\Omega^n\)Z_N(\Omega).
\plabel{Zl}
\ee
Note that the parameters
$\lambda_n$ are coupled exactly to the moments $s_n$
of the twisting matrix playing the role of fugacities of vortices.
Therefore, the functional \Ref{Zl} is nothing else but the Legendre
transform of the twisted partition function considered as a function
of the moments.

This statement can be formulated more rigorously with help
of the following identity
\be
\int [d\Omega]_{SU(N)} \exp\(\sum\limits_{n\ne 0}\lambda_n\tr\Omega^n\)
=\exp\(\sum\limits_{n> 0}n\lambda_n\lambda_{-n}\),
\plabel{propio}
\ee
which is valid up to non-perturbative terms $O(e^{-N})$ provided
the couplings do not grow linearly in $N$.
This property of integrals over the unitary groups
shows that in the large $N$
limit the moments $s_n$ can be considered as independent variables
and the measure $[d\Omega]_{SU(N)}$ is expressed in a very simple way
through them. As a result, the generating function \Ref{Zl}
is written as
\be
Z_N[\lambda]=\int\limits_{-\infty}^{\infty}
\prod\limits_{n\ne 0}{d s_n\over \sqrt{\pi}}\,
e^{\sum\limits_{n\ne 0}\(\lambda_n s_n-{1\over 2|n|}s_ns_{-n}\)}Z_N[s].
\plabel{Zls}
\ee

The relation \Ref{propio} is a generating equation for
integrals of products of moments. Among them the following relation
is most important for us
\be
\int [d\Omega]_{SU(N)} \tr\Omega^n\,\tr\Omega^m =|n|\delta_{n+m,0}.
\plabel{twoio}
\ee
It helps to elucidate the sense of the couplings $\lambda_n$ which
is hidden in the diagrammatic representation of $Z_N[\lambda]$.
This representation follows from the expansion of
the twisted partition function if to perform the integration over
$\Omega$. From \Ref{twoio} one concludes that there will be three kinds
of contributions. The first one is a trivial factor given by the r.h.s of
\Ref{propio}. It comes from the coupling of two moments from
the measure in \Ref{Zl}. The second contribution arises when a
moment from the measure is coupled with
the factor ${1\over N}\tr\Omega^{w_I}$
in \Ref{MQMexpO} associated with a vortex of vorticity $w_I$.
It results in substitution
of the coupling $\lambda_{w_I}$ in place of the trace of the twisting
matrix. Thus, whenever it appears, $\lambda_n$ is always associated with
a vortex of winding number $n$. Hence, it plays the role of the coupling
constant of the operator creating the vortices.
Finally, there is the third contribution related with the coupling of two
moments from \Ref{MQMexpO}. However, it was argued that it vanishes
in the double scaling limit \cite{KKK}.

To summarize, the double scaling limit of the free energy of \Ref{Zl}
coincides with the partition function
of the $c=1$ theory perturbed by vortex operators with the coupling
constants proportional to $\lambda_n$. Thus, we obtain
a matrix model realization of the CFT \Ref{PERTS} with $t_n=0$.
Since the couplings $\lambda_n$ are explicitly introduced from
the very beginning, it is much easier to work with $Z_N[\lambda]$
than with the twisted partition function. Moreover, it turns out
that in terms of $\lambda$'s the system becomes integrable.

\subsection{The partition function as $\tau$-function of Toda hierarchy}

To reveal the relation of the partition function \Ref{Zl} to
integrable systems, one should do two things. First, one should pass
to the grand canonical ensemble
\be
\CZ_{\mu}[\lambda]=\sum\limits_{N=0}^{\infty}e^{-\beta \mu N}Z_N[\lambda].
\plabel{Zlgc}
\ee
One can observe that $-\beta\mu$ plays the role of the ``zero time''
$\lambda_0$ which appears if one includes $n=0$ into the sum in \Ref{Zl}.
Therefore, the necessity to use the grand canonical ensemble
goes in parallel with the change from $\Omega$ to $\lambda$'s
and it is natural in this context.

Second, we use the result \Ref{ZNO} for the twisted partition function
found in the double scaling limit \cite{BULKA}.
Combining \Ref{ZNO}, \Ref{Zl} and \Ref{Zlgc} and integrating out
the angular part of the twisting matrix, one obtains
\be
\CZ_{\mu}[\lambda]=\sum\limits_{N=0}^{\infty}{ e^{-\beta \mu N}\over N!}
\oint \prod\limits_{k=1}^N \( {dz_k\over 2\pi i z_k}
{e^{u(z_k)}\over q^{1/2}-q^{-1/2}}\)
\prod\limits_{i\ne j} {z_i-z_j\over q^{1/2}z_i-q^{-1/2}z_j},
\plabel{intZl}
\ee
where $q=e^{i\beta}$, $z_k$ are eigenvalues of $\Omega$, and
$u(z)=\sum\limits_n \lambda_n z^n$ is a potential associated with
the perturbation.
Initially, the eigenvalues belonged to the unit circle. But
due to the holomorphicity, the integrals in \Ref{intZl} can be
understood as contour integrals around $z=0$.
Finally, using the Cauchy identity
\be
{\Delta(x)\Delta(y)\over \prod\limits_{i,j}(x_i-y_j)}=
\mathop{\det}\limits_{i,j}\( {1\over x_i-y_j}\),
\plabel{Cauchy}
\ee
we rewrite the product of different factors as a determinant what gives
the following representation for the grand canonical partition function
of 2D string theory perturbed by vortices
\be
\CZ_{\mu}[\lambda]=\sum\limits_{N=0}^{\infty}{ e^{-\beta \mu N}\over N!}
\oint \prod\limits_{k=1}^N {dz_k\over 2\pi i }
\mathop{\det}\limits_{i,j}
\( {\exp\left[ \hf (u(z_i)+u(z_j))\right]
\over  q^{1/2}z_i-q^{-1/2}z_j}\).
\plabel{intZd}
\ee

Relying on this representation one can prove
that the grand canonical partition function coincides with
a $\tau$-function of Toda hierarchy \cite{KKK}.
In this case it is convenient to establish the equivalence
with the fermionic representation of $\tau$-function \Ref{glfer}.
We claim that if one chooses the matrix determining the operator
of $GL(\infty)$ rotation as follows
\be
A_{rs}=\delta_{r,s}q^{i\mu+r},
\plabel{AMM}
\ee
the $\tau$-function is given by
\be
\tau_l[t] =e^{-\sum\limits_{n>0} n t_n t_{-n}}\CZ_{\mu-il}[\lambda],
\plabel{tauz}
\ee
where the coupling constants are related to the Toda times through
\be
\lambda_n=2i\sin(\pi n R)\, t_n.
\plabel{tlam}
\ee
Indeed, with the matrix \Ref{AMM} the operator $\bfg$ is written as
\be
\bfg\equiv\exp(q^{i\mu}\hat A)=\exp\( e^{-\beta\mu}\oint {dz\over 2\pi i}
\psi(q^{-1/2}z)\psi^*(q^{1/2}z)\).
\plabel{gMM}
\ee
The expansion of the exponent gives the sum over $N$ and factors
$e^{-\beta\mu N}$ in \Ref{intZd}. Then in the $N$th term of the expansion
one should commute the exponents
$e^{H_{\pm}[t]}$ associated with perturbations between each other and
with $\hat A^N$. The former commutation gives rise to the trivial factor
appearing in \Ref{tauz}. It is to be compared with the similar contribution
to $Z_N[\lambda]$ coming from \Ref{propio}. The commutator with $\hat A^N$
is found using the relations \Ref{ferham}. As a result, one obtains
\be
\oint \prod\limits_{k=1}^N \({dz_k\over 2\pi i }
\exp\left[ \sum\limits_{n\ne 0}  (q^{n/2}-q^{-n/2}) t_n z_k^n \right]\)
\langl l\left|  \prod\limits_{k=1}^N \psi(q^{-1/2}z_k)\psi^*(q^{1/2}z_k)
\right|l \rangl.
\plabel{fact}
\ee
Comparing this expression with \Ref{intZd} we see the necessity to
redefine the coupling constants according to \Ref{tlam} to match
the potentials.
Finally, the quantum average in the vacuum of charge $l$ produces the
same determinant as in \Ref{intZd} and additional factor $q^{lN}$
(see \Ref{twofer}).
The latter leads to the shift of $\mu$ shown in \Ref{tauz}.

Actually, this result is not unexpected because, as we mentioned in the end of
section \ref{chMQM}.\ref{VerCorFun}, a similar result has been obtained for
the generating functional of tachyon correlators.
The tachyon and winding perturbations are related by T-duality.
Therefore, both the tachyon and winding perturbations of 2D string
theory should be described by the same $\tau$-function with $T$-dual
parameters.

Nevertheless, it is remarkable that one can obtain an explicit matrix
representation of this $\tau$-function which can be directly interpreted
in terms of discretized surfaces with vortices. Therefore, one can use
the powerful matrix technique to solve some problems which
may be inaccessible even by
methods of integrable systems. For example, while the tachyon and winding
perturbations are integrable when they are introduced separately,
the integrability disappears as only both of them are present.
In such situation the Toda hierarchy does not work anymore, but
the matrix description is still valid.

The Toda description can be used exploiting its hierarchy of equations.
To characterize their unique solution we should provide either a string
equation or an initial condition.
The string equation can be found in principle \cite{IK}
(and we will show how it appears in the dual picture of tachyon
perturbations), but it is not so evident.
In contrast, it is clear that the initial condition is given by the partition
function with vanishing coupling constants,
{\it i.e.}, without vortices. It corresponds
to the partition function of the compactified MQM in the singlet sector.
It is well known and its expansion is given by \Ref{FRexp}.
Thus, we have the necessary information to use the equations of
Toda hierarchy.

These equations are of the finite-difference type. Therefore,
usually one represents them as a series of partial differential equations.
We associated this expansion in section \ref{chMM}.\ref{TODA} with
an expansion in the Planck constant which is the parameter measuring
the lattice spacing. What is this parameter in our case?
On the string theory side one has the genus expansion.
The only possibility is to identify these two expansions.
From \Ref{FRexp} we see that the parameter playing the role
of the string coupling constant,
which is the parameter of the genus expansion, is $\gc\sim \mu^{-1}$.
Thus, one concludes that the role of the spacing parameter is played
by $\mu^{-1}$ and to get the dispersionless limit of Toda hierarchy
one should investigate the limit of large $\mu$.
Note that this conclusion is in the complete agreement with
the consequence of \Ref{tauz} that $\mu$ is associated
with the discrete charge of $\tau$-function.

\newpage

\section{Matrix model of a black hole}
\label{MMbh}

\subsection{Black hole background from windings}

Due to the result of the previous section, the Toda hierarchy
provides us with equations on the free energy as a function of $\mu$ and
the coupling constants.
Any result found for finite couplings $\lambda_n$ would already correspond
to some result in string theory with a non-vanishing condensate
of winding modes.
In section \ref{chSTR}.\ref{twback} we discussed that such winding
condensates do not have a local target space interpretation. In other words,
there is no special field in the string spectrum describing them.
Therefore, the effect of winding condensation should be seen in another
characteristics of the background: dilaton and metric.
Thus, it is likely that considering MQM with non-vanishing
$\lambda_n$, we actually
describe 2D string theory in a curved background.

Let us consider the simplest case when only $\lambda_{\pm 1}\ne 0$.
Without lack of generality one can take them equal
$\lambda_1=\lambda_{-1}\sim\lambda$.
Then the corresponding string theory is described by the so called
{\it Sine--Liouville CFT}
\be
\SSL={1\over4\pi}\int d^2 \sigma\,
\left[ (\partial X)^2+(\partial \phi)^2
-Q{\hat\CR}\phi + \mu e^{\gamma\phi}+\lambda e^{\rho\phi}\cos(R\tX) \right],
\plabel{SLCFT}
\ee
where $\tX$ is T-dual to the field $X(\sigma)$ which is compactified
at radius $R$. The requirement that the perturbations
are given by marginal operators, leads to the following conditions on the
parameters
\be
\gamma=-Q+\sqrt{Q^2-4}, \qquad \rho=-Q+\sqrt{R^2+Q^2-4}.
\plabel{SLpar}
\ee
The central charge of this theory is $c=2+6Q^2$.
Therefore, to get $c=26$, as always in matrix models,
one should take $Q=2$.

As we discussed above, 
the theory \Ref{SLCFT} with the vanishing cosmological constant $\mu$
was suggested to be dual to 2D string theory
in the black hole background described in section \ref{chSTR}.\ref{curback}.
The exact statement of this conjecture was presented in section 
\ref{chSTR}.\ref{conjFZZ}.
In particular, it was shown that the parameters of the model
should be identified with the level $k$ of the gauge group
as follows
\be
R=\sqrt{k}, \qquad Q={1\over \sqrt{k-2}}.
\plabel{bhpar}
\ee
The condition on $R$ comes from the matching of the asymptotic
radii of the cigar geometry describing the Euclidean black hole and
of the cylindrical target space of the Sine--Liouville CFT.
The value of $Q$ is fixed by matching the central charges.

Due to these restrictions on the parameters, there is only one
point in the parameter space of the two models where they intersect.
It corresponds to the following choice
\be
Q=2, \qquad R={3/ 2}, \qquad \mu=0.
\plabel{valpar}
\ee
Since for these values of the parameters the Sine--Liouville CFT
can be obtained as a matrix model constructed in the previous section,
on the one hand, and is dual to the coset CFT, on the other hand,
at this point we have a matrix model description of string theory in the
black hole background \cite{KKK}.
The string partition function is given by
the free energy of \Ref{Zl} or its grand canonical
counterpart \Ref{Zlgc} where one puts $\mu$ and all couplings except
$\lambda_{\pm 1}$ to zero as well as $R=3/2$.

This remarkable correspondence opens the possibility to study
the black hole physics using the matrix model methods.
Of course, the most interesting questions are related to
the thermodynamics of black holes. In particular, any theory of
quantum gravity should be able to explain the microscopic
origin of the black hole entropy and to resolve the information paradox.
One might hope that the matrix model will allow to identify
the fundamental degrees of freedom of this system to solve both these
problems.

\subsection{Results for the free energy}
\label{KAKsolut}

Before to address the question about the entropy, it is much more easy
to get some information about another thermodynamical quantity ---
the free energy. The grand canonical free energy is given by the logarithm
of the partition function \Ref{Zlgc} and, hence, by the logarithm of
the $\tau$-function. Thus, one can use the integrable structure
of Toda hierarchy to find it.

The problem is that we cannot work directly in the black hole point of the
parameter space. Indeed, it puts $\mu$ to zero
whereas the dispersionless limit
of Toda hierarchy, which allows to write differential equations on the free
energy, requires to consider large $\mu$. The solution is to study
the theory with a large non-vanishing $\mu$ treating $\lambda$ as
a perturbation. Then, one should try to make an analytical continuation
to the opposite region of small $\mu$. In the end, one should also fix
the radius of the compactification.

In fact, in the matrix model it is very natural to turn on the cosmological
constant $\mu$ and to consider an arbitrary radius $R$. The values
\Ref{valpar} (except $Q=2$) are not distinguished anyhow. Even $\mu=0$
is not the most preferable choice because there is another value
which is associated with a critical point where the theory acquires some
special properties (see below).
Moreover, to analyze thermodynamical issues, one should be able
to vary the temperature of the system
what means for the black hole to vary the radius $R$.
Thus, it is strange that only at the values
\Ref{valpar} MQM describes a black hole. What do other values
correspond to? It is not clear at the moment, but it would be quite natural
that they describe some deformation of the initial black hole background.
Therefore, we will keep $\mu$ and $R$ arbitrary in the most of calculations.

Let us use the Toda integrable structure to find the free energy
$\CF(\mu,\lambda)=\log \CZ_{\mu}(\lambda)$ where $t_{1}t_{-1}=\lambda^2$
and all other couplings vanish. Due to the winding number conservation,
it depends only on the product of two couplings and not on them separately.
The identification \Ref{tauz} allows to conclude that
the evolution along the first times
is governed by the Toda equation \Ref{todaqq}.
Since the shift of the discrete charge $l$ is equivalent to an imaginary
shift of $\mu$, in terms of the free energy the Toda equation becomes
\be
{\p^2 \CF(\mu, \lambda)\over \p t_1 \p t_{-1}}+
\exp\left[\CF(\mu+i,\lambda)+\CF(\mu-i,\lambda)-2\CF(\mu,\lambda)\right]=1.
\plabel{todaf}
\ee
Rewriting the finite shifts of $\mu$ as the result of action
of a differential operator, one obtains
\be
{1\over 4}\lambda^{-1}\p_{\lambda}\lambda\p_{\lambda}\CF(\mu, \lambda)+
\exp\left[-4\sin^2\(\hf{\p\over \p\mu}\)\CF(\mu,\lambda)\right]=1.
\plabel{todafd}
\ee

The main feature of this equation is that it is compatible with the scaling
\be
\lambda\sim \mu^{2-R\over 2}
\plabel{scal}
\ee
which can be read off from the Sine--Liouville
action \Ref{SLCFT}. Due to this, the free energy can be found order by
order in its genus expansion which has the usual form of the expansion in
$\mu^{-2}$
\be
\CF(\mu,\lambda)=\lambda^2+
\mu^2\left[-{R\over 2}\log\mu+\tlf_0(\zeta)\right]
+\left[ - 
{R+R^{-1} \over 24}\log\mu+\tlf_1(\zeta)\right]+
\sum\limits_{g=2}^{\infty}\mu^{2-2g}\tlf_g(\zeta),
\plabel{genexpCF}
\ee
where $\zeta=(R-1)\lambda^{2}\mu^{R-2}$ is a dimensionless parameter.
The first term is not universal and can be ignored. It is intended to
cancel $1$ in the r.h.s. of \Ref{todafd}.
The coefficients $\tlf_g(\zeta)$ are smooth functions near $\zeta=0$.
The initial condition given by \Ref{FRexp} fixes them at the origin:
$\tlf_0(0)=\tlf_1(0)=0$ and $\tlf_g(0)= R^{1-g} \fs_{g}(R)$ with
$\fs_{g}(R)$ from \Ref{fnR}.

It is clear that one can redefine the coefficients in such a way
that the genus expansion will be associated with an expansion in
$\lambda$. More precisely, if one introduces the following scaling variables
\be
w=\mu\xi, \qquad \xi=(\lambda\sqrt{R-1})^{-{2\over 2-R}},
\plabel{scp}
\ee
the genus expansion of the free energy reads
\be
\CF(\mu,\lambda)=\lambda^2+
\xi^{-2}\left[{R\over 2}w^2\log\xi+ f_0(w)\right]
+\left[ 
{R+R^{-1} \over 24}\log\xi+f_1(w)\right]+
\sum\limits_{g=2}^{\infty}\xi^{2g-2}f_g(w).
\plabel{gnexpCF}
\ee
Thus, the string coupling constant is identified as $\gc\sim \xi$.
This is a simple consequence of the scaling \Ref{scal}.
It is clear that the dimensionless parameters $\zeta$ and $\w$ are
inverse to each other: $\zeta=w^{R-2}$.
We included the factor $(R-1)$ in the definition of the scaling variables
for convenience as it will become clear from the following formulae.
This implies that $R>1$. It is the region we are interested in
because it contains the black hole radius. However,
the final result can be presented in the form avoiding this restriction.

Plugging \Ref{gnexpCF} into \Ref{todafd}, one obtains a system of
ordinary differential equations for $f_g(w)$. Each equation is associated
with a definite genus. At the spherical level there is a closed
non-linear equation for $f_0(w)$
\be
{R-1\over (2-R)^2}\(w\p_w-2\)^2 f_0(w)+e^{-\p^2_w f_0(w)}=0.
\plabel{eqfo}
\ee
Its solution is formulated as a non-linear algebraic equation for
$X_0=\p^2_w f_0$ \cite{KKK}
\be
w=\efo-\ego.
\plabel{wf}
\ee
In terms of the solution of this equation, the spherical free energy
itself is represented as
\be
\CF_0(\mu,\lambda)=\hf\mu^2\(R\log\xi+X_0\)+
\xi^{-2}\( {3\over 4}{R}\eao -{R^2-R+1\over R-1}\ebo+
{3\over 4}{R\over R-1}\eco \).
\plabel{sphCF}
\ee

Let us rewrite the equation \Ref{wf} in terms of the susceptibility
$\chi=\p_{\mu}^2\CF$, more precisely, in terms of its spherical part
\be
\chi_0=R\log\xi+X_0.
\plabel{sphchi}
\ee
The result can be written in the following form
\be
\mu e^{\oR\chi_0}+(R-1)\lambda^2 e^{{2-R\over R}\chi_0}=1.
\plabel{eqchi}
\ee
From this it is already clear why we included the factor $(R-1)$ in
the scaling variables.
Note that this form of the answer is not restricted to $R>1$
and it is valid for all radii. However, it shows that the limit $\mu\to 0$
exists only for $R>1$. Otherwise the susceptibility becomes imaginary.
For $R>1$, a critical point where the equation \Ref{eqchi} does not have
real solutions anymore also exists and it is given by
\be
\mu_c=-(2-R)(R-1)^{R\over 2-R}\lambda^{2\over 2-R}.
\plabel{crp}
\ee
This critical value of the cosmological constant was found previously
by Hsu and Kutasov \cite{HSU}. The result \Ref{crp}
shows that the vanishing value of $\mu$
is inaccessible also for $R>2$. Actually, in this region the situation
is even more dramatic because $\xi$ becomes an increasing
function of $\lambda$
and the genus expansion breaks down in the limit of large $\lambda$.
This is related to the fact that the vortex perturbation in \Ref{SLCFT}
is not marginal for $R>2$ and is non-renormalizable because it
grows in the weak coupling region.
Thus, the analytical continuation to the black hole point
$\mu=0$ is possible only in the finite interval of radii $1<R<2$.
Fortunately, the needed value $R=3/2$ belongs to this interval
and the proposal survives this possible obstruction.

The equation \Ref{eqchi} can be used to extract expansion of the spherical
free energy either in $\lambda$ or in $\mu$. In particular, the former
expansion reproduces the $2n$-point correlators of vortex operators
\be
\langl \tV_R^n\tV_{-R}^n\rangl_0= -n!\mu^2 R^{2n+1}\((1-R)\mu^{R-2}\)^n
{ \Gamma(n(2-R)-2)\over \Gamma(n(1-R)+1)}.
\plabel{cMoore}
\ee
For small values of $n$ they have been found and for other values
conjectured by Moore in \cite{MOORE}.
These correlators should coincide with
the coefficients in the $\lambda$-expansion
of $\CF_0$ because they can be organized into the partition function
as follows
\be
\CF(\mu,\lambda)=\langl e^{\tl \tV_R+\tl \tV_{-R}} \rangl_{\rm gr.c.} =
\CF(\mu,0)+
\sum\limits_{n=1}^{\infty}{\tl^{2n}\over (n!)^2 }
 \langl \tV_R^n\tV_{-R}^n\rangl_{\rm gr.c.},
\plabel{pfMoore}
\ee
where the expectation value is evaluated in the grand canonical ensemble.
The comparison shows that the correlators do coincide if one identifies
$\lambda =R\tl$.\footnote{In fact, the correlators \Ref{cMoore} differ
by sign from the coefficients in the expansion of $\CF_0$. This is
related to that $\CF_0$ is the grand canonical free energy whereas
the paper \cite{MOORE} considered the canonical ensemble.}
This indicates that the correct relation between the Toda times $t_n$
and the CFT coupling constants $\tilde t_n$ in \Ref{PERTS} is the
following
\be
t_n =iR\tilde t_n, \qquad t_{-n} =-iR\tilde t_{-n}, \quad (n>0).
\plabel{ttn}
\ee
The appearance of the factor $R$ will be clear when we consider
the dual system with tachyon perturbations.

The black hole limit $\mu=0$ corresponds to $X_0=0$. Then the free energy
\Ref{sphCF} becomes
\be
\CF_0(0,\lambda)=-{(2-R)^2\over 4(R-1)}(\sqrt{R-1}\lambda)^{4\over 2-R}.
\plabel{bhCF}
\ee
Note that at the point $R=3/2$ the free energy is proportional to
an integer power of the coupling constant $\sim\lambda^8$.
Therefore, the spherical contribution seems to be non-universal.
However, for general $R$ it is not so and this is crucial for
thermodynamical issues.

At the next levels the equations obtained from \Ref{todafd}
form a triangular system so that
the equation for $f_g(w)$ is linear with respect to this function
and contains all functions of lower genera as a necessary input \cite{KKK}
\be
\left( {R-1\over (2-R)^2}\(w\p_w+2g-2\)^2 -e^{-X_0}\p_w^2\right) f_g=
-\left[\xi^{2-2g}\exp\( -4\sin^2 \( {\xi\over 2}\p_w\)
\sum\limits_{k=0}^{g-1} \xi^{2k-2} f_k \)\right]_0,
\plabel{eqfn}
\ee
where $[\cdots]_0$ means the terms of zero order in $\xi$-expansion.
Up to now, only the solution for the genus $g=1$ has been obtained
\cite{KKK}
\be
\CF_1(\mu,\lambda)={R+R^{-1} \over 24}\(\log\xi+\oR X_0\)-
{1\over 24}\log\( 1-(R-1)e^{{2-R\over R} X_0} \).
\plabel{torCF}
\ee
For the genus $g=2$ the differential operator of the second order contains
4 singular points and the solution cannot be presented in terms
of hypergeometric functions \cite{AKnon}.

\subsection{Thermodynamical issues}
\label{bhterm}

An attempt to analyze thermodynamics of the black hole relying on
the result \Ref{bhCF} was done in \cite{KKK} and \cite{KZ}.
However, no definite conclusions have been obtained. First of all,
it is not clear whether the free energy of the black hole vanishes or not.
The ``old'' analysis in the framework of dilaton gravity predicts
that it should vanish \cite{GP,NP}. However, the matrix model
leads to the opposite conclusion. One can argue \cite{KKK}
that since for $R=3/2$ the leading term is non-universal, it can be
thrown away giving the vanishing free energy. But if
the matrix model realizes string theory in a black hole background
for any $R$, this would be quite unnatural.

Moreover, even in the framework of the dilaton
gravity the issue is not clear. The value of the free energy depends
on a subtraction procedure which is to be done to regularize diverging
answers. There is a natural reparameterization invariant procedure
which leads to a non-vanishing free energy \cite{KZ}
in contradiction with the previous results.
However, in this case it is not clear how to get the correct expressions
for the mass and entropy.

The related problem which prevents to clarify the situation is what quantity
should be associated with the temperature. At the first glance this is the
inverse radius.
In particular, if one follows this idea and uses the reparameterization
invariant subtraction procedure, one arrives at reasonable results but the
mass of the black hole differs by factor 2 from the standard
expression \Ref{mass} \cite{SAnon}.
Note that the possibility to get this additional factor was emphasized
in \cite{LVA}. It is related to the definition of energy in dilaton
gravity.

However, from the string point of view the radius is always fixed.
Therefore, in the analysis of the black hole thermodynamics,
the actual variations of the temperature were associated
with the position of a ``wall'' which is introduced to define the subtraction
\cite{GP,NP,KZ}.
But there is no corresponding quantity in the matrix model.

In the next chapter, relying on the analysis of a dual system,
we will argue that
it is $R^{-1}$ that should be considered as the temperature.
Also we will shed some light on the puzzle with the free energy.

\newpage

\section{Correlators of windings}

After this long introduction, finally we arrived at the point where
we start to discuss the new results of this thesis.
The first of these results concerns correlators of winding operators
in the presence of a winding condensate.
According to the proposal of \cite{KKK} reviewed in the previous section,
they give the correlators of winding modes in the black hole background.
The calculation of these correlators represents the next step in
exploring the Toda integrable structure describing the winding sector
of 2D string theory.
For the one- and two-point correlators in the spherical
approximation this task has been fulfilled in the work \cite{AK}.

Due to the identification \Ref{tauz}, the generating functional of
all correlators of vortices is the $\tau$-function of Toda hierarchy.
For the Sine--Liouville theory where only the first couplings
$\lambda_{\pm 1}$  are non-vanishing, the correlators are defined as follows
\be
\CK_{i_1 \cdots i_n}=\left.
\frac{\partial^{n}}{\partial \lambda_{i_1} \cdots \partial
\lambda_{i_n}} \log\tau_0\right|_{\lambda_{\pm 2}=\lambda_{\pm 3}=\cdots =0},
\plabel{correlat}
\ee
where the coupling constants $\lambda_n$ are related to the Toda times
$t_n$ by \Ref{tlam}.
Whereas to find the free energy it was enough to establish the evolution
law along the first times, to find the correlators one should know
how the $\tau$-function depends on all Toda times, at least near $t_n=0$.

The evolution law for the first times $t_{\pm 1}$ 
was determined by the Toda equation. It is the first
equation in the hierarchy of bilinear differential Hirota equations
\Ref{toda}. The idea of \cite{AK} was to use these equations to find
the correlators.

\subsection{Two-point correlators}
\label{twopoint}

The first step was to identify the necessary equations because
not the whole hierarchy is relevant for the problem.
After that we observe that the extracted equations
are of the finite-difference type.
To reduce them to differential equations, one should plug in the
ansatz \Ref{gnexpCF}, where now the coefficients $f_g$ are functions of
all dimensionless parameters: $w$ and $s=(s_{\pm 2},s_{\pm 3},\dots)$.
The first parameter was defined in \Ref{scp} and the parameters $s_n$
are related to higher times
\be
s_n=i\left(-\frac{t_{-1}}{t_{1}}\right)^{n/2}
\xi^{\Delta[t_n]}t_n,
\plabel{sss}
\ee
where $\Delta[t_n]$ is the dimension of the coupling with respect to $\mu$
\be \Delta[t_n]=1-\frac{R|n|}{2}.
\label{dim}
\ee

The spherical approximation corresponds to the dispersionless limit
of the hierarchy. It is obtained as $\xi\to 0$. Thus, extracting the
first term in the small $\xi$ expansion, we found the spherical approximation
of the initial equations. In principle, they could mix different correlators
and of different genera. However, it turned out that in the spherical limit
the situation is quite simple.
In the equations we have chosen only second derivatives of the spherical part of 
the free energy survive.
We succeeded to rewrite the resulting equations as equations on
the generating functions of the two-point correlators.
There are two such functions: for correlators
with vorticities of the same and opposite signs
\beq
F^{\pm}(x,y)&=&\sum\limits_{n,m=0}^{\infty} x^n y^m
\tX^{\pm}_{\pm n,\pm m},
\plabel{genf}  \\
G(x,y)&=&\sum\limits_{n,m=1}^{\infty} x^n y^m \tX_{n,-m},
\plabel{genfd}
\eeq
where
\beq
& \tX^{\pm}_{0,m} := \mp i
\frac{1}{|n|}\frac{\partial^2 }{\partial t_n \partial \mu}\CF_0,
 \quad n\ne 0, &
\plabel{X1pm}  \\
& \tX^{\pm}_{n,m} := 2\frac{1}{|n||m|}\frac{\partial^2 }{\partial t_n
\partial t_m} \CF_0, \quad n,m\ne 0.
  &
\plabel{X2pm}
\eeq
The equations for $F^{\pm}(x,y)$ and $G(x,y)$, respectively, read
\beq
& \frac{x+y}{x-y}\left(e^{\frac12 F^{\pm}(x,x)} - e^{\frac12 F^{\pm}(y,y)} \right)
= x\partial_x F^{\pm}(x,y) e^{\frac12 F^{\pm}(y,y)} +
y\partial_y F^{\pm}(x,y) e^{\frac12 F^{\pm}(x,x)}, & \label{eqMain} 
\\
& A\left[ y\partial_y (G(x,y)-2F^-(y,0))-2\right] e^{\frac12 F^+(x,x)+F^+(x,0)} =
\frac{1}{y}\partial_x G(x,y) e^{\frac12 F^-(y,y)-F^-(y,0)}, & \label{eqMainn}
\eeq
where
\be
A=\exp\left( -\partial^2_{\mu} \CF_0\right)=
\xi^{-R}e^{-X_0}.
\label{constA}
\ee 
The equations \Ref{eqMain} come from the Hirota bilinear identities \Ref{toda}
taking $i=0$, extracting the coefficients in front of $y_{\pm n}y_{\pm m}$, 
$n,m>0$, multiplying by $x^n y^m$ and summing over all $n$ and $m$. 
The similar procedure with $i=1$ and $y_n y_{-m}$ gives the equation \Ref{eqMainn}.
In fact, the difference between $F^+$ and $F^-$ is not essential and it disappears
when one chooses $t_1=-t_{-1}\ (\lambda_1=\lambda_{-1})$.
Therefore, in the following we will omit the inessential sign label in $F(x,y)$.

Since the dependence of $\mu$ is completely known and
the cosmological constant plays a distinguished role, the quantities
$\tX^{\pm}_{0,m}$ can be actually considered as one-point correlators.
We define their generating function as
\be
h(x)=F(x,0).
\plabel{genonp}
\ee
The equations \Ref{eqMain} and \Ref{eqMainn} 
for the functions $F(x,y)$ and $G(x,y)$
have been explicitly solved in terms of this generating function $h(x)$
\beq
F(x,y)&=&\log\left[ \frac{4xy}{(x-y)^2}{\rm sh}^2\left(\frac{1}{2}
(h(x)-h(y)+\log\frac{x}{y})\right)\right],
\plabel{Fexp}  \\
G(x,y)&=& 2\log\left(1-Axy e^{h^+(x)+h^-(y)}\right).
\label{Dsolution}
\eeq
These solutions are universal in the sense that they are valid for any
system describing by Toda hierarchy. In other words, their form
does not depend on the potential or another initial input. All dependence
of particular characteristics of the model enters through the
one-point correlators and the free energy.
In a little bit different form these solutions appeared in
\cite{kkvwz,Zabrodin} and resemble the equations for the
two-point correlators in 2MM found in \cite{DKK}.

\subsection{One-point correlators}
\label{onepoint}

The main difficulty is to find the one-point correlators.
The Hirota equations
are not sufficient to accomplish this task. One needs to provide
an additional input. It comes from the fact that we know the dependence of
the free energy of the first times $t_{\pm 1}$ and of the cosmological
constant $\mu$. Due to this one can write a relation between the one-point
correlators entering $h(x)$ and two-point correlators of the kind
$\tX_{n,\pm 1}$. (Roughly speaking, one should integrate over $\mu$
and differentiate with respect to $t_{\pm 1}$.)
The latter are generated by two functions, $\p_y F(x,0)$
and $\p_y G(x,0)$. As a result, we arrive at the following two equations
\beq
\p_y F(x,0)&=&\hat K^{(+)} h(x)+\tX^+_{1,0},
\plabel{eqh+}\\
\p_y G(x,0)&=&\hat K^{(-)} h(x),
\plabel{eqh+-}
\eeq
where $\hat K^{(\pm)}$ are linear integral-differential operators.
These operators are found from the explicit expressions
for the free energy and the scaling variables and have the following form
\beq
\hat K^{(+)}&=&-a
\left[\w-(1+(R-1)x\frac{\partial}{\partial x} )
\int\limits^{w} d\w\right],
\label{opK+} \\
\hat K^{(-)}&=&a
\left[\w-(1+x\frac{\partial}{\partial x} )
\int\limits^{\w} d\w\right],
\label{opK-}
\eeq 
where $a=2\frac{\sqrt{R-1}}{2-R}\xi^{-\frac{R}{2}}$.
On the other hand, the generating functions $F$ and $G$
are known in terms of $h(x)$ from \Ref{Fexp} and \Ref{Dsolution}.
Substituting them into \Ref{eqh+} and \Ref{eqh+-}  
and taking the derivative with respect to $\w$,
we obtain two equations for one function $h(x)$
\beq
\left[ -a
\left( (R-1)x\frac{\partial}{\partial x} - \w
\frac{\partial}{\partial \w} \right) +\frac{2}{x}e^{-h(x;\w)}
\frac{\partial}{\partial \w}\right] h(x;\w)
&=&2\frac{\partial}{\partial \w}\tX^+_{0,1},
\label{p2+}  \\
\left[ a
\left( x\frac{\partial}{\partial x} - \w
\frac{\partial}{\partial \w} \right) -2Ax e^{h(x;\w)}
\frac{\partial}{\partial \w}\right] h(x;\w)
&=&2x e^{h(x;\w)}\frac{\partial}{\partial \w}A.
\label{p2-}
\eeq

In \cite{AK} we succeeded to solve these differential equations. 
As it must be, they turned out to be compatible. 
The solution was represented in terms of
the following algebraic equation
\be
e^{\frac{1}{R}h}-ze^{h}=1,
\plabel{hsol}
\ee
where $z=x\,\frac{\xi^{-R/2}}{\sqrt{R-1}}\, e^{-\frac{R-1}{R}X_0}=
x\lambda\, e^{-\frac{R-1}{R}\chi_0}$. Note that if we take different
$t_{\pm 1}$, one would have two equations for $h^{\pm}$ with
the parameters $z^{\pm}$ where $\lambda$ is replaced by $t_{\mp 1}$,
respectively.

The equation \Ref{hsol} represents the main result of the work \cite{AK}. 
It was used to find the explicit expressions for the correlators
which are given by the coefficients of the expansion of
the generating functions $h(x)$, $F(x,y)$ and $G(x,y)$ multiplied by 
the factors relating $t_n$ with $\lambda_n$.
The resulting expressions are 
\beq
\CK_n&=&
\frac{1}{2\sin \pi n R}
\frac{\Gamma(nR+1)}{n!\Gamma(n(R-1)+1)}
\frac{\xi^{-\frac{nR+2}{2}}}{(R-1)^{n/2}}
\left(\frac{e^{-\frac{n(R-1)+1}{R}X_0}}{n(R-1)+1}
-\frac{e^{-(n+1)\frac{R-1}{R}X_0}}{n+1}\right),
\label{onepointcor} \\
\CK_{n,-m}&=& -
\frac{\Gamma(nR+1)}{2\sin \pi n R}\frac{\Gamma(mR+1)}{2\sin \pi m R}
\frac{\xi^{-(n+m)R/2}}{(R-1)^{(n+m)/2}} e^{-(n+m)\frac{R-1}{R}X_0}
\times \nonumber \\
&\times &
\sum\limits_{k=1}^{{\rm min}(n,m)}\frac{k(R-1)^k
e^{k\frac{R-2}{R}X_0}}
{(n-k)!(m-k)!\Gamma(n(R-1)+k+1)\Gamma(m(R-1)+k+1)}.
\label{pm2pc} \\
\CK_{n,m}&=& -
\frac{\Gamma(nR+1)}{2\sin \pi n R}\frac{\Gamma(mR+1)}{2\sin \pi m R}
\frac{\xi^{-(n+m)R/2}}{(R-1)^{(n+m)/2}} e^{-(n+m)\frac{R-1}{R}X_0}
\times \nonumber \\
&\times &
\sum\limits_{k=1}^{n}\frac{k}
{(n-k)!(m+k)!\Gamma(n(R-1)+k+1)\Gamma(m(R-1)-k+1)}.
\label{pp2pc}
\eeq
The correlators are expressed as functions of $X_0$ and $\xi$.
The former contains all the non-trivial dependence
of the coupling constants $\mu$ and $\lambda$, whereas the latter provides
the correct scaling.
The results for the black hole point $\mu=0$ are obtained when one considers
the limit $X_0=0$.

We associated the factor $\(2\sin(\pi n R)\)^{-1}$, coming from the change of
couplings \Ref{tlam}, with each index $n$ of the correlators. 
Exactly at the black hole radius $R=3/2$, it
becomes singular for even $n$. Besides, it leads to negative answers
and breaks the interpretation of the correlators as probabilities to
find vortices of a given vorticity. Most probably, one should not attach
these factors to the correlators because they are part of the leg-factors
which always appear when comparing the matrix model and CFT results
(see section \ref{chMQM}.\ref{VerCorFun}).
This point of view is supported by the fact that these
factors appear as the same wave-function renormalization 
in all multipoint correlators and disappear if one considers
the normalized correlators like ${\CK_{n,m}\over \CK_n\CK_{m}}$.
Besides, one can notice 
that one does not attach this factor to $\lambda=\sqrt{t_1t_{-1}}$
and, nevertheless, one finds agreement for the free energy
with the results of \cite{MOORE} and \cite{HSU}.

\subsection{Comparison with CFT results}
\label{compCFT}

 Due to the FZZ conjecture (see section \ref{chSTR}.\ref{conjFZZ})  
we expect that our one-point
correlators should contain information about the amplitudes of emission
of winding modes by the black hole, whereas the two-point correlators
describe the S-matrix of scattering of the winding states from the tip
of the cigar (or from the Sine-Liouville wall). 
In \cite{FZZ} and \cite{BasFat} some two-
and three-point correlators were computed in the CFT approach to this
theory. It would be interesting to compare their results with the
correlators calculated here from the MQM approach. However,
there are immediate obstacles to this comparison.

First of all, these authors do not give any results for the one point
functions of windings. In the conformal theory such functions are
normally zero since the vortex operators have the dimension one.  But
in the string theory we calculate the averages of a type 
$\langl \int d^2\sigma\,\hat V_n(\sigma)\rangl$ 
integrated over the parameterization space. They are
already quantities of zero dimension, and the formal integration leads
to the ambiguity $0*\infty$ which should in general give a finite
result. Another possible reason for vanishing of the one-point
correlators could be the additional infinite W-symmetry found in
\cite{Fateev} for the CFT (\ref{SLCFT})
at $R=3/2, \ \mu=0$. The generators of this
symmetry do not commute with the vortex operators $\hat V_n, \ n\ne\pm
1$. Hence its vacuum average should be zero, unless there is a singlet
component under this symmetry in it.

Note that the one-point functions were calculated \cite{BasFat} in a
non-conformal field theory with the action
\be
S={1\over4\pi}\int d^2\sigma
\left[ (\partial X)^2 +(\partial\phi)^2-2\hat \CR\phi
+m \left( e^{-\hf\phi} +e^{\hf\phi}\right)\cos({3\over 2} X)\right].
\label{corka:NCFT}
\ee
This theory coincides with the Sine--Liouville CFT (again at $R=3/2$ and
$\mu=0$) in the limit $m\to 0$, $\phi_0\to -\infty$,
with $\lambda=m e^{-\hf\phi_0}$
fixed, where $\phi_0$ is a shift of the zero mode of the Liouville
field $\phi(z)$.
So we can try to perform this limit in the calculated one-point
functions directly. As a result we obtain
$\CK_n^{(m)} \sim m^2 \lambda^{3n-4}$.
The coefficient we omitted is given by a complicated
integral which we cannot perform explicitly. It is important
that it does not depend on the couplings and is purely numerical.
We see that, remarkably,
the vanishing mass parameter enters in a constant power
which is tempting to associate with the measure $d^2 \sigma$ of
integration. Moreover, the scaling in $\lambda$ is the same as
in (\ref{onepointcor}) (at $R=3/2$) up to the $n$-independent factor
$\lambda^{-8}$. All these $n$-independent factors
disappear if we consider the
correlators normalized with respect to $\CK_1^{(m)}$
which is definitely nonzero.
They behave like $\sim \lambda^{3(n-1)}$ what coincides with
the MQM result.

In fact, there is still a possibility for agreement of the matrix model 
results with the CFT prediction that the one-point correlators should vanish.
It involves a mixing of operators where not only primary
operators appear. Probably, after the correct identification,
one-point correlators will vanish in the matrix model too.
But we do not see any reason why this should be the case.

The possibility to introduce leg-factors and the possible mixing of 
operators make difficult to compare also our results 
for the two-point correlators with ones obtained in the Sine--Liouville
or coset CFT and given in \Ref{twobh}.
There only the two-point correlators of opposite and equal by modulo
vorticities have been calculated. They cannot be identified with our
quantity $\CK_{n,-n}$ while the normalization of the vortex operators
in the matrix model with respect to the vortex operators in CFT is not
established. This could be done with help of the one-point correlators
but they are not known in the CFT approach. 

Let us mention that from the CFT side some three-point correlators are also
accessible \cite{FZZ}. However, we did not study the corresponding problem
in the matrix model yet. The calculation of these correlators
would provide already sufficient information to make the comparison
between the two theories.

Note that the situation is that complicated because the two theories
have only one intersection point, so that the correlators
to be compared are just numbers.
We would be in a better position if we are able to extend the FZZ correspondence
to arbitrary radius or cosmological constant, for example.
Such extension may give answers to many questions which are not
understood until now.


\chapter{Tachyon perturbations of MQM}
\label{chTACH}

In this chapter we consider the second way to obtain
a non-trivial background in 2D string theory, which is
to perturb it by tachyon sources.
Of course, the results should be T-dual to ones obtained by
a winding condensation.
However, although expected, the T-duality of tachyons and windings
is not evident in Matrix Quantum Mechanics. We saw in chapter
\ref{chMQM} that they appear in a quite different way. And
this is a remarkable fact that the two so different pictures do agree.
It gives one more evidence that MQM provides the correct description of
2D string theory and their equivalence can be extended to include
the perturbations of both types.

Besides, the target space interpretations of the winding and tachyon
perturbations are different. Therefore, although they are described by the
same mathematical structure (Toda hierarchy), the physics is not the same.
In particular, as we will show, it is impossible to get a curved background
using tachyon perturbations, whereas the winding condensate was conjectured
to correspond to the black hole background. At the same time,
the interpretation in terms of free fermions allows to obtain
a more detailed
information about both the structure of the target space and
thermodynamical properties of tachyonic backgrounds.

Since the introduction of tachyon modes does not require
a compactification of the time direction, as the winding modes do,
we are not forced to work with Euclidean theory.
In turn, the free fermionic representation is naturally formulated
in spacetime of Minkowskian signature.
Therefore, in this chapter we will work with the real Minkowskian
time $t$. Nevertheless, we are especially interested in the case
when the tachyonic momenta are restricted to values of the Euclidean
theory compactified at radius $R$. It is this case that should be dual
to the situation considered in the previous chapter and we expect to find
that it is exactly integrable.

\section{Tachyon perturbations as profiles of Fermi sea}
\label{tachprof}

First of all, one should understand how to introduce the tachyon sources
in Matrix Quantum Mechanics. The first idea is just to follow the CFT
approach: to add the vertex operators realized in terms of matrices
to the MQM Hamiltonian. However, this idea fails by several reasons.
The first one is that although the matrix realization
of the vertex operators is well known \Ref{Tpm}, this form of the
operators is valid only in the asymptotic region. When approaching
the Liouville wall, they are renormalized in a complicated way.
Thus, we cannot write a Hamiltonian that determines the dynamics
everywhere. The second reason is that such perturbations introduced
into the Hamiltonian disappear in the double scaling limit
where only the quadratic part of the potential is relevant.
This is especially obvious for the spectrum of perturbations corresponding
to the self-dual radius of compactification $R=1$. Then
the perturbing terms do not differ from the terms of the usual potential
which are all inessential.

The last argument shows that one should do something with the system
directly in the double scaling limit. There the system is universal
being always described by the inverse oscillator potential. This means
that {\it we should change not the system but its state}. Indeed, we
established the correspondence with the linear dilaton background only
for the ground state of the singlet sector of MQM.
In particular, in the spherical approximation this state is described by
the stationary Fermi sea \Ref{grstsol}.
The propagation of
small perturbations above this ground state was associated with
the scattering of tachyons \cite{POLCHINSKI}. Therefore, it is natural
to expect that a tachyon condensation is obtained when one considers
excited states associated with non-perturbative deformations
of the Fermi sea. This idea has got a concrete realization in the work
\cite{AKK}.

\subsection{MQM in the light-cone representation}

A state containing a non-perturbative source of particles is,
usually, a coherent state obtained by action of the exponent of
the operator creating the particles on a ground state. Of course,
the easiest way to describe such states is to work in the representation
where the creation operator is diagonal. In our case, there
are two creation operators of right and left moving tachyons.
They are associated with powers of the following matrix operators
\be
\Xpm = {M\pm P\over \sqrt{2}}.
\plabel{XpXm}
\ee
Their eigenvalues $\xpm$ can be considered as light-cone like variables
but defined in the phase space of the theory rather than in the target
space. Thus, we see that to describe the tachyon perturbations,
one should work in the light-cone representation of MQM.

Such light-cone representation was constructed in \cite{AKK}.
With $\Xp$ and $\Xm$ we associate the right and left Hilbert spaces, respectively,
whose elements are functions of one of these variables. The scalar product is defined as
\be\label{prof:SCPR}
\langle \Phipm|\Phipm'\rangle
=\int d\Xpm \, \overline{ \Phipm(\Xpm)}\Phipm'(\Xpm).
\ee
Since the matrix operators $\Xpm$ obey the canonical commutation relation
\be\label{prof:ccrYY}
[(\Xp)^i_j,(\Xm)^k_l]=-i  \ \delta^{i}_{l}\delta^k_j,
\ee
the operator of coordinate in the right Hilbert space is
the momentum operator in the left one and the wave functions
in the two representations are related by a Fourier transform.
In the double scaling limit the dynamics of the system is governed
by the inverse oscillator matrix Hamiltonian. It is written in the variables 
\Ref{XpXm} as
\be\label{prof:hammat}
H_0 = -  \hf \Tr ( \Xp \Xm+  \Xm\Xp)
\ee
so that the second-order Schr\"odinger equation associated with it
in the usual representation
becomes a first-order one when written in the $\pm$-representations
\be\label{prof:SCHRPM}
 \p_t \Phipm(\Xpm,t)= \mp \Tr \left(\Xpm {\p\over\p \Xpm}
 +{N \over 2} \right)\Phipm(\Xpm,t).
\ee
The general solution is
\be\label{prof:GENSL} \Phipm (\Xpm , t)  =
 e^{\mp \hf N^2 t}  \Phipm(e^{\mp t}\Xpm ).
\ee

Similarly to the usual representation \Ref{MQMdecom},
the right and left Hilbert spaces are decomposed into a direct
sum of subspaces labeled by irreducible representations of $SU(N)$,
and the functions $ \vec \Phi^{(r)}_\pm=\{\Phi^{(r, a)}_\pm\}_{a=1}^{\dR}$
belonging to given irreducible representation $r$ 
depend only on the $N$ eigenvalues $\xipm{1},\dots,\xipm{N}$.
It is important that all these subspaces are invariant under 
the action of the Hamiltonian \Ref{prof:hammat}, which means that 
the decomposition is preserved by dynamics.
Remarkably, the Hamiltonian \Ref{prof:hammat}
reduces on the functions $\vec \Phi^{(r)}_\pm(\xpm)$ to its radial part only
\be\label{prof:ANGH}
H_0 = \mp i \sum_k (\xkpm\p_{\xkpm}+ {N\over 2}).
\ee
This property is a potential advantage of the light-cone approach in comparison
with the usual one where  
the Hamiltonian does contain an angular part, 
which induces a Calogero-like interaction.

In the scalar product \Ref{prof:SCPR}, the angular part can also be integrated
out, leaving only the trace over the representation indices and
the square of the Vandermonde determinant $\Delta(\xpm)$ as in \Ref{MQMsc}.
Therefore, we do the usual redefinition to remove this determinant
\be\label{prof:FWF}
\vec  \Pspm^{(r)}(\xpm) =\Delta(\xpm)\vec\Phpm{(r)}(\xpm).
\ee
For the new functions $\vec \Pspm^{(r)}(\xpm)$ the scalar product reads
\be\label{prof:SCPROo}
\langle \vec \Pspm^{(r)}|\vec \PPspm{(r)}\rangle=\sum\limits_{a=1}^{\dR}
\int \prod\limits_{k=1}^N d\xkpm 
\overline{\Psi^{(r,a)}_\pm(\xpm)}\Psi'^{(r,a)}_\pm(\xpm),
\ee
and the Hamiltonian 
takes the same form as in \Ref{prof:ANGH}, but with a different constant term
\be\label{prof:ANGHD}
H_0 = \mp i \sum_k (\xkpm\p_{\xkpm}+ {1/2}).
\ee

 In the singlet representation we reproduce the known result:
the wave functions $\Psi_\pm^{({\rm singlet} )}$ are completely
antisymmetric and  
the singlet sector describes a system of $N$ free fermions.
In what follows we will concentrate on this sector of the Hilbert space.
As we know form section \ref{chMQM}.\ref{dasjev}, it is sufficient to
describe tachyon excitations of 2D string theory.
We will start from the properties of the ground state of the
model, representing the unperturbed 2D string background and then go
over to the perturbed fermionic states describing the (time-dependent)
backgrounds characterized by nonzero expectation values of some vertex
operators.

\subsection{Eigenfunctions and fermionic scattering}

A wave function of a free fermionic system is represented by
the Slater determinant of one-particle wave functions. 
Therefore, according to \Ref{prof:ANGHD}
it is enough to study the problem of scattering of one fermion 
described by the one-particle Hamiltonian
\be
H_0= -\hf (\hat \xp\hat \xm +\hat \xm\hat \xp).
\plabel{oneph}
\ee
As we saw, in the light-cone variables the Hamiltonian \Ref{oneph}
is represented by a linear differential operator of the first order. 
Therefore, it is easy to write its eigenfunctions,
which are given by the simple power functions
\be
\psepm(\xpm)={1\over\sqrt{2\pi}} \xpm^{\pm iE-\hf}.
\plabel{wavef}
\ee
Given this simple result, one can ask how it is able to incorporate 
information about the scattering?  

To answer this question, note that there are two light-cone representations, right and
left. It turns out that 
{\it all the non-trivial dynamics of fermions in the inverse oscillator
potential is hidden in the relation between these two representations.}
This relation is just the usual unitary transformation between two quantum
mechanical representations. In the given case it is described by the Fourier
transform
\be
[\hat S \psip](\xm)=  \int
d\xp \, \KK(\xm,\xp) \psip(\xp).
\plabel{Fourtr}
\ee
The exact form of the kernel $K(\xm,\xp)$ depends on the non-perturbative
definition of the model. There are two possible definitions (theories 
of type I and II \cite{MPR}).
In the first model the domain of definition of the wave functions
is restricted to the positive half-lines $\xpm>0$ and the kernel has one 
of the two possible forms\footnote{The fact
that there are two choices for the kernel can
be explained as follows. In order to define the theory of type I
for the original second-order Hamiltonian in the usual representation, 
we should  fix the boundary condition at $\eig=0$, and there are
two linearly independent boundary conditions. The difference between 
them is seen only at the non-perturbative level.}
\be\label{prof:ker}
\KK(\xm,\xp)=\sqrt{2\over \pi}\cos(\xm\xp)\ \  {\rm or} \ \
\KK(\xm,\xp)=i\sqrt{2\over \pi}\sin(\xm\xp).
\ee
In the second model the domain coincides with
the whole line and the kernel is ${1\over \sqrt{2\pi}}e^{i\xm\xp}$. 
The choice of the model corresponds to that either we consider
fermions from one side of the inverse oscillator potential or from both
sides. The two sides of the potential are connected only by
tunneling processes which are non-perturbative in the cosmological constant
being proportional to $e^{-2\pi\mu}$. Therefore, the choice of the model
does not affect the perturbative (genus) expansion of the free energy.
In the following, we prefer to work with the first model 
avoiding the doubling of fermions and choose the cosine kernel.
The description of the light-cone quantization of the second model
can be found in appendix of \cite{AKK}.

Calculating the transformation \Ref{Fourtr} on the eigenfunctions
\Ref{wavef}, one finds that it is diagonal
\be
[\hat S^{\pm 1} \psepm ](\xmp)= \CR(\pm E) \psemp (\xmp),
\plabel{Smtrxx}
\ee
where the coefficient is given by
\be
\CR(E) = \sqrt{2\over \pi}
\cosh\left({\pi\over 2} (i/2-E)\right)  \Gamma( iE + 1/2).
\plabel{rfactor}
\ee
The coefficient is nothing else but the scattering coefficient of fermions
off the inverse oscillator potential. It can be seen as fermionic $S$-matrix
in the energy representation. Since $\CR(E)$ is a pure phase the
$S$-matrix is unitary. Thus, one gets a non-perturbatively defined
formulation of the double scaled MQM or, in other terms, of 2D string theory.

From the above discussion it follows that the scattering amplitude
between two arbitrary in and out states is given by the integral with
the Fourier kernel \Ref{prof:ker}
\beq\label{prof:scpr}
& \langle\psim|\hat S\psip\rangle  =\langle\hat S^{-1}
\psim|\psip\rangle =
\langle\psim|K|\psip\rangle & \nonumber \\
& \langle\psim|K|\psip\rangle
\equiv
\int_{0}^\infty d\xp d\xm\,
\overline{\psim(\xm)} \   \KK(\xm,\xp) \psip(\xp). &
\eeq
The integral \Ref{prof:scpr} can be interpreted as a scalar product between the
in and out states.  
Since the in and out eigenfunctions \Ref{wavef} form two complete systems of
$\delta$-function normalized orthonormal states and the $S$-matrix is diagonal
on them, they satisfy the orthogonality relation
\be\label{prof:scprEE}
\langle\psem|K|\peep\rangle= \CR(E)\delta(E-E').
\ee

Note that the fermionic $S$-matrix has been calculated in \cite{MPR}
from properties of the parabolic cylinder functions defined by
equation \Ref{SEQdbs}.
In our case it appears from the usual Fourier transformation and
it does not involve a solution of complicated differential equations.
Thus, the light-cone representation does crucially simplify the problem
of scattering in 2D string theory.

\subsection{Introduction of tachyon perturbations}

But the main advantage of the light-cone representation becomes clear
when one considers the tachyon perturbations. As we discussed,
they should be introduced as coherent states of tachyons. Following this
idea, we consider the one-fermion wave functions of the form
\be
\Pepm(\xpm)=e^{\mp i \vp_\pm(\xpm;E)}
\psepm(\xpm),
\plabel{asswave}
\ee
where the expansion of the phase $\vp_\pm(\xpm;E)$
in powers of $\xpm$ in the asymptotics $\xpm\to \infty$
is fixed and gives the spectrum of tachyons.
The exact form of $\vp_\pm$
is determined by the condition that $S$-matrix \Ref{Fourtr}
remains diagonal on the perturbed wave functions
\be
\hat S \ \Pep= \Pem,
\plabel{PepSm}
\ee
what means that
$\Pep$ and $\Pem$ are two representations of the same physical state.
This condition can also be expressed as the orthonormality of in and
out eigenfunctions \Ref{asswave}  
\be\label{prof:orth}
\langle  \Pim{E_{{-}}}|K|  \Pip{E_{{+}}}\rangle =\delta(E_+-E_-)
\ee
with respect to the scalar product \Ref{prof:scpr}. The normalization to 1
fixes the zero mode of the phase.  

Two remarks are in order. First, the wave functions \Ref{asswave}
are not eigenfunctions of the Hamiltonian \Ref{oneph}. Nevertheless,
they can be promoted to solutions of the Schr\"odinger equation
by replacement $\xpm\to e^{\mp t}\xpm$ and multiplying by the overall
factor $e^{\mp \hf t}$. As we will show, this leads
to a time-dependent Fermi sea and the corresponding string background.
But the energy of the whole system remains constant.
In principle, one can introduce a perturbed Hamiltonian with respect to
which $\Pepm$ would be eigenfunctions \cite{AKK}.
In the $\pm$-representations it is defined as solutions 
of the following equations
\be\label{prof:Hper}
H_\pm=  H_0^\pm+ \xpm \p \vp_{\pm}(\xpm;H_\pm).
\ee
The orthonormality condition \Ref{prof:orth} can be equivalently
rewritten as the
condition that the Hamiltonians $H_\pm$ define the action of the
same self-adjoint operator $H$.
However, such Hamiltonian has nothing
to do with the physical time evolution.
In particular, with respect to the time defined by $H$ the Fermi sea
is stationary and its profile coincides with the classical 
trajectory of the fermion with the highest energy.
Nevertheless, this Hamiltonian contains all information
about the perturbation and may be a useful tool to investigate
the perturbed system.

The second remark is that introducing the tachyon perturbations according
to \Ref{asswave}, we change the Hilbert space of the system. Indeed,
such states cannot be created by an operator acting in the initial
Hilbert space formed by the non-perturbed eigenfunctions \Ref{wavef}.
Roughly speaking, this is so because the scalar product of a perturbed
state \Ref{asswave} with an eigenfunction \Ref{wavef} diverges.
Therefore, in contrast to the infinitesimal perturbations considered
in \cite{POLCHINSKI},
the coherent states \Ref{asswave} are not elements of the Hilbert space
associated with the fermionic ground state.
Thus, our perturbation is intrinsically non-perturbative and we arrive at
the following picture. With each tachyon background one can associate
a Hilbert space. Its elements describe propagation of small tachyon
perturbations over this background. But the Hilbert spaces associated with
different backgrounds are not related to each other.

An explicit description of these backgrounds can be obtained in the
quasiclassical limit $\mu\to \infty$, which is identified
with the spherical approximation of string theory. In this limit
the state of the system of free fermions is described as an incompressible
Fermi liquid and, consequently, it is enough to define the region of the
phase space filled by fermions to determine completely the state.
Assuming that the filled region is connected, the necessary data
are given by one curve representing the boundary of the Fermi sea.
In a general case the curve is defined by a multivalued function.
For example, for the ground state in the coordinates $(x,p)$ it is given
by the two-valued function $p(x)$ \Ref{grstsol}.

In \cite{AKK} we found the equations determining 
the profile of the Fermi sea for the perturbation \Ref{asswave}.
They arise as saddle point equations for the double integral in the left
hand side of \Ref{prof:orth}. 
Each of the two equations defines a curve in the phase
space on which the corresponding integral is localized.
To produce $\delta$-function these curves should coincide.
As a result, we arrive at the condition that
the following two equations should be compatible at $E_+=E_-=-\mu$
\be
\xp\xm =M_\pm(\xpm)\equiv \mu +\xpm \p \vp_\pm(\xpm;-\mu).
\plabel{TOD}
\ee
One of these equations is most naturally written as $\xp=\xp(\xm)$ where
the function $\xp(\xm)$ is single-valued in the asymptotic region
$\xm\to \infty$.
The other, in turn, is determined by the function $\xm(\xp)$
with the same properties in the asymptotics $\xp\to \infty$.
Their compatibility means that the functions
$\xp(\xm)$ and $\xm(\xp)$ are mutually inverse. This condition
imposes a restriction on the perturbing phases $\vp_\pm$. 
It allows to restore the full phases
from their asymptotics at infinity. The resulting curve coincides with
the boundary of the Fermi sea.

Thus, we see that the tachyon perturbations are associated with
changes of the asymptotic form of the Fermi sea of free fermions of
the singlet sector of MQM. Given the asymptotics, the exact form
can be found with help of equations \Ref{TOD} which express
the matching condition of in-coming and out-going tachyons.
Note that the replacement $\xpm\to e^{\mp t}\xpm$ does lead to
a time-dependent Fermi sea.

\subsection{Toda description of tachyon perturbations}

Up to now, we considered perturbations of tachyons of arbitrary momenta.
Let us restrict ourselves to the case which is the most interesting for us:
when the momenta are imaginary and
form an equally spaced lattice as in the compactified
Euclidean theory or as in the presence of a finite temperature.
Thus, the perturbations to be studied are given by the phases
\be
\vp_\pm  (\xpm;E)
= V_\pm(\xpm) +\hf \phi(E) + v_\pm(\xpm;E),
\plabel{pot}
\ee
where the asymptotic part has the following form
\be
V_\pm(\xpm)
= \sum\limits_{k\ge 1} t_{\pm k} \xpm^{k/R} .
\plabel{Vbig}
\ee
The rest contains the zero mode $\phi(E)$ and the part
$v_\pm$ vanishing at infinity. They are to be found from the
compatibility condition \Ref{PepSm} (or \Ref{TOD}) and expressed through
the parameters of the potentials \Ref{Vbig}. These parameters
are the parameter $R$ measuring the spacing of the momentum lattice,
which plays the role of the compactification radius in the corresponding
Euclidean theory, and the coupling constants $t_{\pm n}$ of the tachyons.

In the work \cite{AKK} we demonstrated that with each coupling $t_n$
one can associate a flow generated by some operator $H_n$.
These operators are commuting in the sense of \Ref{zerocur}.
Moreover, they have the same structure as the Hamiltonians
from the Lax formalism of Toda hierarchy. Namely, one can introduce
the analogs of the two Lax operators $L_\pm$ and
the Hamiltonians $H_n$ are expressed through
them similarly to \Ref{laxh}
\be
 H_{ \pm n} = \pm (L_\pm^{n/R}   ) _{^{>}_{<} } \pm\hf (L_\pm^{n/R}   ) _{0} ,
\qquad n>0.
\label{hashka}
\ee
Thus, the perturbations generated by \Ref{Vbig} are integrable and
described by Toda hierarchy.

This result has been proven by explicit construction of the representation
of all operators of the Lax formalism of section \ref{chMM}.\ref{laxform}.
The crucial fact for this construction is that in the basis
of the non-perturbed functions \Ref{wavef}
the operators of multiplication by $\xpm$ coincide
with the energy shift operator
\be
\hat\xpm\psepm(\xpm)  = \ho^{\pm 1}\psepm(\xpm),  \qquad \ho=e^{-i\p_E}.
\plabel{bareREP}
\ee
Due to this property the perturbed wave functions \Ref{asswave}
can be obtained from the non-perturbed ones by action of some
operators $\CW_\pm$ in the energy space
\be
\Pepm\equiv e^{\mp i \vp_\pm(\xpm) }\psepm=\CW_\pm \psepm.
\plabel{Drss}
\ee
The operators $\CW_\pm$ are constructed from $\ho$ and can be represented
as series in $\ho^{1/R}$
\be\label{prof:DessOp}
\CW_\pm =e^{\mp i\phi/2}\
\left( 1+ \sum\limits_{k\ge 1} w_{\pm k} \hat \o^{\mp k/R}\right)
\exp\(\mp i \sum\limits_{k\ge 1} t_{\pm k} \hat \o ^{\pm k/R}\).
\ee
This shows that if one starts from a wave function of a given energy,
for instance $E=-\mu$, then the perturbed function is a linear combination
of states with energies $-\mu+in/R$. Of course, they do not belong to the
initial Hilbert space, but this is not important for the construction.
The important fact is that only a discrete set of energies appears.
Therefore, one can identify this set of imaginary energies
(shifted by $-\mu$ which plays the role of the Fermi level)
with the discrete lattice $\hb s$ of the Lax formalism.

It is easy to recognize the operators $\CW_\pm$ as the dressing
operators \Ref{dress}. The coupling constants $t_{\pm n}$ play the role
of the Toda times.
And the wave function $\Pep(\xp)$ appears as the Baker--Akhiezer function
\Ref{eigprob}.
Since the Lax operators act on the Baker--Akhiezer function
as the simple multiplication operators, in our case they are just represented
by $\hat \xpm$. Their expansion in terms of $\ho$ is given
by the representation of $\hat \xpm$ in the basis of
$\Pipm{-\mu+in/R}$ and can be obtained by dressing
$\ho^{\pm 1}$. Similarly, the Orlov--Shulman operators \Ref{ORSH}
are the dressed version of the energy operator $-\hat E$
\beq
L_\pm &=& \CW_\pm \, \ho^{\pm 1}\, \CW^{-1}_\pm 
= e^{\mp i\phi/2} \ho^{\pm 1}e^{\pm i \phi/2}
\left(1 +\sum_{k\ge 1} a_{\pm k}   \ho^{\mp k/R}\right),
\plabel{LpLmG} \\
M_\pm &=& -  \CW_\pm \,\hat E\, \CW^{-1}_\pm
= {1\over R} \sum _{k\ge 1}  k t_{\pm k}   L_\pm^{ k/R}- \hat E +
{1\over R}\sum_{k\ge 1} v_{\pm k}  L_\pm^{-k/R}.
\plabel{OrlSh}
\eeq

All the relations determining the Toda structure can be easily established.
The only difficult place is the connection between the right and left
representations. In section \ref{chMM}.\ref{laxform}
we mentioned that to define Toda hierarchy, the dressing operators must be
subject of some condition. Namely, $\CW_+^{-1}\CW_-$ should not depend
on the Toda times $t_{\pm n}$. It turns out that this condition
is exactly equivalent to the requirement \Ref{PepSm}
which we imposed on the perturbations.
Indeed, in terms of the dressing operators
it is written as
\be
\CW_-=\CW_+ \hat \CR ,
\plabel{TakTakT}
\ee
where $\hat \CR$ is the operator corresponding to \Ref{rfactor}.
This operator is independent of the couplings, 
so that the necessary condition is fulfilled.

Besides, this framework provides us with the string equations in a very
easy way. They are just consequences of the trivial
relations\footnote{We neglect the subtleties related with the necessity to
insert the $S$-matrix operator passing from the right to left representation
and back. The exact formulae can be found in \cite{AKK}.}
\be
[\hat \xp, \hat\xm]=-i, \qquad
\hat\xp \hat\xm+\hat\xm \hat\xp=-E.
\plabel{HamOp}
\ee
Relying on the defining equation \Ref{TakTakT}, it was shown that
one obtains the following string equations
\be
  L_+ L_- = M_+ +i/2,
\qquad
    L_- L_+=  M_- -i/2,
\qquad
M_-=M_+.
\plabel{STREQ}
\ee
They resemble the string equations of the two-matrix model
\Ref{strMM} and \Ref{strMMM}. Similarly to that case, only two of them
are independent.

Actually, we should say that $L_\pm$ and $M_\pm$ are not exactly the
operators that appear in the Lax formalism of Toda hierarchy.
The reason is that they are series in $\ho^{1/R}$ whereas
the standard definition of the Lax operators \Ref{Lax} involves
series in $\ho$. This is because our shift operator is $R$th power of the
standard one what follows from the relation between
$s$ and the energy $E$.
Therefore, the standard Lax and Orlov--Shulman operators should be defined
as follows
\be
 L=L_+^{1/R}, \quad \bL = L_-^{1/R}, \quad
M=RM_+, \ \ \bM=RM_-.
\plabel{standLx}
\ee
In terms of these Lax operators the string equation \Ref{STREQ}
is rewritten as
\be
[L^R,\bL^R]=i.
\plabel{strLLR}
\ee
This equation was first derived in \cite{IK} for the dual Toda hierarchy,
where $R\to 1/R$, describing the winding perturbations of MQM
considered in the previous chapter.\footnote{Before the work \cite{IK}, 
only the string equation
with integer $R$ appeared in the literature \cite{EK,Nakatsu}.}
Thus, indeed there is a duality between windings and tachyons:
both perturbations are described by the same integrable structure
with dual parameters.
Finally, the standard Toda times are related with the coupling constants
as $t_{\pm n}\to \pm t_{\pm n}$
(see footnote \ref{foottn} on page \pageref{foottn})
and the Planck constant is pure imaginary $\hb=i$.

\subsection{Dispersionless limit and interpretation of the Lax formalism}

Especially simple and transparent formulation is obtained in
the dispersionless limit of Toda hierarchy.
In this limit the state of the fermionic system is described by the Fermi
sea in the phase space of free fermions.
Therefore, we expect that the Toda hierarchy
governs the dynamics of this sea.

Indeed, as we know from section
\ref{chMM}.\ref{dispers}, the shift operator becomes a classical variable
and together with the lattice parameter defines the symplectic form
\be
 \{  \o, E  \} = \o.
\plabel{ccrom}
\ee
Remembering that the Lax operators $L_\pm$ coincide with
the light-cone variables $\xpm$, the classical limit of the string equations
reads
\beq
&  \{ \xm,  \xp\}= {1}, &
\plabel{cSEQ} \\
&\xp\xm=M_\pm(\xpm)=
 {1\over R}\sum\limits_{k\ge 1}  k t_{\pm k} \  \xpm^{ k/R}  +\mu  +
{1\over R}\sum\limits_{k\ge 1} v_{\pm k}\  \xpm^{-k/R}. &
\plabel{conteq}
\eeq
The first equation is nothing else but the usual symplectic form
on the phase space of MQM.
The second equation coincides with the compatibility
condition \Ref{TOD} where the explicit form of the perturbing phase was
substituted. Thus, this is the equation that determines the exact form of the
Fermi sea.
We note also that for the self-dual case $R=1$, the deformation described
by \Ref{conteq} is similar to the deformation of the ground ring
\Ref{relgrdef} suggested by Witten in \cite{WITTENGR}.

We conclude that for the tachyon perturbations
all the ingredients of the Lax formalism have a clear
interpretation in terms of free fermions:
\begin{itemize}
\item
the discrete lattice on which the Toda hierarchy is defined is the
set of energies given by the sum of the Fermi level $-\mu$
and Euclidean momenta of the compactified theory;
\item
the Lax operators are the light-cone coordinates in the phase space
of the free fermions;
\item
the Orlov--Shulman operators define asymptotics of the profile equation
describing deformations of the Fermi level;
\item
the Baker--Akhiezer function is the perturbed one-fermion function;
\item
the first string equation describes the canonical transformation form
the light-cone coordinates $\xpm$ to the energy $E$ and $\log\o$.
\item
the second string equation is the equation for
the profile of the Fermi sea.
\end{itemize}

\subsection{Exact solution of the Sine--Liouville theory}
\label{solSL}

As in the case of the winding perturbations, the Toda integrable structure
can be applied to find the exact solution of the Sine--Liouville theory
dual to the one considered in \Ref{SLCFT}. But in the present case
we possess a more powerful tool to extract the solution: the string equation.
In the dispersionless limit, it allows to avoid any differential
equations and gives the solution quite directly.
 
In fact, the procedure leading to the solution is quite general and works
for any potential of a finite degree. This procedure was suggested in
\cite{IK} and can be summarized as follows. Let all $t_{\pm k}$ with $k>n$ vanish.
Then in the dispersionless limit the representation \Ref{LpLmG} implies
\be\label{prof:zoom}
\xpm(\o, E)= e^{-{1\over 2R}\chi}\o^{\pm 1}
\left(1+\sum_{k= 1}^n a_{\pm k}(E)\ \o^{\mp k/R}\right),
\ee
where $\chi=\p^2_E \log\tau$.\footnote{From now on, we omit the index $0$
indicating the spherical approximation.}
The problem is to find the coefficients $a_{\pm k}$.
For this it is enough to substitute the expressions \Ref{prof:zoom},
with $E=-\mu$, in the
profile equations \Ref{conteq} and compare the coefficients in front of
$\o^{\pm k/R}$. Thus, the problem reduces to a finite triangular system 
of algebraic equations.

For the case of the Sine--Liouville theory, when there are only 
the first couplings $t_{\pm 1}$, the result of this procedure is the following \cite{AKK}
\beq
&\xpm = e^{-{1\over 2R}\chi}  \o^{\pm 1} (1+ a_\pm \o ^{\mp {1\over R}} ),&
\plabel{zomSG} \\
&\mu e^{ {1\over R} \chi} -{1\over R^2}
\left(1-{1\over R}\right)\tp\tm e^{{2R-1\over R^2} \chi} =1,
\qquad
a_\pm = {\tmp\over R} \, e^{{R-1/2\over R^2} \chi}.&
\plabel{coefzom}
\eeq
This solution reproduces both the free
energy and the one-point correlators. The former is expressed trough the
solution of the first equation in \Ref{coefzom} and the latter
is contained in \Ref{zomSG} where one should identify
\be
\o(\xpm)=e^{{1\over 2R} \chi}\xpm e^{-R h_\pm(\xpm)}.
\plabel{omh}
\ee
 Here $h_\pm(\xpm)$ are the generating functions of the one-point
correlators similar to \Ref{genonp}.

Comparing \Ref{coefzom} and \Ref{zomSG} with \Ref{eqchi} and \Ref{hsol},
respectively, one finds that the former equations are obtained
from the latters by the following duality
transformation\footnote{Performing this transformation, one should take into
account that the susceptibility transforms as
$\chi\to R^{-2}\chi-R^{-1}\log{R}$. This result can be established, for
example, from \Ref{sphchi}.}
\be
R\to 1/R, \qquad \mu\to R\mu,\qquad
t_n\to R^{-nR/2}t_n.
\plabel{dualtran}
\ee
Note that it would be more natural to transform the couplings as
$t_n\to R^{1-nR/2}t_n$ adding additional factor $R$.
Then the scaling parameter $\lambda^{2}\mu^{R-2}$ would not change.
Moreover, in the CFT formulation this factor is naturally associated with
the Liouville factor $e^{-2\phi}$ of each marginal operator.
Its absence in our case is related to that we identified the winding
couplings in incorrect way. The correct couplings $\tilde t_n$
are given in \Ref{ttn}. Thus, whereas the couplings
of the tachyon perturbations are exactly the Toda times, for the winding
perturbations they differ by factor $R$.

\lfig{Profiles of the Fermi sea ($\xpm=x\pm p$)
in the theory of type I at $R=2/3$.  The first picture contains several
profiles corresponding to $\tp=\tm=2$ and values of $\mu$ starting
from $\mu_c=-1$ with step $40$.  For comparison, the
unperturbed profile for $\mu = 100$ is also drawn.  The second picture 
shows three moments of
the time evolution of the critical profile at $\mu = -1$.}
{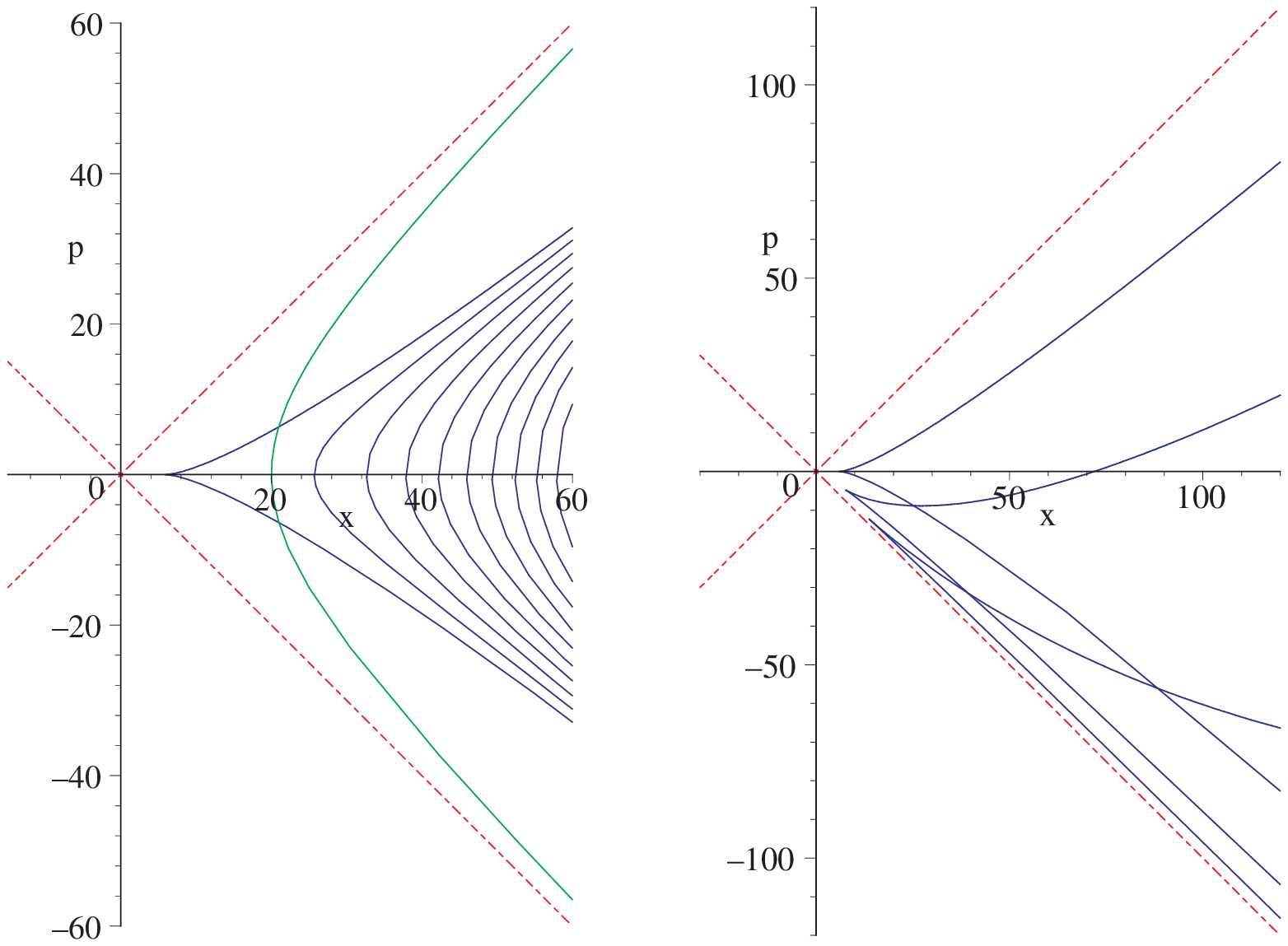}{11cm}{profile}

The explicit solution \Ref{zomSG} determines the boundary of the Fermi sea,
describing a condensate of tachyons of momenta $\pm 1/R$, in
a parametric form. For each set of parameters one can draw the corresponding
curve. 
The general situation is shown on fig. \ref{profile}. The
unperturbed profile corresponds to the hyperbola asymptotically
approaching the $\xpm$ axis, whereas the perturbed curves deviate from
the axes by a power law. We see that there is a critical
value of $\mu$, where the contour forms a spike.
It coincides with (T-dual of) $\mu_c$ given by \Ref{crp}.
At this point the quasiclassical description breaks down
and our results provide a geometric interpretation for this.
On the second picture the physical time evolution of the critical 
profile is demonstrated.

It is interesting that only the case $t_{\pm 1}>0$ has a good interpretation.
In all other cases at some moment of time the Fermi sea begins to penetrate
into the region $x<0$. This corresponds to the transfusion of the fermions
through the top of the potential to the other side. Such processes are
forbidden at the perturbative level.

This can be understood for $t_1t_{-1}<0$ because the corresponding CFT is
not unitary and one can expect some problems. On the other hand, the
case $t_{\pm 1}<0$ is well defined from the Euclidean CFT point of view, 
as is the case of positive couplings, because they differ just by a shift
of the Euclidean time. It is likely that the problem is intrinsically
related to the Minkowskian signature. In terms of the Minkowskian time
the potential is given by $\cosh (t/R)$ and
it is crucial with which sign it appears in the action.
The two possibilities lead to quite different pictures as it is clear
for our solution.

In the same way one can find the solution in the classical limit of
the theory of type II. In this case one can
introduce two pairs of perturbing potentials describing the
asymptotics of the wave functions at $\xpm \to\infty$ and $\xpm
\to-\infty$. For sufficiently large $\m$ the Fermi sea consists of two
connected components and the theory decomposes into two theories of
type I. However, in contrast to the previous case, there are no
restrictions on the signs of the coupling constants. When $\mu$
decreases, the two Fermi seas merge together at some critical value
$\m^*$. This leads to interesting (for example, from the point of view
of the Hall effect) phenomena. 
Here we will only mention that, depending on the choice of
couplings, it can happen that for some interval of $\mu$ around the
point $\mu^*$, the Toda description is not applicable.

\newpage

\section{Thermodynamics of tachyon perturbations}

\subsection{MQM partition function as $\tau$-function}

In the previous section we showed that the tachyon perturbations
with momenta as in the compactified Euclidean theory are described
by the constrained Toda hierarchy. Hence, they are characterized by
a $\tau$-function. What is the physical interpretation of this
$\tau$-function? In \cite{DMP} it was identified as the generating
functional for scattering amplitudes of tachyons. On the other hand,
since the tachyon spectrum coincides with that of the theory at finite
temperature,
it is tempting to think that the theory possesses a thermodynamical
interpretation. Then the $\tau$-function could be seen as the partition
function of the model.

Although expected, the existence of the thermodynamical interpretation is
not guaranteed because the system is formulated in the Minkowskian time
and except the coincidence of the spectra there is no reference to
a temperature. However, we will show that it does exist at least for the
case of the Sine-Liouville perturbation \cite{AKTBH}.
In the following we will accept this point of view and will show that
the $\tau$-function is indeed the grand canonical
partition function at temperature $T=1/\beta$.
In the spherical limit this was done in \cite{AKK,AKTBH}
and we will present that derivation in the next paragraph.
Here we prove the statement to all orders in perturbation theory 
following the work \cite{AKKNMM} 
which will be discussed in detail in chapter \ref{chNORM}.

The grand canonical partition function is defined as follows
\be
\CZ(\mu,t)=\exp\left[ \int_{-\infty}^\infty
d E\, \rho(E)\log\(1+e^{-\bh\beta(\mu+E)}\)\right],
\plabel{FRENO}
\ee
where $\rho(E)$ is the density of states.
It can be found by confining the system in a box of size
$\sim\sqrt{\Lambda}$ similarly as it was done in section
\ref{chMQM}.\ref{doubscl}. The difference is that now we work in the light
cone representation. Therefore, one should generalize the
quantization condition \Ref{WKBq}. The generalization is given by
\be
[\hat S\Psi](\ctf) =\Psi(\ctf)
\plabel{bcctff}
\ee
so that one identifies the scattered state with the initial one at
the wall.
Then from the explicit form of the perturbed wave function \Ref{asswave}
with \Ref{pot} one finds \cite{AKKNMM}
\be
\rho(E)={\log \Lambda\over 2 \pi}-{1\over 2\pi } {d\phi(E) \over d E}.
\plabel{DEN}
\ee
Dropping out the $\Lambda$-dependent non-universal contribution,
integrating by parts in \Ref{FRENO}, closing the contour
in the upper half plane
and taking the integral by residues of the thermal factor, 
one obtains\footnote{Actually, the function $\phi(E)$ has
logarithmic cuts which contribute to the integral.
However, their contribution reduces
to the sum of integrals along these cuts of the form
${\beta\over 2\pi }\int_\Delta dE\,
{\phi(E+\e)-\phi(E-\e)\over 1+e^{\beta(\mu+E)}}$.
The discontinuity on the cut of $\phi(E)$
is constant so that we remain only with the integral of the thermal factor.
As a result the $n$th cut gives the contribution
$\sim\log\left(1+e^{-\beta\left(\mu+i(n+\hf)\right)} \right)=O(e^{-\beta\mu})$,
which is non-perturbative and can be neglected in our consideration.}
\cite{AKKNMM}
\be
\CZ(\mu,t)=\prod\limits_{n \ge 0}
\exp\left[ {i\over\hb}\phi\left(i\hb{n+\hf\over R}-\mu\right)\right].
\plabel{FRENooo}
\ee

On the other hand, the zero mode of the perturbing phase
is actually equal to the zero mode
of the dressing operators \Ref{dress}. Hence it is expressed through
the $\tau$-function as in \Ref{taur}. Since the shift in the discrete
parameter $s$ is equivalent to an imaginary shift of the chemical potential
$\mu$, this formula reads
\be
e^{{i\over \hb} \phi(-\mu)}= {\taum_{0}\(\mu +i{\hbar\over
2R}\)\over \taum_{0}\(\mu -i{\hbar\over 2R}\)}.
\plabel{phitau}
\ee
Comparing this result with \Ref{FRENooo}, one concludes that
\be
\CZ (\mu,t)=\taum_{0}(\mu,t).
\plabel{Ztauu}
\ee

\subsection{Integration over the Fermi sea: free energy and energy}

As always, in the dispersionless limit one can give to all formulae
a clear geometrical interpretation. Moreover, this limit allows to obtain 
additional information about the system. Namely, in this paragraph
we show how such thermodynamical quantities as
the free energy and the energy of 2D string theory perturbed
by tachyon sources are restored from the Fermi sea of the singlet sector
of MQM. This was done in the work \cite{AKTBH}.

Since in the classical limit the profile of the Fermi sea uniquely
determines the state of the free fermion system, it encodes
all the interesting information. Indeed, in the full quantum theory
the expectation value of an operator is given by its quantum average
in the given state. Its classical counterpart is the integral
over the phase space of the corresponding function multiplied by
the density of states.
For the free fermions the density of states equals 1 or 0 depending on either
the phase space point is occupied or not. As a result, the classical value
of an observable $\CO$ is represented by its integral over the Fermi sea
\be
<\CO>=\dpi\mathop{\int\int}\limits_{\Fsea} d\xp d\xm\,
\CO(\xp,\xm).
\plabel{integV}
\ee

Now the description in terms of the dispersionless Toda hierarchy comes
into the game. We saw that the tachyon perturbation gives rise to the canonical
transformation from $\xpm$ to the set of variables $(E,\log\o)$.
The explicit map is given by \Ref{prof:zoom}.
Due to this the measure in \Ref{integV} can be rewritten in terms of $E$
and $\o$. As a result, one arrives at the following formula
\be
\p_{\mu}<\CO>=-\dpi \int\limits_{\bom(\mu)}^{\bop(\mu)}
{d\o \over \o}\,  \CO(\xp(\o,\mu),\xm(\o,\mu)).
\plabel{avV}
\ee
The limits of integration are defined by the cut-off.
We choose it as two walls at
$\xp=\sqrt{\Lambda}$ and $\xm=\sqrt{\Lambda}$.
Then the limits can be found from the equations
\be\label{tbh:grcon}
\xpm(\bopm(\mu),-\mu)=\sqrt{\Lambda}.
\ee
In the following we will need the solution of these equations to the second order
in the cut-off. From \Ref{prof:zoom} one can obtain the following result 
\be\label{tbh:grom}
\bopm\approx\oo^{\pm 1}(1\mp a_{\pm}\oo^{-1/R}), 
\qquad \oo=\sqrt{\Lambda e^{\chi/R}}.
\ee

The result \Ref{avV} is rather general. It is valid for any observable.
We applied it to the main two observables which
are the number of particles and the energy.
They were already defined in \Ref{FSen} and \Ref{FSnorm} for the case
of the ground state. In a more general case they are written as
integrals \Ref{integV} with $\CO=1$ and $\CO=-\xp\xm$, respectively.

In the first case the integral \Ref{avV} is easily calculated for 
the perturbing potential of any degree. For the energy, in principle, 
it can also be calculated for any finite $n$. 
However, the expressions become quite complicated and involve
higher order terms of the expansion of the limits $\bopm$ in $\Lambda$.
Therefore, we restrict the calculation of the energy to the case of 
the Sine--Liouville perturbation. Then, taking into account \Ref{tbh:grom}, 
one finds \cite{AKTBH}
\beq
\p_{\mu}N&=&
-\dpi \log\Lambda- {1 \over 2\pi R} \chi,
\plabel{frener} \\
\p_{\mu}E&=&
{1\over 2 \pi } e^{-{1\over R}\chi}\left\{ (1+a_+a_-)
\left(\log\Lambda+\oR\chi\right) -2a_+a_-\right\}+
{R\over 2\pi}(\tp+\tm)\Lambda^{1/2R},
\plabel{freeres}
\eeq
where the first equation is valid for any $t_{\pm k}$.
Using \Ref{coefzom}, which implies $a_+a_-={R\over 1-R} e^{{2R-1\over R^2}X}$, 
one can integrate \Ref{freeres} and obtain 
the following expression for the energy of the system
\beq
2\pi E & =&
\xi^{-2}\left(
{1\over 2R} \ea +{2R-1 \over R(1-R)}\eb- {1\over 2(1-R)}\ec \right)
(\chi+R\log\Lambda)
\nonumber \\
 & + &\xi^{-2} \left({1\over 4} \ea-{R \over
1-R}\eb+ {R(4-5R)\over 4(1-R)^2}\ec\right)+R(\tp+\tm)\mu
\Lambda^{1/2R}.  \label{tbh:valen}
\eeq
We observe that the last term is non-universal since it does not contain
a singularity at $\mu=0$. However, $\log \Lambda$ enters in a
non-trivial way.
It is combined with a non-trivial function of $\mu$ and $\xi$.

To reproduce the free energy, note that in the grand canonical ensemble
it is related to the number of particles through
\be
N = \p \CF/\p \mu,
\plabel{dFdm}
\ee
where we changed our notations assuming the usual thermodynamical
definition of the free energy at the temperature $T$
\be
\CF = -T\log \CZ.
\plabel{FRZ}
\ee
Comparing \Ref{dFdm} and \Ref{frener}, one obtains the expected
result
\be
\CF=- {1\over\beta }\log \tau.
\plabel{frenercom}
\ee
In particular, the temperature is associated with $1/\beta$.
Integrating the equation \Ref{coefzom} for the susceptibility $\chi$, one can get
an explicit representation for the free energy. 
For example, in the Sine--Liouville case one obtains
\be\label{tbh:valfe}
2\pi \CF=-{1\over 2R} \mu^2 (\chi+R\log\Lambda)- \xi^{-2}\left(
{3\over 4} \ea-{R^2-R+1 \over 1-R}\eb+{3R\over 4(1-R)}\ec\right).
\ee
This result coincides with the T-dual transform of \Ref{sphCF}.

\subsection{Thermodynamical interpretation}

In \cite{AKTBH} we proved that the derived expressions for the
energy and free energy allow a thermodynamical interpretation.
This is not trivial because there are two definitions of the macroscopic
energy. One of them is the sum of microscopic energies of the individual
particles which is expressed as an integral over the phase space.
We used this definition in the previous paragraph. Another definition
follows from the first law of thermodynamics which relates
the energy, free energy and entropy
\be
S=\beta (E-F).
\plabel{diff}
\ee
It allows to express the energy and entropy as derivatives of the free
energy with respect to the temperature
\be
E={\p (\beta F) \over \p \beta}, \qquad
S=-{\p F \over \p T}.
\plabel{termrel}
\ee
To have a consistent thermodynamical interpretation, the two definitions
must give the same result.
It is a very non-trivial check, although very simple from the technical point
of view. All that we need is to differentiate the free energy, which is given
in \Ref{tbh:valfe}, and check that the result coincides with 
\Ref{tbh:valen}.

However, there are two important subtleties.
The first one is that the first law is formulated in terms of
the canonical free energy $F$ rather than for its grand canonical
counterpart $\CF$. This is in agreement with the fact that
it is the canonical free energy that is interpreted as the partition function
of string theory. Thus, it is $F$ that carries an information about
properties of the string background and should be differentiated. 
Due to this one should pass to the
canonical ensemble
\be
F=\CF-\mu{\p \CF \over \p \mu}. 
\plabel{legtr}
\ee
Then from \Ref{tbh:valfe}, \Ref{tbh:valen} and \Ref{diff}, one finds the final
expressions for the canonical free energy and entropy to be used in the
thermodynamical formulae \Ref{termrel}
\beq
2\pi F&=&{1\over R}\int^{\mu} s\chi(s)ds  
= {1\over 2R} \mu^2 (\chi+R\log\Lambda)  \nonumber \\
&+& \xi^{-2}\left(
{1\over 4} \ea-R\eb+{R\over 4(1-R)}\ec\right), \plabel{valfren} \\
S&=&\xi^{-2} \left(
{R\over 1-R}\eb-{1\over 2(1-R)}\ec \right)
(\chi+R\log\Lambda)
 \nonumber \\
& + &\xi^{-2}\left( -{R^3\over 1-R}\eb +
{R^2(3-4R)\over 4(1-R)^2}\ec\right) +
{R^2\over 2\pi}(\tp+\tm)\mu\Lambda^{1/2R}. \label{tbh:valent}
\eeq

If the first subtlety answers to the question {\it ``what to differentiate?''},
the second one concerns the problem {\it ``how to differentiate?''}.
The problem is what parameters should be held fixed when one differentiates
with respect to the temperature.
First, this may be either $\mu$ or $N$.
It is important to make the correct choice because they are 
non-trivial functions of each other.
Since we are working in the canonical ensemble,
it is natural to take $N$ as independent variable.
Besides, one should correctly identify the coupling $\lambda$
and the cut-off $\Lambda$. Their definition can involve $R$ and, thus,
contribute to the result.

It turns out that the coupling and the cut-off that
we have chosen are already the correct ones. In \cite{AKTBH} 
it was shown by direct calculation
that the thermodynamical relations \Ref{termrel},
where the derivatives are taken with $N$, $\lambda$ and $\Lambda$ fixed,
are indeed fulfilled. Thus, the 2D string theory perturbed
by tachyons of momenta $\pm i/R$ in Minkowskian spacetime has a consistent
interpretation as a thermodynamical system at temperature $T=1/(2\pi R)$.

This result also answers to the question
risen in section \ref{chWIND}.\ref{bhterm}:
what variable should be associated with the temperature?
Our analysis definitely says that this is the parameter $R$.
We do not need to introduce such notion
as ``temperature at the wall'' \cite{GP,NP,KZ}. The differentiation is done
directly with respect to the compactification radius. Also this
supports the idea that in the dual picture a black hole background
should exist for any compactification radius, at least in the interval
$1<R<2$.

We calculated entropy \Ref{tbh:valent} using the standard thermodynamical relations.
It vanishes in the absence of perturbations
when $\lambda=0$ but it is a complicated function in general case.
It would be quite interesting to understand the microscopic origin
of this entropy. In other words, we would like to find the microscopic
degrees of freedom giving rise to the non-vanishing entropy. However, we
have not found the solution yet. The problem is that a state of the system
is uniquely characterized by the profile of the Fermi sea and there is only
one profile described by our solution for each state. The only possibility
which we found to obtain different microscopic states is to
associate them with different positions in time of the same Fermi sea.
Although the Fermi sea is time-dependent, all macroscopic thermodynamical
quantities do not depend on time. Thus, different microscopic states
would define the same macroscopic state. This idea is supported also by
the fact that the entropy vanishes only if the Fermi sea is stationary.
However, we have not succeeded to get the correct result for the entropy from
this picture.

It is tempting to claim that the obtained thermodynamical quantities
describe after the duality transformation \Ref{dualtran} the
thermodynamics of winding perturbations and their string backgrounds.
This is, of course, true for the free energy.
However, it is not clear whether the energy and entropy 
are dual in the two systems. 
For example, it is not understood even how to define 
the energy of a winding condensate.
Nevertheless, if we assume that our results can be related to
the backgrounds generated by windings, this gives a
plausible picture. For example, the non-vanishing entropy is compatible 
with the existence of a black hole.

Our results concern arbitrary radius $R$ and cosmological constant $\mu$.
When the latter goes to zero, which corresponds to
the black hole point according to the FZZ conjecture 
(section \ref{chSTR}.\ref{conjFZZ}),
the situation becomes little bit special. In this limit we find
\beq
2\pi F&=&{(2R-1)^2\over 4(1-R)}\tl^{4R\over 2R-1} , \nonumber\\
2\pi E&=&{2R-1\over 2R(1-R)}\left( \log(\Lambda\xi)
-{R\over 2(1-R)}\right) \tl^{4R\over 2R-1} ,\label{tbh:valallbh} \\
S&=&{2R-1\over 2(1-R)}\left(\log(\Lambda\xi)-{R^2(3-2R)\over 2(1-R)}\right)
\tl^{4R\over 2R-1} , \nonumber
\eeq
where $\tl=\left({1-R\over R^3}\right)^{1/2}\lambda$.
One observes that
the logarithmic term $\xi^{-2}\log(\Lambda\xi)$
in the free energy disappears, whereas it is present
in the energy and the entropy.
This term is the leading one. Therefore, both the energy and the entropy
are much larger than the free energy. This can explain the puzzle
that, on the one hand, the dilaton gravity predicts the vanishing
free energy and, on the other hand, the matrix model gives a non-vanishing
result. Our approach shows that it does not vanish but it is negligible in
comparison with other quantities so that in the main approximation
the law $S=\beta E$ is valid.

\newpage

\section{String backgrounds from matrix solution}

\subsection{Collective field description of perturbed solutions}

We introduced the tachyon perturbations as one of the ways to change the
background of 2D string theory. In particular, we expect that the perturbations
change the value of the tachyon condensate.
To confirm this expectation, one should extract a target space picture
from the matrix model solution describing the perturbed system.
As we saw in the unperturbed case, for this purpose it is convenient to use
the collective field theory of Das and Jevicki
(see section \ref{chMQM}.\ref{dasjev}). Whereas
the fermionic representation of section \ref{tachprof}
is suitable for the solution of many problems,
the relation to the target space phenomena is hidden in 
this formulation and it becomes clear
in the collective field approach. We tried to understand it
relying on the Das--Jevicki theory in the work \cite{ALEX}.

We restrict ourselves to the spherical approximation.
Then the string background is uniquely determined by the profile of the Fermi sea
made of free fermions of the singlet sector of MQM.
Thus, we should find the background using a given
profile as a starting point.
This was already done for the simplest case of the linear
dilaton background which was obtained from the ground state of the fermionic 
system. In particular, we related
the tachyon field to fluctuations of the density
of matrix eigenvalues around the ground state.
We expect that this identification
holds to be true also in more complicated cases.
All that we need to obtain another background is to replace the background
value of the density by a new function determined by the exact form of
the deformed profile of the Fermi sea.
In this way one can find an effective action for the field describing
the density fluctuations.

It appears after the substitution \Ref{expf} into the background independent
effective action of Das and Jevicki \Ref{Scol}. 
However, in contrast to the previous case,
the background value of the density $\bff_0$ now depends on time $t$,
so we write 
\be\label{metr:expf}
\bff(x,t)={1\over \pi}\ff(x,t)+{1\over \sqrt{\pi}}\p_x\eta(x,t).
\ee
This time-dependence leads to additional terms. Extracting only
the quadratic term of the expansion in the fluctuations $\eta$,
one finds the following result
\be
S_{(2)}=\hf \int dt \int {dx\over \ff}\, \left[
(\p_t \eta)^2 -2  {\int dx \p_t \ff \over \ff}\p_t\eta \p_x \eta
-\left( \ff^2 -\left( {\int dx \p_t \ff \over \ff}\right)^2
\right) (\p_x\eta)^2  \right].
\plabel{metr:Sd}
\ee

The crucial property of this action is that for any $\bff_0(x,t)$
the determinant of the matrix coupled to the derivatives of $\eta$
equals $-1$. Besides, there are no terms without derivatives.
As a result, this quadratic part can be represented as the usual action for
a massless scalar field in a curved metric $g_{\mu\nu}$.
\be
S_{(2)}=-\hf \int dt \int dx\, \sqrt{-g} g^{\mu\nu}\p_\mu \eta \p_\nu\eta.
\plabel{Sg}
\ee
In a more general case we would have to introduce a dilaton dependent
factor coupled to the kinetic term.
The metric $g_{\mu\nu}$ in the coordinates $(t,x)$ is fixed by \Ref{metr:Sd} 
up to a conformal factor. 
For example, we can choose it to coincide with the matrix
we were talking about, so that $\det g=-1$.

The action \Ref{Sg} is conformal invariant. Therefore, one can always redefine
the coordinates to bring the metric to the usual Minkowski form
$\eta_{\mu\nu}=\diag(-1,1)$.
We are able to find the transformation to such flat coordinates for a large 
class of functions $\ff(x,t)$, which are solutions of the classical equations following from
\Ref{Scol}. In particular, it includes the
integrable perturbations generated by the potential \Ref{Vbig}.

First, let us extract the background value of the density $\ff$ from an MQM solution,
which is usually formulated in terms of the exact form of the Fermi sea
of the MQM singlet sector.
The form of the sea can be described by the two chiral fields $p_\pm(x,t)$
introduced in \Ref{chf}. They are two branches of the function $p(x,t)$
representing the boundary and coincide with its upper and lower 
components. 

\lfig{The Fermi sea of the perturbed MQM.
Its boundary is defined by a two-valued function with two branches
parameterized by $p(\o,t)$ and $\pp(\o,t)$.
The background field $\ff$ coincides with the width of
the Fermi sea.}{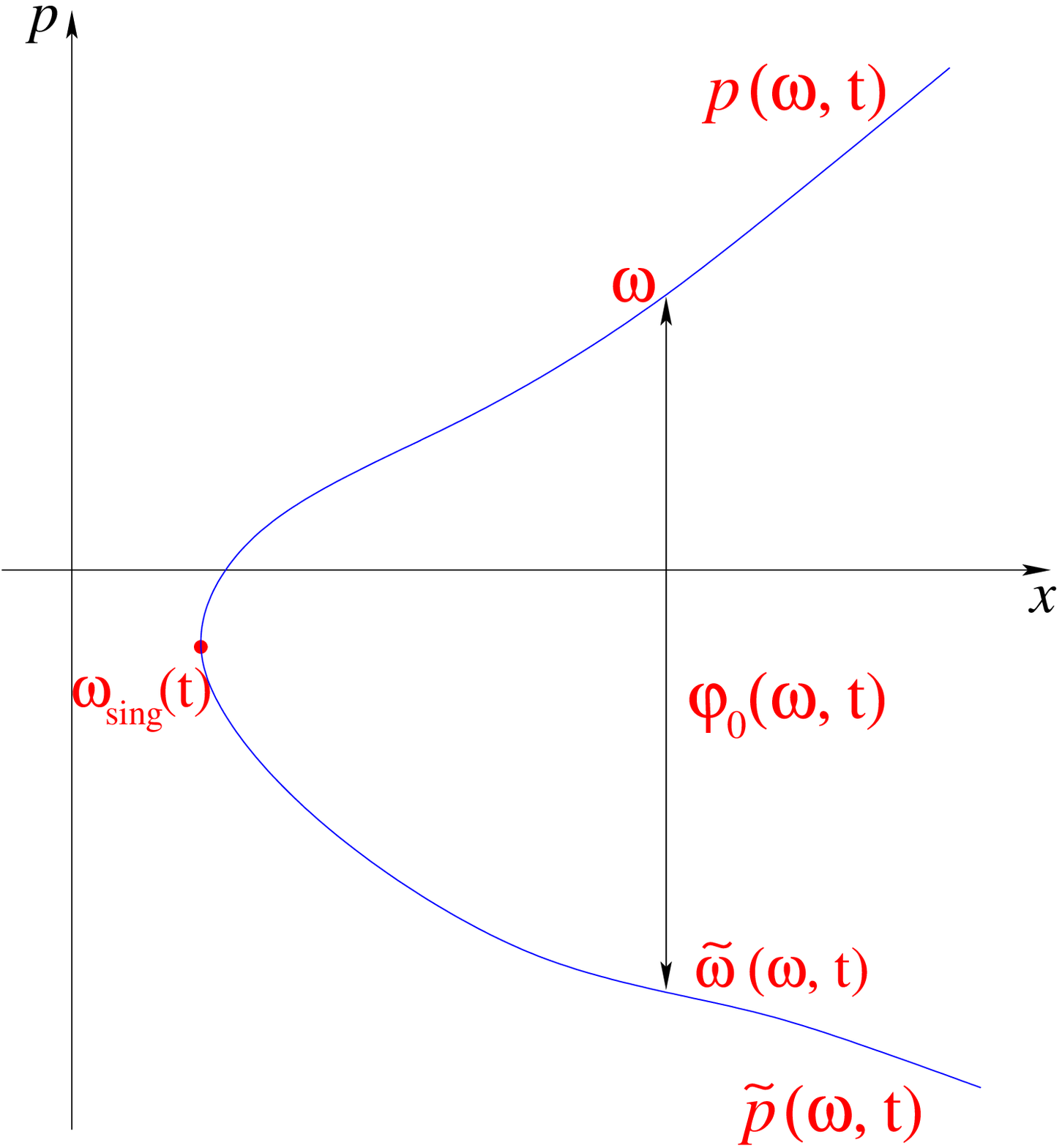}{5cm}{fermsch}

Let $\o$ be a parameter along the boundary. Then the position of the boundary 
in the phase space is given in the parametric form by two functions, $x(\o,t)$ and $p(\o,t)$.
From the equation of motion \Ref{Hopfeq} with the inverse oscillator potential,
it is easy to derive that these functions should satisfy
\be
{\p x\over \p\o} \({\p p \over \p t} -x\)={\p p\over \p \o} \({\p x \over \p t} -p\).
\plabel{Hopfoo}
\ee
It is convenient also to introduce the``mirror" parameter $\op(\o,t)$ such that
(see fig. \ref{fermsch}).
\be\label{metr:ooooo}
x(\op(\o,t),t)=x(\o,t), \ \ \op\ne \o.
\ee
Then $p_+$ can be identified with $p(\o,t)$ and
$p_-$ with $p(\op(\o,t),t)$.
The solution for the background field is given by their difference
and therefore it is represented in the parametric form
\be\label{metr:solf}
\ff(\o,t)=\hf(p(\o,t)-\pp(\o,t)),
\ee
where we denoted $\pp(\o,t)=p(\op(\o,t),t)$.
Due to \Ref{cleq}, \Ref{chf} and \Ref{metr:solf},
the effective action \Ref{metr:Sd} can now be rewritten as
\be\label{metr:Sdfpp}
S_{(2)}=\int dt \int {dx\over p-\pp}\, \left[
(\p_t \eta)^2 +  (p+\pp)\p_t\eta \p_x \eta
+p\pp (\p_x\eta)^2  \right].
\ee

A solution of \Ref{Hopfoo} can be easily constructed if one takes  
\be
{\p x\over \p\o} =p- {\p x \over \p t}.
\plabel{Hopfsol}
\ee
Then the equation \Ref{Hopfoo} implies 
\be
{\p p\over \p\o} =x- {\p p \over \p t}.
\plabel{Hopfsoll}
\ee
An evident solution of these two equations is
\bea{rcl}{SLgen}
p(\o,t)&=&\sum\limits_{k=0}^{\infty} a_k
\sinh\left[\(1-b_k\)\o+b_k t+\alpha_k\right],\\
x(\o,t)&=&\sum\limits_{k=0}^{\infty} a_k
\cosh\left[\(1-b_k\)\o+b_k t+\alpha_k \right],
\eea
for any set of $a_k$, $b_k$ and $\alpha_k$. In principle, the solution can also
contain a continuous spectrum.

Relying only on the property \Ref{Hopfsol} and the relations following from
the definition of the mirror parameter
\be\label{metr:porper}
{\p \op\over \p \o}={{\p x/\p\o} \over \pp -\xtp}, \qquad
{\p \op\over \p t}={p-\xtp \over \pp -\xtp}-{\p \op\over \p \o},
\ee
where $\xtp\equiv {\p_t x}(\op,t)$,
we showed  \cite{ALEX} that the following coordinate transformation
brings the action \Ref{metr:Sdfpp} to the standard form with the kinetic term
written in the Minkowski metric~$\eta_{\mu\nu}$
\be
\tf=t-{\o+\op\over 2}, \qquad \xf={\o-\op\over 2}.
\plabel{flcor}
\ee
The change of coordinates \Ref{flcor} is remarkably simple and has
a transparent interpretation. It associates the 
light-cone coordinates $\tf\pm \xf$ with the parameters
of incoming and outgoing tachyons, $t-\o$ and $t-\op$.
For the ground state given by the solution \Ref{cmf}, the mirror
parameter is $\op=-\o$ so that we return to the simple situation described
in section \ref{chMQM}.\ref{identld}.

In a particular case when
$b_k=k/R$, the solution \Ref{SLgen} reproduces the profile of the Fermi sea
corresponding to the tachyon perturbations of the two previous sections.
Indeed, in that case the description in terms of Toda hierarchy
implies the representation \Ref{prof:zoom}. Returning from the light-cone 
to the $(x,p)$ coordinates, one obtains \Ref{SLgen},
where $\o$ is now the logarithm of the shift operator from the
previous sections.

\subsection{Global properties}

The field $\eta$ in coordinates $(\tf,\xf)$ satisfies the simple 
Klein--Gordon equation. Thus, the transformation \Ref{flcor} trivializes
the dynamics and makes the integrability explicit.
It is interesting that the spacetime in coordinates $(\tf,\xf)$ can still be non-trivial.
Although the metric is flat and have the Minkowski form, we still have a possibility
to have a non-trivial global structure because the image of the initial $(t,x)$-plane 
(or, more precisely, $(t,\o)$-plane, on which the initial solution is defined)
under the coordinate transformation \Ref{flcor} can cover only a subspace 
of the plane of the new coordinates. If we identify this subspace as the physical region
to be considered, the global structure
of this space will be non-trivial. Depending on boundary conditions,
either boundaries will appear or a compactification will take place.

The explicit form of the transformation \Ref{flcor} allows to study
the exact form of the resulting spacetime. This was done in \cite{ALEX} for
the simplest case of the Sine--Liouville perturbation corresponding to 
the following parameters in  \Ref{SLgen}
\bea{rclrclrclc}{metr:param}
a_0&=&\sqrt{2}e^{-{1\over 2R}\chi},\qquad
&a_1&=&{\sqrt{2}{\lambda}\over R}\, e^{{R-1\over 2R^2}\chi}, \qquad 
&a_k&=&0, \quad & k>1, \\
b_0&=&0, \qquad 
& b_1&=&1/R, \qquad 
& \alpha_k&=&0.
\eea
The result crucially depends on the parameter $R$ playing the role of the
compactification radius.
There are 3 different cases. 

When $R\ge 1$ nothing special happens
and the image of the $(t,\o)$-plane coincides with the whole plane of $\tf$ and $\xf$.

\lfig{Flat spacetime of the perturbed theory for
the case $R<1$.}{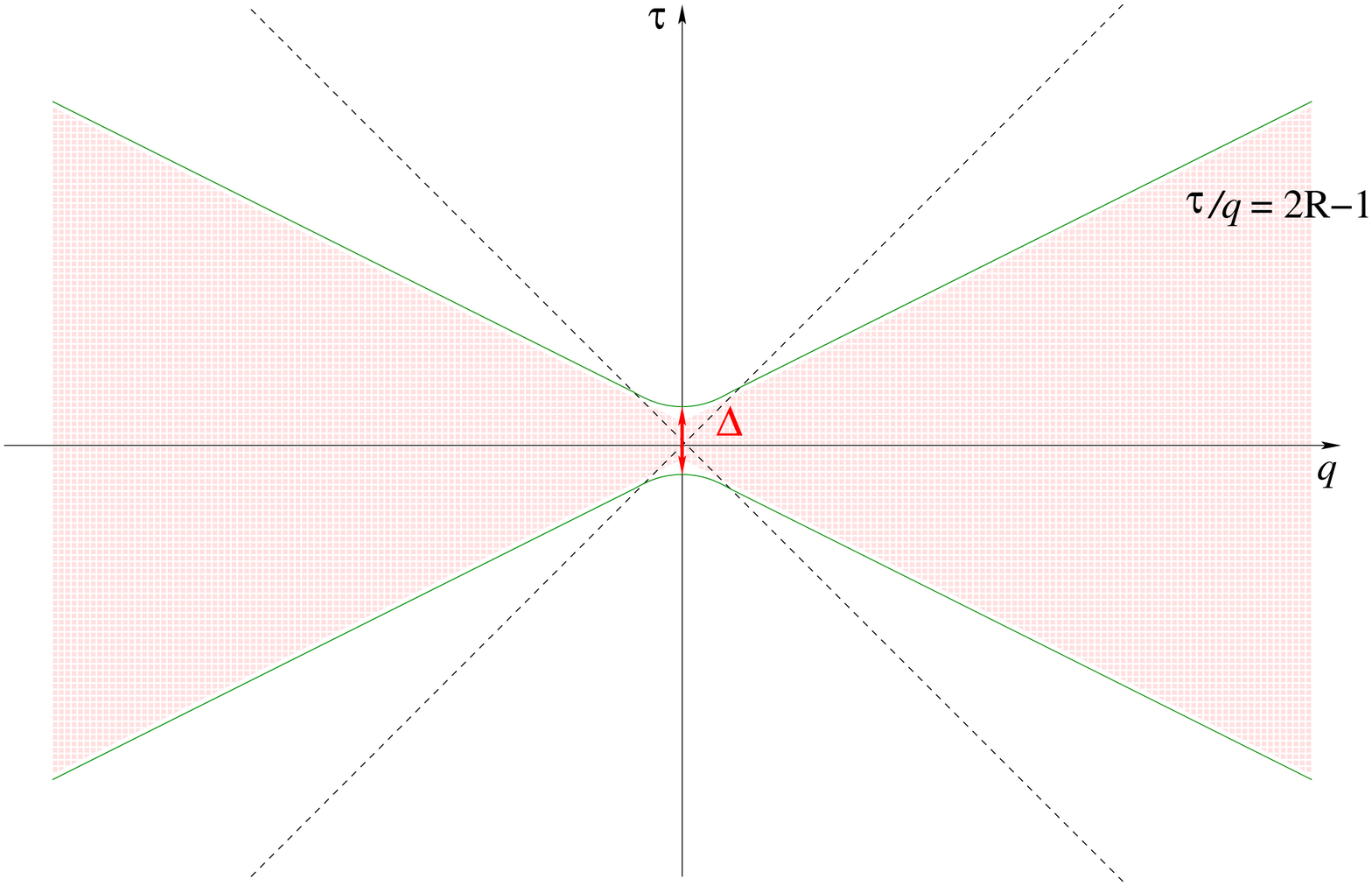}{9cm}{space}

For $1/2<R<1$ the resulting space is deformed
so that asymptotically it can be considered as two conic regions
$|\tf/\xf|<2R-1$ (fig. \ref{space}). 
When one approaches the origin, one finds that the two cones
are smoothly glued along a finite interval of length
\be
\Delta={2R} \log\left[{R\over 1-R}{a_0\over a_1}\right].
\plabel{delt}
\ee
From this result it is easy to understand what happens when we switch off
the perturbation. This corresponds to the limit $a_1\to 0$.
Then the interval $\Delta$ (the minimal distance
between the upper and lower boundaries) logarithmically diverges and the boundaries
go away to infinity. In this way we recover the entire Minkowski space.
The similar picture emerges in the limit $R\to 1$ when we return to
the case $R\ge 1$.

For $R<1/2$ the picture is similar to the previous case, but the time
and space coordinates are exchanged so that the picture should be rotated by
$90^{\circ}$.

For special value $R=1/2$, the space is also deformed and has the form of a strip.

It is interesting that the deformed spacetime shown on fig. \ref{space} 
exhibits the same singularity
as the one we found in the analysis of the free energy 
and the perturbed Fermi sea.
The singularity appears when the length $\Delta$ vanishes so that
the two conic regions separate from each other.
The corresponding critical value of $\mu$ is exactly the same as \Ref{crp}
after the duality transformation \Ref{dualtran}.

The obtained conic spacetimes possess boundaries.
Therefore, some boundary conditions should be imposed on the fields
propagating there. The most natural ones are 
vanishing and periodic boundary conditions.
However, we have not been able to make a definite choice which conditions 
are relevant.

Another unsolved problem is related to the thermodynamical
interpretation. We saw that the perturbed 2D string theory
in the Minkowskian spacetime can be considered as a theory at finite
temperature $T=1/(2\pi R)$. It would be very interesting
to reproduce this result from the analysis of quantum field theory
in the spacetime obtained above.
It is well known that moving boundaries or a compactification with varying radius
can give rise to a thermal spectrum of observed particles \cite{BirDav}.
Therefore, our picture seems to be reasonable. However, the problem is
technically difficult due to the complicated form of the boundary near the origin.

\subsection{Relation to string background}

According to the analysis of section \ref{chMQM}.\ref{identld},
a massless scalar field of the collective theory in the flat coordinates
can be identified with the tachyon field redefined by a dilaton factor. 
This was shown for the case of the linear dilaton background, but 
it is valid also in the perturbed case.
Indeed, let us compare the action \Ref{Sg} 
to the low-energy effective action for
the tachyon field \Ref{Stach} restricted to two dimensions.
They coincide if one makes the usual identification $T=e^{\dl}\eta$
and requires the following property
\be
\mt^2=(\dd\dl)^2-\dd^2\dl-{4\pal}=0,
\plabel{mas}
\ee
which ensures that $\eta$ is a massless field. This condition appears as
an additional constraint for the equations of motion on the background
fields. Its appearance is directly related to the restriction on the form
of the action coming from the Das--Jevicki formalism. In particular,
if the determinant of the matrix in \Ref{metr:Sd} was arbitrary,
we would not have this condition.

In \cite{ALEX} we argued that the constraint \Ref{mas} selects
a unique solution of the equations of motion similarly
to an initial condition. It ensures that 
the metric and dilaton are fixed as in the usual linear dilaton
background \Ref{ldl}. As we know from section  \ref{chSTR}.\ref{intach}, 
the tachyon can not be fixed by equations written
in the leading order in $\alp$. We expect that it does not vanish and modified
by perturbations from the Liouville form \Ref{soltach}.
However, we cannot expect that the dilaton or the metric are modified too.
Thus, the introduction of arbitrary tachyon perturbation cannot change
the local structure of the target space: it always remains flat.

Note that this result illustrates that the T-dual theories
on the world sheet are not the same in the target space. Whereas
the CFT perturbed by windings looks similar to that of perturbed by
tachyons, the former was supposed to correspond to the black
hole background and the latter lives always in the target space of
the vanishing curvature.

We conclude that the field $\eta$ representing the fluctuations of the density
of the matrix model and the coordinates \Ref{flcor} 
can be seen as the tachyon and coordinates
of the string target space, respectively.
However, as it was discussed in the end of section \ref{chMQM}.\ref{identld},
this identification is valid only in the free asymptotic region where one can neglect
the Liouville and other interactions. In particular, in this region the change 
of coordinates \Ref{flcor} becomes inessential.

The exact relation between the collective field theory and the target space of string theory
requires identification of the tachyon with
the loop operator rather than with the density (see \Ref{WLT}).
The former is the Laplace transform of the latter, so that the relation between
the density and the string tachyon is non-local.
For the simplest case of the ground state, it was shown 
that this integral transform maps the Klein--Gordon equation
to the Liouville equation.
In our case we expect to obtain the Liouville equation 
perturbed by the higher vertex operators.
Let us check whether this is the case.

The loop operator $W(l,t)$ is related to the density fluctuations $\eta(x,t)$
by \Ref{WLphi}. Therefore, to derive an equation on $W(l,t)$, we should
make the Laplace transform of the equation of motion following from \Ref{metr:Sdfpp}.
It is convenient to rewrite this equation in the following form
\be
\p_t^2 \eta+\p_x\( p_+p_-\p_x\eta\)+x\p_x\eta+\p_x\( (p_++p_-)\p_t \eta\)=0.
\plabel{eqmpp}
\ee
To obtain this result we used the equation \Ref{Hopfeq}
on $p_{\pm}(x,t)$ and the condition \Ref{metr:ooooo}.
Now we differentiate \Ref{eqmpp} with respect to $x$ and substitute
\be
\p_x\eta(x,t)\longrightarrow \int\limits_{-i\infty}^{i\infty} dl\, e^{lx} W(l,t).
\plabel{etaxW}
\ee
The resulting equation reads
\be 
\[\p_t^2 +l^2 p_+(-\p_l,t)p_-(-\p_l,t) -l\p_l+l^2 \(p_+(-\p_l,t)+p_-(-\p_l,t)  \){1\over l} \p_t \]W(l,t)=0.
\plabel{eqW}
\ee
Since, in general, $p_+(x,t)$ and $p_-(x,t)$ are very complicated functions, 
the derived equation does not look like the Liouville equation perturbed by vertex operators.
Moreover, the operator in the left hand side is pseudodifferential. 
This becomes clear if we consider a limit of \Ref{eqW} where it takes a more 
explicit form.

Let us study the case of the Sine--Liouville perturbation. Then the function $p(x,t)$ is represented
in the parametric form \Ref{SLgen} with \Ref{metr:param}.
When $\lambda=0$, \Ref{eqW} reduces to the Liouville equation
and one returns to the case of the linear dilaton background.
We are interested in the first $\lambda$ correction to this result. This is equivalent to the
linear approximation in $a_1$.
To this order, one has
\beq
\op(\o,t)&\approx&  -\o-{2a_1\over a_0}{\sinh{t\over R} \sinh\(\(1-{1\over R}\)\o\) \over \sinh\o}, \\
x(\o,t)&\approx& a_0\cosh\o +a_1 \cosh\(\(1-{1\over R}\)\o+{t\over R}\), \\ 
p(\o,t)&\approx& a_0\sinh\o +a_1 \sinh\(\(1-{1\over R}\)\o+{t\over R}\), \\ 
\pp(\o,t)&\approx& -a_0\sinh\o -a_1 \[\cosh\(\(1-{1\over R}\)\o-{t\over R}\)
+ {2\sinh{t\over R} \sinh\(\(1-{1\over R}\)\o\) \over \tanh\o}\]. 
\eeq
Substituting these expansions into \Ref{eqW}, one obtains
\be
\[p_t^2-(l\p_l)^2+a_0^2l^2 +2a_1l^2\left\{ a_0\cosh{t\over R}\,\,\cosh{\hat\o\over R}+
\sinh{t\over R}\,\,\,{\sinh{\hat\o\over R}\over\sinh\hat\o}{1\over l}\p_t \right\}\]W(l,t)=0,
\plabel{eqWl}
\ee 
where $\hat\o$ is considered as a differential operator $\hat\o={\rm arccosh}\(-{1\over a_0}\p_l\)$.
This shows that even the first term in the $\lambda$-expansion of the equation
on the loop operator of the matrix model does not have a simple form.

The usual identification of the Liouville coordinate $\phi$ suggests that $l=e^{-\phi}/\sqrt{2}$.
This does not give any simplifications in the equation \Ref{eqWl}. 
May be they appear at least in the limit $\phi\to \infty$?
Then $\hat\o\sim \phi+\log\p_\phi$ and \Ref{eqWl}
reduces to
\be
\[p_t^2-(\p_\phi)^2+{a_0^2\over 2}\, e^{-2\phi} +
{a_0a_1 \over 2}\,e^{-2\phi}\( {\sqrt{2}\over a_0}e^{\phi}\p_{\phi} \)^{1\over R}
\left\{ \cosh{t\over R}+
\sinh{t\over R}\( \p_{\phi} \)^{-1}\p_t \right\}\]W(l(\phi),t)=0.
\plabel{eqWphi}
\ee 
We see that although the last term scales as $e^{\({1\over R}-2\) \phi}$,
similarly to the Sine--Liouville term,
it contains also pseudodifferential operators like $(\p_\phi)^{1/R}$.
As a result, the loop operator does not satisfy any equations of the Liouville type 
even in the asymptotic region and in the weak coupling regime. 

This result suggests that the tachyon field of string theory is related to
the collective field of MQM in a more complicated way than by the transformation
\Ref{WLT}. The exact relation can require the Laplace transform with respect to
a different variable and a more complicated relation between the momentum
and the Liouville coordinate.
We have not succeeded to find it. Moreover, it seems that the analysis of the 
leading order in $\eta$ is not enough to find the string background.
For example, we showed how to relate a matrix model solution to a
solution of the Klein--Gordon equation.
As we know from section \ref{chMQM}.\ref{identld}, the latter can be transformed to
a solution of the Liouville equation. Therefore, one could conclude that 
even a perturbed solution of the matrix model corresponds to the linear dilaton
background, what is, of course, not true. Thus, one should involve an additional 
information to uniquely fix the background.

Note that all these results concern 2D string theory
with tachyon condensation, but they are not related (at least directly)
to the most interesting problem of 2D string theory in curved backgrounds.
The latter should be obtained by winding perturbations.
Although from the CFT point of view they are not much different from the
tachyon perturbations, in the MQM framework their description is much more
complicated. For the tachyon modes there is a powerful free fermion
representation and, as a consequence, the Das--Jevicki collective field
theory. It is this formulation that allows to
get some information about the string target space.
For windings there is no analog of this formalism and how to show,
for example, that the winding perturbations of MQM correspond to
the black hole background is still an open problem.


\chapter{MQM and Normal Matrix Model}
\label{chNORM}

In this chapter we present the next work included in this thesis
\cite{AKKNMM}.
It stays a little bit aside the main
line of our investigation. Nevertheless it opens one more aspect
of 2D string theory in non-trivial backgrounds and its Matrix Quantum
Mechanical formulation, in particular. This work establishes
an equivalence of the tachyonic perturbations of MQM described
in the previous chapter with the so called Normal Matrix Model (NMM).
This model appears in the study of various physical
and mathematical problems.
Therefore, before to discuss our results, we will briefly describe
the main features of NMM and the related issues.

\section{Normal matrix model and its applications}
\label{normap}

\subsection{Definition of the model}

The Normal Matrix Model was first introduced in \cite{ChauYu,ChauZ}.
It is a statistical model of random complex
matrices which commute with their conjugates
\be
[Z,\bZ]=0.
\plabel{defnm}
\ee
As usual, its partition function is represented as the matrix integral
\be
Z_N=\int d\nu(Z, \bZ)\, \exp\[ -{\bh} W(Z,\bZ)\] ,
\plabel{normint}
\ee
where the measure $d\nu$ is a restriction of the usual measure
on the space of all $N\times N$ complex matrices to
those that satisfy the relation \Ref{defnm}.
For our purposes we introduced explicitly
the Planck constant at the place of $N$. They are supposed to be related
in the large $N$ limit which is obtained as $N\to \infty,\ \hb\to 0$
with $\hb N$ fixed.

The potentials which can be considered are quite general. We will be
interested especially in the following type of potentials
\be
W_{R}(Z,\bZ)=\tr (Z\bZ)^R -\hb\gamma \tr \log (Z\bZ) -
\tr V(Z)-\tr \bV(\bZ).
\plabel{potNMM}
\ee
where
\be
V(Z)=\sum_{k\ge 1} t_k Z^k ,
\qquad
\bV= \sum_{k\ge 1}t_{-k}{\bZ}^k.
\plabel{normpot}
\ee
The probability measure with such a potential depends on several parameters.
These are $R$, $\gamma$ and two sets of $t_{n}$ and $t_{-n}$.
The latter are considered as coupling constants because the dependence
on them of the partition function contains an information about
correlators of the matrix operators $\tr Z^k$.
The other two parameters $R$ and $\gamma$ are just
real numbers characterizing the particular model.

The partition function \Ref{normint} resembles the two-matrix model studied
in section \ref{chMM}.\ref{twomm}. Similarly to that model, one can do
the reduction to eigenvalues. The difference with respect to 2MM
is that now the eigenvalues are complex numbers and the measure takes the form
\be
d\nu(Z, \bZ)={1\over N!}[d\Omega]_{SU(N)}
\prod\limits_{k=1}^N d^2 z_k\,|\Delta(z)|^2.
\plabel{normmes}
\ee
Thus, instead of two integrals over real lines one has one integral over
the complex plane. Despite this difference, one can still introduce
the orthogonal polynomials and the related fermionic representation.
Then, repeating the arguments of section \ref{chMM}.\ref{tmmtau},
it is easy to prove that the partition function \Ref{normint}
is a $\tau$-function of Toda hierarchy as well.
For the case of $R=1$ this was demonstrated in
\cite{kkvwz} and for generic $R$
the proof can be found in appendix A of \cite{AKKNMM}.
In fact, for the particular case $R=1$
(and $\gamma=0$) it was proven \cite{KM} that NMM and 2MM coincide
in the sense that they possess the same free energy as function of
the coupling constants. Nevertheless, their interpretation remains different.

The eigenvalue distribution of NMM in the large $N$ limit is also similar to
the picture arising in 2MM and shown on fig. \ref{eigtwo}.
The eigenvalues fill some compact spots on a two-dimensional plane.
The only difference is that earlier this was the plane formed by real
eigenvalues of the two matrices, and now this is the complex $z$-plane.
Therefore, the width of the spots does not have anymore a direct
interpretation in terms of densities. 
Instead, the density inside the spots for a generic potential
can be non-trivial. For example, for the potential \Ref{potNMM} it is given by
\be
\rho(z,\bz)={1\over \pi}
\p_{z}\p_{\bz} W_{R}(z,\bz)={R^2\over \pi}(z\bz)^{R-1}.
\plabel{normdens}
\ee
For $R=1$ where NMM reduces to 2MM, we return to the constant density.

\subsection{Applications}

Recently, the Normal Matrix Model found various physical applications
\cite{Znorm,WZn}. Most remarkably,
it describes phenomena whose characteristic scale
differs by a factor of $10^9$. Whereas some of these phenomena
are purely classical, another ones are purely quantum.

\subsubsection{Quantum Hall effect}

First, we mention the relation of NMM to the Quantum Hall effect
\cite{WIEGAGAM}.
There one considers electrons on a plane in a strong magnetic field $B$.
The spectrum of such system consists from Landau levels.
Even if the magnetic field is not uniform, the lowest level is highly
degenerate. The degeneracy is given by the integer part of the total magnetic
flux ${1\over 2\pi \hb}\int B(z) d^2z$, and the one-particle wave functions
at this level have the following form
\be
\psi_n(z)=P_n(z)\exp\(- {W(z)\over 2\hb}\).
\plabel{QHEf}
\ee
Here $W(z)$ is related to the magnetic field through $B(z)=\hf \Delta W(z)$
and $P_n(z)$ are holomorphic polynomials of degree $n$ with the first
coefficient normalized to $1$.

Usually, one is interested
in situations when all states at the lowest level are occupied.
Then the wave function of $N$ electrons is the Slater determinant of
the one-particle wave functions \Ref{QHEf}.
Hence, it can be represented as
\be
\Psi_N(z_1, \dots,z_N)={1\over \sqrt{N!}}\Delta(z)\exp\(-{1\over 2\hb}
\sum\limits_{k=1}^N W(z_k)\).
\plabel{QHEF}
\ee
Its norm coincides with the probability measure of NMM.
Therefore, the partition function \Ref{normint} appears in this picture as
a normalization factor of the $N$-particle wave function
\be
Z_N=\int\prod\limits_{k=1}^N d^2 z_k \, |\Psi_N(z_1, \dots,z_N)|^2.
\plabel{QHEZ}
\ee
Similarly, the density of electrons can be identified with the eigenvalue
density and the same is true for their correlation functions.

Due to this identification, in the quasiclassical limit the study
of eigenvalue spots is equivalent to the study of electronic droplets.
In particular, varying the matrix model potential one can investigate
how the shape of the droplets changes with the magnetic field.
At the same time, varying the parameter $\hb N$ one examines its evolution
with increasing the number of electrons.

Note that although we discussed the semiclassical regime, the system remains
intrinsically quantum. The reason is that all electrons
under consideration occupy
the same lowest level, whereas the usual classical limit implies
that higher energy levels are most important.

\subsubsection{Laplacian growth and interface dynamics}

It was shown \cite{WIEGAGAM} that when one increases the number of
electrons the semiclassical droplets from the previous paragraph
evolve according to the so called Darcy's law which is also known as
Laplacian growth. It states that
the normal velocity of the boundary of a droplet occupying a simply
connected domain $\CD$ is proportional to the gradient of a scalar function
\be
{1\over \hb}{\delta \vec n \over \delta N}\sim \vec\nabla \vp(z),
\plabel{Darcy}
\ee
which is harmonic outside the droplet and vanishes at its boundary
\bea{rcl}{harpot}
\Delta\vp(z,\bz)&=&0, \quad z\in \Cb\backslash \CD,
\\
\vp(z,\bz)&=&0, \quad z\in \p\CD.
\eea
In the matrix model this function appears as the following correlator
\be
\vp(z,\bz)=\hb \langl \tr\(\log(z-Z)(\bz-Z^{\dagger}) \)\rangl.
\plabel{matphi}
\ee

It turns out that exactly the same law governs the dynamics of viscous flows.
This phenomenon appears when an incompressible fluid with negligible
viscosity is injected into a viscous fluid. In this case the harmonic
function $\vp$ has a concrete physical meaning. It is identified with
the pressure in the viscous fluid $\vp=-P$.
In fact, the Darcy's law is only an approximation to a real evolution.
Whereas the condition $P=0$
in the incompressible fluid is reasonable, the vanishing of the pressure
at the interface is valid only when the surface tension can be neglected.
This approximation fails to be true when the curvature of the boundary
becomes large. Then the dynamics is unstable and the incompressible fluid
develops many fingers so that its shape looks as a fractal. This is known
as the Saffman--Taylor fingering.

NMM provides a mathematical
framework for the description of this phenomenon.
From the previous discussion it is clear that it describes the
interface dynamics in the large $N$ limit.
In this approximation the singularity corresponding to the described
instabilities arises when the eigenvalue droplet forms a spike
which we encountered already in the study of the Fermi sea of MQM
(section \ref{chTACH}.\ref{solSL}). We know that at this point
the quasiclassical approximation is not valid anymore.
But the full quantum description still exists.
Therefore, it is natural to expect that the fingering, which is a feature
of the Laplacian growth, can be captured by including next orders of
the $1/N$ expansion of NMM.

\subsubsection{Complex analysis}

The Darcy's law \Ref{Darcy} shows that there is a relation between NMM
and several problems of complex analysis. Indeed, on the one hand, it
can be derived from NMM as the evolution law of eigenvalue droplets and,
on the other hand, it gives rise to a problem to find a harmonic function
given by the domain in the complex plane.
The latter problem appears in different contexts
such as the conformal mapping
problem, the Dirichlet boundary problem, and the 2D inverse potential
problem \cite{Zabrodin}.
For instance, if we fix the asymptotics of $\vp(z,\bz)$ requiring that
\be
\vp(z,\bz)\under{\sim}{z\to\infty} \log|z|,
\plabel{assphi}
\ee
the solution of \Ref{harpot} is unique and given by the holomorphic
function $\o(z)$
\be
\vp(z,\bz)=\log|\o(z)|,
\plabel{solphi}
\ee
which maps the domain $\CD$ onto the exterior of the
unit circle and has infinity as a fixed point.
To find such a function is the content of the conformal mapping problem.

Further, the Dirichlet boundary problem, which is to find
a harmonic function $f(z,\bz)$ in the exterior domain given a function
$g(z)$ on the boundary of $\CD$, is solved in terms of the above
defined $\o(z)$. The solution is given by
\be
f(z,\bz)=-{1\over \pi i}\oint_{\p\CD}g(\zeta)\p_{\zeta}G(z,\zeta)d\zeta,
\plabel{solDir}
\ee
where the Green function is
\be
G(z,\zeta)=\log\left| {\o(z)-\o(\zeta) \over
\o(z)\overline{\o(\zeta)}-1}\right|.
\plabel{Greenf}
\ee
In turn, $\o(z)$ is obtained as the holomorphic part of $G(z,\infty)$.

Finally, the inverse potential problem can be formulated as follows.
Let the domain $\CD$ is filled by a charge spread with some density.
The charge creates an electrostatic potential
which is characterized by two functions
$\vp_+$ and $\vp_-$ defined in the interior and exterior domains,
respectively. They and their derivatives are continuous at the boundary
$\p\CD$. The problem is to restore the form of the charged domain
given one of these functions.
At the same time, when both of them are known the task can be trivially
accomplished. Therefore, the problem is equivalent to the question
how to restore $\vp_-$ from $\vp_+$.
Its relation to the Dirichlet boundary problem becomes now evident because,
since $\vp_-$ is harmonic, it is given (up to a logarithmic singularity
at infinity) by the formula \Ref{solDir} with $g=\vp_+$.

Thus, we see that the Normal Matrix Model provides a unified description for
all these mathematical and physical problems. The main lesson which
we learn from this is that all of them possess a hidden integrable
structure revealed in NMM as Toda integrable hierarchy.

\newpage

\section{Dual formulation of compactified MQM}

\subsection{Tachyon perturbations of MQM as Normal Matrix Model}

In section \ref{chTACH}.\ref{tachprof} we showed how to introduce
tachyon perturbations into the Matrix Quantum Mechanical
description of 2D string theory.
Although it is still the matrix model framework, we have done it
in an unusual way. Instead to deform the matrix model potential,
we have reduced MQM to the singlet sector and
changed there the one-fermion wave functions.
One can ask: can the resulting partition function be represented
directly as a matrix integral?

In fact, the tachyon perturbations of MQM are quite similar to
the perturbations of the two-matrix model. 
First, they are both described by Toda hierarchy.
Second, the phase space of MQM looks as the eigenvalue $(x,y)$ plane
of 2MM. In the former case the fermions associated with the time
dependent eigenvalues fill the non-compact Fermi sea,
whereas in the latter case they fill some spots. If the 2MM potential
is unstable like the inverse oscillator potential $-\xp\xm$,
the spots will be non-compact as well. Thus, it is tempting to
identify the two pictures.

Of course, MQM is much richer theory than 2MM and one may
wonder how such different theories could be equivalent. The answer is
that we have restricted ourselves just to a little sector of MQM.
First, we use the restriction to the singlet sector and, second,
we are interested only in the scattering processes. This explains
why only two matrices appear. They are associated with in-coming
and out-going states or, in other words,
with $M(-\infty)$ and $M(\infty)$.

But it is easy to guess that the idea to identify MQM perturbed by
tachyons with 2MM does not work. Indeed, the partition function of 2MM
is a $\tau$-function of Toda hierarchy when it is considered in the
canonical ensemble. Therefore, one should find a representation of
the grand canonical partition function of MQM which
has the form of a canonical one. Such a representation does exist
and is given by \Ref{FRENooo}. But it implies a discrete equally spaced
energy spectrum. It is evident for the system without perturbations
where the partition function is a product of $\CR$-factors \Ref{rfactor}
corresponding to $E_n=-\mu+i\hb {n+\hf\over R}$.
However, it is difficult to obtain such a spectrum from two
hermitian matrices. Moreover, one can show that
their diagonalization would produce
Vandermonde determinants of monomials of incorrect powers.

All these problems are resolved if one considers another model of
two matrices which is the NMM. In the work \cite{AKKNMM}
we proved that the grand canonical partition function of MQM
with tachyon perturbations coincides with a certain analytical continuation
of the canonical partition function of NMM. Thus, NMM can be regarded
as a new realization of 2D string theory perturbed by tachyons.

Our proof is based on the fact that the two partition functions are
$\tau$-functions of Toda hierarchy. Therefore, it is enough to show that
they are actually the same $\tau$-function. This fact follows
from the coincidence of either string equations or the initial
conditions given by the non-perturbed partition functions.
Finally, one should correctly identify the parameters of the two models.
We suggested two ways to identify the parameters.

\subsubsection{Model I}

First, let us consider NMM given by the integral \Ref{normint}
with the potential \Ref{potNMM} where $\gamma=\hf(R-1)+{\alpha\over \hb}$.
We denote its partition function by $\ZN_{\hb}(N,t,\alpha)$.
Then there is the following identification
\be
\ZM _{\hb}(\mu,t)= \lim\limits_{N\to \infty}
\ZN _{i\hb}(N,t,R\mu-i\hb N).
\plabel{PARTFU}
\ee
We proved this result by direct comparison of the two non-perturbed partition
functions. Then the coincidence \Ref{PARTFU} follows from the fact that
both $\ZM$ and $\ZN$ are $\tau$-functions.

According to \Ref{FRENooo}, the non-perturbed partition function of MQM is 
given by
\be
\ZM_{\hb}(\mu,0)=\prod\limits_{n \ge 0}
\CR\left(i\hb{n+\hf\over R}-\mu\right),
\plabel{FRENooooo}
\ee
where $\CR$ is the reflection coefficient \Ref{rfactor}. (Recall that 
$\CR$ is related to the zero mode $\phi$ of the perturbing phase
through $\log \CR(E)={i\over \hb}\phi(E)|_{t_n=0}$.)
On the other hand, the partition function of NMM similarly to 2MM can be represented
as a product \Ref{intpl} of the normalization coefficients of orthogonal polynomials. 
When all $t_{\pm k}=0$, the orthogonal polynomials
are simple monomials and the normalization factors $h_n$ are given by
\be\label{norm:hn}
h_n(\alpha)={1\over 2\pi i}
 \int_{\Cb} d^2 z \,
e^{-\bh(z\bz)^R}  (z\bz)^{(R-1)/2+{\alpha\over \hb}+n}.
\ee
Up to some inessential factors, the integral produces the same $\Gamma$-function 
which appears in the expression for the reflection coefficient $\CR$.
Therefore, up to non-perturbative corrections, one can identify
\be
h_n (\alpha)\sim 
\Gamma\left( {\alpha\over \hb R} +{n+\hf\over R}+\hf\right) \sim
\CR\left( -i\hb {n+\hf\over R}-i{\alpha\over R} \right).
\plabel{hnRfac}
\ee
Given this fact, it is trivial to establish the relation \Ref{PARTFU}
in the non-perturbed case:
\beq
& \lim\limits_{N\to \infty} \ZN_{i\hb}(N,0,R\mu-i\hb N)=
 \lim\limits_{N\to \infty}  \prod\limits_{n=0}^{N-1} 
\left. h_n(R\mu-\hb N)\right|_{\hb\to i\hb}
\nonumber \\
&=
\lim\limits_{N\to \infty}  \prod\limits_{n=0}^{N-1}
 \CR\left(i\hb  {N-n-\hf \over R}-\mu\right)
=\prod\limits_{n=0}^{\infty}
\CR\left( i\hb  {n+\hf \over R}-\mu\right)=\ZM_{\hb}(\mu,0). &
\label{norm:provefs}
\eeq
Besides, it is easy to show that a shift of the discrete charge $s$
of the $\tau$-function associated with $\ZN$ is equivalent
to an imaginary shift of $\mu$:
\be\label{norm:taumu}
\taum_{s}(\mu,t)=
\taum_{0}\(\mu+i\hb {s\over R},t\).
\ee
As we know, this is the characteristic property of the $\tau$-function of MQM.
This completes the proof of \Ref{PARTFU}.

Note that although the relation \Ref{PARTFU}
involves the large $N$ limit,
it is valid to all orders in the genus expansion.
This is because $N$ enters non-trivially through the parameter $\alpha$.
Actually, $N$ appears always in the combination with $\mu$ like in \Ref{PARTFU}. 
This can be understood from the fact that 
the discrete charge of the $\tau$-function is identified,
on the one hand, with $N$ (see \Ref{Ztau} for the 2MM case) and,
on the other hand, with $-{i\over \hb}R\mu$ (see \Ref{norm:taumu}).

\subsubsection{Model II}

This fact hints that there should exist
another way to match the two models where these parameters
are directly identified with each other 
\be\label{norm:numb}
N=-\ibh R\mu.
\ee
This gives the second model proposed in \cite{AKKNMM}, which
relates the two partition functions as follows 
\be
 \ZM_{\hb}(\mu,t)=\ZN_{i\hb}(-\ibh R\mu,t,0).
\plabel{identZZ}
\ee
This second model is simpler than the first one
because it does not involve the large $N$
limit and allows to compare the $1/N$ expansion of NMM
directly with the $1/\mu$ expansion of MQM.

The equivalence of the two partition functions is proven in the same way
as above.
As for the first model, they are both given by
$\tau$-functions of Toda hierarchy.  After the identification \Ref{norm:numb}, 
the charges of these $\tau$-functions are identical.  
Therefore, it only remains to show that without the perturbation
$\ZN_{i\hb}(N,0)$ is equal to the unperturbed
partition function \Ref{FRENooooo}.  
In this case the method of orthogonal polynomials together with \Ref{hnRfac}
gives
\be\label{norm:matrwww}
\ZN_{i\hb}(N,0)=\prod\limits_{n=0}^{N-1}\CR\left(-i\hb {n+\hf\over
R}\right).
\ee
Then we represent the finite product as a ratio of two infinite products
\be\label{norm:zzzzz}
 \ZN_{i\hb}(N,0)=\Xi (0)/\Xi (N),\qquad {\rm where} \quad
\Xi (N)=\prod\limits_{n=N}^{\infty}\CR\left(-i\hb (n+\hf)/R\right).
\ee
$\Xi (0)$ is a constant and can be neglected, whereas $\Xi (N)$ can be
rewritten as
\be\label{norm:zzzz}
 \Xi(N)=\prod\limits_{n=0}^{\infty}\CR\left(-i\hb N/R-i\hb
(n+\hf)/R\right).
\ee
Taking into account the unitarity of the $\CR$-factor,
\be\label{norm:unitrtyr}
 \overline{\CR(E) }
\CR(E) = \CR(-E) \CR(E)=1,
\ee
and substituting $N$ from \Ref{norm:numb}, we obtain
\be\label{norm:ZZZ}
\ZN_{i\hb}(-\ibh R\mu,0)\sim \Xi ^{-1}(-\ibh
R\mu)=\prod\limits_{n=0}^{\infty}\CR \left(\mu+i\hb (n+\hf)/R\right)=\ZM_{\hb}(\mu,0).
\ee
Note that the difference in
the sign of $\mu$ from \Ref{FRENooooo}
does not matter since the partition function is an even
function of $\mu$ (up non-universal terms). Since the two partition functions
are both solutions of the Toda hierarchy, the fact that they coincide
at $t_{k}=0$ implies that they coincide for arbitrary perturbation.

\subsection{Geometrical description in the classical limit and duality}

The relation of the perturbed MQM and NMM is a kind of duality.
Most explicitly, this is seen in the classical limit where
the both models have a geometrical description in terms of incompressible
liquids. In the case of MQM, it describes the Fermi sea in the phase space
parameterized by two real coordinates $\xpm$, whereas in the case of NMM
the liquid corresponds to the compact eigenvalue spots
on the complex $z$-plane.
Thus, the first conclusion is that the variables of one model are
obtained from the variables of the other
by an analytical continuation. The exact relation is the following
\be
\xp \leftrightarrow z^R,
\qquad \xm \leftrightarrow \bz^R.
\plabel{idvar}
\ee
It relates all correlators in the two models if simultaneously
one substitutes $\hb\to i\hb$ and $N=-iR\mu/\hb$.

In particular, the analytical continuation \Ref{idvar} replaces
the non-compact Fermi sea of MQM by a compact eigenvalue spot of NMM.
We have already discussed in the context of MQM that
the profile of the Fermi sea is determined by the solution $\xm(\xp)$
(or or its inverse $\xp(\xm)$ depending on what asymptotics is considered)
of the string equation \Ref{conteq}.
The same is true for the boundary of the NMM spot.
Since the two models coincide,
the string equations are also the same up to the change \Ref{idvar}.
Nevertheless, they define different profiles.
The MQM equation gives a non-compact curve and
the NMM equation leads to a compact
one. For example, when all $t_n=0$ the two equations read
\be
\xp\xm=\mu, \qquad (z\bz)^R=\hb N/R
\plabel{streqqq}
\ee
and describe a hyperbola and a circle, correspondingly.
This explicitly shows how the analytical continuation
relates the Fermi seas of the two models.

A more transparent way to present this relation is to consider
a complex curve associated with the solution in the classical limit.
We showed in section \ref{chMM}.\ref{cctwo} how to construct such a curve
for the two-matrix model. NMM does not differ from 2MM in this respect
and the construction can be repeated in our case.
The only problem is that for generic $R$ the potential \Ref{potNMM}
involves infinite branch singular point.
Therefore, the curve given by the Riemann
surface of the function $\bz(z)$, which describes the shape of the eigenvalue
spots, is not rational anymore and the results of \cite{KM} cannot
be applied. However, the complex curve constructed as a ``double'' still
exists and has the same structure as for the simple $R=1$ case
shown in fig. \ref{double}.

In the work \cite{AKKNMM} we considered the situation
when there is only one simply connected domain
filled by eigenvalues of the normal matrix. This restriction corresponds to the fact
that the dual Fermi sea of MQM is simply connected. If we give up this
restriction, it would correspond to excitations of MQM which break
the Fermi sea to several components.
They represent a very interesting issue to study but
we have not considered them yet.

\lfig{Complex curve associated with both models, NMM and MQM.
The regions filled by eigenvalues of the two models coincide
with two real sections of the curve. The duality exchanges the
$A$ and $B$ cycles which bound the filled regions.}
{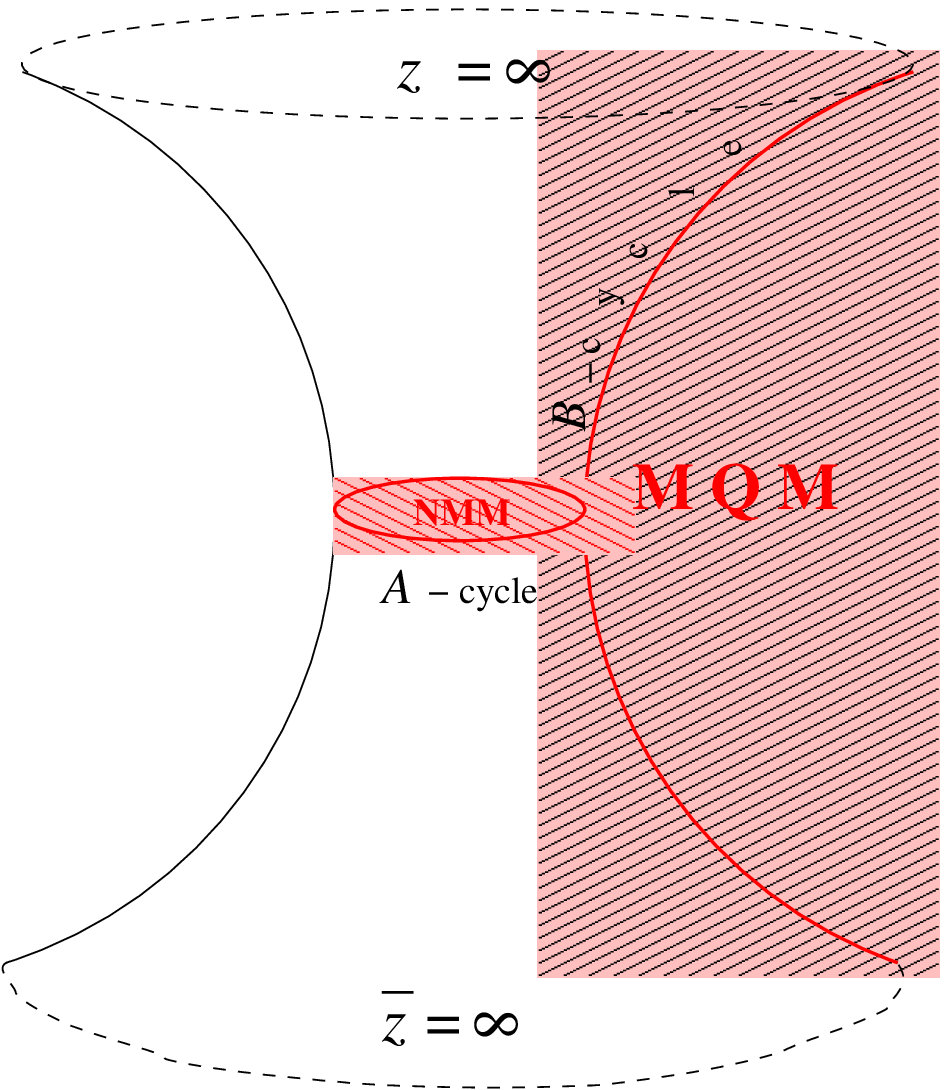}{5cm}{nmmmqm}

When there is only one spot, one gets the curve shown in fig. \ref{nmmmqm}.
It is convenient to think about it as a curve embedded into $\Cb^2$.
Let the coordinates of $\Cb^2$ are parameterized by $z$ and $\bz$.
Then the embedding is defined by the function $\bz(z)$ (or its
inverse $z(\bz)$) which is a solution of the string equation.
But we know that when $z$ and $\bz$ are considered as complex conjugated
this function defines also the boundary of the eigenvalue spot.
At the same time if one takes $z$ and $\bz$ to be real, the same
function gives the profile of the Fermi sea of MQM.
Thus, the two models are associated with two real sections of
the same complex curve: its intersection with the planes
\be
z^*=\bz \quad {\rm and } \quad  z^*=z,\ \bz^*=\bz
\plabel{plane}
\ee
coincides with the boundary of the region filled by eigenvalues
of either NMM or MQM, respectively.
It is clear that these sections can be thought also as the pair of  
non-contractible cycles $A$ and $B$ on the curve.

Moreover, the integrals over these cycles, which give the
moduli of the curve, also exhibit duality.
To discuss them, one should define a holomorphic differential to be
integrated along the cycles.
In \cite{AKKNMM} we derived it using the interpretation of the
large $N$ limit of NMM in terms of the inverse potential problem discussed
in the previous section. 
This interpretation supplies us with the notion of electrostatic
potential of a charged domain, which appears to be very natural in our context.

Such potential is a harmonic function outside the domain $\CD$ and it
is a solution of the Laplace equation with the density
\Ref{normdens} inside the domain
\be\label{norm:phiout}
\vp (z,\bar z)=\cases{ \vp (z )+\bar\vp ( \bar z),& $z\not\in \CD$,
\cr (z\bz)^R, & $z\in \CD$.}
\ee
To fix the potential completely, we should also impose some asymptotic
condition at infinity. This asymptotics is determined by the
coupling constants $t_{\pm n}, n=1,2,...$ and can be considered as the
result of placing a dipole, quadrupole etc. charges at infinity.

The solution of this electrostatic problem is obtained as follows.
The  continuity of the potential $\vp(z,  \bar z)$ and its first derivatives
leads to the following  conditions to be satisfied on the boundary $\g=\p \CD$
 \be\label{norm:dirichl}
\vp (z)+ \bar\vp (\bz)=(z\bz)^R ,
\ee
\be\label{norm:streq} z\p_z\vp (z)= \bar z \p_{\bz} \bar\vp
(\bz)=R z^R\bar z^R.
\ee
Each of two equations \Ref{norm:streq} can be interpreted as an equation 
for the contour $\gamma$. Since we obtain two equations for one curve,
they should be compatible. It is clear that these equations are nothing else but
the string equation.
Therefore, solutions for the chiral fields $\vp$ and $\bar\vp$ can
be found comparing \Ref{norm:streq} with \Ref{conteq}.
In this way we have
\be\label{norm:vppZ}
\begin{array}{rcl}
\vp(z)&= & \hb N \log z + \hf \phi  + \sum\limits_{k\ge 1}t_k z^k-
\sum\limits_{k\ge 1}{1\over k}v_{k} z^{-k}, \\
\bar \vp(\bz)&= &\hb N \log \bz + \hf \phi  +
\sum\limits_{k\ge 1}t_{-k} \bz^k-
\sum\limits_{k\ge 1}{1\over k}v_{-k} \bz^{-k} .
\end{array}
\ee
The zero mode $\phi$ is fixed by the condition \Ref{norm:dirichl}. However, 
we already know its relation to the $\tau$-function given by \Ref{taur}. Thus, in the 
dispersionless limit $\hb\to 0$, we find
\be\label{norm:zeromod}
\phi =-\hb {\p \over \p N}\log \ZN_{\hb}.
\ee

Now we can construct the holomorphic differential on the curve described above.
It is given by
\be
\vvp  \mathop{=}\limits^{\rm def}
\cases{\vvp_+(z)= \vp (z)-\hf (z\bz(z))^R
& in the north hemisphere, \cr
\vvp_-(\bz) = -\bar\vp (\bz)+\hf (z(\bz)\bz)^R
& in the south hemisphere.}
\plabel{chirf}
\ee
The field $\vvp$ gives rise to a closed (but not exact) holomorphic
1-form $d\vvp$, globally defined on the complex curve. 
Its analyticity  follows from equation \Ref{norm:dirichl}, which now holds
on the entire curve, since $z$ and $\bz$ are no more considered
as conjugated to each other.
With this definition and using the relation of the zero mode of the
electrostatic potential to the $\tau$-function, it is easy to calculate the integrals
of $d\vvp$ around the cycles:
\be
{1\over 2\pi i} \oint_A d\vvp=\hb N=R\mu, \qquad
 \int_B d\vvp=\hb {\p \over \p N}\log \ZN_{\hb}=-{1\over R}{\p \CF\over \p \mu}.
\plabel{cycls}
\ee
Here the first integral is obtained by picking up the pole
and the second integral  is given by the zero mode $\phi$ and
the diverging non-universal contribution, which we neglected.
This result can also be found by means of the procedure of
transfer of an eigenvalue from a point belonging to the boundary 
$\gamma$ of the spot to infinity. This procedure was first applied
to 2MM in \cite{KM} and then generalized to the present case in \cite{AKKNMM}.

As the relations \Ref{cycls} are written, the duality between MQM and NMM
is not seen. It becomes evident if to remember that the free energies
are taken in different ensembles. If one changes the ensemble, the
cycles are exchanged.
For example, for the canonical free energy of MQM
defined as $F=\CF+\hb R\mu M$, where $M=-{1\over \hb
R}{\p\CF\over\p\mu}$ is the number of eigenvalues, the
relations \Ref{cycls} take the following form
\be
{1\over 2\pi i} \oint_A d\vvp=\bh {\p
F\over \p M}, \qquad \int_B d\vvp= \hb M.
\plabel{intMQMF}
\ee
Thus, the duality between the two models is interpreted as
the duality with respect to
the exchange of the conjugated cycles on the complex curve.

We see that the fact that one compares the grand canonical partition
function of one model with the canonical one of the other is very crucial.
Actually, such kind of dualities can be interpreted as an electric-magnetic
duality which replaces a gauge coupling constant by its inverse \cite{vafa}.
This idea may get a concrete realization in supersymmetric gauge
theories. Their relation to matrix models was established recently
in \cite{DVc,Dvd}. This may open a new window for application of matrix
models already in critical string theories.



\chapter{Non-perturbative effects in matrix models and D-branes}
\label{chNONP}

\section{Non-perturbative effects in non-critical strings}
\label{nonperef}

In all previous chapters we considered matrix models as a tool
to produce perturbative expansions of two-dimensional gravity
coupled to matter and of string theory in various backgrounds.
However, it is well known that these theories exhibit non-perturbative
effects which play a very important role. In particular,
they are responsible for the appearance of D-branes and dualities
between different string theories (see section \ref{chSTR}.\ref{Mth}).
Are matrix models able to capture such phenomena?

We saw that in the continuum limit matrix models can have
several non-perturbative completions. For example,
in MQM one can place a wall either at the top
of the potential or symmetrically at large distances from both sides
of it. In any case, the non-perturbative definition is always
related to particularities of the regularization which is done
putting a cut-off. Therefore, it is non-universal.

Nevertheless, it turns out that the non-perturbative completion
of the perturbative results of matrix models is highly restricted.
Actually, we expect that it is determined, roughly speaking, up to
a coefficient. For example, if we consider the perturbative expansion of
the free energy
\be
F_{\rm pert}=\sum\limits_{g=0}^{\infty} \gst^{2g-2}f_g,
\plabel{pertfr}
\ee
the general form of the leading non-perturbative corrections
is the following 
\be
F_{\rm non-pert}\sim C\gst^{f_A}e^{-{f_D\over \gst}},
\plabel{nonpertfr}
\ee
where $f_A$ and $f_D$ can be found from the asymptotic behaviour
of the coefficients $f_g$ as the genus grows.
The overall constant $C$ is undetermined and reflects the
non-universality of the non-perturbative effects.

On the other hand, the matrix model free energy should reproduce
the partition function of the corresponding string theory.
More precisely, its perturbative part describes the partition function
of closed strings. At the same time, the non-perturbative corrections are
associated with open strings with ends living on a D-brane.
A particular source of non-perturbative terms of order $e^{-{1/\gst}}$
was identified with D-instantons --- D-branes localized in spacetime.
It was shown \cite{PolchinD} that
the leading term in the exponent of \Ref{nonpertfr} should be given
by a string diagram with one hole, which is a disk for the
spherical approximation \cite{PolchinD}.
The boundary of the disk lives on a D-instanton
implying the Dirichlet boundary conditions for all string coordinates.

Thus, in the two approaches one has more or less clear qualitative picture
of how the non-perturbative corrections to the partition function arise.
However, to find them explicitly, one should
understand which D-instantons are to be taken into account and how
to calculate the corrections using these D-instantons.
This is not a hard problem in the critical string theory,
whereas it remained unsolved for a long time for non-critical strings.
The obstacle was that the D-instantons can be localized in the
Liouville direction only in the strong coupling region, because
this is the region where the minimum of the energy of the brane,
which goes like $1/\gst$, is realized.
As a result, the perturbative expansion breaks down together with the
description of these D-instantons.

A clue came with the work of A. and Al. Zamolodchikov \cite{Zamol}
where the necessary D-branes were constructed in Liouville field theory.
This opened the possibility to study non-perturbative effects
in non-critical strings and to compare them with the matrix model results.
In fact, such results existed only for some class of minimal ($c<1$)
models \cite{EynZ} described in the matrix approach, for example, using 2MM
(see section \ref{chMM}.\ref{twomm}).

In the work \cite{KAK} we extended the matrix
model results on non-perturbative corrections
to the case of the $c=1$ string theory perturbed by windings. 
This was done using the Matrix Quantum
Mechanical description of section \ref{chWIND}.\ref{MMbh}.
Namely, the leading non-perturbative contribution
described by $f_D$ in \Ref{nonpertfr} was calculated. 
In our case this coefficient
is already a function of a dimensionless parameter composed from $\mu$
and the Sine--Liouville coupling $\lambda$.
Also it was verified that in all cases, including 
both the minimal unitary models
and the small coupling limit of the considered $c=1$ string,
these results can be reproduced from conformal field theory calculations
where some set of D-branes is appropriately chosen.

In the following we describe the matrix models results on the non-perturbative 
corrections in non-critical string theories and then we reproduce some of them
in the CFT framework. 



\newpage

\section{Matrix model results}

\subsection{Unitary minimal models}

To understand how the non-perturbative corrections to the partition function
appear in matrix models, let us start with the simplest case,
of pure gravity. 
It corresponds to the unitary minimal $(p,q)$ model with $p=2$, $q=3$. 
The partition sum of the model
is given by the solution of the Painleve-I equation for the matrix model free energy 
$F(\mu)$, where $\mu$ is the cosmological constant. 
More precisely, the equation is written for its second
derivative $u(\mu)=-\p_\mu^2 F(\mu)$ and reads as follows
\be
\plabel{kak:PAIN}
u^2(\mu)-{1\over 6} u''(\mu)=\mu.
\ee
In the considered case, string perturbation theory is an expansion in even powers of
$\gst=\mu^{-5/4}$:
\be
\plabel{kak:PAINs}
u(\mu)=\mu^{1/2}\sum_{h=0}^\infty c_h \mu^{-5h/2},  
\ee
where $c_0=1,c_1= -1/48,\ldots$, and $c_h\under{\sim}{h\to\infty}- 
a^{2h}\G(2h-1/2)$,
with $a=5/8\sqrt3$.
The series \Ref{kak:PAINs} is asymptotic, and hence non-perturbatively ambiguous. 
The size of the leading non-perturbative ambiguities can 
be estimated as follows. Suppose
$u$ and $\tilde u$ are two solutions of \Ref{kak:PAIN} which share 
the asymptotic behavior
\Ref{kak:PAINs}. Then, the difference between them, $\eps=\tilde u-u$, 
is exponentially small
in the limit $\mu\to\infty$, and we can treat it perturbatively. 
Plugging $\tilde u=u+\eps$
into \Ref{kak:PAIN}, and expanding to first order in $\eps$, we find that
\be
\plabel{kak:epsueps}
\eps''=12u\eps
\ee
which can be written for large $\mu$ as
\be
\plabel{kak:REPS}
{\eps'\over\eps}=r\sqrt{u}+b{u'\over u}+\cdots
\ee
with $r=-2\sqrt3$, $b=-1/4$. 
Using \Ref{kak:PAINs}, $u=\sqrt\mu+\cdots$, one finds that
\be
\plabel{kak:EXPCOR}
\eps\propto\mu^{-{1\over8}}e^{-{8\sqrt{3}\over5}\mu^{5\over4}}.
\ee
As we mentioned, the constant of proportionality in \Ref{kak:EXPCOR} 
is a free parameter of  the solution
and cannot be determined from the string equation \Ref{kak:PAIN} 
without further physical input.

This example demonstrates the general procedure to extract non-perturbative
corrections in the matrix model framework. All that we need is to know a differential
equation on the free energy (string partition function). Then the leading
behaviour of the corrections follows from the expansion around a perturbative
solution.

Another lesson is that it is quite convenient to look for the answer in the form of the
quantity $r$ defined as in \Ref{kak:REPS}:
\be
\plabel{kak:rdeff}
r={\p_{\mu}\log\eps \over\sqrt{u}},
\ee
where $\eps$ is again the leading non-perturbative ambiguity in $u=-F''$.
It is clear that this quantity is directly related to the constant 
$f_A$ appearing in \Ref{nonpertfr}. 
Its main advantage in comparison with $f_A$ is that it is a pure number 
and does not depend on normalization of the string coupling $\gst$.
 
In \cite{EynZ} the analysis above was generalized to the case of $(p,p+1)$ 
minimal models coupled to gravity which correspond to the unitary series. 
It is not surprising that the authors of \cite{EynZ} found it convenient
to parameterize the results in terms of $r$ \Ref{kak:rdeff}.
It was found that for general $p$ there is in fact
a whole sequence of different solutions for $r$ labeled by two integers
$(m,n)$ which vary over the same range as the Kac indices labeling
the degenerate representations of the Virasoro algebra or 
the primary operators in the minimal models:
\be
m=1,2,\dots, p-1, \qquad
n=1,2,\dots, p \quad  {\rm and} \quad (m,n)\sim(p-m,p+1-n).
\plabel{kacind}
\ee
The result for $r_{m,n}$ was found to be:
\be
\plabel{kak:MMR}
r_{m,n}=-4\sin{\pi m\over p}\sin{\pi n\over p+1}. 
\ee
%

\subsection{$c=1$ string theory with winding perturbation}
\label{conemat}

In the work \cite{KAK} we performed the similar analysis for the compactified 
$c=1$ string theory perturbed by windings with the Sine--Liouville potential.
In the CFT framework this theory is described by the action 
\Ref{SLCFT}.\footnote{In fact, the CFT couplings can differ 
from the corresponding matrix model quantities 
by multiplicative factors. Therefore, we will distinguish between $\mu$ and $\mul$
for Liouville theory.}
Its matrix counterpart is represented by the model considered in section 
\ref{chWIND}.\ref{windmodel}.
Its solution was presented in section \ref{chWIND}.\ref{KAKsolut}.
In particular, it was shown that the Legendre transform $\FSL$ 
of the string partition sum $F$ satisfies the Toda differential equation \Ref{todafd}.
The initial condition for this equation is supplied by the 
unperturbed $c=1$ string theory on a circle. The perturbative part of its 
partition partition function was given in \Ref{FRexp}.
The full answer contains also non-perturbative corrections
which can be read off its integral representation
\be
\FSL(\mu,0)  = 
{R\over 4} \Re \int_{\Lambda^{-1}}^\infty {ds\over s}
{ e^{-i\mu s}\over \sinh{s\over 2}\sinh{s\over 2R}}=
\FSL_{\rm pert}(\mu,0) +O(e^{-2\pi\mu})+O(e^{-2\pi R \mu}).
\plabel{kak:FrenO}
\ee

The result \Ref{kak:FrenO} shows that at $\lambda=0$
there are two types of non-perturbative corrections associated 
with the poles of the integrand. These occur at $s=2\pi ik$ and
$s=2\pi R ik$, $k\in \Zb$, and give rise to the non-perturbative
effects $\exp(-2\pi\mu k)$ and $\exp(-2\pi R\mu k)$, respectively.

At finite $\lambda$, the situation is more interesting. 
In general, the corrections can become dependent on the Sine--Liouville
coupling $\lambda$.
However, the series of non-perturbative corrections
\be
\plabel{kak:dirnonpert}
\Delta \FSL=\sum_{n=1}^\infty C_n  e^{ -2\pi n\mu}
\ee
gives rise to an exact solution of the {\it full} Toda equation \Ref{todafd}. 
Due to this the corresponding instantons are
insensitive to the presence of the Sine--Liouville perturbation. 
We will return to this fact, and explain
its interpretation in Liouville theory, in the next section.

The second type of corrections, which starts at $\lambda=0$ like
$\Delta \FSL=e^{ -2\pi R\mu} $, does not solve the full equation  \Ref{todafd},
and does get $\lambda$ dependent corrections. To study these corrections,
we proceed in a similar way to that described in the previous paragraph.
Namely, we expand the differential equation on the free energy 
of the matrix model, which is in our case the Toda equation \Ref{todafd}, 
around some perturbative solution. The difference with respect to the previous case
is that now we have a partial differential equation instead of the ordinary one.
As a result, the final equation on a quantity measuring the strength of the
non-perturbative correction will be differential, whereas it was algebraic for 
the $c<1$ case.

Indeed, the linearization of the Toda equation around some solution $\FSL$,
gives
\be
\plabel{kak:todasph}
{1\over 4}\lambda^{-1}\p_{\lambda}\lambda\p_{\lambda}\eps(\mu, \lambda)
-4e^{-\p^2_{\mu}\FSL_0(\mu,\lambda)}
\sin^2\(\hf{\p\over \p\mu}\)\eps(\mu,\lambda)=0, 
\ee
where in the exponential in the second term we approximated
\be
\plabel{kak:sphfr}
4\sin^2\(\hf{\p\over \p\mu}\)\FSL(\mu,\lambda)\simeq
\p^2_{\mu}\FSL_0= R\log\xi+X(y). 
\ee
This is similar to the fact that in the discussion of the Painleve equation,
one can replace $u$ in \Ref{kak:epsueps} by its spherical limit $\sqrt{\mu}$.
This also means that to find the leading correction one should know
explicitly only the spherical part of the perturbative expansion.
After the  change of variables from $(\lambda, \mu)$ to $(\xi,y)$ defined in
\Ref{scp} (we change the notation from $w$ to $y$ to follow the paper \cite{KAK}), 
equation \Ref{kak:todasph} can be written as
\be
\plabel{kak:eqep}
\alpha \xi^2 (y\p_y +\xi\p_{\xi})^2\eps(\xi,y)
=4 e^{-X(y)}
\sin^2\( {\xi\over 2}\p_y \)\eps(\xi,y),
\ee
where $\alpha\equiv {R-1\over (2-R)^2}$. As in section \ref{chWIND}.\ref{MMbh},
we will work only with radii $1<R<2$ which include the black hole radius $R=3/2$.
 
To proceed further, we should plug in some ansatz for $\eps$ into equation
\Ref{kak:eqep}. We expect that the leading non-perturbative correction
has the exponential form \Ref{nonpertfr}. Since in the $c=1$ theory the string coupling 
is proportional to $1/\mu$,
we use the following ansatz 
\be
\plabel{kak:anz}
\eps(\xi,y)= P(\xi,y) e^{-\mu f(y)}. 
\ee
Here $P(\xi,y)$ is a power-like prefactor in $\gst$, and $f(y)$ is the
function we are interested in (the analogue of $r$ in the minimal models).
Substituting \Ref{kak:anz} into \Ref{kak:eqep} and keeping only the leading terms in
the $\xi\to 0$  limit, one finds the following first order  differential
equation
\be
\plabel{kak:eqg}
\sqrt{\alpha}\,e^{\hf X(y)} (1- y\p_y) g(y)
=\pm \sin\[\p_y g(y)\],  
\ee
where we introduced
\be
\plabel{kak:ggffyy}
g(y)=\hf yf(y).
\ee
The $\pm$ in \Ref{kak:eqg} is due to the fact that one actually finds
the square of this equation. Below we will show that the solution with the minus sign
is in fact unphysical. But for a while we keep both signs.

Equation \Ref{kak:eqg} is a first order differential equation in $y$, and to
solve it we need to specify boundary conditions. As discussed earlier
for the perturbative series, it is natural to specify these boundary
conditions at $\lambda\to 0$, or $y\to\infty$. We saw
above that there are two solutions,
$f(y\to\infty)\to 2\pi$ or $2\pi R$. This implies via \Ref{kak:ggffyy} that
$g(y\to\infty)\simeq \pi y$ or $\pi Ry$. We already saw that
$g(y)=\pi y$ gives an exact solution, and this is true
for \Ref{kak:eqg} as well (as it should be). Thus, to study non-trivial
non-perturbative effects, we must take the other boundary condition
\be
\plabel{kak:incon}
g(y\to\infty)\simeq\pi R y. 
\ee

Remarkably, the non-linear differential 
equation \Ref{kak:eqg} is exactly solvable. 
For the initial condition \Ref{kak:incon},
the solution can be written as \cite{KAK}
\be
\plabel{kak:gphi}
g(y)=y\phi(y) \pm {1\over \sqrt{\alpha}}e^{-\hf X(y)}\sin \phi(y),
\ee
where $\phi(y)=\partial_y g$ satisfies the equation
\be
\plabel{kak:zzz}
e^{{2-R\over 2R}X(y)}=\pm \sqrt{R-1}\, { \sin \({1\over R} \phi\)
\over \sin \( {R-1\over R} \phi\) }.  
\ee
Equations \Ref{kak:gphi}, \Ref{kak:zzz} are the main result of this subsection.
They provide the leading non-perturbative correction for all couplings $\mu$
and $\lambda$. As we will see, they contain much more information
than it is accessible in the CFT framework.
We next discuss some features of the corresponding non-perturbative
effects.

\subsubsection{Small coupling limits}

Consider first the situation for small $\lambda$, or large $y$, when the Sine--Liouville 
term can be treated perturbatively.
The first three terms in the expansion of $\phi(y)$ are
\be
\plabel{kak:corr}
\phi(y) \approx
\pi R\pm{R \sin(\pi R)\over \sqrt{R-1}}\, y^{-{2-R\over 2}}
+{R\over 2}\sin(2\pi R)\, y^{-(2-R)}. 
\ee
This gives the following result for $f(y)$ \Ref{kak:anz}:
\beq
f(y)&=&
2\pi R \pm {4\sin(\pi R)\over\sqrt{R-1}}\, y^{-{2-R\over 2}}
+{R\sin (2\pi R)\over R-1} y^{-(2-R)} +O(y^{-3(2-R)/2})
\nonumber \\
&=& 2\pi R \pm {4\sin(\pi R)}\, \mu^{-{2-R\over 2}}\, \lambda +
R\sin(2\pi R)\, \mu^{-(2-R)}\,\lambda^2+O(\lambda^3).
\plabel{kak:corrf}
\eeq
We see that for large $y$, the expansion parameter is $y^{-{(2-R)\over2}}\sim
\lambda$, as one would expect.

Another interesting limit is  $\mu\to 0$ at fixed $\lambda$, \ie\
$y\to 0$, which leads to the Sine-Liouville model with $\mu=0$.
In this limit $X\to 0$ and
the first two terms in the expansion of $\phi$ around this point are
\be
\plabel{kak:bhphi}
\phi(y)=\phi_0+
{R\over 2}
\((R-1)\cot\({R-1\over R}\phi_0\)-\cot\({1\over R}\phi_0\)\)^{-1}y +O(y^2),
\ee
where $\phi_0$ is defined by the equation
\be
\plabel{kak:zzzbh}
 { \sin \({1\over R} \phi_0\)
\over \sin \( {R-1\over R} \phi_0\) }=\pm {1\over \sqrt{R-1}}.  
\ee
The function $f(y)$ is given in this limit by the expansion
\be
\plabel{kak:gphibh}
f(y)= \pm {2(2-R)\over y\sqrt{R-1}}\sin \phi_0
+ 2\phi_0 + O(y).
\ee
Note that the behavior of $f$ as $y\to 0$, $f\sim 1/y$, leads
to a smooth limit as $\mu\to 0$ at fixed $\lambda$. The
non-perturbative correction \Ref{kak:anz} goes like $\exp(-\mu f(y))$, so
that as $y\to 0$ the argument of the exponential goes like
$\mu/y=1/\xi$, and all dependence on $\mu$ disappears.

For $R=3/2$, which is supposed to correspond to the Euclidean black hole,
the equations simplify. One can explicitly find $\phi_0$
because \Ref{kak:zzzbh} gives
\be
\plabel{kak:eqbhR}
 \cos {\phi_0\over 3}
= \pm {1\over\sqrt{2}} \Rightarrow \phi_0={3\pi\over 4}\ {\rm or\ }
\phi_0={9\pi\over 4}.  
\ee
As a result, one finds at this value of the radius a simple result
\be
\plabel{kak:gbh}
\mu f(y)= \pm {\mu \over y}+{(6\mp3)\pi\over 2}\mu+\cdots
=\pm \hf \lambda^{4}+{(6\mp3)\pi\over 2}\mu+\cdots. 
\ee
Note that the solution with the minus sign leads to
a growing exponential, $e^{\hf \lambda^4}$.
Therefore, it can not be physical, as mentioned
above.
In fact, for the particular case $R=3/2$ one can find the whole function
$f(y)$ explicitly.
The result is \cite{KAK}
\be
f(y)=6\,\arccos\left[\pm \( 1+\sqrt{1+4y}\)^{-1/2}\right]
\pm {1\over 2y}(1+4y)^{1/4}(3-\sqrt{1+4y})~.
\plabel{kak:gggr}
\ee

\subsubsection{$c=0$ critical behaviour}

A nice consistency check of our solution is to study the RG flow from $c=1$
to $c=0$ CFT coupled to gravity. Before
coupling to gravity, the Sine-Gordon model associated to \Ref{SLCFT}
describes the following RG flow. In the UV, the Sine-Gordon coupling
effectively goes to zero, and one approaches the standard CFT of a compact
scalar field. In the IR, the potential given by the Sine-Gordon interaction
gives a world sheet mass to $X$, and the model approaches a trivial $c=0$
fixed point. This RG flow manifests
itself after coupling to gravity in the dependence of the physics on $\mu$.
Large $\mu$ corresponds to the UV limit; in it, all correlators approach
those of the $c=1$ theory coupled to gravity. Decreasing $\mu$ corresponds
in this language to the flow to the IR, with the $c=0$ behavior recovered
as $\mu$ approaches a critical value $\mu_c$. 
In fact, this critical value coincides with \Ref{crp} found studying the partition
function obtained from the matrix model.

The non-perturbative contributions to the partition function computed in this
section must follow a similar pattern. In particular, $f(y)$ must exhibit
a singularity as $y\to y_c$, with
\be
\plabel{kak:yyccrr}
y_c=-(2-R)(R-1)^{R-1\over 2-R}
\ee
and furthermore, the behavior of $f$ near this singularity should reproduce
the non-perturbative effects of the $c=0$ model coupled to gravity discussed
in the previous paragraph. Let us check whether this is the case.

Near the critical point the relation \Ref{wf} between $y$ and $X$ degenerates:
\be
\plabel{kak:yXrel}
{y_c-y\over y_c}\simeq {R-1\over 2R^2}(X-X_c)^2+O\((X-X_c)^3\).
\ee
Solving it for the critical point, one finds that
\be
\plabel{kak:xxccrr}
e^{-{2-R\over 2R}X_c}=\sqrt{R-1}.
\ee
Substituting \Ref{kak:xxccrr} in \Ref{kak:zzz} we find
\be
\plabel{kak:zzzww}
{ \sin \({1\over R} \phi\)\over \sin \( {R-1\over R} \phi\)}
={1\over R-1}.
\ee
Thus, the $c=0$ critical point corresponds
to $\phi\to 0$.\footnote{Note that if we chose
the minus sign in \Ref{kak:zzz}, we would find a more complicated solution
for $\phi$. One can show that it would lead to a wrong critical
behavior. This is an additional check of the fact that the physical
solution corresponds to the plus sign in \Ref{kak:zzz}.}
The first two terms in the expansion of $\phi$ around the singularity are
\be
\plabel{kak:crphi}
\phi(y)=\sqrt{3}(X_c-X)^{1/2}-{\sqrt{3}(R^2-2R+2) \over 20 R^2}(X_c-X)^{3/2}
+O\((X_c-X)^{5/2}\).
\ee
Substituting this in \Ref{kak:gphi} one finds
\be
\plabel{kak:gcrit}
g(y)=-y_c {2\sqrt{3}(R-1)\over 5 R^2}(X_c-X)^{5/2}+O\((X_c-X)^{5/2}\)
\ee
or, using \Ref{kak:ggffyy}:
\be
\plabel{kak:fcrit}
f(y)\approx - {8\sqrt{3}\over 5 }\({2 R^2\over R-1}\)^{1/4}
\({\mu_c-\mu \over \mu_c}\)^{5/4}.
\ee
The power of $\mu-\mu_c$ is precisely right to describe the leading
non-perturbative effect in pure gravity. It is interesting to
compare also the coefficient in \Ref{kak:fcrit} to what is expected in pure gravity.
It is most convenient to do this by again computing the quantity $r$ \Ref{kak:rdeff}
because it does not depend on the relative normalization of the $c=0$ 
cosmological constant and the critical parameter in the $c=1$ theory.
$u$ is computed by evaluating the leading singular term as $\mu\to\mu_c$
in $\p^2_{\mu}\FSL_0=R\log\xi+X(y)$. One finds
\be
\plabel{kak:rcr}
r=-2\sqrt{3}\({2 R^2\over R-1}\)^{1/4}
\({\mu_c-\mu \over \mu_c}\)^{1/4} ( X_c-X)^{-1/2}=
-2\sqrt{3}
\ee
in agreement with the result \Ref{kak:MMR} for pure gravity. This provides another
non-trivial consistency check of our solution.

\newpage

\section{Liouville analysis}

In this section we study the non-perturbative effects in non-critical strings
from the CFT point of view. As we discussed in section \ref{nonperef},
they are associated with D-instantons and given by the string disk amplitudes
with Dirichlet boundary conditions corresponding to a given instanton.
The first question that we need to address is which D-branes
should be considered for this analysis?
In other words, which D-branes contribute to the leading non-perturbative effects?

If a conformal field theory is a coupling of 
some matter to Liouville theory, all its correlations functions,
and the partition function itself, are factorized to the product of contributions
from the matter and from the Liouville part.
Due to this property we can discuss the boundary conditions in the two theories
independently from each other. The possible boundary conditions in the matter
sector will be discussed in the following subsections.
Now we will be concentrating on the Dirichlet boundary conditions in Liouville theory
discovered by Zamolodchikovs.

In the work \cite{Zamol} they constructed 
boundary states appearing as quantizations of a classical
solution for which the Liouville field $\phi$ goes to the strong coupling
region on the boundary of the world sheet.
In fact, it was shown that there is a two-parameter family of consistent quantizations.
Thus, one can talk about $(m',n')$ branes of Liouville theory.
Which of these branes should be taken in
evaluating instanton effects?

The analysis of \cite{Zamol} shows that open strings
stretched between the $(m',n')$ and $(m'',n'')$
Liouville branes belong to one of a finite number
of degenerate representations of the Virasoro algebra
with a given central charge. 
The precise set of degenerate
representations that arises depends on $m',n',m'',n''$.
Degenerate representations at $c>25$ 
(the case of the minimal models coupled to Liouville) 
occur at negative values of
world sheet scaling dimension, except for the simplest degenerate operator,
$1$, whose dimension is zero. One finds \cite{Zamol} 
that in all sectors
of open strings, except those corresponding to $m'=n'=m''=n''=1$
there are negative dimension operators. It is thus natural to conjecture
that the only stable D-instantons correspond to the case $(m',n')=(1,1)$.
We will assume this in the analysis below both for the minimal models
and for the $c=1$ string theory.

\subsection{Unitary minimal models}

First, let us briefly describe the CFT formulation of the minimal models.
In the conformal gauge they are represented by the Liouville action 
\be
\plabel{kak:LIOU}
S_L= \int {d^2\s\over 4\pi} \left( (\p\phi )^2 - Q\hat \CR\phi
+\mul e^{-2b \phi} \right),
\ee
where the central charge of the Liouville model is
\be
\plabel{kak:cliouv}
c_L=1+6Q^2
\ee
and the parameter $b$ is related to Q via the relation
\be
\plabel{kak:bQ}
Q=b+{1\over b}.
\ee
In general, $b$ and $Q$ are determined by the requirement that the 
total central charge of matter, which is given in \Ref{mincc}, 
and Liouville is equal to $26$. In our case, 
\Ref{mincc} and \Ref{kak:cliouv} imply that
\be
\plabel{kak:bminmod}
b=\sqrt{p\over q}.
\ee
An important class of conformal primaries in Liouville 
theory corresponds to the operators
\be
\plabel{kak:opalph}
V_\alpha(\phi)=e^{-2\alpha\phi}
\ee
whose scaling dimension is given by $\Delta_\alpha=
\bar\Delta_\alpha=\alpha(Q-\alpha)$.
The Liouville interaction in \Ref{kak:LIOU} is $\delta\CL=\mul V_b$.
Finally, we mention that the unitary models correspond to the series 
with $q=p+1$. In the following we restrict our attention to this 
particular case. In any case, only these models were analyzed
in the matrix approach.

Now we turn to the discussion of non-perturbative effects.
Minimal model D-branes are well understood. They were constructed and analyzed
in  \cite{CARDY}. These D-branes are in one to one
correspondence with primaries of the Virasoro algebra and, therefore, 
they are labeled by the indices from the Kac table \Ref{kacind}.
For our purposes, the main property that will be important
is the disk partition sum (or boundary entropy) corresponding
to the $(m,n)$ brane, which is given by
\be
\plabel{kak:cormn}
Z_{m,n}=\({8 \over p(p+1)}\)^{1/4}
{\sin{\pi m \over p}\sin{\pi n \over p+1} \over
\( \sin{\pi  \over p}\sin{\pi  \over p+1}\)^{1/2}}. 
\ee

The minimal model part of the background can be thought
of as a finite collection of points. All D-branes
corresponding to it are localized and therefore should
contribute to the non-perturbative effects.
Taking into account that only the (1,1) Liouville D-brane is supposed to contribute,
we conclude that the D-instantons to be considered 
in $c<1$ minimal models coupled to gravity have the form:
$(1,1)$ brane in Liouville $\times$ $(m,n)$ brane in the minimal model.
We next show that these D-branes give rise to the correct leading
non-perturbative effects \Ref{kak:MMR}.

As we explained in the previous section, 
the quantity $r$ \Ref{kak:rdeff} is very convenient to compare 
results of two theories. It turns out that that it is a
natural object to consider in the continuum approach as well.
Indeed, from the continuum point of view, $r$ is represented as follows
\be
\plabel{kak:CFTr}
r={{\p \over \p\mul} Z_{\rm disk}\over \sqrt{-{\p_{\mul}^2 F}}}, 
\ee
where in the numerator we used the fact that $\log \eps$ is the disk
partition sum corresponding to the D-instanton (see \Ref{nonpertfr}).
Thus we see that $r$ is the ratio between the one point function of 
the cosmological
constant operator $V_b$ on the disk, and the square root of its two 
point function on the
sphere. This is a very natural object to consider since it is known 
in general in CFT
that $n$ point functions on the disk behave like the square roots
of $2n$ point functions on the sphere. This is actually the reason why 
we do not have to worry about the multiplicative factor 
relating $\mu$ and $\mul$,
as it drops out in the ratio leaving just a number.

To compute \Ref{kak:CFTr}, we start with the numerator in \Ref{kak:CFTr}. We have
\be
\plabel{kak:DISKP}
{\p\over \p\mul}Z_{\rm disk}= Z_{m,n} \times \langle V_b\rangle_{(1,1)},
\ee
where we used the fact that the contribution of the minimal model is simply the
disk partition sum \Ref{kak:cormn}, and the second 
factor is the one point function of
the cosmological constant operator \Ref{kak:LIOU} on the disk with 
the boundary conditions
corresponding to the $(1,1)$ Liouville D-brane.
$Z_{m,n}$ is given by eq. \Ref{kak:cormn}, whereas the one point function
of $V_b$ can be computed through the boundary wave function 
constructed in \cite{Zamol}
\be
\plabel{kak:wavef}
\Psi_{1,1}(P)={ 2^{3/4} 2\pi i P [\pi\mul \gamma(b^2)]^{-iP/b}
\over \Gamma(1-2ibP)\Gamma(1-2iP/b)},
\ee
where
\be
\plabel{kak:ggaamm}
\gamma(x)={\Gamma(x)\over\Gamma(1-x)}.
\ee
The wave function $\Psi_{1,1}(P)$ can be interpreted as an overlap 
between the $(1,1)$ boundary state and the state with Liouville momentum $P$.
Therefore, $\Psi_{1,1}(P)$ is proportional to the one point function
on the disk, with $(1,1)$ boundary conditions, of the Liouville operator
$V_\alpha$, with
\be
\plabel{kak:alphaP}
\alpha={Q\over2}+iP.
\ee
The proportionality constant is a pure number (independent of $P$ and $Q$).
We will not attempt to calculate this number precisely. Instead 
we deduce it by matching any one of the matrix model predictions.
Then we can use it in all other calculations.
As a result, we obtain
\be
\plabel{kak:wavefP}
\langle V_b\rangle_{(1,1)}=
-C \, { 2^{1/4} \sqrt{\pi}  [\pi\mul \gamma(b^2)]^{\hf(1/b^2-1)}
\over b \Gamma(1-b^2)\Gamma(1/b^2)}.
\ee

We next move on to the denominator of \Ref{kak:CFTr}. 
This is given by the two point function
$\langle V_bV_b\rangle_{\rm sphere}$. This quantity was calculated in
\cite{DornXN,ZamolodchikovAA}. It is convenient
to first compute the three point function  
$\langle V_bV_bV_b\rangle_{\rm sphere}$ and then
integrate once, to avoid certain subtle questions regarding 
the fixing of the ${\rm SL}(2,\Cb)$
Conformal Killing Group of the sphere. The final result is \cite{KAK}
\be
\plabel{kak:twob}
\langle V_bV_b\rangle_{\rm sphere}=
{1/b^2-1\over \pi b} \left[\pi\mul\gamma(b^2)\right]^{1/b^2-1}
\gamma(b^2)\gamma(1-1/b^2).
\ee

We are now ready to compute $r$. 
Plugging in \Ref{kak:cormn}, \Ref{kak:wavefP} and
\Ref{kak:twob} into \Ref{kak:CFTr}, we find
\be
\plabel{kak:RFIN}
r_{m,n}=-2C\sin{\pi m \over p}\sin{\pi n \over p+1},
\ee
which agrees with the matrix model result \Ref{kak:MMR} if we set $C=2$.
Since $C$ is independent of $m$, $n$ and $p$,
we can fix it by matching to any one
case, and then use it in all others. Thus, we conclude that the Liouville analysis
gives the same results for the leading non-perturbative corrections
as the matrix model one.

\subsection{$c=1$ string theory with winding perturbation}

In this subsection we will discuss the Liouville interpretation of the
matrix model results presented in section \ref{conemat}. 
Following the lesson of the previous paragraph, we expect 
that the non-perturbative effects in the $c=1$ string theory are
all associated with the $(1,1)$ Liouville brane.
Thus, it remains to reveal the D-brane content of the matter sector
and to perform calculation of the corresponding correlation functions.

First, let us consider the unperturbed $c=1$ theory corresponding to $\lambda=0$. 
As we saw, in the matrix model analysis one finds two
different types of leading non-perturbative effects (see \Ref{kak:FrenO}),
$\exp(-2\pi\mu)$, and $\exp(-2\pi R\mu)$. It is not difficult to guess
the origin of these non-perturbative effects from the CFT point of
view. The $\exp(-2\pi\mu)$ contribution is due to a Dirichlet brane in the $c=1$ CFT,
\ie a brane located at a point on the circle parameterized by $X$, whereas
the $\exp(-2\pi R\mu)$ term comes from a brane wrapped around
the $X$ circle.

This identification can be verified in the same way as we did for
the minimal models in the previous paragraph. To avoid normalization issues, one can
again calculate the quantity $r$ \Ref{kak:CFTr}. The matrix model prediction
for the Neumann brane\footnote{Similar formulae can be written for the
Dirichlet brane.} is
\be
\plabel{kak:rmu}
r=-{2\pi \sqrt{R} \over \sqrt{\log{\Lambda\over\mu}}},
\ee
where we used the sphere partition function $\FSL_0(\mu,0)$ of the
unperturbed compactified $c=1$ string given by the first term in
\Ref{FRexp}.

The CFT calculation is similar to that performed in the $c<1$ case.
The partition function of $c=1$ CFT on a disk with Neumann boundary
conditions is well known and is given by
\be
\plabel{kak:pfdisc}
Z_{\rm Neumann}=2^{-1/4}\sqrt{R}.
\ee
The disk amplitude corresponding to the $(1,1)$ Liouville brane,
and the two-point function on the sphere are computed using equations
\Ref{kak:wavefP} and \Ref{kak:twob}, 
in the limit $b\to 1$. The limit is actually singular,
but computing everything for generic $b$ and taking the limit at the end
of the calculation leads to sensible, finite results. The leading behavior
of \Ref{kak:wavefP} as $b\to 1$ is
\be
\plabel{kak:psii}
\langle V_b\rangle_{(1,1)}\approx - 
{2^{5/4}\sqrt{\pi}\over \Gamma(1-b^2)},
\ee
with the constant $C$ in \Ref{kak:wavefP} chosen to be equal to $2$
according to the minimal model analysis.
The two point function \Ref{kak:twob} approaches
\be
\plabel{kak:twomu}
\p^2_{\mul} \FSL_0\simeq
-{ \log\mul\over \pi\Gamma^{2}(1-b^2)}.
\ee
Substituting \Ref{kak:pfdisc}, \Ref{kak:psii} and \Ref{kak:twomu}
into \Ref{kak:CFTr} gives precisely the result \Ref{kak:rmu}. This provides
a non-trivial check of the statement that
the constant $C$ is a pure number independent
of all the parameters of the model. 

The agreement of \Ref{kak:rmu} with the Liouville analysis supports
the identification of the Neumann D-branes as the source
of the non-perturbative effects $\exp(-2\pi R\mu)$. A similar analysis
leads to the same conclusion regarding the relation between the Dirichlet
$c=1$ branes and the non-perturbative effects $\exp(-2\pi\mu)$ (the two
kinds of branes are related by T-duality).

Having understood the structure of the unperturbed theory, we next turn to
the theory with generic $\lambda$. In the matrix model we found
that the non-perturbative effects associated with 
the Dirichlet brane localized
on the $X$ circle are in fact independent of $\lambda$ 
(see \Ref{kak:dirnonpert}
and the subsequent discussion). In the continuum formulation this 
corresponds to the claim that the disk partition sum with the $(1,1)$
boundary conditions for Liouville, and Dirichlet boundary conditions
for the matter field $X$ is $\lambda$-independent.  In other words,
all $n$-point functions of the Sine--Liouville operator given by the last term in 
\Ref{SLCFT} on the disk vanish
\be
\plabel{kak:vanishcor}
\langl \left(\int d^2z e^{(R-2)\phi}
\cos (R\tX)\right)^n\rangl_{(1,1)\times{\rm Dirichlet}}=0.
\ee
Is it reasonable to expect \Ref{kak:vanishcor} to be valid from the world
sheet point of view?  For odd $n$ \Ref{kak:vanishcor} is trivially zero
because of winding number conservation.  Indeed, the Dirichlet
boundary state for $X$ breaks translation invariance, but preserves
winding number. The perturbation in \Ref{SLCFT} carries winding number,
and for odd $n$ all terms in \Ref{kak:vanishcor} have non-zero winding
number. Thus, the correlator vanishes.

For even $n$ one has to work harder, but it is still reasonable to
expect the amplitude to vanish in this case. Indeed, consider the
T-dual statement to \Ref{kak:vanishcor}, that the $n$ point functions of the
momentum mode $\cos(X/R)$, on the disk with $(1,1)$ $\times$ Neumann
boundary conditions, vanish. This is reasonable since the operator
whose correlation functions are being computed localizes $X$ at the
minima of the cosine, while the D-brane on which the string ends is
smeared over the whole circle.  It might be possible to make this
argument precise by using the fact that in this case the D-instanton
preserves a different symmetry from that preserved by the perturbed
theory, and thus it should not contribute to the non-perturbative
effects.

To summarize, the matrix model analysis predicts that \Ref{kak:vanishcor} is
valid. We will not attempt to prove this assertion here from the
Liouville point of view (it would be nice to verify it even for the
simplest case, $n=2$), and instead move on to discuss the
non-perturbative effects due to the branes wrapped around the $X$
circle.

The solution given by \Ref{kak:gphi} and \Ref{kak:zzz} should
correspond from the Liouville point of view to the disk partition sum
with the $(1,1)$ boundary conditions on the Liouville field and
Neumann conditions on the $X$ field. The prediction is that
\be
\plabel{kak:nonvanishcor}
\langl \left(\int d^2z e^{(R-2)\phi}
\cos (R\tX)\right)^n\rangl_{(1,1)\times{\rm Neumann}}
\ee
are the coefficients in the expansion of $f(y)$ \Ref{kak:anz} in a power
series in $\lambda$, the first terms of which are given by \Ref{kak:corrf}. It
would be very nice to verify this prediction directly using Liouville
theory, but in general this seems hard given the present state of the
art. A simple check that can be performed using results of
\cite{Zamol} is to compare 
the order $\lambda$ term in \Ref{kak:corrf}
with the $n=1$ correlator \Ref{kak:nonvanishcor}. 

Like in the other cases studied earlier, to make this comparison 
it is convenient to define a
dimensionless quantity given by the ratio of the one point
function on the disk \Ref{kak:nonvanishcor} and the square root of the
appropriate two point function on the sphere,
\be
\plabel{kak:rcone}
\rho=\left. {{\p\over \p \lambda} \log \eps
\over \sqrt{-\p_{\lambda}^2\FSL_0}}\right|_{\lambda=0}. 
\ee
The matrix model result for this quantity is
\be
\plabel{kak:rconeMM}
\rho=-\left.{\mu {\p\over \p \lambda} f \over
\sqrt{-\p_{\lambda}^2\FSL_0}}\right|_{\lambda=0}
= -2\sqrt{2}\sin(\pi R).  
\ee
In the Liouville description, $\rho$ is given by
\be
\plabel{kak:CFTrho}
\rho= {B_\CT \langle V_{b-{R\over2}}\rangle_{(1,1)}
\over \sqrt{-\langle\CT^2\rangle}},
\ee
where $\CT=V_{b-{R\over2}}\cos (R\tX)$ and
$B_{\CT}$ is the one point function of $\cos (R\tX)$ on the disk.
It has the same value as \Ref{kak:pfdisc} 
\be
\plabel{kak:bt}
B_{\CT}=2^{-1/4}\sqrt{R}. 
\ee
The one-point function of the operator of $V_{b-{R\over2}}$ is related
to the wave function $\Psi_{1,1}(P)$ \Ref{kak:wavef} 
with momentum $iP=b-R/2-Q/2$ and is given by
\be
\plabel{kak:wavefPR}
\langle V_{b-{R\over2}}\rangle_{(1,1)}=
-{ 2^{5/4} \sqrt{\pi}  [\pi\mul \gamma(b^2)]^{\hf(1/b^2-1+R/b)}
\over b \Gamma(1-b^2+Rb)\Gamma(1/b^2+R/b)}. 
 \ee
The two-point function of $\CT$ on the sphere
is computed as above from the three point function.
One finds \cite{KAK}
\be
\plabel{kak:twol}
\langl \CT^2\rangl ={\( {1\over b^2}+{R\over b}-1\)\over 2\pi b}
\left[\pi\mul \gamma(b^2)\right]^{{1\over b^2}+{R\over b}-1}
\gamma\(b^2-Rb\) \gamma\(1-{1\over b^2}-{R\over b}\).
\ee
Substituting these results into \Ref{kak:CFTrho} leads,
in the limit $b=1$, to \Ref{kak:rconeMM}. We see that the 
Liouville results are again in complete agreement with
the corresponding matrix model calculation.

Thus, we found that whenever it is possible to compare matrix model
and CFT predictions for non-perturbative effects they always coincide.
All these agreements also support our proposal 
that only the $(1,1)$ Liouville D-brane is responsible for the leading 
non-perturbative corrections. 
Unfortunately, these results say nothing about other 
$(m',n')$ Zamolodchikov's D-branes.
(See, however, the recent work \cite{SAmn} about the role of $(1,n)$ branes
in the $c=1$ string theory.)

It seems to be a very remarkable fact that matrix models, whose connection  
with string theory relies only on a perturbative expansion and even is not completely
understood, are able to describe correctly also the non-perturbative physics.
This should give a promising direction for future research and for new developments
both in matrix models and string theory itself.

\begin{chapternon}{Conclusion}

We conclude this thesis by summarizing the main results achieved
here and giving the list of the main problems,
which either were not solved or not
addressed at all, although their understanding would shed light on
important physical issues.

\section{Results of the thesis}

\begin{itemize}

\setlength{\itemsep}{1pt plus 1pt minus 1pt}

\item
The two- and one-point correlators of winding modes at the spherical level
in the compactified Matrix Quantum Mechanics in
the presence of a non-vanishing winding condensate (Sine--Liouville
perturbation) have been calculated \cite{AK}.

\item It has been shown how the tachyon perturbations can be
incorporated into MQM. They are realized by changing
the Hilbert space of the one-fermion wave functions of the singlet
sector of MQM in such way that the asymptotics of the phases contains
the perturbing potential. At the quasiclassical level these
perturbations are equivalent to non-perturbative deformations of the
Fermi sea which becomes time-dependent. The equation determining the
exact form of the Fermi sea has been derived \cite{AKK}.

\item
When the perturbation contains only tachyons of discrete momenta as in
the compactified Euclidean theory, it is integrable and described by
the constrained Toda hierarchy. Using the Toda structure, the exact
solution of the theory with the Sine-Liouville perturbation
has been found \cite{AKK}. The grand canonical partition
function of MQM has been identified as a $\tau$-function of Toda
hierarchy \cite{AKKNMM}.

\item
For the Sine-Liouville perturbation the energy, free energy and entropy
have been calculated. It has been shown that they satisfy the standard
thermodynamical relations what proves the interpretation of the parameter
$R$ of the perturbations in the Minkowski spacetime as
temperature of the system \cite{AKTBH}.

\item
A relation of the perturbed MQM solution to a free field satisfying
the Klein--Gordon equation in the flat spacetime has been established.  
The global structure of this spacetime and its relation to 
the string target space were discussed \cite{ALEX}.

\item
MQM with tachyon perturbations with equidistant spectrum
has been proven to be equivalent to
certain analytical continuation of the Normal Matrix Model.
They coincide at the level of the partition functions and all correlators.
In the quasiclassical limit this equivalence has been interpreted as a
duality which exchanges the conjugated cycles of a complex curve associated
with the solution of the two models. Physically this duality is of the 
electric-magnetic type (S-duality) \cite{AKKNMM}.

\item
The leading non-perturbative corrections to the partition function of
2D string theory perturbed by a source of winding modes have been
found using its MQM description. In particular, from this result some
predictions for the non-perturbative effects of string theory in the
black hole background have been extracted \cite{KAK}.

\item
The matrix model results concerning non-perturbative corrections
to the partition function of the $c< 1$ unitary minimal models 
and the $c=1$ string theory have been verified from the string theory side
where they arise from amplitudes of open strings attached to D-instantons.
Whenever this check was possible it showed excellent agreement
of the matrix model and CFT calculations \cite{KAK}.

\end{itemize}

\newpage

\section{Unsolved problems}

\begin{itemize}

\setlength{\itemsep}{1pt plus 1pt minus 1pt}

\item
The first problem is the disagreement of the calculated (non-zero) one-point
correlators with the CFT result that they should vanish.
The most reasonable scenario is the existence of an operator mixing
which includes also some of the discrete states.
However, if this is indeed the case, by comparing with the
CFT result one can only find the coefficients of this mixing.
But it was not yet understood how to check this coincidence independently.

\item
Whereas we have succeeded to find the correlators of windings in 
presence of a winding condensate and to describe the T-dual picture
of a tachyon condensate, we failed to calculate tachyon correlators
in the theory perturbed by windings and {\it vice versa}.
The reason is that the integrability seems to be lost when the two types
of perturbations are included. Therefore, the problem
is not solvable anymore by the present technique.

\item
On this way it would be helpful to find a matrix model
incorporating both these perturbations. Of course, MQM does this, but
we mean to represent them directly in terms of a matrix integral
with a deformed potential. Such representation for windings was constructed as
a unitary one-matrix integral, whereas for tachyon perturbations this task is
accomplished by Normal Matrix Model. However, there is no matrix integral
which was proven to describe both perturbations simultaneously.

Nevertheless, we hope that such matrix model exists. For example, in the
CFT framework at the self-dual radius of compactification there is a nice
description which includes both winding and tachyon modes.
It is realized in terms of a ground ring found by Witten.
A similar structure should arise in the matrix model approach.

In fact, in the end of the paper \cite{AKK} a 3-matrix model was 
proposed, which is supposed to incorporate both tachyon and winding
perturbations. However, the status of this model is not clear up to now.
The reason to believe that it works is based on the expectation that
in the case when only one type of the perturbations is present,
the matrix integral gives the corresponding $\tau$-function of MQM.
This is obvious for windings, but it is difficult to prove this statement
for tachyons. It is not clear whether these are technical difficulties
or they have a more deep origin.

\item
Studying the Das--Jevicki collective field theory, we saw that
the discrete states are naturally included into the MQM description
together with the tachyon modes. However, we realized only how to introduce
a non-vanishing condensate of tachyons.
We did not address the question how the discrete states can also be
incorporated into the picture where they appear as a kind of
perturbations of the Fermi sea.

\item
Also we did not consider seriously how the perturbed Fermi sea consisting
from several simply connected components can be analyzed.
Although a qualitative picture is clear, the exact mathematical description
is not known yet. In particular, it would be interesting to generalize
the duality of MQM and NMM to this multicomponent case.

\item
The next unsolved problem is to find the exact relation between the collective field of 
MQM and the tachyon of string theory.
The solution of this problem can help to understand the correspondence, 
including possible leg-factors, of the vertex operators of the matrix model 
to the CFT operators.

\item
It is not clear whether the non-trivial global structure of the 
spacetime on which the collective field of MQM is defined has a physical meaning.
What are the boundary conditions?
What is the physics associated with them?
All these questions have no answers up to now.

Although it seems to be reasonable that the 
obtained non-trivial global structure can give rise to a finite temperature,
this has not been demonstrated explicitly.
This is related to a set of technical problems. However,
the integrability of the system, which has already led to a number of
miraculous coincidences, allows to hope that these
problems can be overcome.

\item
One of the main unsolved problems is how to find 
the string background obtained by the winding condensation.
In particular, one should reproduce the black hole target space metric
for the simplest Sine--Liouville perturbation.
Unfortunately, this has not been done.
In principle, some information about the metric should be contained
in the mixed correlators already mentioned here. But they have neither 
been calculated.

For the case of tachyon perturbations, the crucial role in establishing
the connection with the target space physics is played by the collective
field theory of Das and Jevicki. There is no analogous theory
for windings. Its construction could lead to a real breakthrough 
in this problem.

\item
The thermodynamics represents one of the most interesting issues
because we hope to describe the black hole physics.
We have succeeded to analyze it in detail for the tachyon perturbations
and even to find the entropy. However, we do not know yet
how to identify the degrees of freedom giving rise to the entropy.
Another way to approach this problem would be to consider
the winding perturbations. But it is also unclear how to extract
thermodynamical quantities from the dynamics of windings.

\item
All our results imply that it is very natural to consider the theory
where all parameters like $\mu$, $\lambda$ and $R$ are kept arbitrary.
At the same time, from the CFT side a progress has been made only either for
$\lambda=0$ (the $c=1$ CFT coupled to Liouville theory) or for $\mu=0, \ R=3/2$
(the Sine--Gordon theory coupled to gravity at the black hole radius).
This is a serious obstacle for the comparison of results of the matrix model and CFT
approaches.

In particular, we observe that from the matrix model point of view
the values of the parameters
corresponding to the black hole background of string theory are not
distinguished anyhow. Therefore,
we suppose that for other values the corresponding string background
should have a similar structure.
But the explicit form of this more general background
has not yet been found.

\item
Finally, it is still a puzzle where the Toda integrable structure is hidden in
the CFT corresponding to the perturbed MQM.
In this CFT there are some infinite symmetries indicating the presence
of such structure. But
this happens only at the self-dual radius, whereas MQM does not give
any restrictions on $R$.
Probably the answer is in the operator mixing mentioned above
because the disagreement in the one-point correlators found by the two approaches
cannot be occasional.
Until this problem is solved, the understanding of the relation
between both approaches will be incomplete.

\end{itemize}

We see that many unsolved problems wait for their solution.
This shows that, in spite of the significant progress, 2D string theory
and Matrix Quantum Mechanics continue to be a rich field
for future research.
Moreover, new unexpected relations with other domains of theoretical
physics were recently discovered. And maybe some manifestations of the universal
structure that describes these theories are not discovered yet and
will appear in the nearest future.

\end{chapternon}


\renewcommand{\baselinestretch}{1} \normalsize
\renewcommand{\bibname}{References}

\clearemptydoublepage

\bibliographystyle{osa}
\addcontentsline{toc}{chapter}{References}
\bibliography{biblio}

\end{document}